\documentclass[aps,preprint,preprintnumbers,superscriptaddress,nofootinbib,aps,floatfix,prd]{revtex4-2}
\pdfoutput = 1
\usepackage{graphicx, xcolor}
\usepackage{amsthm} 
\usepackage{amsmath,amssymb,physics}
\usepackage{url,hyperref}
\definecolor{nicered}{rgb}{0.4,0.2,0.}
\definecolor{nicegreen}{rgb}{0.,0.5,0.}
\definecolor{niceblue}{rgb}{0.,0.,0.5}
\definecolor{darkpink}{rgb}{0.8,0.47,0.47}
\definecolor{darkblue}{rgb}{0.0,0,0.5}
\definecolor{blue-violet}{rgb}{0.54, 0.17, 0.89}
\hypersetup{colorlinks,citecolor=nicegreen,linkcolor=nicered,urlcolor=niceblue}
\usepackage{enumitem, array,multirow,booktabs} 
\usepackage[normalem]{ulem}
\setlist{nolistsep} 
 \usepackage{setspace}

\usepackage{float}
\usepackage{setspace}

\begin{document}
\preprint{ANL-182798, DESY-23-068, FERMILAB-PUB-23-276-T,
MSUHEP-23-016, SMU-HEP-23-02, PITT-PACC-2315}
\title{Quantifying the interplay of experimental constraints in analyses of parton distributions}
\author{Xiaoxian Jing} 
\affiliation{Department of Physics, Southern Methodist University, Dallas, TX 75275-0181, USA}
\author{Amanda Cooper-Sarkar}
\affiliation{Department of Physics, University of Oxford, Oxford, OX1 3RH,
  UK\looseness=-1}
 \author{Aurore Courtoy}
\affiliation{Instituto de F\'isica,
  Universidad Nacional Aut\'onoma de M\'exico, Apartado Postal 20-364,
  01000 Ciudad de M\'exico, Mexico\looseness=-1}
\author{Thomas Cridge}
\affiliation{Deutsches Elektronen-Synchrotron DESY, Notkestr. 85, Hamburg, 22607, Germany \looseness=-1}
\author{Francesco Giuli}
\affiliation{CERN, CH-1211 Geneva, Switzerland\looseness=-1}
\author{Lucian Harland-Lang}
\affiliation{ Department of Physics and Astronomy, University College, London,
WC1E 6BT, UK}
\author{T.~J.~Hobbs}
\affiliation{High Energy Physics Division, Argonne National Laboratory, Argonne, IL 60439, USA}
\author{Joey Huston}
\affiliation{Department of Physics and
  Astronomy, Michigan State University, East Lansing, MI 48824,
  USA\looseness=-1}
\author{Pavel Nadolsky} 
\email{Corresponding author: nadolsky@smu.edu}
\affiliation{Department of Physics, Southern Methodist University,
  Dallas, TX 75275-0181, USA}
\author{Robert S. Thorne}
\affiliation{ Department of Physics and Astronomy, University College, London,
WC1E 6BT, UK}
\author{Keping Xie}
\affiliation{Pittsburgh Particle Physics, Astrophysics, and Cosmology Center,\\ 
  Department of Physics and
  Astronomy, University of Pittsburgh, Pittsburgh, PA 15260,
USA \looseness=-1}
\author{C.-P. Yuan} 
\affiliation{Department of Physics and Astronomy, Michigan State University, East Lansing, MI 48824, USA\looseness=-1}

\date{June 6, 2023}

\begin{abstract}
  Parton distribution functions (PDFs) play a central role in calculations for the Large Hadron Collider (LHC). To gain a deeper understanding of the emergence and interplay of constraints on the PDFs in the global QCD analyses, it is important to examine the relative significance and mutual compatibility of the experimental data sets included in the PDF fits. Toward this goal, we discuss the $L_2$ sensitivity, a convenient statistical indicator for exploring the statistical pulls of individual data sets on the best-fit PDFs and identifying tensions between competing data sets. Unlike the Lagrange Multiplier method, the $L_2$ sensitivity can be quickly computed for a range of PDFs and momentum fractions using the published Hessian error sets. We employ the $L_2$ sensitivity as a common metric to study the relative importance of data sets in the recent ATLAS, CTEQ-TEA, MSHT, and reduced PDF4LHC21 PDF analyses at NNLO and approximate N3LO. We illustrate how this method can aid the users of PDFs to identify data sets that are important for a PDF at a given kinematic point, to study quark flavor composition and other detailed features of the PDFs, and to compare the data pulls on the PDFs for various perturbative orders and functional forms. We also address the feasibility of computing the sensitivities using Monte Carlo error PDFs. Together with the article, we present a companion interactive website with a large collection of plotted $L_2$ sensitivities for eight recent PDF releases and a C++ program to plot the $L_2$ sensitivities.
\end{abstract}

\keywords{parton distribution functions, Large Hadron Collider, quantum chromodynamics, uncertainty quantification}

\maketitle
\newpage
\tableofcontents
\newpage

\section{Introduction}
\label{sec:Intro}
Parton distribution functions (PDFs) used for predictions at the Large
Hadron Collider (LHC) and elsewhere are determined by multivariate
fits to a selection of precise experimental measurements from deeply
inelastic scattering, vector boson, jet, top-quark production,
and other processes, including data from the LHC.
The number of experimental data sets
included in the recent PDF fits ranges from a few most precise ones to
several tens in the most comprehensive global fits.
The contemporary next-to-next-to-leading-order (NNLO) PDF fits may either
use the Hessian method  \cite{Pumplin:2001ct,Pumplin:2002vw} or the
Monte Carlo method  \cite{Giele:1998gw} to both determine the
central PDF and estimate the uncertainty on the PDF parameters. For
data from an experimental measurement to influence the PDF fit in a
particular region of $x$ and $Q^2$, two conditions usually must be
met: ({\it i}) the parton-level dynamics underlying the measurement
must substantially depend on a particular PDF ({\it e.g.}, that of the
gluon), as  manifest via a statistical correlation between the PDF in that kinematic region and the experimental observable \cite{Nadolsky:2008zw}; and ({\it ii}) the measurement must have sufficient
resolving power to nontrivially contribute to the likelihood function of a QCD analysis. The latter
depends on the experimental errors of the data set, both statistical and systematic. 

In general, all experimental data sets included in a PDF fit have some influence on the determination of a given PDF flavor for a given $x,Q^2$, with the amount of influence typically varying over a wide range. 
In this article we focus on the examination of the statistical pulls that the experimental data sets impose on the PDFs at the best fit, determined from the dependence of the goodness-of-fit function $\chi^2$ on the PDF values.  In numerical efforts to quantify the PDF pulls of commonly-fitted experiments through figures-of-merit~\cite{Wang:2018heo,Hobbs:2019gob},
some data sets are found to be particularly influential, with the aggregated PDF sensitivity of the full data set typically dominated
by a handful of measurements and a power law-like falloff in the pulls of less sensitive experiments. Detailed investigations of these
sensitivities may indicate ways for increasing the collective impact of the full data set, either for currently available measurements,
from the perspective of alternative implementations of new measurements, or in anticipation of possible data that might be recorded at upcoming facilities like the high-luminosity LHC, Electron-Ion Collider, or neutrino experiments.

For many experiments, their pulls on the PDFs are indistinguishable from those due to statistical fluctuations in the data samples. Such pulls may either reflect a good agreement of the experiment with the best-fit PDF model (in which case such an experiment may nevertheless impose essential constraints on the PDF uncertainty); or it may be that the constraints from the experiment are just weak. In addition,
due to imperfections in the data measurements and/or theory, there can be tensions among data sets and even within a single experiment or data set ({\it e.g.}, among differing rapidity bins in hadroproduction experiments), resulting in significant opposing pulls on the PDFs. The existence of disagreements between some available data sets has been noticed since the early days of PDF fitting (see, e.g., \cite{Collins:2001es}). Tensions among the experiments may lead to a smaller reduction in the PDF uncertainty than might have been expected based on the nominal constraining power of the individual data sets and have motivated introduction of tolerance \cite{Pumplin:2001ct,Martin:2009iq,Kovarik:2019xvh} on the final uncertainty.

The complex inner workings of a PDF fit may leave an impression of a black box, especially to end-users of the PDFs. For this reason, there is substantial interest in numerical methods to quantify the pulls among fitted experiments {\it inside the full fit} in terms of the respective log-likelihood variations for these experiments under systematic shifts in the PDFs. Such tools 
can thereby clarify the influences on the extracted PDFs by the various data sets as well as by variations in the assumed theoretical formalism, such as the perturbative order or deployment of nonperturbative corrections. One such technique is the Lagrange Multiplier (LM) scan~\cite{Stump:2001gu}, which provides robust (not dependent on
the Gaussian approximation) information on the constraints on a particular PDF or observable due to the data sets. This
technique can, in practice, only be applied by the PDF authors, and in addition it is computationally expensive and limited to fixed values of the parton momentum fraction, $x$, and the factorization scale, $Q$.  
Other popular approximate techniques include Hessian profiling and updating \cite{Paukkunen:2014zia, Schmidt:2018hvu,Hou:2019gfw} as well as Monte Carlo PDF
reweighting~\cite{Giele:1998gw, Ball:2010gb, Ball:2011gg,
  Sato:2013ika}, which however depend either on the choice of statistical weights or of tolerance, as well are limited to using static parametric forms. 

Another technique without the above drawbacks is based on the $L_2$ sensitivity measure~\cite{Hobbs:2019gob}, employed together with the LM scans in the recent CT18
global analysis~\cite{Hou:2019efy} and in the PDF4LHC21 benchmarking study~\cite{PDF4LHCWorkingGroup:2022cjn}. The $L_2$ sensitivity technique maps the influences of the data sets on a given PDF or PDF-dependent quantity by taking into account both
the correlation of each data set with the PDF and the degree to which
the data set is influencing the determination of the PDF. 
The $L_2$ sensitivity can be plotted against $x$ in a PDF for a given $Q$ or, in the case of the parton-parton luminosities,  as a function of the final-state invariant mass $M_X$ for a given $\sqrt{s}$. In this form, the $L_2$ sensitivity quantifies the pull of each experiment on the PDF at a given $x$ and $Q$.

The $L_2$ sensitivity is calculated using the log-likelihood ($\chi^2$) values for the fitted experiments and the error PDF sets -- the readily available outputs of the PDF fits. As such, it streamlines comparisons among independent PDF analyses. A few years ago, the CTEQ-JLab and CTEQ-TEA groups performed such comparisons at the NLO accuracy in the QCD coupling strength with the goal to understand the role of large-$x$ and nuclear experimental measurements \cite{Accardi:2021ysh}. 

In this article, we expand the $L_2$ sensitivity comparisons to the next-to-next-to-leading order (NNLO) PDFs by the ATLAS~\cite{ATLAS:2021vod}, CTEQ-TEA \cite{Hou:2019gfw}, and MSHT~\cite{Bailey:2020ooq} groups. We include the recent CT18As\_Lat NNLO analysis with lattice QCD constraints \cite{Hou:2022onq} and the approximate N3LO analysis by the MSHT group, MSHT20aN3LO \cite{McGowan:2022nag}.

Aside from the utility of quantifying the statistical pulls of fitted data
on the PDFs for the sake of practical phenomenology, there is another fundamental motivation for applying the LM or $L_2$ sensitivity methods.  They both explore the parametric dependence of $\chi^2$ in the immediate vicinity of the global minimum of $\chi^2$, i.e., the best fit. They therefore contain rich information regarding the multidimensional geometry of the likelihood function, which is closely connected to the ultimate PDF
uncertainty and its interpretation. For example, the $L_2$ method can elucidate the complicated correlations among PDF flavors or regions of $x$ or $Q$.
Interpretation of the $L_2$ or similar methods, especially when contrasting distinct fits, invokes a range of subtleties in the precise definition of the PDF uncertainty, use of tolerance criteria, and relationships between methods based on Hessian or Monte Carlo uncertainties. We discuss these more formal aspects of uncertainty quantification and their relation to the $L_2$ method before proceeding to the numerical comparisons of the fits by three groups.

The outline of this paper is as follows. In Sec.~\ref{sec:Theory}, we first give a general description of the $L_2$ sensitivity and explain how it can be calculated for both Hessian-based and Monte Carlo replica-based PDF fits. In Sec.~\ref{sec:Particulars} we summarize the PDF sets that will be considered in this analysis, discussing the data and theory settings, as well as $\chi^2$ and PDF error definitions. In Sec.~\ref{sec:non-global} the sensitivities for the non--global PDF sets, namely dedicated ATLAS and reduced benchmarking fits, are presented. In Sec.~\ref{sec:SFL2CT} the sensitivities for the global CT family of PDF sets are presented. In Sec.~\ref{sec:SFL2MSHT} the sensitivities for the MSHT20 fits are presented. In Sec.~\ref{sec:SFL2global} direct comparisons are made of the senstivities for a range experiments and parton flavors between the different PDF sets. Finally, in Sec.~\ref{sec:Conclusions} we conclude. The appendix summarizes a computation of the sensitivities using Monte-Carlo replicas that results in a close agreement with the Hessian approach.

\section{Definitions and basic properties of the $L_2$ sensitivity 
\label{sec:Theory}}

\subsection{The Hessian method \label{sec:HessianMethod}}
Error PDFs are widely used to estimate probability distributions for PDF-dependent quantities according to two common methods.

 The ATLAS, CTEQ-TEA, and MSHT groups adopt the Hessian format \cite{Pumplin:2001ct,Pumplin:2002vw} as the default to publish their PDF error sets. An ensemble of $D$ Hessian error PDFs estimates the uncertainty by assuming that the probability  distribution is approximately Gaussian. In a notation adopted from Ref.~\cite{Hou:2016sho}, a function $X(\vec{R})$ of the parameters $R_i$ in the vicinity of the minimum of the global $\chi^2$ corresponding to $\vec R=\vec 0$ and $X(\vec 0)\equiv X_0$ can be estimated as the Taylor series expansion, 
\begin{equation}
X(\vec{R})=X_{0}+\sum_{i=1}^{D}\left. \frac{\partial X}{\partial R_{i}}\right|_{\vec R =\vec 0}R_{i}+\frac{1}{2}\sum_{i,j=1}^{D}\left. \frac{\partial^{2}X}{\partial R_{i}\partial R_{j}}\right|_{\vec R=\vec 0} R_{i}R_{j}+...\,.
\label{Taylor}
\end{equation}
Given $X_{\pm i} \equiv X(0,...,R_i=\pm 1,0,...) $ for a pair of PDF displacements $R_{\pm i} \equiv \pm 1$ at the 68\% confidence level (C.L.) along the eigenvector direction $i$, the first-order derivative in this direction is estimated by a symmetrized finite difference,
\begin{equation}
\left. \frac{\partial X}{\partial R_i}\right|_{\vec R=\vec 0} \approx  \frac{X_{+i}-X_{-i}}{2}. \label{TaylorI}
\end{equation}
A symmetric estimate of the 68\% C.L.~PDF uncertainty \cite{Pumplin:2002vw} then follows as the maximal variation of $X(\vec R)$ within a hypersphere of unit radius centered at the global minimum, called the ``tolerance hypersphere": 
\begin{equation}
\delta_{\rm H}X=\left|\vec \nabla X\right|=\frac{1}{2}\sqrt{\sum_{i=1}^{D}\left[X_{+i}-X_{-i}\right]^{2}} .\label{dHessian}
\end{equation}

The second-order Taylor terms in Eq.~(\ref{Taylor}) are important when the probability distribution is asymmetric. The full description of the second-order terms, while possible in principle \cite{Hou:2016sho}, would require having additional Hessian eigenvector sets that are not provided in the published PDF ensembles. Contributions from diagonal second-order derivatives, $\partial^{2}X/\partial R_{i}^2$, can be estimated by using the asymmetric PDF uncertainties \cite{Nadolsky:2001yg} with the usual Hessian PDFs. The linear approximation captures the essential features of the uncertainties, while the complete description of the non-linear terms involves many subtleties \cite{Hou:2016sho}.

We will thus restrict ourselves entirely to the linear approximations and will use symmetrized finite-difference formulae like in Eq.~(\ref{TaylorI}) to minimize non-linear terms in subsequent derivations. In this spirit, the Pearson correlation between two quantities $X(\vec R)$ and $Y(\vec R)$, interpreted as the cosine of the correlation angle for $X$ and $Y$ in the PDF parameter space, can be computed as ~\cite{Pumplin:2001ct,Nadolsky:2008zw}
\begin{align}
C_{\rm H}(X,Y)  =\frac{1}{4\delta_{\rm H} X\ \delta_{\rm H} Y}\sum_{i=1}^{D}\left(X_{+i}-X_{-i}\right)\left(Y_{+i}-Y_{-i}\right). 
\label{CorrHessian}
\end{align}

The PDF sensitivity is a statistical indicator that visualizes constraints from
the included experiments on the PDFs. In the Hessian representation, the $L_2$ sensitivity for some $f(\vec R)$ reads \cite{Hobbs:2019gob}
\begin{align}
	S_{f,L2}^{\rm H}(E)\, &\equiv\, \frac{\vec{\nabla} \chi^2_E \cdot \vec{\nabla} f}{\delta_{\rm H} f} \nonumber \\
	&
	= \left( \delta_{\rm H} \chi^2_E \right)\,\, C_{\rm H}(f,\chi^2_E)\ ,
\label{SFL2Hessian}
\end{align}
where $C_{\rm H}(f,\chi^2_E)$ represents the cosine of the correlation angle between $f$ and the $\chi^2$ for experiment $E$, evaluated over the $2D$ Hessian eigenvector sets.  Thus, if the direction of decreasing $\chi^2_E$ of data set $E$ is also the direction of decreasing values of the PDF $f$ (at a given $x$ value), the two quantities are positively correlated, and $S_{f,L2}^{\rm H}(E)$ is positive. This indicates that the data from 
 this data set would like to pull the PDF downward. If the two quantities are anticorrelated, $S_{f,L2}^{\rm H}(E)$ is negative, and the data would like to pull the PDF upwards. 

The name ``$L_2$ sensitivity'' reflects its reliance on the $\chi^2$ to quantify the
pulls of experimental data on the PDFs, i.e., on  the quadratic, or
$L_2$, norm of the vectors of the statistical residuals 
between theoretical predictions and experimental measurements. This emphasizes its
distinction from the alternative definitions of the PDF sensitivity
that are also possible. One alternative definition operates with the absolute
values of the residuals, i.e., the $L_1$ norm, and was employed 
in the first practical studies of the sensitivities using the \texttt{PDFSense} program \cite{Wang:2018heo,Hobbs:2019gob}.

The other connotation is that the $L_2$ sensitivity
serves as the ``second Lagrangian technique'' that complements
the classical conditional optimization with multiplier
terms invented by J.-L. Lagrange. This technique is a fast approximation to the Lagrange Multiplier scan \cite{Stump:2001gu},  now realized using the published eigenvector sets outside of the PDF fit. The LM scan and $L_2$ sensitivity both visualize the probability in the multidimensional PDF parameter space.

\subsection{The Monte-Carlo method \label{sec:MCMethod}}

The Monte-Carlo (MC) method \cite{Giele:1998gw} for PDF
uncertainties is adopted by default by NNPDF, while all groups can
convert between the Hessian and MC error PDFs in both directions
\cite{Watt:2012tq,Gao:2013bia,Carrazza:2015aoa,Carrazza:2016htc}. By
analogy with the Hessian approach, the $L_2$ sensitivity can be
introduced in the MC method \cite{Wang:2018heo}. The MC
case involves subtleties that are absent in the Hessian method. The MC
method provides an ensemble of PDF replicas ${f_a^{(k)}(x,Q)} \equiv
f^{(k)}$ with $k\in[1,..,N_{{\rm rep}}]$, i.e., stochastically
generated error PDF sets. By evaluating a PDF-dependent quantity
$X(f^{(k)}_a(x,Q)) \equiv X_k$ with these replica PDFs, the
probability distribution for $X$ might in principle be reconstructed with
arbitrary accuracy for a sufficiently large $N_{\rm rep}$, including
the non-Gaussian features. However, high dimensionality of the PDF parameter
space presents a salient impediment and requires careful
implementation of the sensitivity.  

As in the Hessian case, we find that accounting for the asymmetries of the probability, while possible in principle, substantially complicates the analysis. We will therefore work with the MC formulae that average over the asymmetries in the probability, such as the standard formulae for the central value and PDF uncertainty of $X$ given respectively by the expectation value and standard deviation:
\begin{align}
    & \langle X\rangle = \frac{1}{N_{{\rm rep}}}\sum_{k=1}^{N_{{\rm rep}}}X_k,
    \label{X0MC}\\
    & \delta_{\rm MC} X = \sqrt{\frac{1}{N_{{\rm rep}}-1}\sum_{k=1}^{N_{{\rm rep}}} \left(X_k-\langle X\rangle\right)^2} \label{dMC}.
\end{align}

Similarly, the Pearson correlation is represented as 
\begin{equation}\label{CorrMC}
    C_{\rm MC}(X,Y) = \frac{\langle XY \rangle - \langle X\rangle \langle Y \rangle}{\delta_{\rm MC} X \cdot \delta_{\rm MC} Y}.
\end{equation}

The MC analog of the Hessian $L_2$ sensitivity in Eq.~(\ref{SFL2Hessian})
is written as
\begin{equation}\label{SFL2MC}
    S_{f,L2}^{\rm MC}(E) = \eta(D)\,\, \left(\delta_{\rm MC} \chi^2_E\right) \,\, C_{\rm MC}(f,\chi^2_E),
\end{equation}
in terms of the MC estimates (\ref{dMC}) and (\ref{CorrMC}) for $f$ and $\chi^2_E$.
We have also introduced a normalization constant $\eta(D)$ that depends on the number of PDF parameters $D$ and may differ from unity, depending on how the MC ensemble samples the space of PDF solutions. 

The Hessian and Monte-Carlo definitions of the $L_2$ sensitivity are equivalent in the linear approximation, which in turn is justified when displacements of the PDF error sets from the global minimum are small. In practice, the linearity condition for $\chi^2_E$ does not hold with the standard MC replicas from NNPDF or other groups, and hence the direct estimation of $S_{f,L2}^{\rm MC}(E)$ according to Eq.~(\ref{SFL2MC}) may be vulnerable to errors. The reason is that, in a typical MC ensemble, the majority of PDF replicas lie several standard deviations away from the global minimum as a consequence of the high dimensionality of the parameter space \cite{Hou:2016sho}.  As $\chi^2_E$ includes quadratic and higher powers of the PDF parameters, such far displacements introduce large non-linearities, which lead to the accuracy loss in  $S_{f,L2}^{\rm MC}(E)$. 

Thus, to obtain a numerically stable estimate of $S_{f,L2}^{\rm MC}(E)$ using the MC replicas, one must avoid large parameter displacements from the global minimum, which in many dimensions requires unconventional sampling. Appendix~\ref{sec:Hessian2MC} presents such an example, in which a Hessian ensemble is converted into an MC one so that the MC replicas are distributed uniformly on the surface of a hypersphere rather than over the whole parameter space. With this ensemble, probability integrations in the angular (radial) directions are performed numerically (analytically), and the radial integration contributes a normalization constant $\eta(D)=\sqrt{D}$ in $S_{f,L2}^{\rm MC}(E)$ in Eq.~(\ref{SFL2MC}). With this procedure, the MC sensitivity in Eq.~(\ref{SFL2MC}) closely agrees with the sensitivity (\ref{SFL2Hessian}) of the progenitor Hessian PDFs, as demonstrated in the appendix.

In this article, we primarily compare the Hessian sensitivities, which are simpler to compute. 
In the next subsection we illustrate the meanings of the $L_2$ sensitivities in the Hessian and MC frameworks using a toy example. 

\subsection{A one-dimensional example \label{sec:SFL2OneDim}}

Let us take $f$ in Eqs.~(\ref{SFL2Hessian}) and (\ref{SFL2MC}) to be a PDF, $f_a(x,Q)$, at some value of $\{x,Q\}$. The minimum $\chi^2_0$ of the total $\chi^2$ is obtained at $f=f_0$. In the Hessian representation, we have $ \vec \nabla f=\delta_{\rm H} f \,\hat{e}_f$, with $\delta_{\rm H} f$ being the one-sigma uncertainty on the PDF $f$, and $\hat{e}_f$ the unit vector along $ \vec \nabla f$  \cite{Pumplin:2001ct,Pumplin:2002vw}.  
Similarly,  $\vec \nabla \chi_E^2 \cdot \hat{e}_f =\partial \chi_E^2/\partial f$, and hence
$S_{f,L_2}^H=\delta_{\rm H} \chi^2_E(\hat{e}_f)$ is the variation of $\chi_E^2$ from the best-fit $\chi^2_{E,0}$ along direction $\hat{e}_f$. 
We can approximate $S_{f,L_2}^H$ in Eq.~(\ref{SFL2Hessian}) using a symmetric finite-difference derivative for $\Delta \chi^2_E(f) \equiv \chi^2_E(f) - \chi^2_{E,0}$:
\begin{eqnarray}
S_{f,L_2}^H & = & \left. \frac{\partial \chi_E^2(f)}{\partial f} \right|_{f=f_0} \, \delta_{\rm H} f,   \nonumber \\
    &\approx & \frac{\Delta\chi_E^2(f+\delta_{\rm H} f)-\Delta\chi_E^2(f-\delta_{\rm H} f)}{2}.
\label{SFL2finitedifference}
\end{eqnarray}

\begin{figure*}[t]
\centering
\includegraphics[width=0.75\textwidth]{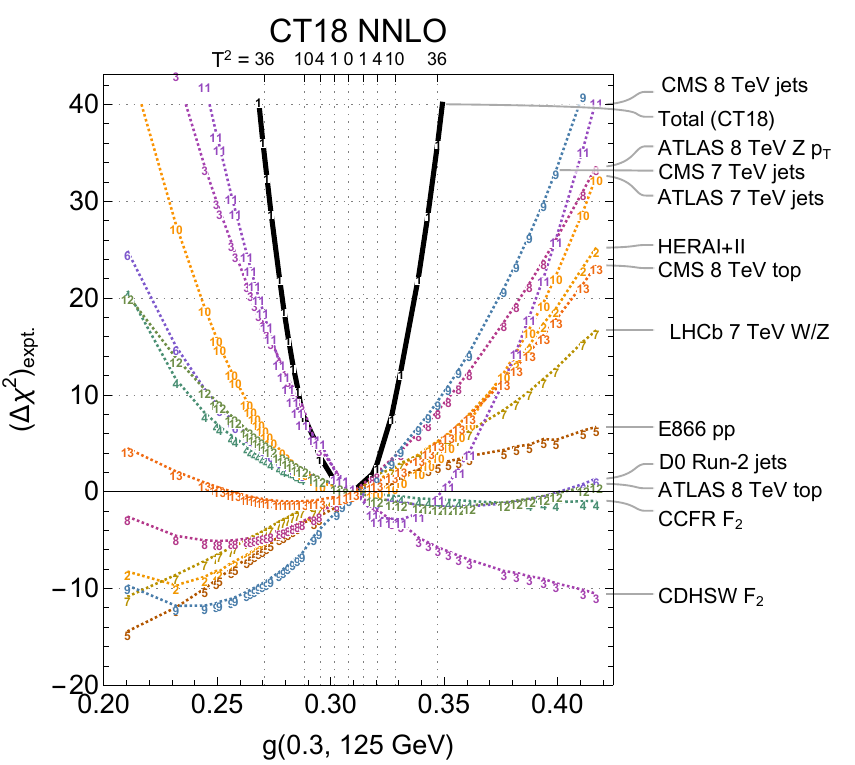}
\caption{A Lagrange Multiplier scan for the gluon PDF, $ g(0.3,125\mbox{ GeV})$, in the CT18 NNLO analysis. The tolerance $T^2$, computed from the total $\Delta\chi^2$ curve, is displayed on the top of the plot. The total chi-squared distribution corresponds to the thick black curve. Each curve is marked by a unique numerical ID.}
\label{fig:LMScan1D}
\end{figure*}

Taking $f=g(0.3, 125\mbox{ GeV})$ (the gluon PDF) from CT18 NNLO as an example, in Fig.~\ref{fig:LMScan1D} we plot $\Delta \chi^2(f)$ versus $f$ for several leading experimental data sets and for all data sets together. We perform a series of fits with a condition $f=f_i$ for a set of discrete $f_i$ and interpolate $\chi^2(f)$ in between the $f_i$ values. In this LM scan, we find that the minimum of the total $\chi^2$ corresponds to $f=f_0\approx 0.31$. This minimum optimizes the total $\chi^2$ among competing pulls of the individual experiments. 

The total $\chi^2$ enters the likelihood probability, 
\begin{equation}
P(f) \propto e^{-(\chi^2(f)-\chi^2_0)/(2 T^2)},
\end{equation}
which also depends on the chosen tolerance $T^2$. The latter in turn fixes the value of  $\delta_{\rm H} f$, as illustrated by the vertical dashed lines in Fig.~\ref{fig:LMScan1D}. The tolerance should be such that the linear approximation dominates for the leading $\chi^2_E$ curves in the interval $-\delta_{\rm H} f \leq f -f_0 \leq +\delta_{\rm H} f$. In this example, we choose $T^2=10$.

In the MC approach, we first generate $N_{\rm rep}$ instances of $f$ by stochastically sampling them according to $P(f)$. Then, we compute the $L_2$ sensitivity through Eq.~(\ref{SFL2MC}), where  $\eta(D)=1$ in one dimension.

\begin{figure*}[t]
\centering
\includegraphics[width=0.45\textwidth]{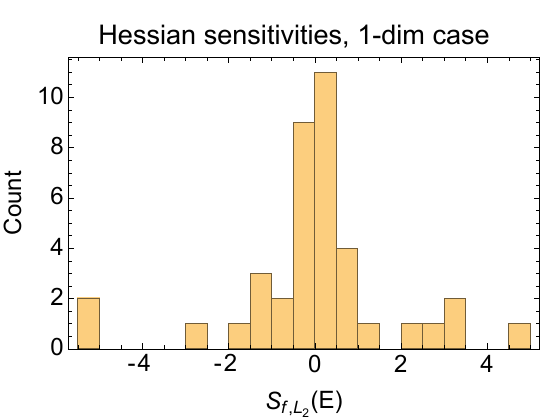}
\includegraphics[width=0.45\textwidth]{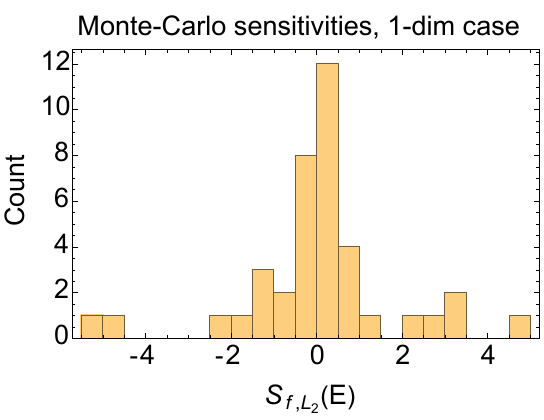}
\caption{The $S_{f,L_2}$ counts for the $N_E$ experiments for a) the Hessian case and b) $N_{\rm rep}=1000$ MC replicas generated from the distribution of the total $\chi^2$ from the LM scan.}
\label{fig:Histo_1D}
\end{figure*}

Figure~\ref{fig:Histo_1D} shows histograms of the Hessian and MC $L_2$ sensitivities computed for the $\chi^2_E$ curves in Fig.~\ref{fig:LMScan1D}. The histograms agree with one another, confirming that both the Hessian and MC methods are compatible in the neighborhood of the minimum. The trend of the MC histogram does not critically depend on the number of replicas, $N_{\rm rep}$, as long as $N_{\rm rep} > 100$.
The experiments ``CDHSW $F_2$" and ``CMS 7 TeV jets", which have the largest negative and positive $\partial\Delta \chi^2_E(f)/\partial f$ at $f=f_0$ in Fig.~\ref{fig:LMScan1D}, contribute to $S_{f,L_2}^H$ values of about -5 and 4.5 in Fig.~\ref{fig:Histo_1D}, respectively.

On the other hand, since the first derivative of the total $\chi^2$ vanishes at the global minimum, the sum of $S_{f,L_2}^H$ over all experiments must be zero within uncertainties. For both the Hessian and MC representations, we find in this example that
\begin{eqnarray}
\sum_E S_{f, L_2} \ll T^2 <\sum_E |S_{f, L_2}|\,,
\label{sumSFL2}
\end{eqnarray}
with $\sum S^{\rm H}_{f, L_2}\lesssim \sum S^{\rm MC}_{f, L_2}$.
 
In fact, an informative validation test of the Hessian approximation also
in a $D$-dimensional case consists in checking that Eq.~(\ref{sumSFL2}) is satisfied. The equation reflects the existence of the global minimum of the total $\chi^2$ for any dimensionality of the fit. In our studies, we have observed that a non-negligible number of published ensembles in the LHAPDF library do not automatically satisfy this condition, which can happen with older PDF sets or poorly constrained eigenvector sets that don't comply with the stated $T^2$, especially when $T^2$ is of order 10 or less. In these cases, the $\sum S^{\rm H}_{f, L_2}$ estimates may be biased; one can correct the deficient eigenvector sets by rescaling their displacements along the respective EV directions, as summarized in the appendix of the CT-CJ NLO comparative study \cite{Accardi:2021ysh}, in which such deviation was first observed and corrected. We apply the test in our study as well to validate the accuracy of the examined Hessian PDFs. The plots of the summed sensitivities can be viewed on our website \cite{L2website}.

\subsection{How to interpret $L_2$ sensitivities
\label{sec:MoreLML2Relation}
}
This article presents the $L_2$ sensitivities in two forms, chosen to illustrate either the leading sensitivities of the experiments to a given PDF or the sensitivities of a given experiment to a collection of PDF flavors or PDF combinations. 

\begin{figure*}[p]
\centering
\includegraphics[height=0.42\textheight]{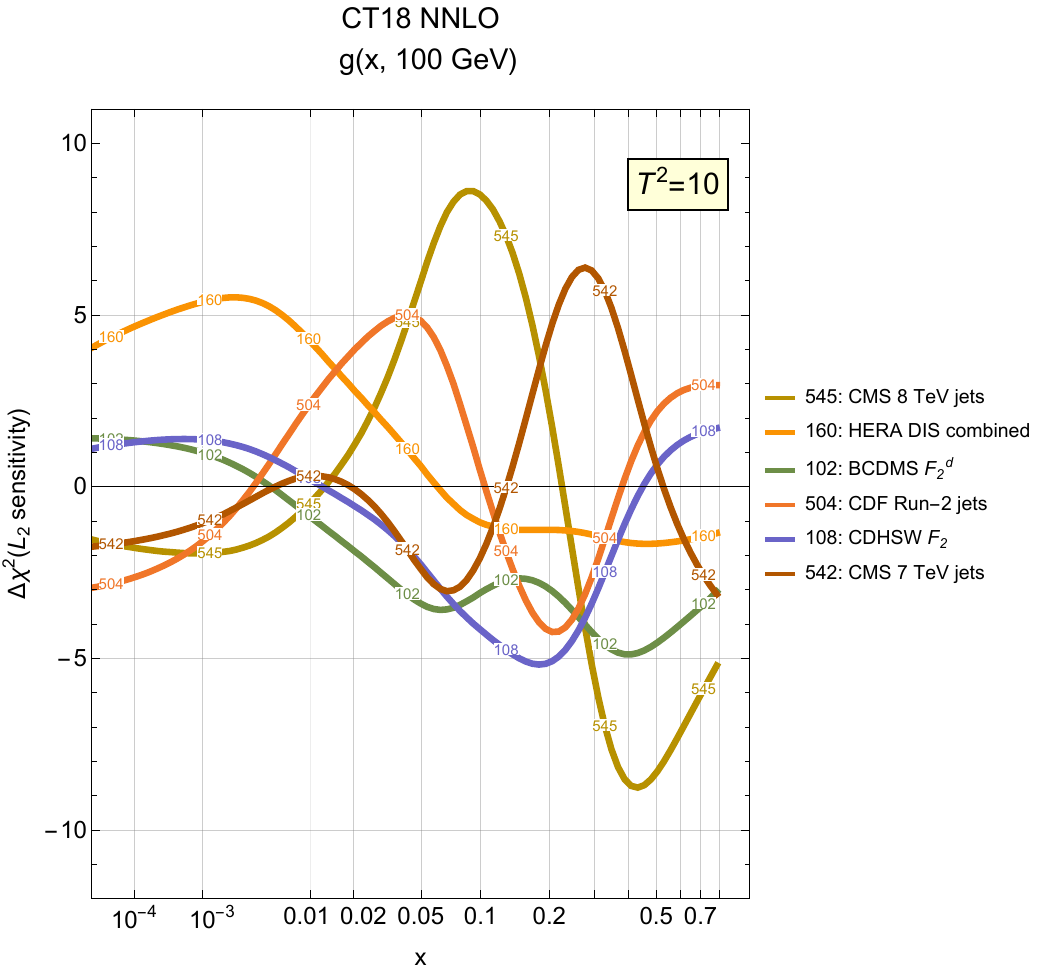}
\caption{The $L_2$ sensitivity to
  the CT18 NNLO gluon distribution at $Q=100$ GeV of the six experiments with the greatest pulls, taking $T^2=10$. Each curve is marked by the numerical ID of the corresponding experiment provided in Tables~\ref{tab:DISdataCTMSHT}-\ref{tab:JetdataCTMSHT}.}
\label{fig:L2gluonCT18}
\end{figure*}

{\bf Cumulative sensitivities to PDF flavors.} As an example from the first category, Fig.~\ref{fig:L2gluonCT18} shows the $L_2$ sensitivities to the gluon distribution $g(x,Q)$ for CT18 NNLO, as a function of the partonic momentum fraction $x$ at a $Q$ value of 100 GeV. Only the $L_2$ sensitivities from the six most significant experiments are plotted, for purposes of clarity.\footnote{On the companion website \cite{L2website}, such plots can show 4, 6, or 8 most sensitive experiments, or alternatively all experiments that have $|S_{f,L2}|>3$ in some range of $x$.} 

\begin{figure*}[p]
\centering
\includegraphics[height=0.42\textheight]{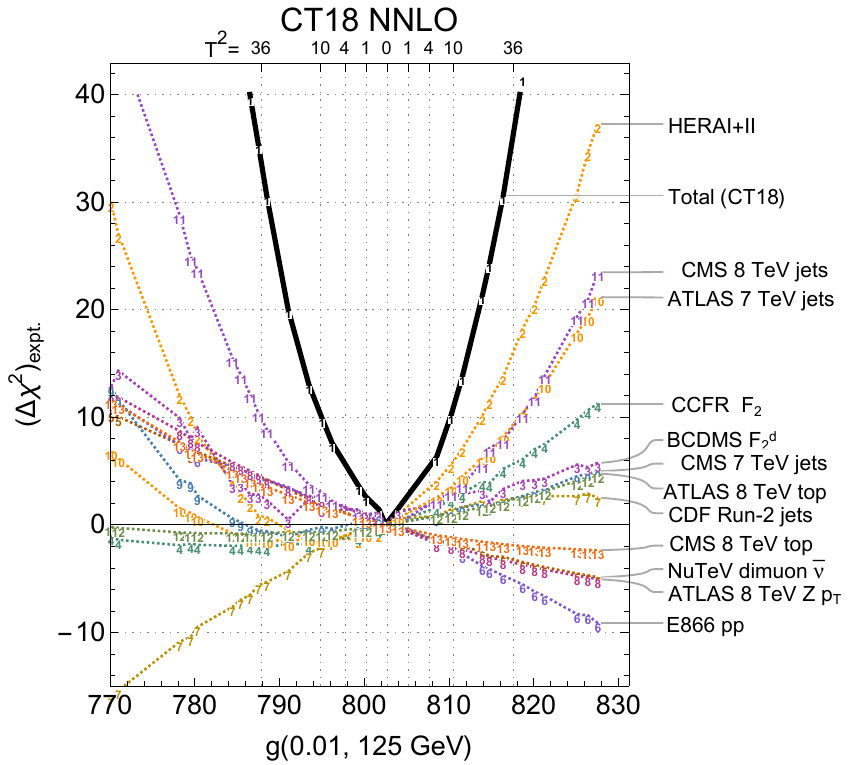}
\caption{A LM scan for the gluon, $ g(x=0.01,Q=125\mbox{ GeV})$, in the CT18
  NNLO fit.}
\label{fig:LMScan2}
\end{figure*}

One way to understand more intuitively the meaning of the $L_2$ sensitivity is to compare it to the  more familiar LM scans, either the already discussed case of the LM scan on the high-$x$ gluon, $g(0.3,125\mbox{ GeV})$, in Fig.~\ref{fig:LMScan1D} or an analogous scan on $g(0.01,125\mbox{ GeV})$ as relevant for $gg\!\to\! \mathrm{Higgs}$ production, shown in Fig.~\ref{fig:LMScan2}. In a LM scan, the strength of the constraint provided by a data set determines the narrowness of its corresponding $\chi^2_E$ parabola. For example, in Fig.~\ref{fig:LMScan2}, the CMS experiment on jet production at 8 TeV (curve 11) imposes such a constraint, similar in magnitude to that provided by the HERA I+II data.
The central gluon at this $x$ agrees with the value preferred by the CMS 8 TeV jet data, so there is no pull from these data. In contrast, the LM scan at $x=0.3$, in Fig.~\ref{fig:LMScan1D}, indicates that the CMS 8 TeV jet data prefers a gluon value of 0.33, larger than the best-fit value of 0.31.
These features of the CMS 8 TeV jet data at $x=0.01$ and $x=0.3$  can be seen in the $L_2$ sensitivity plot of Fig.~\ref{fig:L2gluonCT18}, as explained below~\footnote{Note that sensitivities change very slowly with the scale $Q$, as illustrated in Section~\ref{sec:non-global}}.

At either $x$ value, a number of experiments prefer somewhat different values of the gluon than at the best fit. We can quantify these pulls on the gluon by computing the change, $\Delta\chi^2_E$, for each experiment $E$ when the considered PDF increases by one standard deviation from the total $\chi^2_0$, which in the following comparisons was chosen to correspond to $\Delta \chi^2=10$. We note that $\Delta \chi^2_E$ were defined above Eq.~(\ref{SFL2finitedifference}) in  Sec.~\ref{sec:SFL2OneDim}. The $L_2$ sensitivity $S_{f,L2}$ estimates these $\Delta \chi^2_E$ in the linear approximation for the whole range of $x$. The magnitude of $S_{f,L2}$ depends on the correlation of the $\chi^2_E$  with a particular PDF, the constraining power of the data on that PDF, and the difference between the PDF preferred by that data set and the PDF determined by the full fit. 
{\bf A positive value of $S_{f,L2}$ indicates the preference for a lower value of the PDF at the specified $x$ and $Q$, and vice versa.}

From Fig.~\ref{fig:L2gluonCT18}, we conclude following this rule that the HERA I+II deep inelastic scattering (DIS) data (ID=160) has the largest  sensitivity at $x<0.01$, preferring a lower gluon here than the CDF and CMS jet data sets (504, 542, 545).
The CMS 8 TeV jet data set has little sensitivity at $x=0.01$, even though it is well-constraining at this $\{x,Q\}$ point, in accord with the behavior in the LM scan in Fig.~\ref{fig:LMScan2} noted above. By contrast, we see that at $x=0.3$ the sensitivity is negative, so that this data set prefers a larger gluon at this $x$, as noted in Fig.~\ref{fig:LMScan1D}. The CMS 7 and 8 TeV inclusive jet data have the largest $|L_2|$ sensitivity at $x>0.05$.\footnote{It is interesting to note that, at high $x$, these jet data at the two energies have large $L_2$ sensitivity values and opposite signs. Even though the two measurements are carried out by the same experiment at very similar energies, there is a tension between them, representative of the situations encountered in a global PDF fit. Ref.~\cite{Kovarik:2019xvh} discusses consequences of such tensions for the interpretation of PDF uncertainties.} We note that no single experiment dominates over the entire $x$ range, and the $L_2$ sensitivity for many experiments will switch sign as a function of $x$. 

Again, these trends are consistent with the LM scans at two fixed $x$ values shown in 
Figs.~\ref{fig:LMScan1D} and \ref{fig:LMScan2}. 
If an experiment has a major influence on the best-fit PDF at a
particular $x$ value,  it will normally have a large absolute value of $S_{f,L_2}$. But, if the reference PDF value already agrees with that preferred by the data set, it may have
only a relatively small value of $S_{f,L_2}$  (since the gradient of the
$\chi^2$ is small).\footnote{This agreement may be an accident, or an indication of the constraining power of that experiment.}

It is also possible that an experiment is constraining enough for a particular PDF over a wide $x$ region and has a low value of $S_{f,L_2}$  across that entire $x$  region, as  the PDF is forced  to be close to that preferred  by that  experiment. In reality, this rarely occurs: the combined weight of the other  experiments tends to provide a counter-constraint  to the  dominant experiment. See,  for example, that the HERA  DIS  combined data in Fig.~\ref{fig:L2gluonCT18} -- the most dominant data  set in  all global PDF fits -- strongly prefers a smaller  gluon distribution at $x<0.02$, but even then is counteracted by the CDF and CMS inclusive jet data. 

The sum of the $S_{f,L_2}$ values for all experiments for a given kinematic point
should be close to zero, as the pulls have to balance out to produce the central PDF
at that point. One can also sum up all of the positive values of $S_{f,L_2}$, as well as 
all of the negative values, for each parton $x$ value. The sums (both positive and negative) tend to be relatively flat as a function of $x$, and are roughly equal to the number of experiments, i.e. each experiment contributes on the order of one to the sum. There is no obvious correlation between the value of the sum and the size of the PDF uncertainty at that point. 

\begin{figure*}[h]
\centering
\includegraphics[height=160pt,trim={ 0 0 1.6cm 0},clip]{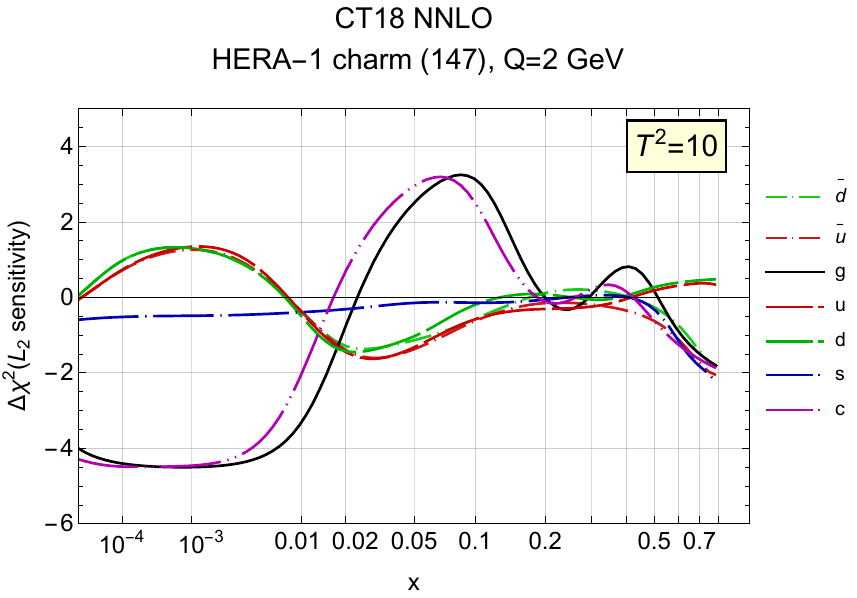}
\includegraphics[height=160pt,trim={ 1.cm 0 0 0},clip]{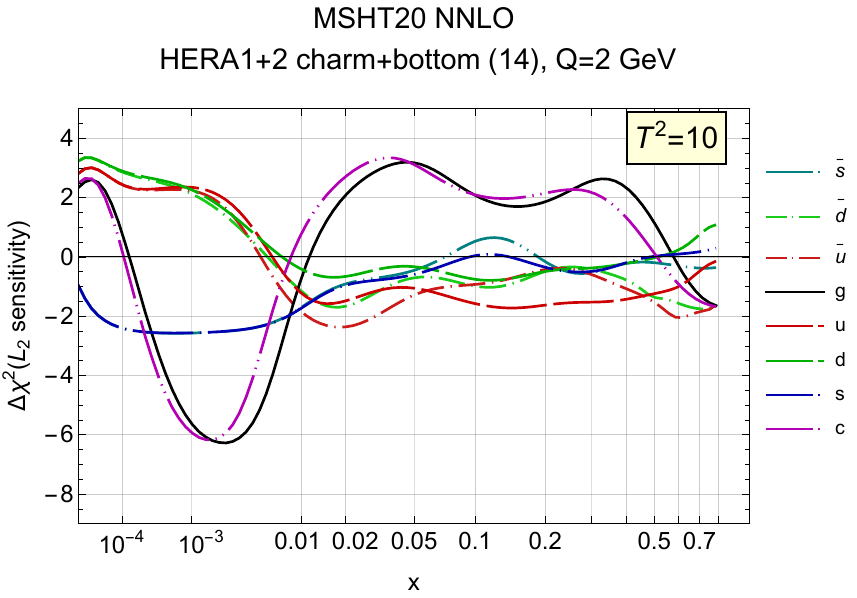}
\caption{
Sensitivities for the HERA I charm data set in the CT18 NNLO fit and HERA I+II charm+bottom data set in the MSHT20 NNLO fit at $Q=2$ GeV.}
\label{fig:compar_HERAcharm}
\end{figure*}

{\bf Cumulative sensitivities of individual data sets.} Figure~\ref{fig:compar_HERAcharm} is an example of the second form of comparisons, in which sensitivities for the indicated PDF flavors are plotted for a given data set indicated in the plot label. In the figure, we compare the sensitivities of the HERA I combined charm data set \cite{H1:2012xnw} included in the CT18 NNLO analysis in the left subpanel, and the combined HERA I+II charm+bottom data set \cite{H1:2018flt} included in the MSHT20 NNLO analysis in the right one. While the more recent HERA I+II data set \cite{H1:2018flt} covers an extended kinematic range and has smaller uncertainties, the CTEQ-TEA group found it difficult to accommodate these data, with its $\chi^2/N_{\rm pts}$ remaining high ($>1.7$) under a variety of explored assumptions \cite{Guzzi:2021rvo}. The HERA I+II data set is included in the MSHT20 analysis, and it is interesting to compare its impact with that of the HERA I charm data in the CT18 NNLO fit. We find that both data sets prefer higher (lower) gluon and charm PDFs at $10^{-4}<x<0.01$ ($0.02 < x <0.2$). In the MSHT20 NNLO case, the preferences for a lower gluon extends to higher $x$ of up to 0.5. In both PDF fits, these heavy-quark data prefer lower $u$ and $d$ (anti)quark PDFs at $x<0.01$. There is mild preference for a higher strangeness PDF at $x<0.1$. This preference is more pronounced in the MSHT20 case. Overall, according to the sensitivities that stay within a few units, the HERA charm data (at $Q$ of tens of GeV) impose moderate constraints on the gluon and other PDFs.

\subsection{$L_2$ sensitivity and the likelihood-ratio test}
The likelihood-ratio test is a classical Bayesian test, together with
the closely related Lagrange Multiplier test and Wald test, that
discriminates between two theoretical hypotheses, $T_1$ and $T_2$,
based on their agreement with a set of observational data $D$. In the
context of PDF fits \cite{Kovarik:2019xvh,Soper:1994km}, it is more
common to formulate the likelihood-ratio test as a comparison of
log-likelihood functions $\chi^2$ for two PDF models, $T_1$ and $T_2$,
related to the augmented likelihood as $P(D|T_i)=\text{const}\cdot
\exp\left({-\chi^2(D,T_i)/2}\right)$. When there is no strong prior
preference for either $T_1$ or $T_2$, but the data $D$ strongly favor
one of them, the ratio of posterior probabilities is dominated by the
ratio of the likelihoods that is related to the difference of $\chi^2$
for $T_1$ and $T_2$: 
\begin{equation}
\left(\frac{P(T_2)}{P(T_1)}\right)_\text{posterior}
=
\frac{P(D|T_2)}{P(D|T_1)} 
\left(\frac{P(T_2)}{P(T_1)}\right)_\text{prior},
\label{LikelihoodRatioTest}
\end{equation}
where 
\begin{equation}
\frac{P(D|T_2)}{P(D|T_1)} = \exp\left({-\frac{\chi^2(D,T_2)-\chi^2(D,T_1)}{2}}\right).
\label{LikelihoodRatio}
\end{equation}
Based on this ratio, the PDFs rendering the lowest $\chi^2$ are the
most likely ones according to the empirical data. The PDFs with a low, but
not the lowest $\chi^2$ can be acceptable with some probability
determined by the tolerance prescription. A LM scan examines the
change in $\chi^2$ as a function of a PDF parameter or PDF-dependent
observable, hence it realizes the ratio test between the best-fit PDF
and nearby PDF solutions. The $L_2$ sensitivity serves the same
purpose. These tests do not need to know how the PDFs are found. Section~\ref{sec:LogLikelihood} reviews the practical implementations of the likelihoods in the fits by three groups.

\section{Particulars of the compared fits
\label{sec:Particulars}
}

\subsection{Overview of the global analyses
\label{sec:OverviewOfFits}}

One of the goals of this study is to explore how the constraints on PDFs emerge as one successively adds new experiments into the PDF analysis. We compare sensitivities in two categories of PDFs:
\begin{enumerate}
    \item {\bf Non-global fits}, which include a small(er) number of sensitive experiments. Examples include the ATLASpdf21 and PDF4LHC21 benchmark reduced fits.
    \item {\bf Global fits}, which include several tens of data sets of varied sensitivity. The examples considered here include NNLO ensembles CT18, CT18As, CT18As\_Lat, MSHT20 as well as the approximate N3LO ensemble MSHT20aN3LO.
\end{enumerate}

This section provides the background for the comparisons of the $L_2$ sensitivities that will be presented in the following sections. The reader broadly familiar with the selection of experiments and methodologies of three groups can skip much of this section. The ensuing discussions of the $L_2$ sensitivities will extensively refer to the selections of the experimental data sets in the fits. Table~\ref{tab:ATLASData} lists the data sets included in the ATLASpdf21 analysis, while Tables~\ref{tab:DISdataCTMSHT}-\ref{tab:JetdataCTMSHT} list the data sets included in the CT18 and MSHT fits. Next to each data set in the tables, we list its numerical ID adopted to mark the corresponding $L_2$ sensitivity curves in the figures. We will see that, while there are large differences among these PDF analyses, the $L_2$ sensitivity elucidates their comparisons as a common metric.

The next three subsections briefly review the selections of experiments and theoretical computations in ATLASpdf21, CTEQ-TEA and MSHT global fits, then moving on to the comparisons of CT18 and MSHT20 data sets in Sec.~\ref{sec:CTMSHTDifferences}, and then to additional discussion of the PDF4LHC21 reduced fits in Sec.~\ref{sec:PDF4LHC21Data}. While the CT18 and MSHT20 NNLO global fits share many key data sets, there are important differences among them. To compare the CT18, MSHT20, and NNPDF3.1 methodologies using (nearly) the same data sets, reduced fits were performed by the PDF4LHC group in the course of 2021 benchmarking study \cite{PDF4LHCWorkingGroup:2022cjn} by including only twelve shared data sets indicated in the rightmost column in Tables~\ref{tab:DISdataCTMSHT}-\ref{tab:JetdataCTMSHT}. The findings from these comparisons guided the construction of the PDF4LHC21 combination of the PDFs from the three NNLO global analyses. 

The value of $\chi^2_E$ generally depends on the approximations made in the likelihood, or ``$\chi^2$ definition". The specific implementations are reviewed in Sec.~\ref{sec:LogLikelihood}. The tolerance conventions for the published PDFs, which determine both the size of PDF uncertainties and $L_2$ sensitivities, are compared for the three groups in Sec.~\ref{sec:ToleranceConventions}. 

\subsubsection{Summary of ATLAS fits 
\label{sec:DataSetsATLAS}
}

\begin{table}[bt]\small
\caption{Summary of all the input data sets considered in the ATLASpdf21 fit. \label{tab:ATLASData}}
\begin{center}
\resizebox{\textwidth}{!}{%
\begin{tabular}{clcccc}
\hline
ID & Data set & $\sqrt{s}$ [ TeV] & Luminosity [fb$^{-1}$] & Decay channel & Observables entering the fit \\
\hline
160 & HERA inclusive DIS ~\cite{Abramowicz:2015mha} & Varied & Varied &  & Reduced cross sections \\
68 & Inclusive $W,Z/\gamma^*$~\cite{ATLAS:2016nqi} & 7 & ~~4.6 & $e,\mu$ combined& $\eta_{\ell}$ ($W$), $y_{\mathit{Z}}$ ($Z$) \\
89 & Inclusive $Z/\gamma^*$~\cite{ATLAS:2017rue} & 8 & 20.2 & $e,\mu$ combined & $\cos\theta^{\ast}$ in bins of $y_{\ell\ell}$, $m_{\ell\ell}$\\
86 & Inclusive $W$~\cite{ATLAS:2019fgb} & 8 & 20.2 & $\mu$ & $\eta_{\mu}$ \\
\multirow{2}{*}{56} & $W^{\pm}$\,+\,jets~\cite{ATLAS:2017irc} & 8 & 20.2 & $e$ & $p_{\mathrm{T}}^{\mathit{W}}$ \\
                    & $Z$\,+\,jets~\cite{ATLAS:2019bsa} & 8 & 20.2 & $e$ & $p_{\mathrm{T}}^{\mathrm{jet}}$ in bins of $|y^{\mathrm{jet}}|$\\
7 & $t\bar{t}$~\cite{ATLAS:2015lsn,ATLAS:2016pal} & 8 & 20.2 & lepton\,+\,jets, dilepton & $m_{t\bar{t}}$, $p_{\mathrm{T}}^t$, $y_{t\bar{t}}$ \\ 
8 & $t\bar{t}$~\cite{ATLAS:2019hxz} & 13~~ & 36~~~ & lepton\,+\,jets & $m_{t\bar{t}}$, $p_{\mathrm{T}}^t$, $y_t$, $y_{t\bar{t}}^{\mathrm{b}}$ \\ 
9 & Inclusive isolated $\gamma$~\cite{ATLAS:2019drj} & 8, 13 & 20.2, 3.2 & - & $E_{\mathrm{T}}^{\gamma}$ in bins of $\eta^{\gamma}$ \\
10 & Inclusive jets~\cite{ATLAS:2017kux}& 8  & 20.2 & - & $p_{\mathrm{T}}^{\mathrm{jet}}$ in bins of $|y^{\mathrm{jet}}|$ \\
\hline
\hline
\end{tabular}}
\end{center}
\end{table}
ATLAS PDF fits concentrate on the impact of ATLAS data on PDFs. However, it is not possible to make
an accurate PDF fit to ATLAS data alone. The HERA DIS combined data~\cite{Abramowicz:2015mha} are used as the backbone of the ATLAS PDF fits, to which ATLAS data are added.
The HERA experiments cover a very broad range of $Q^2$,
the absolute four-momentum transfer squared, 
 from near $1\mbox{ GeV}^2$ to above $10^4\mbox{ GeV}^2$, and of Bjorken $x$ from $\sim0.6$ down to $10^{-4}$.

 ATLASpdf21 \cite{ATLAS:2021vod} --- the most comprehensive  fit from this series -- uses the HERA inclusive DIS data and a broad variety of ATLAS data, while accounting for correlations between the various ATLAS measurements. We also consider several intermediate ATLAS fits leading to ATLASpdf21. Table~\ref{tab:ATLASData} lists these fits together with the included data sets, their numerical IDs, center-of-mass energies, luminosities, decay channels, and observables entering each fit.

The $L_2$ sensitivities for the intermediate fits can be used as a pedagogical example of the effect of adding in sensitive data sets: all these PDFs are obtained starting from the PDF set HERAPDF2.0 that was determined staring with the HERA data  alone~\cite{Abramowicz:2015mha} and by successively adding new ATLAS data sets. First in this series is the ATLASepWZ16 PDF set~\cite{ATLAS:2016nqi}, in which the ATLAS precision measurements of the inclusive differential
$W^{\pm}$ and $Z/\gamma^*$ boson cross sections at 7 TeV were added to HERA data. This PDF set improved on the HERAPDF2.0 set in various respects. Firstly, the strange content of the sea could be fitted rather than assumed to be a fixed fraction of the light sea. Indeed, the strange sea was found to be enhanced at low $x$, $x \lesssim 0.05$ compared to previous determinations. Secondly, the accuracy of the valence quark distributions was considerably improved. 

Further improvement was achieved by adding $t\bar{t}$ differential cross sections at 8 TeV from both the lepton+jets and the dilepton channels. This fit was called ATLASepWZtop18~\cite{ATLAS:2018ttbar}. In the lepton+jets channel, the mass of the $t\bar{t}$ pair, $m_{tt}$, and  the average top-quark transverse momentum, $p^t_{T}$, were fitted simultaneously. In the dilepton channel, the rapidity of the $t\bar{t}$ pair was fitted. Care was taken to include correlated systematic and statistical uncertainties. This fit improved the uncertainties of the high-$x$ gluon.

The final ATLASpdf21 ensemble discussed here was obtained by including all the previous ATLAS data sets mentioned above and adding to them more data constraining the valence quarks and the flavor of the sea (from ATLAS 8 TeV inclusive $W$ and $Z$ data), more data constraining the high-$x$ strange sea and the gluon PDFs (from ATLAS 8 TeV $W$ and $Z$ boson data +jets\footnote{In fact, the ATLASepWZjets20~\cite{ATLAS:2021qnl} PDF fit also uses these data, but it is not further discussed here.}), and more data constraining the gluon (from ATLAS 8 TeV data on inclusive jets, ATLAS 13 TeV data on top-antitop distributions, and ratios of ATLAS 13 and 8 TeV direct photon data).

The increase in the number of data sets from HERAPDF2.0 to ATLASpdf21 has facilitated an increase in the freedom of the parametrizations from 14 to 21 parameters (and in the corresponding number of eigenvectors). Extra freedom  has been added in the high-$x$ valence and gluon PDFs and in the sea parametrization at low $x$, whereby the low-$x$ $\bar{u}$, $\bar{d}$, and $\bar{s}$ were all independently parametrized.  Extra variations due to model assumptions and additional parameters were also considered. Tension between data sets has also led to a consideration of appropriate $\chi^2$ tolerance as $T=3$ as well as $T=1$. See Sec.~\ref{sec:ToleranceConventions} for an explanation of the choice of tolerance for the present $L_2$ study.

Uniquely, the ATLASpdf21 fit included full information both on correlated 
systematic sources of uncertainty of all data sets and on the correlations between the ATLAS data sets, in contrast to the global fits. The largest sources of such correlations among the data sets come from the measurements involving jets: not only from inclusive jet production but also boson+jets and $t\bar{t}$ in the lepton+jets channel. The $L_2$ sensitivities in our comparisons account for the correlations among the experimental data sets.

The theoretical predictions for the ATLASpdf21 fit were computed at NNLO in QCD using programs \texttt{DYNNLO} \cite{Catani:2007vq,Catani:2009sm}, \texttt{FEWZ} \cite{Gavin:2010az,Gavin:2012sy,Li:2012wna}, \texttt{NNLOJET} \cite{Currie:2016bfm,Gehrmann-DeRidder:2017mvr} and results from \cite{Gehrmann-DeRidder:2015wbt, Czakon:2017dip, Czakon:2016olj, Campbell:2016lzl}, and at NLO in electroweak theory using \texttt{DYNNLO}, \texttt{FEWZ}, \texttt{SHERPA} and results from \cite{Czakon:2017wor, Becher:2013zua, Dittmaier:2012kx}.
A full description of the theoretical treatment is given in Ref.~\cite{ATLAS:2021vod}.

\subsubsection{Summary of CT18 NNLO 
\label{sec:CT18DataTheory}}
The CT18 NNLO analysis \cite{Hou:2019efy} constituted a major new release of the CTEQ-TEA family of PDF studies, having included $\sim $700 LHC data points on top of the baseline fit in the previous main release, CT14HERA2 \cite{Hou:2016nqm}. A detailed discussion of the theory used in CT18, selected data sets, and other statistical or methodological choices is presented in Ref.~\cite{Hou:2019efy}; these aspects were further summarized for the purpose of the recent PDF4LHC21 benchmarking study in Ref.~\cite{PDF4LHCWorkingGroup:2022cjn}.

The strategy of the CT18 analysis was to first examine a large group of data sets using preliminary fits and fast Hessian techniques in order to select an ensemble of constraining and maximally consistent data sets. The LM scans and $L_2$ sensitivities were extensively employed to identify such data sets, as documented in Ref.~\cite{Hou:2019efy} and on the CT18 website \cite{CT18L2Sensitivity}.
The published CT18 NNLO fit was performed to this final data set. 
The selected experiments are shown in Tables~\ref{tab:DISdataCTMSHT}-\ref{tab:JetdataCTMSHT} and include neutral-current and charged-current DIS, as well as production of vector bosons, jets, and top quark pairs.

When validating the fit, the mutual agreement of the data sets was examined based on the strong goodness-of-fit criteria \cite{Kovarik:2019xvh}, in addition to requiring a good total $\chi^2$. Out of the newly included LHC data sets, the ATLAS 7 TeV $W, Z$ production data set \cite{ATLAS:2016nqi} (ID=248) was found to be both precise and showing a tension with the NuTeV dimuon and HERA DIS data sets as a result of its preference for a larger strangeness PDF at $x\sim 0.02$. This data set was included in the alternative fits, CT18A and CT18Z, and not in the nominal CT18. 

The size and complexity of the CT18 analysis made it worthwhile to consider a number of variations on the assumptions within the analysis. Together with the default CT18, several complementary PDF ensembles were released to quantify the impact of these assumptions, including different selections in the fitted
data sets (e.g., CT18A, which fitted the 2016 7 TeV ATLAS inclusive $W, Z$ production data); alternative factorization scales (e.g., CT18X, which assigned a different $x$-dependent scale, $\mu_\mathrm{DIS}$, to DIS
data to mimic the effects of low-$x$ resummation); and an amalgamation of these choices (CT18Z, which also
took a slightly enlarged charm mass, $m_c = 1.4$ GeV, relative to the default of $m_c = 1.3$ GeV used in the other CT18 fits).

In subsequent years, a number of follow-up studies have expanded the CT18 framework by 
investigating various physics issues relevant for both high- and medium-energy data sets
fitted in CT.
These have included:
the introduction of an explicit photon PDF and associated
electroweak corrections in the CT18QED analysis for a proton \cite{Xie:2021equ} and neutron \cite{Xie:2023qbn};
evaluation of the S-ACOT-$\chi$ NNLO theory with heavy-quark mass effects with
phenomenology for the EIC and neutrino DIS \cite{Gao:2021fle}, including, in a later study, scattering of cosmic neutrinos at ultra-high energies \cite{Xie:2023suk};
an investigation of the PDF impact of light nuclear corrections, in particular, those
connected to nucleon off-shellness in deuterium \cite{Accardi:2021ysh};
an analysis of the high-$x$ PDF behavior in light of power-counting arguments \cite{Courtoy:2020fex};
a study of fitted charm in light of the full CT18 data set with a range of
nonperturbative models allowing $c\!\neq\!\bar{c}$ \cite{Guzzi:2022rca};
inclusion of high-$x$ lattice QCD constraints for the strange quark and antiquark PDFs in an analysis
allowing for $s\!\neq\!\bar{s}$ \cite{Hou:2022onq};
and an investigation of the impact on the PDFs by the post-CT18 data on vector boson production at the LHC at 8 and 13 TeV \cite{Sitiwaldi:2023jjp}.

In CT18 and follow-up studies, the perturbative QCD theory for the LHC data is evaluated at NNLO accuracy,
typically through the use of fast interpolation tables provided by \texttt{fastNLO} \cite{Kluge:2006xs,Wobisch:2011ij,Britzger:2012bs} and \texttt{APPLGRID} \cite{Carli:2010rw} and calculated
at NLO using \texttt{MCFM} \cite{Campbell:2010ff,Boughezal:2016wmq,MCFM6,MCFM8}, \texttt{NLOJET++} \cite{Nagy:2003tz}, and \texttt{aMCfast} \cite{Bertone:2014zva};
these calculations are then corrected to NNLO via point-by-point NNLO/NLO K-factors based on \texttt{DYNNLO} \cite{Catani:2007vq,Catani:2009sm}, \texttt{FEWZ} \cite{Gavin:2010az,Gavin:2012sy,Li:2012wna},
\texttt{MCFM}, and \texttt{NNLOJET} \cite{Currie:2016bfm,Gehrmann-DeRidder:2017mvr}; an exception to this procedure applies to
top-quark production, which is computed at NNLO directly using \texttt{fastNNLO} \cite{Czakon:2017dip,Czakon:2017wor,fastnnlo:grids} grids.
We note that scale-choice and other theoretical uncertainties were explored in the main CT18 publication; while these
were not quantified systematically in a specialized error treatment, the final CT18 uncertainty was determined to
ensure coverage of variations associated with these uncertainties.
The ultimate CT18 parametrization (given explicitly in App.~C of Ref.~\cite{Hou:2019efy}) resembles that used
in CT14HERA2 and is similarly formulated in terms of Bernstein polynomials, but with slightly more flexibility accorded to the light-quark sea.
In addition, CT18 considered $\mathcal{O}(250)$ alternative parametrization forms.
The nominal CT18 PDF uncertainty was also assessed to encompass the corresponding variations driven by these alternative
nonperturbative forms.
Like earlier studies, CT18 deployed an NNLO implementation of the S-ACOT-$\chi$ scheme to treat heavy-flavor production,
partonic thresholds, and related dynamics.

Computing the $L_2$ sensitivities provided powerful insights into many post-CT18 studies
and was an essential feature of Refs.~\cite{Hou:2022onq,Accardi:2021ysh,Courtoy:2020fex}.
In the current article, we present the $L_2$ sensitivities for CT18 NNLO in the uniform format that facilitates comparisons with ATLASpdf21 and MSHT20. We also apply the sensitivity method to two NNLO PDF fits CT18As and CT18As\_Lat \cite{Hou:2022onq} that determined the magnitude of a non-zero strangeness asymmetry, $s_-(x,Q) \equiv s(x,Q)- \bar s(x,Q)$, by releasing the assumption of $s=\bar s$ made in the CT18 and CT18Z analyses. These fits follow the setup of the CT18A analysis and, in particular, include the high-luminosity ATLAS 7 TeV $W, Z$ data set 248 \cite{ATLAS:2016nqi}. They also examine agreement with the E906/SeaQuest data set on the Drell-Yan $pd/pp$ ratio without fitting it. To constrain the $s_-(x,Q)$ combination at $0.3 < x <0.8$, where no relevant experimental sensitivity currently exists, the CT18As\_Lat analysis includes constraints from 
quasi-PDF matrix elements (extrapolated to physical pion mass) computed in lattice QCD \cite{Zhang:2020dkn}. The lattice QCD input significantly reduces the allowed magnitude of $s_-(x,Q)/s_+(x,Q)$ at $x\to 1$, which otherwise can be very large (approaching 100\%) if only the extrapolations of experimental constraints are included. In the CT18As\_Lat analysis, the lattice QCD input is implemented with the help of Lagrange multipliers. The fitting code reports its contribution to $\chi^2$ in one category with the contributions from the normalization shifts for BCDMS, CDHSW, and CCFR data on the DIS structure functions (ID=101, 102, 108, 109, 110, 111). We compute $S_{f,L2}$ for this ``Lattice + DIS normalizations'' category under ID=701, keeping in mind that the lattice constraints dominate between the two.\footnote{The DIS normalizations are mildly correlated with the gluon PDF in the fixed-target region.}

\begin{table}[tb] 
\resizebox{\textwidth}{!}{%
\small
\begin{tabular}{|l||c|c|c||c|c|c||c|}
\hline 
\multirow{2}{*}{Data set} & \multicolumn{3}{c||}{CT18 NNLO} & \multicolumn{3}{c||}{MSHT20 NNLO} & In the PDF4LHC21\tabularnewline
\cline{2-7} \cline{3-7} \cline{4-7} \cline{5-7} \cline{6-7} \cline{7-7} 
 & Ref. & $N_{{\rm pts}}$ & ID & Ref. & $N_{{\rm pts}}$ & ID & reduced fit?\tabularnewline
\hline 
BCDMS$F_{2}^{p}$ & \cite{Benvenuti:1989rh} & 337 & 101 & \cite{Benvenuti:1989rh} & 163 & 1 & yes\tabularnewline
BCDMS $F_{2}^{d}$ & \cite{Benvenuti:1989rh} & 250 & 102 & \cite{Benvenuti:1989rh} & 151 & 2 & yes\tabularnewline
NMC $F_{2}^{p}$ &  &  &  & \cite{Arneodo:1996qe} & 123 & 3 & \tabularnewline
NMC $F_{2}^{d}$ &  &  &  & \cite{Arneodo:1996qe} & 123 & 4 & \tabularnewline
NMC ratio & \cite{Arneodo:1996qe} & 123 & 104 & \cite{NewMuon:1996uwk} & 148 & 11 & yes\tabularnewline
SLAC $ep$ $F_{2}$ &  &  &  & \cite{Whitlow:1991uw,Whitlow:1990gk} & 37 & 5 & \tabularnewline
SLAC $ed$ $F_{2}$ &  &  &  & \cite{Whitlow:1991uw,Whitlow:1990gk} & 38 & 6 & \tabularnewline
E665 $\mu d$ $F_{2}$ &  &  &  & \cite{E665:1996mob} & 53 & 7 & \tabularnewline
E665 $\mu p$ $F_{2}$ &  &  &  & \cite{E665:1996mob} & 53 & 8 & \tabularnewline
\hline
HERA I+II DIS combined & \cite{Abramowicz:2015mha} & 1120 & 160 & \cite{Abramowicz:2015mha} & 1185 & 160 & yes\tabularnewline
\qquad HERA $e^{+}p$ CC &  &  &  & \cite{H1:2009pze} & 39 & 22 & \tabularnewline
\qquad HERA $e^{-}p$ CC &  &  &  & \cite{H1:2009pze} & 42 & 23 & \tabularnewline
\qquad HERA $e^{+}p$ ${\rm NC}~820$ GeV &  &  &  & \cite{H1:2009pze} & 75 & 24 & \tabularnewline
\qquad HERA $e^{+}p$ ${\rm NC}~920$ GeV &  &  &  & \cite{H1:2009pze} & 402 & 25 & \tabularnewline
\qquad HERA $e^{-}p$ ${\rm NC}~460$ GeV &  &  &  & \cite{H1:2009pze} & 209 & 26 & \tabularnewline
\qquad HERA $e^{-}p$ ${\rm NC}~575$ GeV &  &  &  & \cite{H1:2009pze} & 259 & 27 & \tabularnewline
\qquad HERA $e^{-}p$ ${\rm NC}~920$ GeV &  &  &  & \cite{H1:2009pze} & 159 & 28 & \tabularnewline
HERA charm+bottom &  &  &  & \cite{H1:2018flt} & 79 & 14 & \tabularnewline
\qquad HERA I charm & \cite{H1:2012xnw} & 47 & 147 &  &  &  & \tabularnewline
\qquad H1 bottom & \cite{Aktas:2004az} & 10 & 145 &  &  &  & \tabularnewline
H1 $F_{L}$ & \cite{H1:2010fzx} & 9 & 169 &  &  &  & \tabularnewline
NMC/BCDMS/SLAC/HERA $F_{L}$ &  &  &  & \cite{Arneodo:1996qe,Benvenuti:1989rh,Whitlow:1990gk,H1:2008rkk,H1:2010fzx,ZEUS:2009nwk} & 57 & 15 & \tabularnewline
\hline
CDHSW $F_{2}^{p}$ & \cite{Berge:1989hr} & 85 & 108 &  &  &  & \tabularnewline
CDHSW $x_{B}F_{3}^{p}$ & \cite{Berge:1989hr} & 96 & 109 &  &  &  & \tabularnewline
CCFR $F_{2}^{p}$ & \cite{Yang:2000ju} & 69 & 110 &  &  &  & \tabularnewline
CCFR $x_{B}F_{3}^{p}$ & \cite{Seligman:1997mc} & 86 & 111 &  &  &  & \tabularnewline
CHORUS $\nu N$ $F_{2}$ &  &  &  & \cite{CHORUS:2005cpn} & 42 & 19 & \tabularnewline
CHORUS $\nu N$ $x_B F_{3}$ &  &  &  & \cite{CHORUS:2005cpn} & 28 & 20 & \tabularnewline 
NuTeV $\nu N$ $F_{2}$ &  &  &  & \cite{NuTeV:2005wsg} & 53 & 9 & \tabularnewline
NuTeV $\nu N$ $x_B F_{3}$ &  &  &  & \cite{NuTeV:2005wsg} & 42 & 10 & \tabularnewline
\hline
CCFR dimuon combined &  &  &  & \cite{Goncharov:2001qe} & 86 & 16 & \tabularnewline
\qquad $\nu_{\mu}$ & \cite{Goncharov:2001qe} & 40 & 126 &  &  &  & \tabularnewline
\qquad $\bar{\nu}_{\mu}$ & \cite{Goncharov:2001qe} & 38 & 127 &  &  &  & \tabularnewline
NuTeV dimuon combined &  & 81 & 593 & \cite{Goncharov:2001qe} & 84 & 17 & yes\tabularnewline
\qquad $\nu_{\mu}$ & \cite{Mason:2006qa} & 38 & 124 &  &  &  & \tabularnewline
\qquad $\bar{\nu}_{\mu}$ & \cite{Mason:2006qa} & 33 & 125 &  &  &  & \tabularnewline
\hline 
\end{tabular}
}
\caption{DIS and SIDIS data sets in the CT18 and MSHT20 ensembles. For each data set, we indicate the publication reference, number of data points, and the numerical ID in the figures.  
\label{tab:DISdataCTMSHT}}
\end{table}

\begin{table}[tb]
\resizebox{\textwidth}{!}{%
\small
\begin{tabular}{|l||c|c|c||c|c|c||c|}
\hline 
\multirow{2}{*}{Data set} & \multicolumn{3}{c||}{CT18 NNLO} & \multicolumn{3}{c||}{MSHT20 NNLO} & In the PDF4LHC21\tabularnewline
\cline{2-7} \cline{3-7} \cline{4-7} \cline{5-7} \cline{6-7} \cline{7-7} 
 & Ref. & $N_{{\rm pts}}$ & ID & Ref. & $N_{{\rm pts}}$ & ID & reduced fit?\tabularnewline
\hline 
ATLAS 7 TeV $35\mbox{ pb}^{-1}$ $W, Z$ cross sec., $A_{ch}$ & \cite{Aad:2011dm} & 41 & 268$^{\ddagger\ddagger}$ & \cite{Aad:2011dm} & 30 & 52 & yes\tabularnewline
ATLAS 7 TeV high-mass Drell-Yan &  &  &  & \cite{Aad:2013iua} & 13 & 58 & \tabularnewline
ATLAS 7 TeV high precision $W$, $Z$ & \cite{ATLAS:2016nqi} & 34 & 248$^{\ddagger}$ & \cite{ATLAS:2016nqi} & 61 & 68 & \tabularnewline
ATLAS 8 TeV high-mass Drell-Yan &  &  &  & \cite{ATLAS:2016gic} & 48 & 82 & \tabularnewline
ATLAS 8 TeV $W$ &  &  &  & \cite{ATLAS:2019fgb} & 22 & 86 & \tabularnewline
ATLAS 8 TeV $Z$ $p_{T}$ & \cite{ATLAS:2015iiu} & 27 & 253 & \cite{ATLAS:2015iiu} & 104 & 71 & \tabularnewline
ATLAS 8 TeV double differential $Z$ &  &  &  & \cite{ATLAS:2017rue} & 59 & 89 & \tabularnewline
\hline
CDF Run-1 lepton $A_{ch}$, $p_{T\ell}>25$ GeV & \cite{CDF:1998uzn} & 11 & 225 &  &  &  & \tabularnewline
CDF Run-2 electron $A_{ch}$, $p_{T\ell}>25$ GeV & \cite{CDF:2005cgc} & 11 & 227 &  &  &  & \tabularnewline
CDF Run-2 $W$ asymmetry &  &  &  & \cite{CDF:2009cjw} & 13 & 43 & \tabularnewline
CDF Run-2 $Z$ rapidity & \cite{Aaltonen:2010zza} & 29 & 261 & \cite{Aaltonen:2010zza} & 28 & 37 & \tabularnewline
\hline
CMS 7 TeV lepton asymmetry &  &  &  &  &  &  & \tabularnewline
\qquad $36\mbox{ pb}^{-1}$ &  &  &  & \cite{CMS:2011bet} & 24 & 54 & \tabularnewline
\qquad $840\mbox{ pb}^{-1}$ & \cite{Chatrchyan:2012xt} & 11 & 267 & \cite{Chatrchyan:2012xt} & 11 & 53 & yes\tabularnewline
\qquad $4.7\mbox{ fb}^{-1}$, muon & \cite{Chatrchyan:2013mza} & 11 & 266 &  &  &  & \tabularnewline
CMS 7 TeV $Z\rightarrow e^{+}e^{-}$ &  &  &  & \cite{CMS:2011wyd} & 35 & 57 & \tabularnewline
CMS 7 TeV double diff. Drell-Yan &  &  &  & \cite{CMS:2013zfg} & 132 & 60 & \tabularnewline
CMS 8 TeV muon asymmetry & \cite{CMS:2016qqr} & 11 & 249 & \cite{CMS:2016qqr} & 22 & 64 & \tabularnewline
\hline
E605 Drell-Yan & \cite{Moreno:1990sf} & 119 & 201 &  &  &  & \tabularnewline
E866/NuSea $pd/pp$ Drell-Yan ratio & \cite{Towell:2001nh} & 15 & 203 & \cite{Towell:2001nh} & 15 & 13 & yes\tabularnewline
E866/NuSea $pp$ Drell-Yan & \cite{Webb:2003ps} & 184 & 204 & \cite{Webb:2003bj} & 184 & 12 & \tabularnewline
E906/SeaQuest $pd/pp$ Drell-Yan ratio & \cite{SeaQuest:2021zxb} & 6 & 206$^\star$ & &  &  & 
\tabularnewline
\hline
D{Ø} Run-2 $W\rightarrow\nu\mu$ asymmetry & \cite{D0:2007pcy} & 9 & 234 & \cite{D0:2013xqc} & 10 & 38 & \tabularnewline
D{Ø} Run-2 $W\rightarrow\nu e$ asymmetry & \cite{D0:2014kma} & 13 & 281 & \cite{D0:2008cgv} & 12 & 44 & \tabularnewline
D{Ø} Run-2 $Z$ rapidity & \cite{D0:2007djv} & 28 & 260 & \cite{D0:2007djv} & 28 & 36 & yes\tabularnewline
D{Ø} Run-2 $W$ asymmetry &  &  &  & \cite{D0:2013lql} & 14 & 70 & \tabularnewline
\hline
LHCb 7 TeV $W$ asymmetry $p_{T\ell}>20$ GeV &  &  &  & \cite{Aaij:2012vn} & 10 & 56 & \tabularnewline
LHCb 7 TeV $Z\rightarrow e^{+}e^{-}$ &  &  &  & \cite{LHCb:2012gii} & 9 & 55 & \tabularnewline
LHCb 7 and 8 TeV $W$ and $Z$ & \cite{LHCb:2015okr,LHCb:2015kwa} & 50 & 258 & \cite{LHCb:2015okr,LHCb:2015kwa} & 67 & 61 & yes\tabularnewline
\qquad 7 TeV & \cite{LHCb:2015okr} & 33 & 245 & \cite{LHCb:2015okr} &  &  & \tabularnewline
\qquad 8 TeV & \cite{LHCb:2015kwa} & 17 & 246 &  &  &  & \tabularnewline
LHCb 8 TeV $Z\rightarrow ee$ & \cite{LHCb:2015mad} & 34 & 250 & \cite{LHCb:2015mad} & 17 & 62 & yes\tabularnewline
\hline 
\end{tabular}
}
\caption{Vector boson production data sets in the CT18 and MSHT20 ensembles. The ATLAS 7 TeV $35\mbox{ pb}^{-1}$ $W, Z$ data set, marked by $\ddagger\ddagger$, is replaced by the high-luminosity data set (4.6 fb$^{-1}$), marked by $\ddagger$, in the CT18As and CT18As\_Lat fits. The E906/SeaQuest Drell-Yan ratio (marked by $\star$) is included in the CT18As and CT18As\_Lat comparisons, but not fitted.
\label{tab:DYdataCTMSHT}}
\end{table}

\begin{table}[tb]
\resizebox{\textwidth}{!}{%
\small
\begin{tabular}{|l||c|c|c||c|c|c||c|}
\hline 
\multirow{2}{*}{Data set} & \multicolumn{3}{c||}{CT18 NNLO} & \multicolumn{3}{c||}{MSHT20 NNLO} & In the PDF4LHC21\tabularnewline
\cline{2-7} \cline{3-7} \cline{4-7} \cline{5-7} \cline{6-7} \cline{7-7} 
 & Ref. & $N_{{\rm pts}}$ & ID & Ref. & $N_{{\rm pts}}$ & ID & reduced fit?\tabularnewline
\hline 
ATLAS 7 TeV incl. jets & \cite{ATLAS:2014riz} & 140 & 544 & \cite{ATLAS:2014riz} & 140 & 66 & \tabularnewline
CDF Run-2 $p\bar{p}$ incl. jets & \cite{CDF:2008hmn} & 72 & 504 & \cite{CDF:2007bvv} & 76 & 35 & \tabularnewline
CMS 2.76 TeV incl. jets &  &  &  & \cite{CMS:2015jdl} & 81 & 87 & \tabularnewline
CMS 7 TeV incl. jets & \cite{CMS:2014nvq} & 158 & 542 & \cite{CMS:2014nvq} & 158 & 69 & \tabularnewline
CMS 8 TeV incl. jets & \cite{CMS:2016lna} & 185 & 545 & \cite{CMS:2016lna} & 174 & 73 & yes\tabularnewline
D{Ø} Run-2 $p\bar{p}$ incl. jets & \cite{D0:2008nou} & 110 & 514 & \cite{D0:2011jpq} & 110 & 48 & \tabularnewline
\hline
ATLAS 8 TeV $W+\text{jets}$ & & & & \cite{ATLAS:2017irc} & 30 & 83 &     \tabularnewline
CMS 7 TeV $W+c$ &  &  &  & \cite{Chatrchyan:2013uja} & 10 & 67 & \tabularnewline
\hline
ATLAS 8 TeV single diff $t\bar{t}$ & \cite{ATLAS:2015lsn} & 15 & 580 & \cite{ATLAS:2015lsn} & 25 & 74 & \tabularnewline
ATLAS 8 TeV single diff $t\bar{t}$ dilepton &  &  &  & \cite{ATLAS:2016pal} & 5 & 81 & \tabularnewline
 Tevatron, ATLAS, CMS $\sigma_{t\bar{t}}$ &  &  &  & \cite{CDF:2013hmv,CMS:2013hon} & 17 & 59 & \tabularnewline
CMS 8 TeV double differential $t\bar{t}$ & \cite{CMS:2017iqf} & 16 & 573 & \cite{CMS:2017iqf} & 15 & 84 & \tabularnewline
CMS 8 TeV single differential $ t\bar{t}$ ($y_{t\bar t}$) &  &  &  & \cite{CMS:2015rld} & 9 & 88 & \tabularnewline
\hline 
\end{tabular}
}
\caption{Inclusive jet, boson+jet, and top-quark pair production data sets
in the CT18 and MSHT20 ensembles. \label{tab:JetdataCTMSHT}}
\end{table}

\subsubsection{Summary of MSHT20 NNLO and approximate N3LO 
\label{sec:MSHTData}}

The baseline MSHT20 PDF sets \cite{Bailey:2020ooq} represent the latest in the MRST/MSTW/MMHT line of PDF fits, with substantial improvements made on top of the previous MMHT2014 study~\cite{Harland-Lang:2014zoa} on the experimental, methodological, and theoretical fronts. 

On the experimental side, these include additional LHC data sets on vector boson, top and jet production among others, as well as somewhat augmented and different selections of the fixed-target and Tevatron data sets relative to the previous MMHT2014 PDFs, as can be seen from Tables~\ref{tab:DISdataCTMSHT}-\ref{tab:JetdataCTMSHT}. Simultaneously with the increase in the number and precision of included experimental constraints on the PDFs, methodological improvements were also made. Foremost among these was an extension of the PDF parametrization at the input scale $Q_0 = 1$ GeV. In MSHT the PDFs are parametrized in terms of orthogonal Chebyshev polynomials, with the number of parameters now increased from 4 to 6 for each of the input PDF flavor combinations. This followed previous work \cite{Martin:2012da}, in which it was shown that this would allow a fit of the PDFs to data of better than 1\% accuracy over the entire $x$ range, provided the data and theory allowed. As a result, the number of PDF parameters has increased from 37 in MMHT to 52 in MSHT, with a consequent increase in the number of eigenvectors from 25 to 32 to allow a more detailed reflection of the PDF uncertainties. Further details of these improvements are given in \cite{Harland-Lang:2017ytb,Bailey:2019yze,Thorne:2019mpt,Thorne:2022abv} as well as the MSHT20 paper~\cite{Bailey:2020ooq}.
 The MSHT PDFs also allow a non-zero strangeness asymmetry by default, although in this case the constraints are weaker, so fewer parameters are used, while the charm PDF is generated perturbatively. 
 
 In our baseline MSHT20 study and subsequent work, the theoretical predictions for hadronic cross-sections are made using NNLO QCD theory via grids provided at NLO by \texttt{fastNLO}~\cite{Kluge:2006xs,Wobisch:2011ij,Britzger:2012bs} and \texttt{APPLGRID}~\cite{Carli:2010rw} using calculations from \texttt{MCFM}~\cite{Campbell:2010ff,Boughezal:2016wmq,MCFM6,MCFM8} and \texttt{NLOJET++}~\cite{Nagy:2003tz}. These are supplemented by NNLO/NLO K-factors from a variety of sources including \texttt{DYNNLO}~\cite{Catani:2007vq,Catani:2009sm}, \texttt{FEWZ}~\cite{Gavin:2010az,Gavin:2012sy,Li:2012wna}, \texttt{MCFM}, $N_{\rm jetti}$~\cite{Boughezal:2015dva,Gehrmann-DeRidder:2015wbt} and \texttt{NNLOJET}~\cite{Currie:2016bfm,Gehrmann-DeRidder:2017mvr}; with the K-factors smoothed and an uncertainty applied to account for this - as described futher in \cite{Bailey:2020ooq}. Top production at the LHC is however computed directly at NNLO using provided \texttt{fastNNLO} grids \cite{Czakon:2017dip,Czakon:2017wor,fastnnlo:grids}. These are then supplemented by NLO electroweak corrections where relevant, as described further in \cite{Cridge:2021pxm}.

A variety of follow-up studies have extended the MSHT20 investigations since its publication. A study of the  dependence of the PDFs on the strong coupling $\alpha_s$ and the heavy-quark masses was presented in~\cite{Cridge:2021qfd}, while an update to include QED effects in the DGLAP evolution and a corresponding photon PDF was presented in~\cite{Cridge:2021pxm}, the latter representing a further theoretical improvement and building on previous work~\cite{Nathvani:2018pys}.  The MSHT20 NNLO PDF set was also the PDF set which contributed to the PDF4LHC21 combination~\cite{PDF4LHCWorkingGroup:2022cjn} and the reduced fit studies leading up to it also described in~\cite{Cridge:2021qjj}.
 Most recently, the first global PDF analysis at approximate N3LO (aN3LO) was presented in~\cite{McGowan:2022nag}, extending beyond the current highest NNLO order in QCD of theoretical predictions achieved in contemporary PDF fits. This MSHT20aN3LO study includes for the first time already known N3LO ingredients, as well as the estimates of uncertainties due to the yet unavailable pieces and implicitly some higher-order contributions beyond N3LO. In particular, the splitting functions, transition matrix elements, coefficient functions and $K$-factors for multiple processes are approximated using the exact partial N3LO results and constrained to be consistent with the wide range of already available information about this order \cite{McGowan:2022nag}. Theoretical nuisance parameters are then used to include uncertainties from the missing pieces into the overall PDF uncertainties. 
 
For the present study, we compute the $L_2$ sensitivities using the $\chi^2$ values and MSHT20 Hessian eigenvector sets presented in the NNLO~\cite{Bailey:2020ooq} and the aN3LO~\cite{McGowan:2022nag} analyses.  For this study, we note minor modifications in the precise details from the above publications. In the NNLO case, a correction has been made in the treatment of photon--initated production for the ATLAS 8 TeV double differential $Z$ \cite{ATLAS:2017rue} data, as described in~\cite{Cridge:2021pxm}. For the aN3LO fit, an incorrect application of the N3LO K--factor was corrected in the final version of~\cite{McGowan:2022nag} for two Drell-Yan data sets. \footnote{Since the K--factors are very largely decorrelated from PDF parameters, the publicly available PDFs did not require a refit. A full refit has, however, been performed for the PDFs used in this article and results in a very small change in the PDFs and a reduction in the total $\chi^2$ by a few units.}

The MSHT data sets included in the aN3LO study are identical to the MSHT20 NNLO baseline. In~\cite{McGowan:2022nag}, it was found that the addition of aN3LO theory leads in some places to significant changes and improvements in the fit quality, with evidence that in some cases this is due to the tensions between data sets being alleviated. This therefore motivates studying the pulls of the data sets in the fit in both the MSHT20 NNLO and the aN3LO fit using the $L_2$ sensitivity approach.

\subsubsection{CT18 and MSHT20 data sets side-by-side 
\label{sec:CTMSHTDifferences}}
 Tables~\ref{tab:DISdataCTMSHT}-\ref{tab:JetdataCTMSHT} list the CT18 and MSHT20 data sets side-by-side together with the references, numbers of points, and numerical ID's.
Sometimes one PDF analysis fits several data sets independently, while the other analysis fits
them as a combined data set. There are several such examples in Tables~\ref{tab:DISdataCTMSHT}-\ref{tab:JetdataCTMSHT}. 
To facilitate the comparisons in such cases, 
we occasionally compute the sensitivity using the $\chi^2$ for the combination in both fits and then plot the sensitivities both for the constituent and combined data sets.  Such data sets include:
\begin{enumerate}
    \item HERA DIS (a combined data set 160 in CT18; independent sets 22-28 in MSHT);
    \item HERA charm and bottom production (independent HERA I charm and H1 bottom data sets, 147 and 145, in CT18; a combined data HERA I+II charm and bottom set 14 in MSHT);
    \item CCFR dimuon SIDIS (independent neutrino and antineutrino data sets, 126 and 127, in CT18; a combined data set 16 in MSHT);
   \item NuTeV dimuon SIDIS (independent neutrino and antineutrino data sets, 124 and 125, as well as their a posteriori combination 593, in CT18; a combined data set 17 in MSHT);
      \item LHCb $W$ and $Z$ production at forward rapidities (independent data sets at 7 and 8 TeV, 245 and 246, as well as their a posteriori combination 258, in CT18; a combined data set 61 in MSHT).
\end{enumerate}

The numbers of the data points selected by the groups from a shared data set can be different in reflection of the trade-off between the total number of points and accuracy of the individual points and theoretical predictions. One example are the BCDMS data sets on proton and deuteron DIS structure functions, which include more points in the CT18 fit (as data sets 101 and 102) than in MSHT20 (as data sets 1 and 2).  This difference, however, reflects a different presentation of the same data, with the same information and constraints encoded. In particular in MSHT20 the data are averaged over energy runs~\cite{Benvenuti:1989rh}. In the CT18 analysis, the BCDMS data sets have a particularly pronounced sensitivity to the large-$x$ behavior of the up and down PDFs \cite{Courtoy:2020fex}. Of these, the extracted high-$x$ behavior of the $d$-PDF can be especially influenced by model-dependent prescriptions to correct the deuteron structure function to that of an isoscalar nucleon target~\cite{Accardi:2021ysh}. In the MSHT20 analysis, the BCDMS sensitivities differ from CT18 in reflection of the differences in the obtained PDFs in terms of central values and uncertainties, and from the treatment of nuclear effects.

Another example are the ATLAS 8 TeV data sets on the transverse momentum of $Z$ boson, labeled as 253 and 71 by the two groups, respectively. The CT18 analysis selected 27 data points presented single-differentially in the interval $45\leq p_{T, Z} <150$ GeV, where the fixed-order NNLO theory has best convergence. The MSHT analyses, on the other hand, include 104 points and additional intervals of $30 \leq p_{T, Z} < 45 $ GeV and $p_{T, Z} > 150 $ GeV, while using the cross-sections that are double-differential in $p_{T, Z}$ and $y_Z$ in the $Z$ mass bin. At NNLO, the $Z\ p_T$ data set demonstrates a stronger sensitivity to the gluon than in CT18, which is consistent with the larger number of data points and extended $p_{T, Z}$ range in the MSHT20 fit. Both CT and MSHT groups choose the factorization scale to be the transverse mass of the vector boson. On the other hand, some potential differences in their theoretical cross sections (including electroweak contributions at high $p_{TZ}$) and prescriptions for systematic uncertainties need to be further explored. MSHT20 observe further changes when going to aN3LO, as discussed in Sec.~\ref{sec:SFL2MSHT}.

\subsubsection{PDF4LHC21 reduced fits
\label{sec:PDF4LHC21Data}}
The PDF4LHC21 reduced fits are discussed here only for the two participating Hessian-based analyses, the modified CT18 and MSHT20 NNLO. In that exercise, an agreed-upon list of data sets, common to all groups and marked in Tables~\ref{tab:DISdataCTMSHT}-\ref{tab:JetdataCTMSHT}, was fitted. 
The CTEQ-TEA collaboration contributed CT18$^\prime$ NNLO, which included the ATLAS 7 TeV high-precision $W, Z$ data (ID=68 for ATLAS and MSHT, or 248 for CT) as in CT18A, and differed slightly from the default CT18 in the choice of $m_c^{\rm pole}=1.4$ GeV and small differences in the other data sets. Besides this baseline choice of data sets, common theory settings, like coinciding QCD parameters or setting $s=\bar{s}$, were adapted to minimize discrepancies between all fitting groups. All differences found in the reduced fits were hence mostly due to methodological choices. In reflection of this, the $L_2$ sensitivities for the CT18$^\prime$ and MSHT20 reduced fits discussed in Sec.~\ref{sec:SFL2reduced} quantify the differences due to their methodologies while fitting (practically) the same data set. 

\subsection{Conventions for log-likelihoods
\label{sec:LogLikelihood}
}
The probabilities $P(D|T)$ in the likelihood ratio tests (\ref{LikelihoodRatioTest}) and
(\ref{LikelihoodRatio}) are in fact {\it augmented} likelihoods
\cite{Lepage:2001ym} in the sense that they include prior probability contributions
associated with the nuisance parameters describing correlated systematic
effects.  Normally these prior contributions are included as quadratic
sums of the nuisance parameters that are assumed to obey the standard normal distribution ${\cal N}(0,1)$. Since the experiments provide only a partial information
about correlations among systematic factors, the PDF analyses
use approximate correlation or covariance matrices to estimate these
factors according to a number of prescriptions. The choice of these $\chi^2$
definitions leads to non-negligible differences among the resulting PDFs, with proper modeling of systematic effects presenting a central issue for NNLO PDF determinations \cite[Sec.~5.A in][]{Amoroso:2022eow}. For example, the pulls on the large-$x$ gluon by jet production experiments depend on the $\chi^2$ definition. 

For completeness, here we list the functional forms of the $\chi^2$ that were used to compute the sensitivities. 

In the ATLASpdf21 NNLO analysis \cite{ATLAS:2021vod},
the definition of the $\chi^2$ is
 \begin{equation}
\begin{aligned}
\chi^2 = &\sum_{i,k=1}^{N_{\rm pts}}\left(D_i -
  T_i  +  \sum_{\alpha=1}^{N_{\rm corr}} X_{i,\alpha}
  \beta_{i\alpha}\lambda_\alpha \right)C^{-1}_{\text{stat,uncor},
    ik}(D_i,D_k)\left(D_k - T_k +  \sum_{\alpha^\prime=1}^{N_{\rm corr}}X_{k,\alpha}\beta_{k\alpha^\prime}\lambda_{\alpha^\prime} \right)\\
&+ \sum_{i=1}^{N_{\rm pts}} \log \frac{\delta^{2}_{i,\text{uncor}}T_i^2 + \delta^{2}_{i,\text{stat}}D_i T_i}{\delta^{2}_{i,\text{uncor}}D_i^2 + \delta^{2}_{i,\text{stat}}D_i^2}\\
&+ \sum_{\alpha=1}^{N_{\rm corr}} \lambda_\alpha^2 ,
\label{Chi2ATLAS}
\end{aligned}
\end{equation}
where $D_i$ represent the central values of the measured data, $T_i$ are the corresponding
theoretical predictions,  
$\delta_{i,\text{uncor}}$ and $\delta_{i,\text{stat}}$ are the fractional
uncorrelated systematic uncertainties and the  
statistical uncertainties of $D_i$, respectively, and the correlated systematic
uncertainties described by the correlation matrix $\beta_{i\alpha}$  
are accounted for using the nuisance parameters $\lambda_\alpha$.
The quantity $C_{\text{stat,uncor}, ik}$ is a covariance matrix for both the  
statistical and uncorrelated systematic uncertainties.
Summations over $i$ and $k$ run over all $N_{\rm pts}$  data points, and summations
over $\alpha$ and $\alpha'$ run over all $N_{\rm corr}$ sources of correlated systematic uncertainty. 
For each data set, the first term gives the main contribution to the
partial $\chi^2$ of the data set. The second term is a small bias
correction term, referred to as the {\em log penalty}, which arises
because the diagonal term of the matrix, $C$, is  
given by $C_{ii} =\delta^{2}_{i,\text{uncor}}T_i^2 + \delta^{2}_{i,\text{stat}}D_i T_i$,
with different weighting for
statistical uncertainties and uncorrelated systematic
uncertainties. This form of the $\chi^2$ is used as standard in HERA
and ATLAS PDF fits~\cite{Abramowicz:2015mha,ATLAS:2016nqi,ATLAS:2018ttbar}.

The CTEQ-TEA and MSHT analyses use
\begin{equation}
\chi^2=\sum_{i=1}^{N_{\rm pts}}\left(\frac{D_i -T_i +  \sum_{\alpha=1}^{N_\text{corr}}
 X_{i,\alpha} \beta_{i,\alpha}\lambda_\alpha}{\sqrt{\delta_{i,\text{uncor}}^2 +
  \delta_{i,\text{stat}}^2} D_i}\right)^2 + \sum_{\alpha=1}^{N_{\rm corr}}\lambda_\alpha^2,
\label{Chi2CTMSHT}
\end{equation}
when the individual correlated sources are provided. When only the
final covariance matrix $C_{ij}$ is available, they instead compute 
\begin{equation}
\chi^2 = \sum_{i=1}^{N_{\rm pts}} \sum_{i=j}^{N_{\rm pts}}
(D_i-T_i) (C^{-1})_{ij} (D_j-T_j).
\label{Chi2Cov}
\end{equation}

In all these fits, $\beta_{i,\alpha}$ are published as percentage correlation
matrices that must be multiplied by a reference value $X_i$ for each
data point when included in Eqs.~(\ref{Chi2ATLAS}) and
(\ref{Chi2CTMSHT}). 
In addition, the MSHT20aN3LO fit includes
nuisance parameters and the associated correlation matrices to estimate theoretical uncertainty sources from the modeled aN3LO ingredients -- see~\cite{McGowan:2022nag} for details.
The three groups treat all correlated systematic
uncertainties as multiplicative ones, which amounts to setting
$X_{i,\alpha}=T_i$ for all $\alpha$ in the above $\chi^2$ definitions. While such a prescription
is not unique, it helps to reduce the bias when the uncertainties are
dominated by statistical fluctuations \cite{DAgostini:1993arp, DAgostini:1999gfj,Ball:2009qv,Ball:2012wy,Gao:2013xoa,Courtoy:2022ocu}. It should be kept in mind, however,
that the true $X_{i,\alpha}$ are generally unknown, 
reflecting the broad challenges in modeling of the LHC systematic uncertainties \cite[Sec. 5.A in][]{Amoroso:2022eow},
and other biases can be present,
especially when the statistical errors are small compared to the
systematic ones. See the discussion of this point in Ref.~\cite{Courtoy:2022ocu}.

\subsection{Tolerance conventions
  \label{sec:ToleranceConventions}
}
The $L_2$ sensitivity $S_{f,L_2}^H$ in Eq.~(\ref{SFL2Hessian})
is proportional to the PDF uncertainty $\delta_{\rm H} f$, and hence its magnitude
reflects the tolerance prescription for constructing the Hessian
eigenvector sets. These prescriptions get elaborate and non-uniform in the published
PDF ensembles and may reflect the asymmetric behavior of the
uncertainties, parametrization and scale dependence, 
and other nontrivial features. For the comparisons of
the $L_2$ sensitivities, it is desirable to follow 
a simple and identical tolerance prescription for all compared fits.

In this regard, we choose the global tolerance with $T^2$ of about 10 
for the presented comparisons, as it amounts to generating all eigenvector
sets with $\Delta \chi^2 \approx 10$ and can be easily implemented in
the fit and related to the LM scans, as discussed in
Sec.~\ref{sec:SFL2OneDim}. The value of $\Delta \chi^2 =10$ is low
enough to suppress non-linear deviations from the $L_2$ sensitivity
formula.

Furthermore, the actual tolerances in the ATLASpdf21 and MSHT20 fits are
close on average to the global $\Delta \chi^2 = 10$ tolerance. From
the ATLASpdf21 NNLO analysis, we take the Hessian eigenvector sets with
$t^2=9$. We include only the error sets corresponding to the
experimental uncertainties and not to the model and parametrization
uncertainties, because when tolerance $t=3$ is used, the model and parametrization uncertainties are relatively small. 
Here and in the following,  the lowercase ``$t$'' indicates
that the displacements of the eigenvector sets from the best-fit set
were computed using the idealized Gaussian $\chi^2$ arising
in the diagonalization of the Hessian matrix. The uppercase ``$T$''
indicates that the displacements are computed using  the actual
$\chi^2$,  which includes some non-quadratic terms and is mildly asymmetric. The $T$ and $t$ criteria are generally close but not identical.

The MSHT20  fits  by default apply the ``dynamic tolerance'' procedure, based on a weaker hypothesis-testing criterion and described in more detail in~\cite{Martin:2009iq,Harland-Lang:2014zoa,Bailey:2020ooq}. This enlarges the uncertainties beyond the $\Delta\chi^2=1$ definition to account for data set tensions, as well as potential mismatch of data and theory due to imprecision in the theory or parametrization and experimental measurements. 
This procedure may be too complicated
when calculating the $L_2$ sensitivity, so for this study the MSHT
group provides additional Hessian sets that are computed
with a fixed tolerance of $T^2=10$. The resulting PDF
errors are rather close to the ones obtained by applying the dynamic tolerance of MSHT.

While the aN3LO central PDFs include the estimated N3LO corrections, in the plotted aN3LO uncertainties and sensitivities, we exclude the contributions from the eigenvector sets due to the nuisance parameters that correspond to the theoretical uncertainty from the remaining ambiguities
in the N3LO (and implicitly higher-order) ingredients.
This is done to provide the most like--for--like comparison with 
the other PDF ensembles, which do not include such theoretical uncertainty. 
As observed in~\cite{McGowan:2022nag}, the neglected
error  sets are in general subleading and do not significantly modify the outcomes 
for the purposes of the current study.

Finally, the CT18 error sets are determined using
a two-tier (global+dynamic) tolerance that accounts for experimental, parametrization,
and methodological uncertainties and results in error bands that are
wider than with the $T^2=10$ tolerance. It was easy,
however, to obtain a special set of CT18 eigenvector sets
corresponding to the global $T^2=10$.

\section{Sensitivities for non-global  fits \label{sec:non-global}}

With the background given in the previous sections, we can now analyze the Hessian sensitivities of the selected PDF ensembles and even compare $L_2$ sensitivity patterns obtained with ensembles from various groups.
In this section, the sensitivities will highlight the prominent features of the fits that are based on specific data sets, like the  ATLASpdf21 fit~\cite{ATLAS:2021vod} that is dedicated to HERA+ATLAS data, Sec.~\ref{sec:SFL2ATLAS}. Another class of fits that we consider here is the so-called {\it reduced fits}, Sec.~\ref{sec:SFL2reduced}. Examples of such fits were proposed in the context of the PDF4LHC21 benchmarking exercise~\cite{PDF4LHCWorkingGroup:2022cjn,Cridge:2021qjj}.  

The full collection of the figures is available at the companion website \cite{L2website}, showing $L_2$ sensitivities to \href{http://metapdf.hepforge.org/L2/2023_T210/index2.html}{PDFs and PDF combinations} and for \href{http://metapdf.hepforge.org/L2/2023_T210/index3.html}{individual fitted experiments}.

\subsection{Sensitivities for ATLAS fits \label{sec:SFL2ATLAS}}

We begin by comparing the $\chi^2_E$ values and the $L_2$ sensitivity 
variable for the three classes of data, HERA, ATLAS W, Z 7 TeV and ATLAS $t\bar{t}$  8 TeV, within the three 
fits (HERAPDF2.0NNLO, ATLASepWZ16 and ATLASepWZtop18) summarized in Sec.~\ref{sec:DataSetsATLAS}. 
Table~\ref{tab:sparty} gives the $\chi^2_E$ values and number of data points, $N_{\rm pts}$, for data sets entering each of the ATLAS fits. Note that not all data sets enter every fit.
If a data set is not fitted, its values are printed in italics. Note also that the number of data points for a particular data set may differ according to the PDF fit under consideration.
Comparing the $\chi^2$ values among the fits shows how fitting improves the description of these data. Inclusion of additional data sets may also degrade the description of data already in the fit, revealing tensions among the data sets.
\begin{table}
\begin{center}
\begin{tabular}{ccccc}
\hline
\hline
   PDF   &  data set  &  $\chi^2_E$ &  $N_{\rm pts}$  \\
\hline
  HERAPDF2.0& HERA I+II combined &  1363 & 1145 \\
  ATLASepWZ16 & HERA I+II combined &  1213 & 1056 \\
  ATLASepWZtop18 & HERA I+II combined &  1149   & 1016     \\
\hline
  HERAPDF2.0& ATLAS W,Z 7 TeV  &{\it 384}  & {\it 61}  \\
  ATLASepWZ16 & ATLAS W,Z 7 TeV &  108 & 61 \\
  ATLASepWZtop18 & ATLAS W,Z 7 TeV &  79   &  55     \\
\hline  
  HERAPDF2.0& ATLAS $t\bar{t}$ 8 TeV  & {\it 31}  & {\it 20} \\
  ATLASepWZ16 & ATLAS $t\bar{t}$ 8 TeV & {\it 19} & {\it 20} \\
  ATLASepWZtop18 & ATLAS $t\bar{t}$ 8 TeV &   17  & 20      \\
\hline  
\end{tabular}
\end{center}
\caption{$\chi^2_E$ values, and number of data points, $N_{\rm pts}$,  for the HERA combined, 
ATLAS $W, Z$ 7 TeV and ATLAS $t\bar{t}$ data sets as predicted by
each of the PDFs HERAPDF2.0NNLO, ATLASepWZ16, ATLASepWZtop18. If a data set is not used in the PDF fit, then its values are in italics.\label{tab:sparty}}
\end{table}

The $\chi^2_E$ value of the HERA data (160) within the HERAPDF2.0NNLO fit is higher than ideal, but similar in value to that in the CT18 and MSHT20 fits. However, it is significantly reduced in the ATLASepWZ16 and ATLASepWZtop18 fits because of the harder $Q^2$ cuts imposed on these HERA data: $Q^2 > 7.5\mbox{ GeV}^2$ for the ATLASepWZ16 PDFs and $Q^2 > 10\mbox{ GeV}^2$
for the ATLASepWZtop18 PDF. These cuts avoid data for which low-$x$ higher twist effects or $\ln(1/x)$ 
resummation may be important. 

The $\chi^2_E$ value of the ATLAS $W, Z$ 7 TeV data (68) is very high for the HERAPDF2.0NNLO fit, reduces in the ATLASepWZ16 fit (as the $W, Z$ data set is now a part of the fit), and reduces further in the ATLASepWZtop18 fit, in which the 6 data points from the low-mass Drell-Yan data are removed  due to poor understanding of their electroweak radiative contributions. 

The $\chi^2_E$ values of the ATLAS $t\bar{t}$ 8 TeV data (7) are not very high, even for PDFs which are not fitted to the data.
There is no tension between the $t\bar{t}$ data included in the ATLASepWZtop18 fit 
and the HERA data or the $W, Z$ data.

\begin{figure*}[bht]
  \begin{centering}
    \includegraphics[height=166pt,trim={ 0 0 .1cm 0},clip]{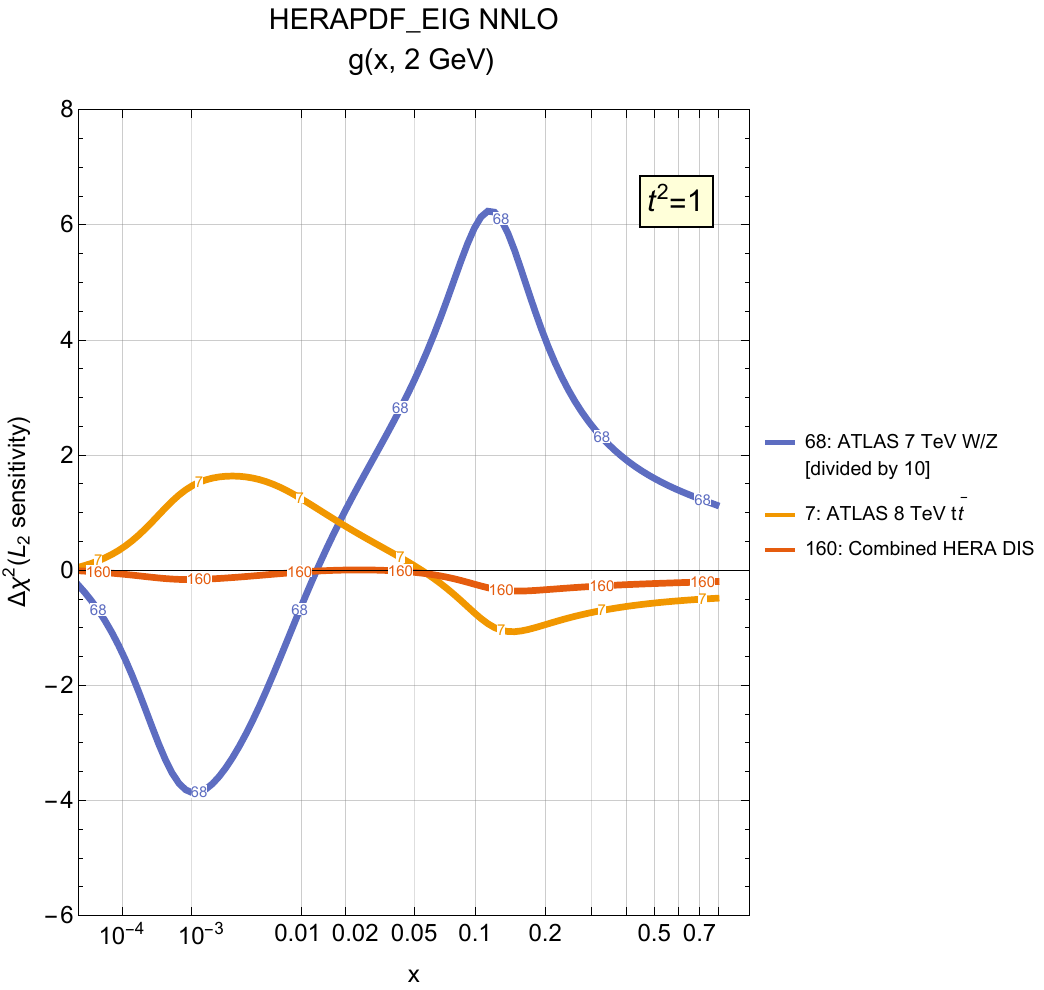}      
   \includegraphics[height=166pt,trim={ .8cm 0 4.9cm 0},clip]{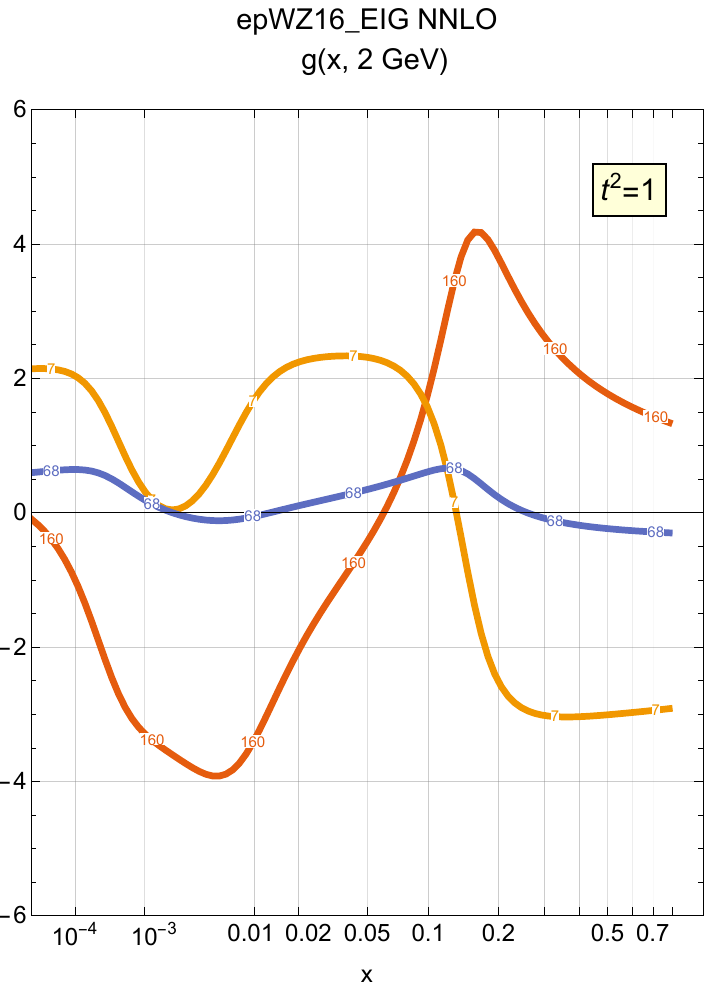}
     \includegraphics[height=166pt,trim={ .8cm 0 0.1cm 0},clip]{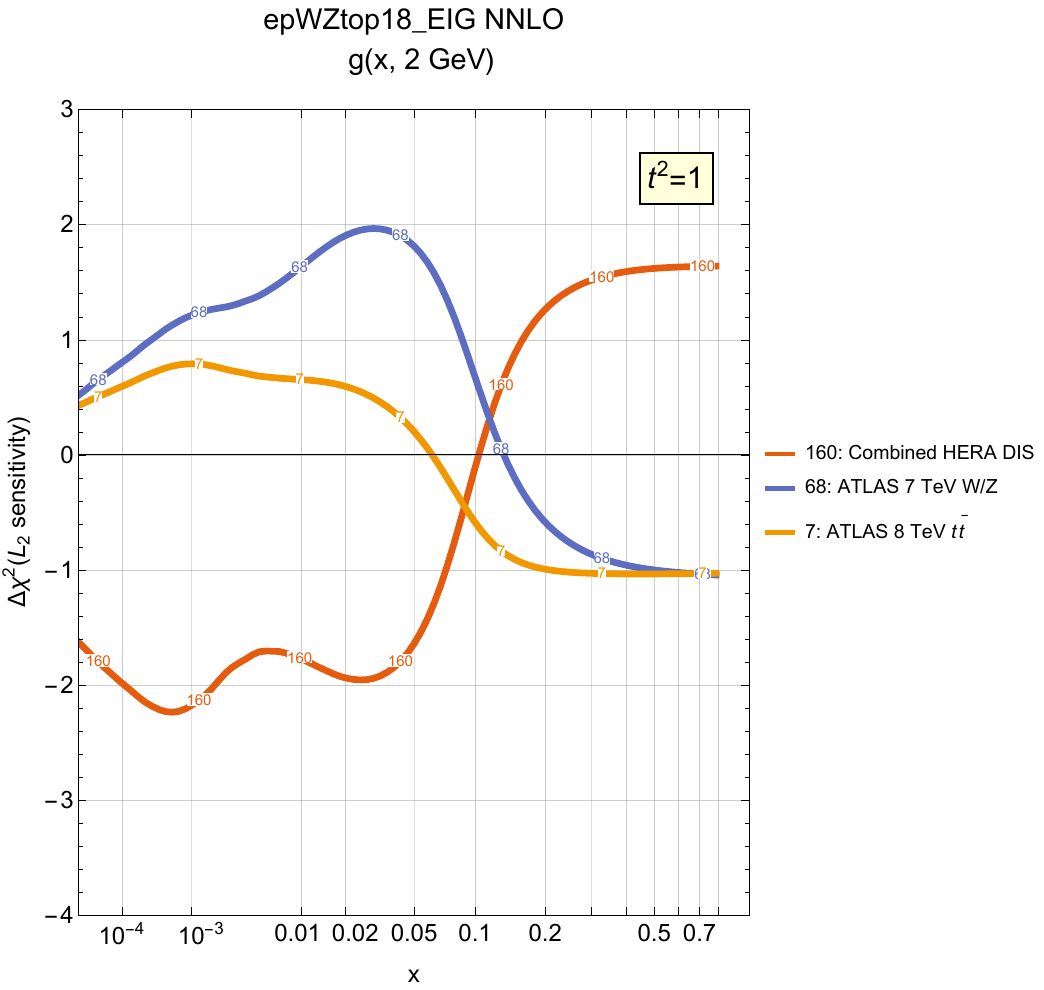}   
    \caption{$L_2$ sensitivity for HERA, ATLAS $W, Z$ 7 TeV and $t\bar{t}$  8 TeV data sets for the gluon at $Q=2$ GeV in the fits: left, HERAPDF2.0; middle, ATLASepWZ16; right, ATLASepWZtop18.
    In the left panel for HERAPDF2.0, the ATLAS $W, Z$ sensitivity is divided by ten.
    \label{fig:L2HERAglue} 
    } 
  \end{centering}
\end{figure*}

The features of the fits revealed above using the $\chi^2_E$ values are much more clearly seen in the $L_2$ sensitivty plots. Fig.~\ref{fig:L2HERAglue} compares the $L_2$ sensitivities for the gluon PDF in the HERAPDF2.0NNLO, ATLASepWZ16 and ATLASepWZtop18 fits for the three data sets: HERA combined inclusive DIS data (ID=160); ATLAS W, Z 7 TeV data (68); ATLAS $t\bar{t}$ 8 TeV data in lepton+jets and dilepton channels (7). Note that the sensitivity to the ATLAS $W, Z$ data for the HERAPDF2.0NNLO fit is very large and has been divided by a factor of 10 for display. These large sensitivities are dramatically reduced when the $W, Z$ data are fitted. Then one can observe that the HERA data and the $t\bar{t}$ data pull against each other in the ATLASepWZ16 fit, with the $t\bar{t}$ data favoring a harder gluon at high $x$.  Once the $t\bar{t}$ data are fitted in the ATLASepWZ18 fit, the gluon becomes harder, and the magnitudes of sensitivity for these data decrease. One can also observe the change in sensitivity to the HERA data between these two fits, where sensitivity to the gluon at low $x$ is decreased when the harder $Q^2$ cut is imposed in the ATLASepWZtop18 fit.
Although there are subtle changes in the pulls of the data sets between the two ATLAS fits,
no significant tensions between the ATLAS data sets and the HERA data set are evident.  As we commented earlier, the sensitivity quantifies such tensions, while the HERA DIS data set is in a good agreement with the PDFs in this case and imposes strong constraints on them.

\begin{figure*}[bht]
  \begin{centering}
    \includegraphics[height=166pt,trim={ 0 0 .1cm 0},clip]{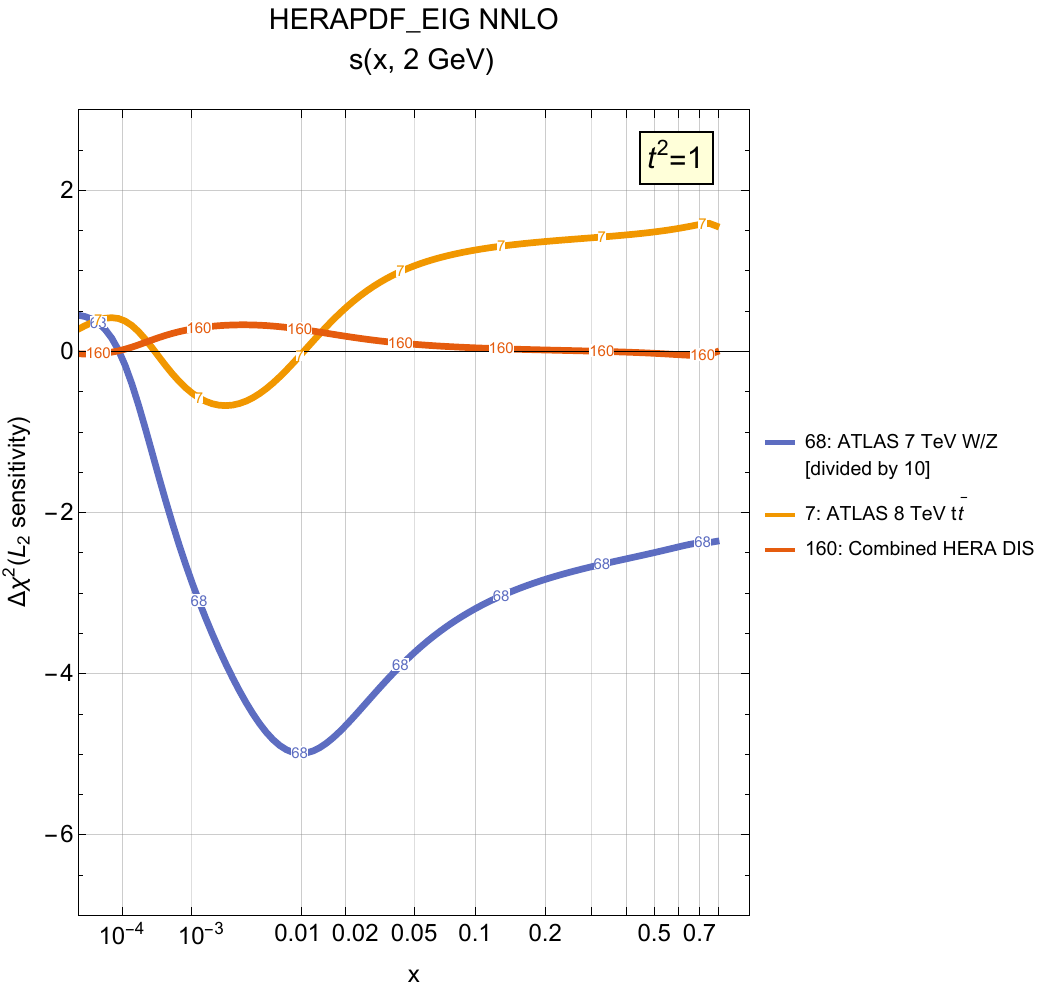}
    \includegraphics[height=166pt,trim={ .8cm 0 4.9cm 0},clip]{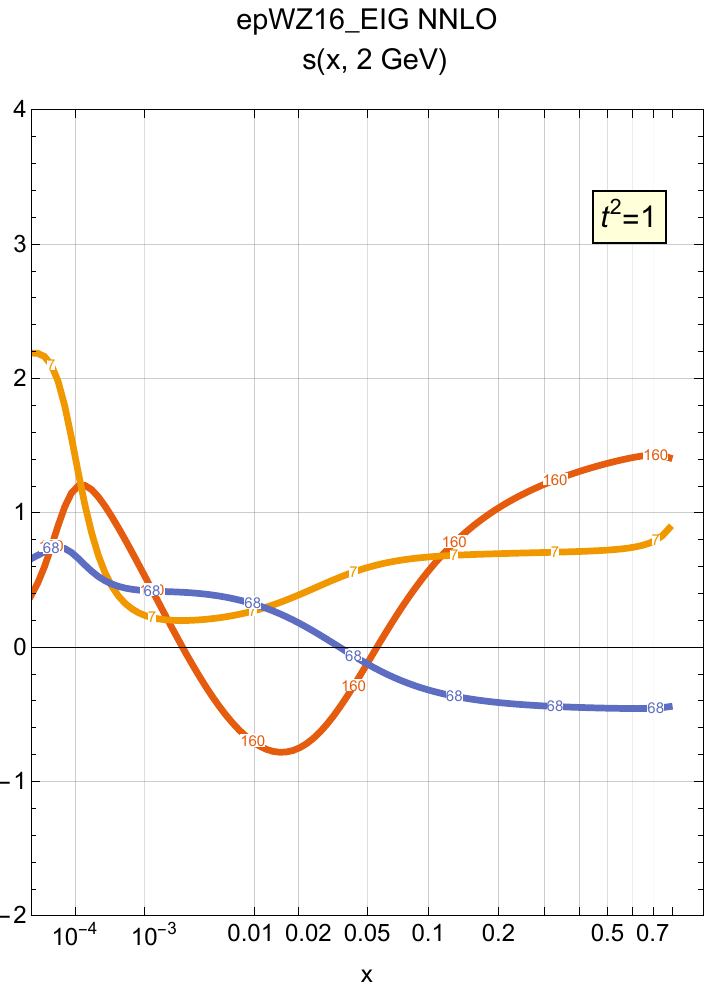}
    \includegraphics[height=166pt,trim={ .8cm 0 .1cm 0},clip]{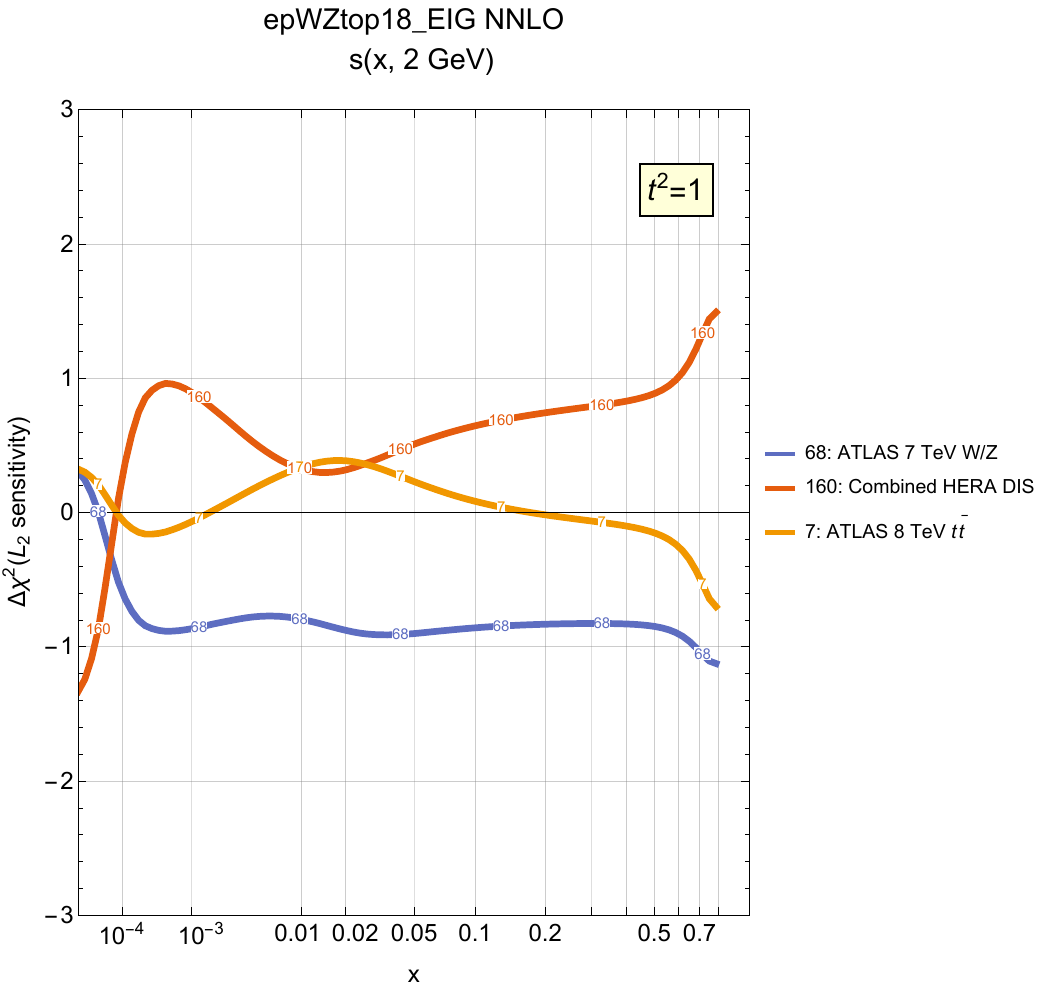}
    \caption{Same as Fig.~\ref{fig:L2HERAglue}, for the strangeness PDF.
    \label{fig:L2HERAstrange} 
    } 
  \end{centering}
\end{figure*}

Figure \ref{fig:L2HERAstrange} compares the $L_2$ sensitivities for the strangeness PDF
in the HERAPDF2.0NNLO, ATLASepWZ16 and ATLASepWZtop18 PDFs for the same data sets. Again the sensitivity to the ATLAS $W, Z$ data in the HERAPDF2.0NNLO fit is reduced by a factor of 10 for display. Clearly the ATLAS $W, Z$ data show preference to increase the strangeness PDF of HERAPDF2.0, particularly at $x\sim 0.01$. This was confirmed when the data were fitted.
There are no large sensitivities remaining in the ATLAS PDFs, although there are subtle changes of sensitivity shape and sign between the two fits.

\begin{figure*}[bht]
  \begin{centering}
    \includegraphics[width=0.48\textwidth]{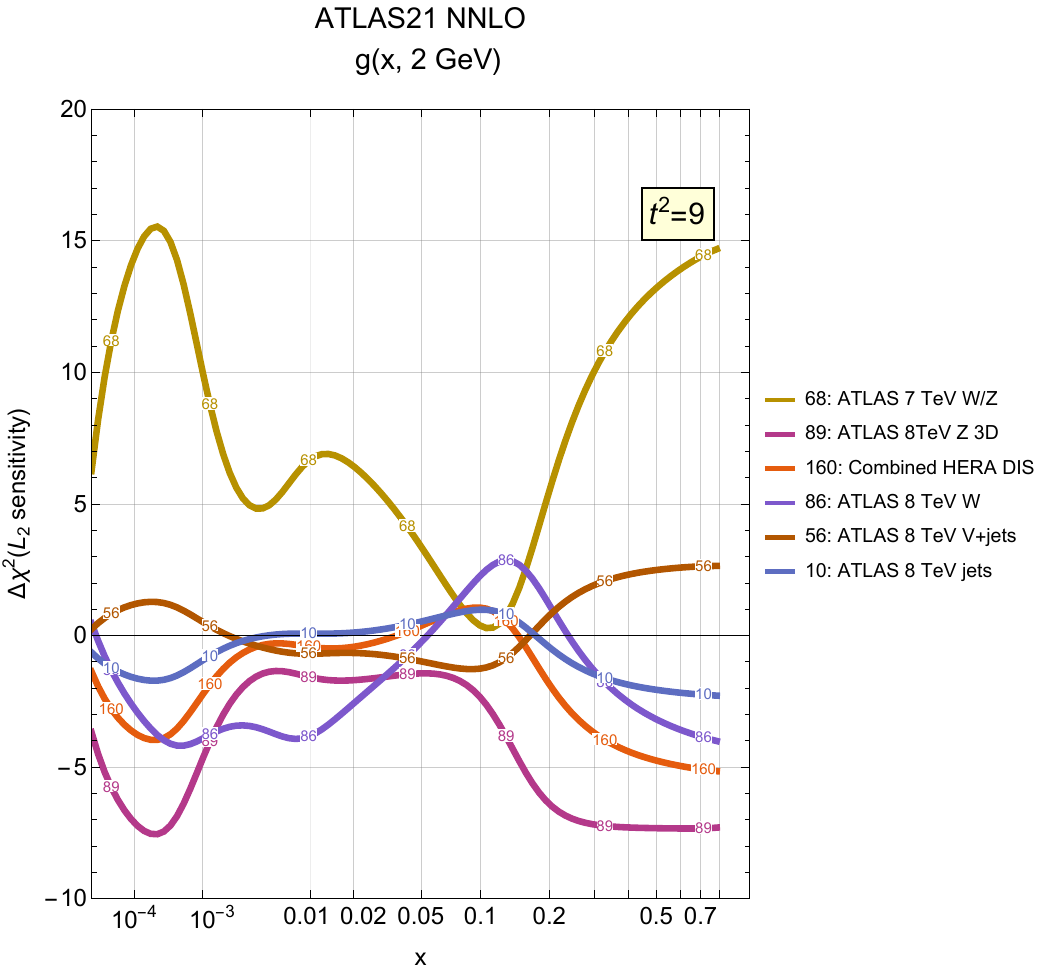}
    \includegraphics[width=0.48\textwidth]{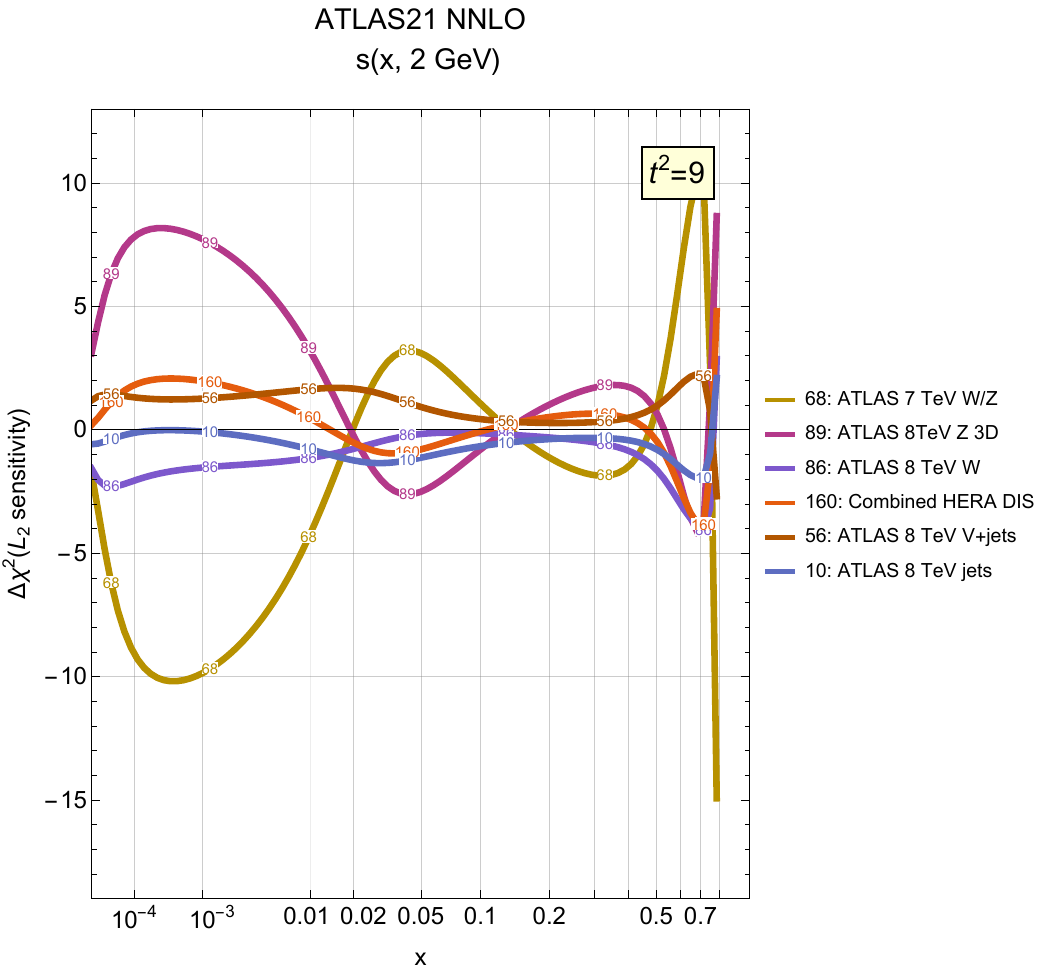}
    \caption{ $L_2$ sensitivity for all data sets in the ATLASpdf21 fit: gluon (left), strange (right).\label{fig:gluonstrall} } 
  \end{centering}
\end{figure*}

Now we consider the $L_2$ sensitivity for the most complete ATLAS fit to date, ATLASpdf21, which includes all the data sets listed in Table~\ref{tab:ATLASData}.
We first plot the sensitivities to the gluon and strange PDFs for six most sensitive experiments in Fig.~\ref{fig:gluonstrall}. It turns out that, whatever the PDF flavor (including $u$ and $d$ flavors, not shown), the most sensitive experiments are the ATLAS 7 TeV $W, Z$, the ATLAS 8 TeV Z3D data, HERA combined data, the ATLAS 8 TeV $W$ data, the ATLAS 8 TeV $V$ + jets data, and the ATLAS 8 TeV inclusive jet data. The $t\bar{t}$ data and the direct photon data have smaller sensitivities and so are not shown.   
For the gluon, the ATLAS 7 TeV $W, Z$ data are the most sensitive, and we observe the opposing, though weaker, tendency of the ATLAS 8 TeV Z3D data. 
For the strangeness, we again see the opposing tendencies of ATLAS 7 TeV $W, Z$ and ATLAS 8 TeV Z3D data, with ATLAS 7 TeV $W, Z$ favoring more strangeness at low $x$, and ATLAS 8 TeV Z3D data favoring less with almost equal weight. It is interesting that these data are in agreement at $x\sim 0.02$ for $Q = 2$ GeV, since this is approximately the $x$ and $Q^2$ values at which ATLAS chose to illustrate the ratio of strangeness to light quarks. Above this $x$ value, the opposition persists but is much weaker until $x > 0.5$, where ATLAS 7 TeV $W, Z$ data and ATLAS 8 TeV $V$ + jets data favor less strangeness, but are opposed by HERA combined data, ATLAS 8 TeV $W$ data and ATLAS 8 TeV inclusive jets data. 

\begin{figure*}[bht]
  \begin{centering}
    \includegraphics[width=0.48\textwidth]{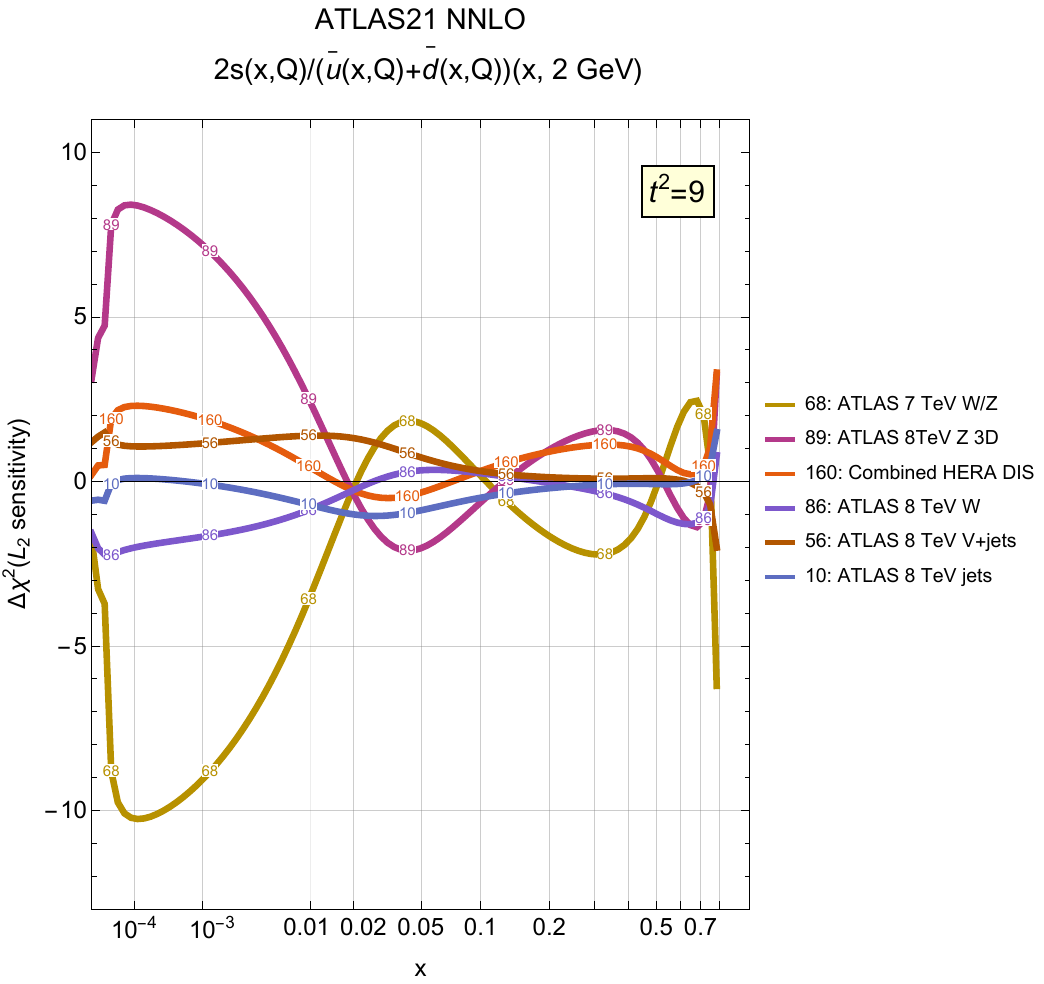}
    \includegraphics[width=0.48\textwidth]{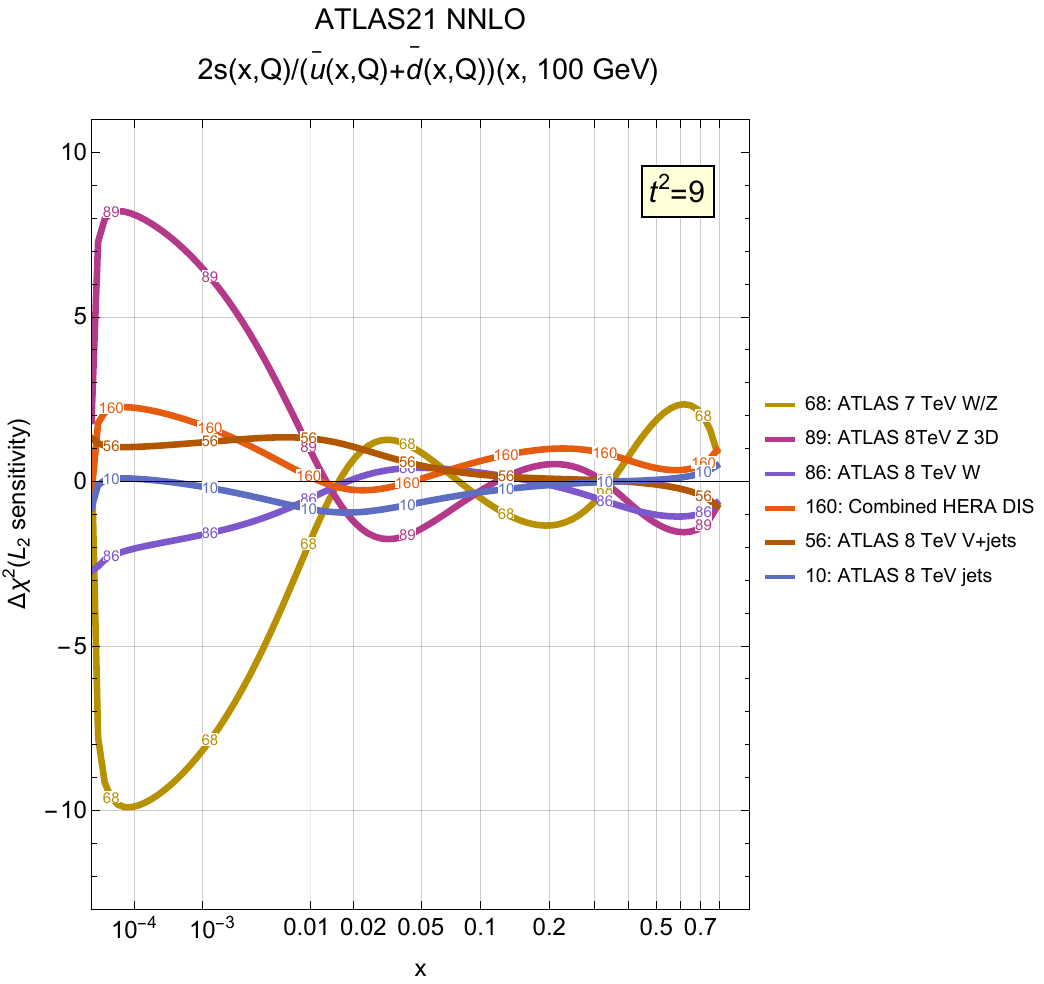}
    \caption{ $L_2$ sensitivity for all data sets in the ATLASpdf21 fit for the strangeness ratio $2s(x,Q)/(\bar{u}(x,Q) +\bar{d}(x,Q))$: $Q=2$ GeV (left), $Q=100$ GeV (right).\label{fig:srat2100} } 
  \end{centering}
\end{figure*}

In Fig.~\ref{fig:srat2100}, we show the sensitivities to the strangeness ratio $2s(x,Q)/(\bar{u}(x,Q) +\bar{d}(x,Q))$ at $Q=2$ GeV and $Q=100$ GeV.  We again see the opposing tendencies of ATLAS 7 TeV $W, Z$ and ATLAS 8 TeV Z3D data, with ATLAS 7 TeV $W, Z$ favoring a higher strangeness ratio at low $x$, and ATLAS 8 TeV Z3D data favoring a smaller ratio with almost equal weight. As we saw for the strangness PDF, these experiments are in agreement for $x\sim 0.02$. The picture at low scale $Q=2$ and at high scale $Q=100$ GeV is very similar, with the shapes moving to lower $x$, as expected by QCD evolution.

\begin{figure*}[p]
  \begin{centering}
    \includegraphics[width=0.48\textwidth]{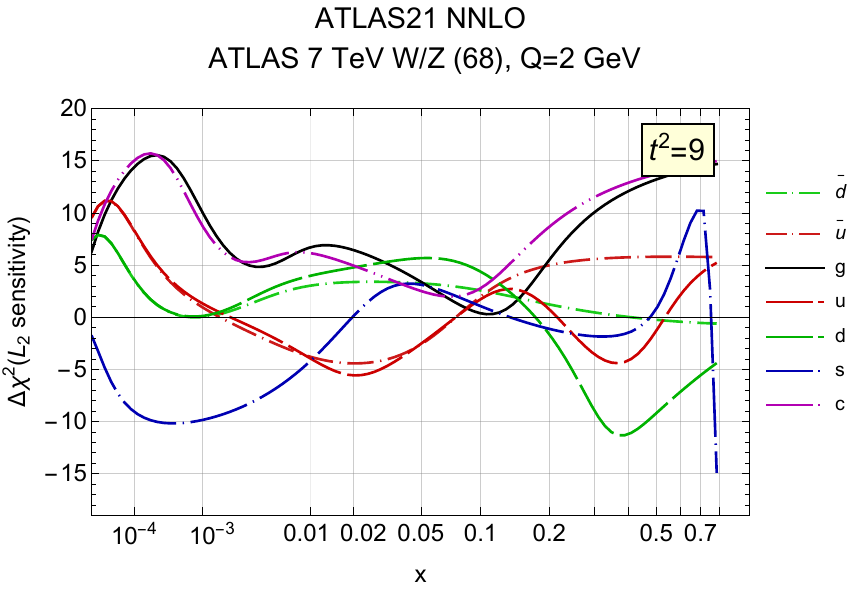}
    \includegraphics[width=0.48\textwidth]{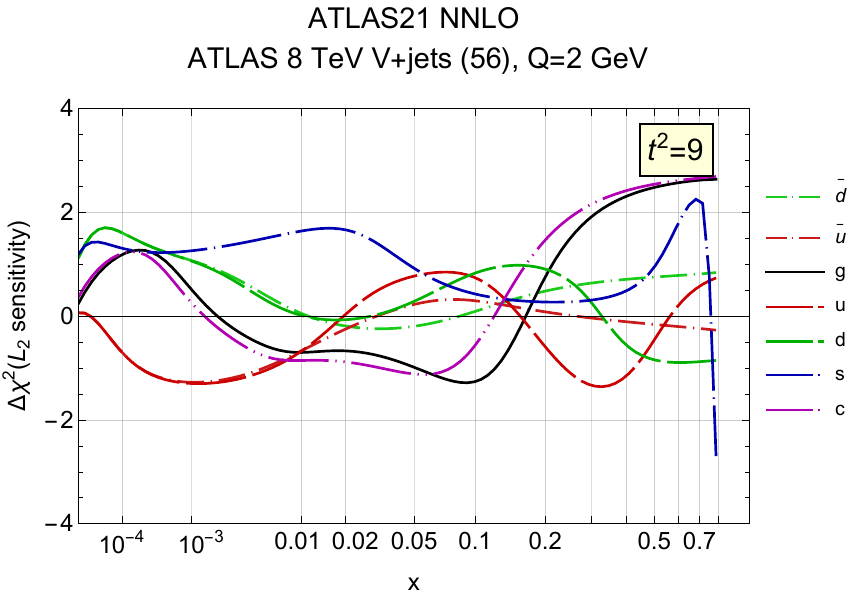}
    \caption{ The $L_2$ sensitivities to all PDF flavors in the ATLASpdf21 fit for the ATLAS 7 TeV $W, Z$ data (left), ATLAS 8 TeV  $V$ + jets data (right).\label{fig:WZ7Z3D} } 
  \end{centering}
\end{figure*}

An alternative way to look at the sensitivities is to plot the sensitivity to each experiment for all the PDF flavors.
This is shown in Fig.~\ref{fig:WZ7Z3D} for the ATLAS 7 TeV $W, Z$ data and the ATLAS 8 TeV $V$ +jets data.

\begin{figure*}[p]
  \begin{centering}
    \includegraphics[width=0.4\textwidth]{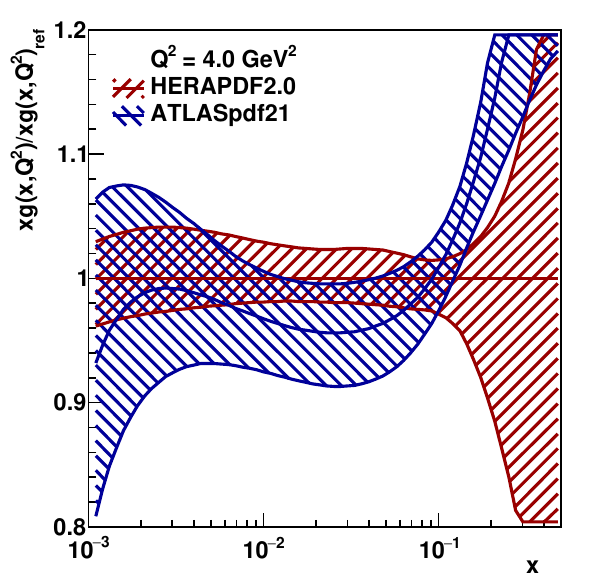}
    \includegraphics[width=0.4\textwidth]{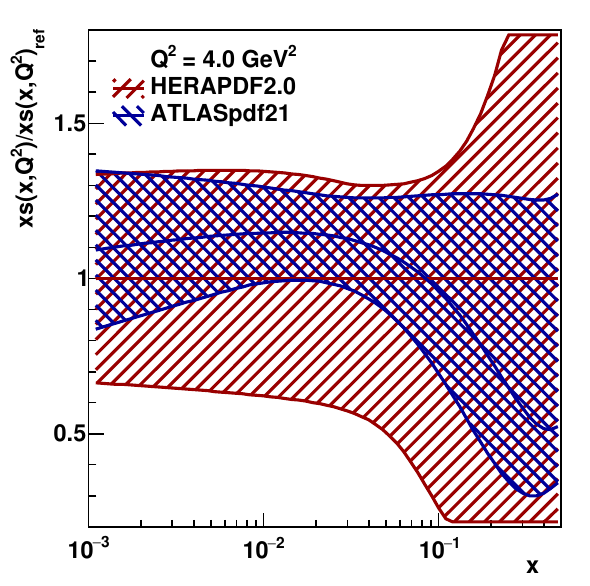} 
    \includegraphics[width=0.4\textwidth]{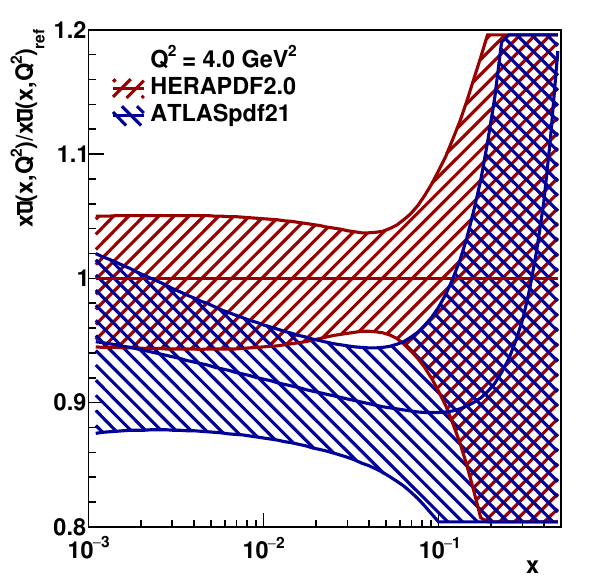}
     \includegraphics[width=0.4\textwidth]{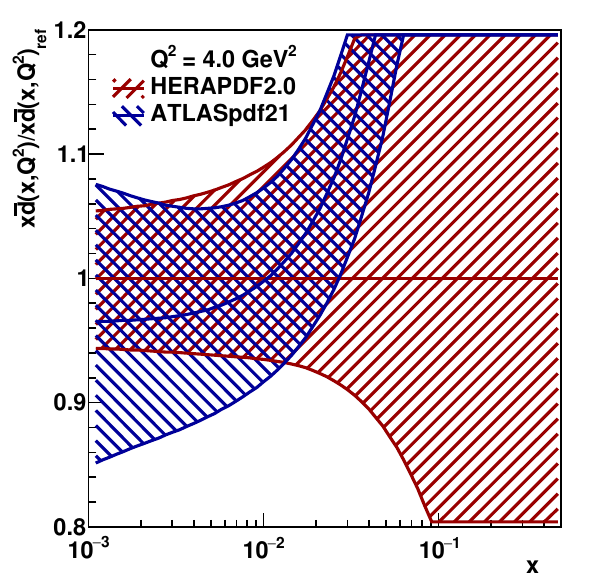}       
    \caption{The ratio of HERAPDF2.0 (red) and ATLASpdf21 (blue) taken to the central PDF of HERAPDF2.0 for $g$, $s$, $\bar{u}$, and $\bar{d}$ PDFs at $Q^2=4\mbox{ GeV}^2$. 
     The HERAPDF band represents the full uncertainty (EIG+VAR) for $T=1$.  ATLASpdf21 is given for the tolerance of $T = 3$.
    \label{fig:HERA-ATLAS-PDFs}  
    } 
  \end{centering}
\end{figure*}

The ATLAS 7 TeV $W, Z$ data are quite strongly sensitive to the gluon as well as to the strange PDF at $x<0.05$, to $d$ and (to a lesser extent) $u$ at high $x$. The ATLAS 8 TeV $V$ +jets data are sensitive both to the gluon (mostly due to $Z$+jets data) and to the strange PDF (mostly due to $W$ + jets data) at high $x$, wanting both smaller high-$x$ gluon and smaller high-$x$ strangeness. A consequence of the smaller high-$x$ strangeness is also a larger high-$x$ $\bar{d}$, as discussed in Ref.~\cite{ATLAS:2021qnl}.

The $L_2$ sensitivities for the ATLAS fits thus illustrate how large sensitivity to a data set is evident before it is fitted, how sensitivity is reduced if a good fit is made to these data, and how smaller residual tensions between fitted data sets can be made evident. More information can be gained by the use of the online plotter \cite{L2website}.
The trends described above, of adding ATLAS data to the HERA data, can be observed in the comparison plot for the HERA and ATLASpdf21 ensembles themselves, Fig.~\ref{fig:HERA-ATLAS-PDFs}.

%

\subsection{Sensitivities for reduced benchmarking  fits \label{sec:SFL2reduced}}

Section~\ref{sec:PDF4LHC21Data} summarized the CT18$^\prime$ and MSHT20 reduced fits that led to the PDF4LHC21 combination~\cite{PDF4LHCWorkingGroup:2022cjn,Cridge:2021qjj}. 
These two analyses based on the Hessian methodology show little differences in their central values as well as in the magnitude of their uncertainties, as can be appreciated in, e.g., Figs.~3.4, 3.5 of Ref.~\cite{PDF4LHCWorkingGroup:2022cjn}.
 The $L_2$ sensitivity technique was used to aid selection of the optimal
 common data set for the benchmarking exercise, and it motivated dedicated
 studies for specific data, see Appendices C and D of Ref.~\cite{PDF4LHCWorkingGroup:2022cjn}. In this section, we illustrate the insights that can be gleaned about the reduced fits from the plots of their sensitivities collected on the online plotter \cite{L2website}.

\begin{figure*}[h]
\centering
\includegraphics[height=160pt,trim={ 0 0 1.6cm 0},clip]{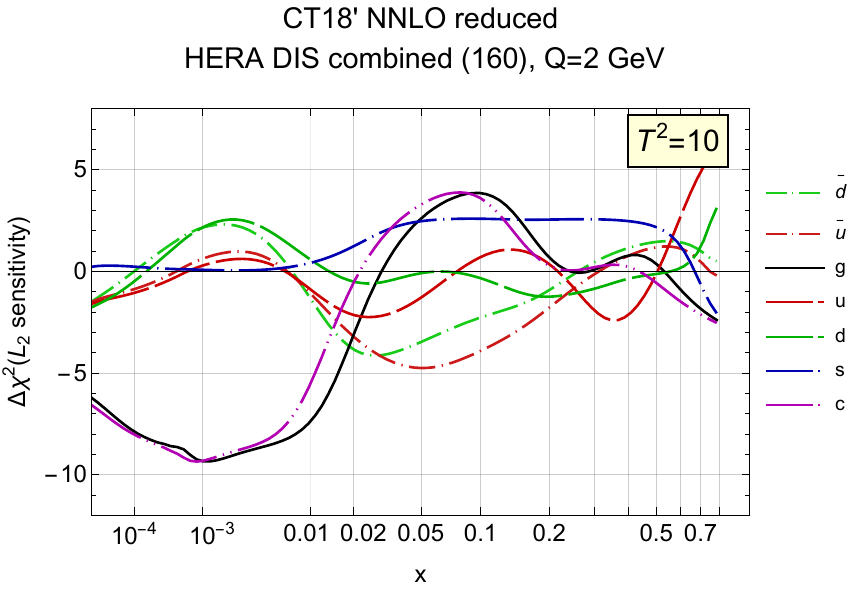}
\includegraphics[height=160pt,trim={ .85cm 0 0 0},clip]{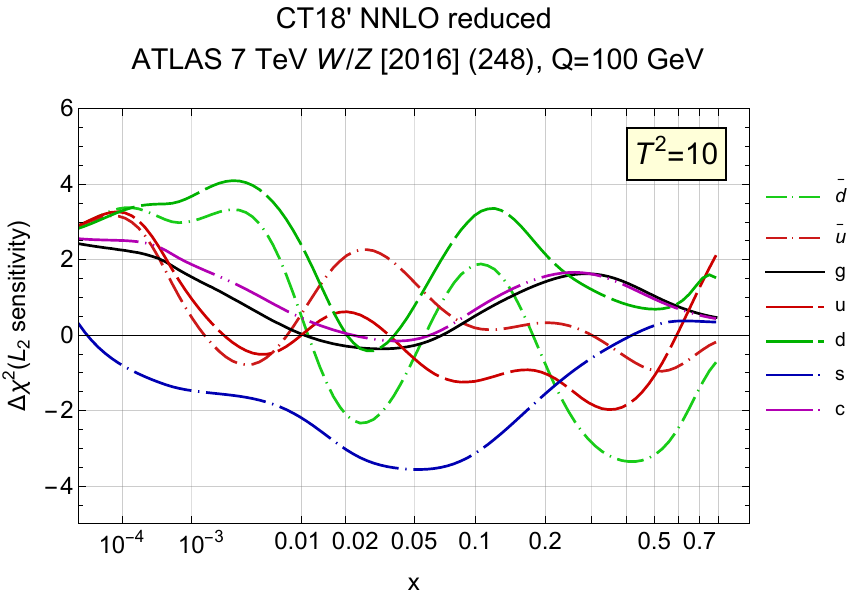}\\
\includegraphics[height=160pt,trim={ 0 0 1.6cm 0},clip]{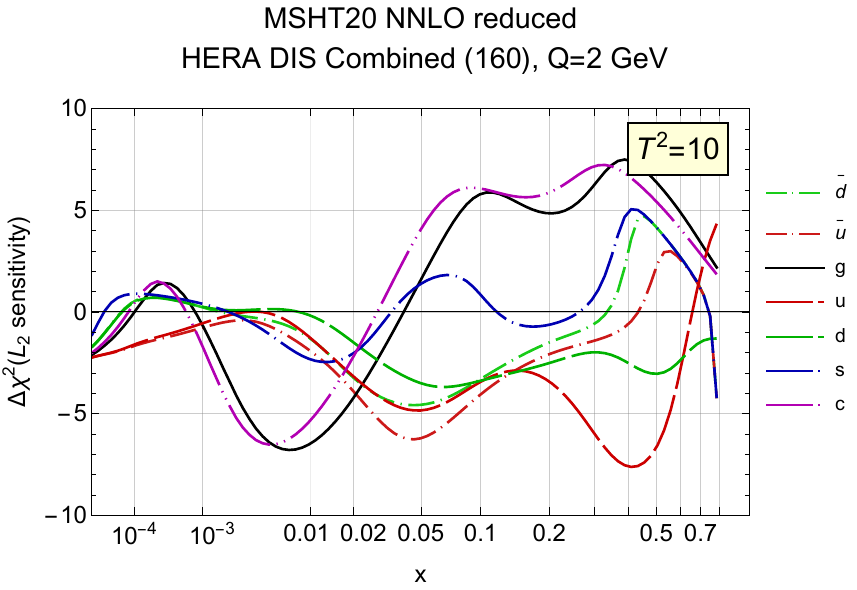}
\includegraphics[height=160pt,trim={ .85cm 0 0 0},clip]{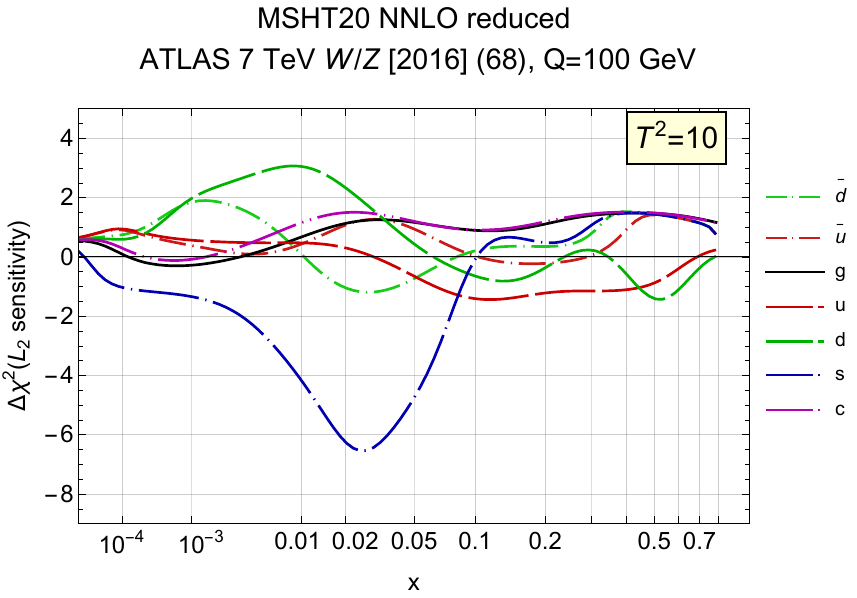}\\
\includegraphics[height=160pt,trim={ 0 0 1.6cm 0},clip]{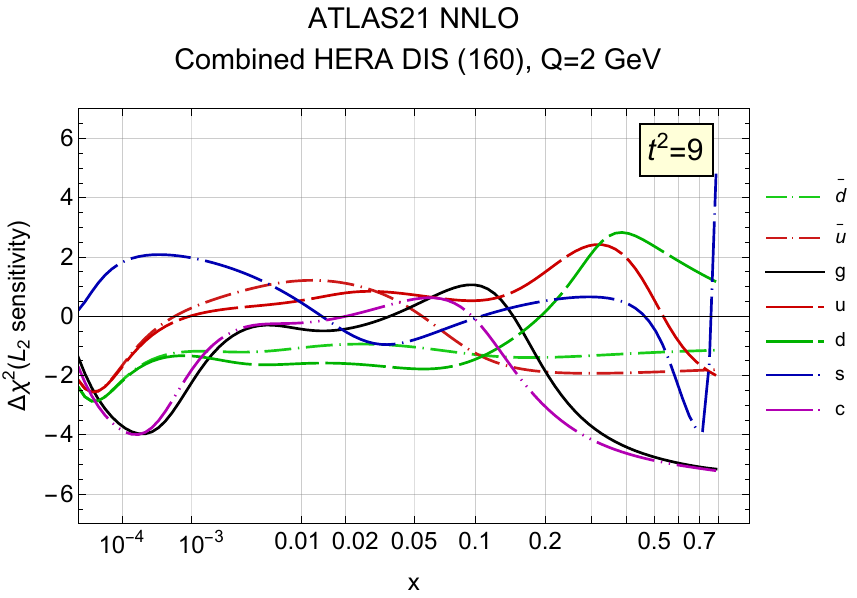}
\includegraphics[height=160pt,trim={ 1.cm 0 0 0},clip]{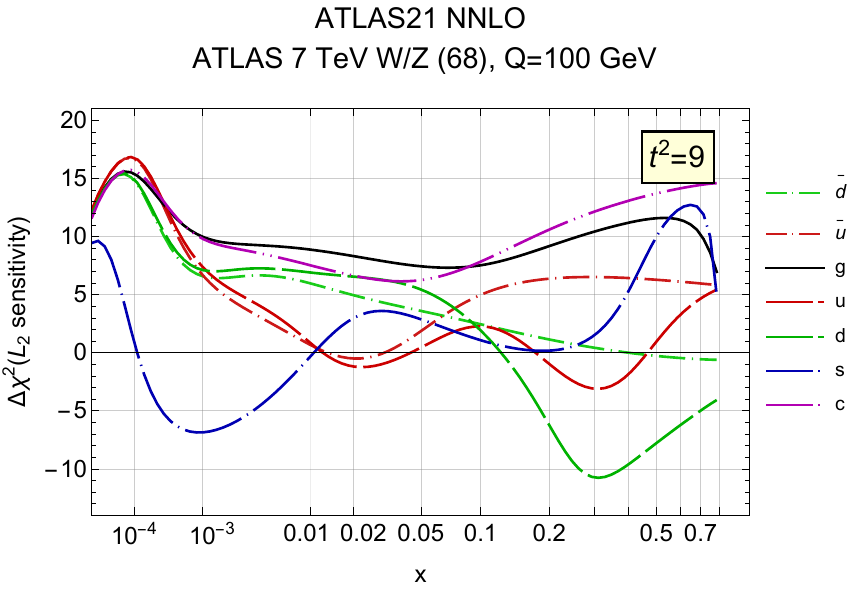}
\caption{
Sensitivities for the CT18$^\prime$,  MSHT20, and ATLASpdf21 non-global fits for HERA DIS data set (left column,  $Q=2$ GeV in $S_{f,L_2}$) and ATLAS 7 TeV $W, Z$ data set (right column,  $Q=100$ GeV).
}
\label{fig:compar_red_bench}
\end{figure*}

\begin{figure*}[h]
\centering
\includegraphics[width=0.49\textwidth]{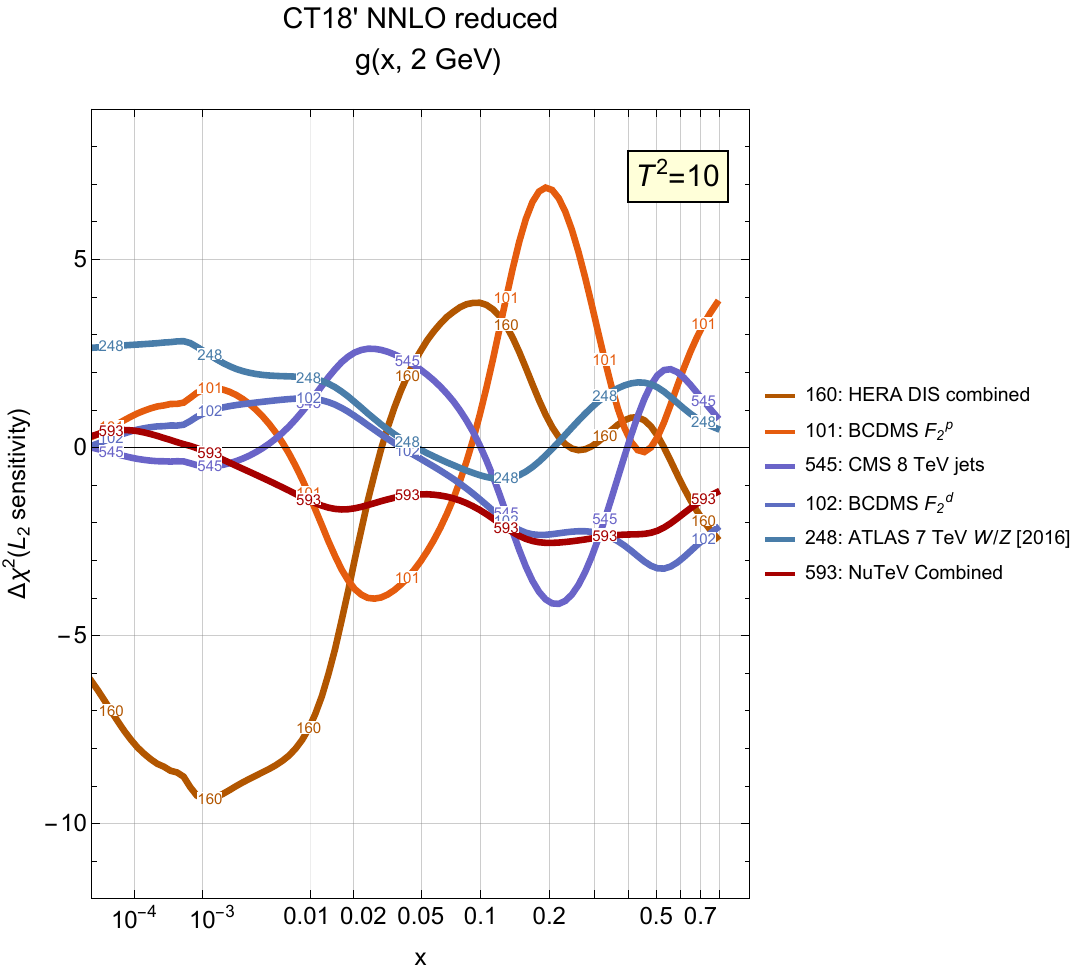}
\includegraphics[width=0.49\textwidth]{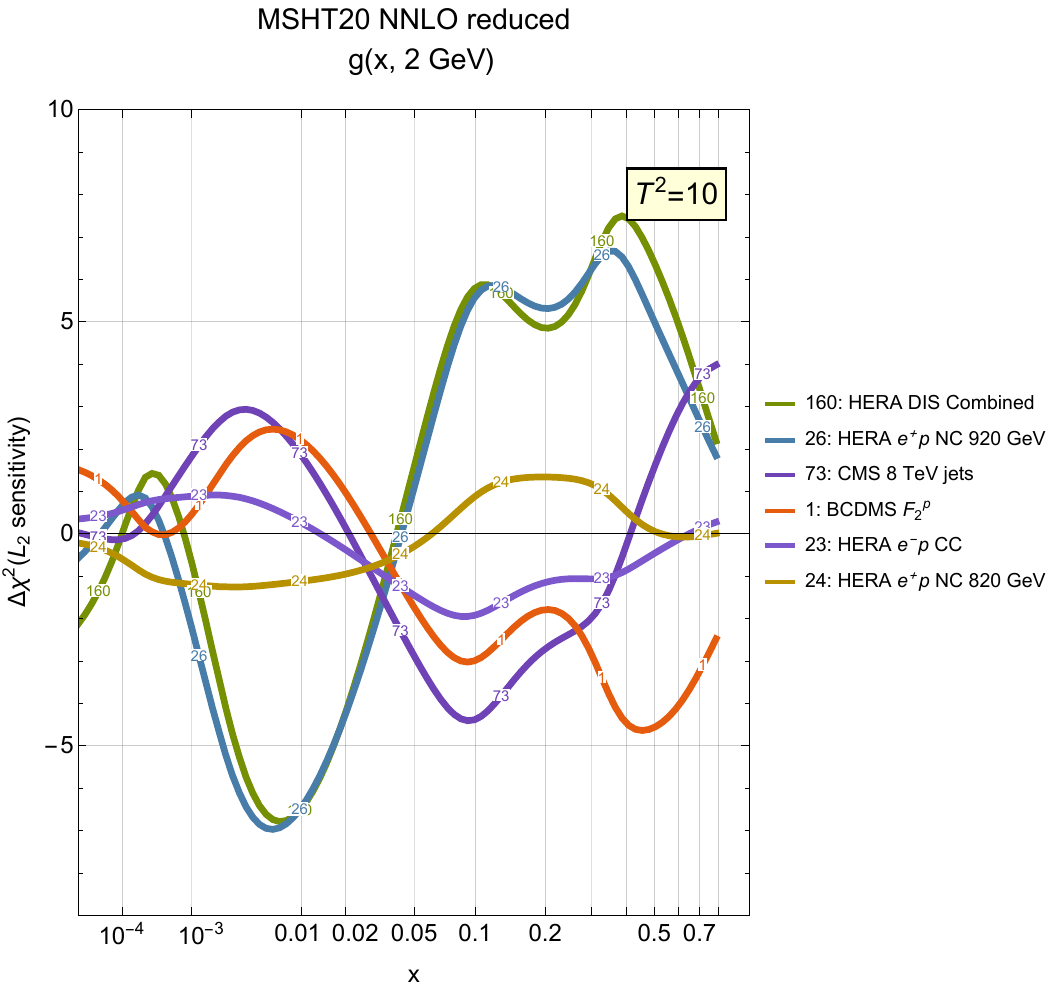}\\
\caption{
Sensitivities to the gluon PDF at $Q=2$ GeV in the CT18$^\prime$ and  MSHT20 reduced fits. The equivalent for ATLASpdf21 is given in Fig.~\ref{fig:gluonstrall}.
}
\label{fig:compar_red_bench2}
\end{figure*}

Figure~\ref{fig:compar_red_bench} shows the $L_2$ sensitivities for the HERA combined DIS and ATLAS 7 TeV $W, Z$ [2016] data sets in the CT18$^\prime$, MSHT20 reduced fits and the ATLASpdf21 fit. The PDF factorization scales in $S_{f,L_2}$ are set to $Q=2$ and $100$ GeV for the HERA DIS and ATLAS $W, Z$ data sets, respectively. Qualitatively, the shapes of the sensitivities are more similar between the CT18$^\prime$ and MSHT20 reduced fits than in the full fits. The differences in the magnitudes of sensitivities are more pronounced. Attention must be paid to the ranges of sensitivities on the vertical scales that are not the same across the figures.

The HERA data ({\it l.h.s.}) impose a pronounced preference for a higher gluon at $x<0.02$ in all three non-global fits. At higher $x$, the ATLASpdf21 fit still mostly prefers a larger gluon, in reflection of having less data sensitive to the gluon at high $x$. 
In both the CT18$^\prime$ and MSHT20 reduced fits, on the other hand, the HERA data prefer a suppressed gluon at $x=0.02-0.2$, and in the MSHT20 reduced fit a smaller gluon is preferred above $x=0.2$ as well, in the pattern that is not dissimilar to the differences in the sensitivities of the HERA heavy-quark cross sections in the full fits in Fig.~\ref{fig:compar_HERAcharm}. As the reduced analyses fit the same HERA data, the behavior of sensitivities may thus reflect methodological differences among the CT18$^\prime$ and MSHT20 reduced fits, such as their different heavy-quark schemes and gluon parametrization forms.

The pulls on the strangeness from ATLAS 7 TeV $W, Z$ data ({\it r.h.s.}) favor larger strangeness at intermediate $x$ for both CT18$^\prime$ and MSHT20 reduced fits. This pull is reduced in the ATLAS fit, in which these data exert the dominant influence on strangeness, and the strangeness PDF has maximally increased to reflect this. For the reduced fits, this  strong pull on the strangeness is opposed by the NuTeV data, as shown in App. D of Ref.~\cite{PDF4LHCWorkingGroup:2022cjn}. 
The variations between the pulls for the large-$x$ gluon and strangeness go hand-in-hand.
\\

In Fig.~\ref{fig:compar_red_bench2}, we show the $L_2$ sensitivities to the gluon at $Q=2$ GeV for the CT18$^\prime$ and MSHT20 reduced fits. The sensitivity to the ATLASpdf21 gluon PDF is shown in Fig.~\ref{fig:gluonstrall}. The orders of magnitude of $S_{f, L2}^{\rm H}$ are comparable for the two reduced ensembles. However, ATLASpdf21 shows slightly larger values for the single sensitivities, the dominant one coming from the high precision ATLAS 7 TeV $W, Z$ measurement.  Global features of the pulls are similar for the three ensembles for, {\it e.g.}, the HERA combined data, for which the  
negative pull at small $x$, favoring a larger gluon PDF, is larger for the reduced sets than for ATLASpdf21, because of the relative dominance of the HERA data in the ATLAS fit, which has already adjusted to reflect this.
Here, both CT18$^\prime$ and MSHT20 reduced fits show counter-balancing pulls on the gluon from the CMS 8 TeV jet data (labels 73 and 545), with the ATLAS $W, Z$ data set 248 exerting an additional downward pull in the case of CT18$^\prime$. 

On the other hand, at $x>0.1$, MSHT20red is still pulled downward mostly by the HERA combined data (160), while BCDMS data on $F_{2,p}$ (1) and CMS 8 TeV jets exert the main opposing upward pulls. 
In the CT18$^\prime$ case, however, the BCDMS $F_{2,p}$ pull (101) at large $x$ is downward and dominates over the weakened HERA DIS pull (160) of the same sign. The pull of the CMS 8 TeV jets continues to be consistently upward for CT18$^\prime$.

The PDF4LHC21 document \cite{PDF4LHCWorkingGroup:2022cjn} noted this pronounced agreement of the pulls by the CMS 8 TeV jets between two reduced fits. The respective plot for the $L_2$ sensitivity was shown at $2$ GeV in Fig.~D1 of \cite{PDF4LHCWorkingGroup:2022cjn}. Already the early \texttt{PDFSense} study \cite{Wang:2018heo} pointed out the prominence of constraints from the CMS jet production revealed by the sensitivity method. Namely, this data set dominates the pulls for the gluon and charm PDFs, resulting in smaller values of both PDFs at $x\sim 10^{-3}$ as well as in the valence region, but larger gluon and charm PDFs  at $x\sim 0.05-0.1$. 
The corresponding plots for the full fit are shown in Sec.~\ref{sec:SFL2global} (see Fig.~\ref{fig:compar_bench_CMS8_TeV}), where we observe the opposite trend between reduced and full fits for both MSHT20 and CT18, nonetheless its pulls remain in excellent agreement also between the full fits.  
The difference between the reduced and the full fits shows that, for the CMS 8 TeV jet data set, the $L_2$ sensitivity is most influenced by the competing pulls of the other input data sets  rather than the functional form. 
A counter example is seen for in the gluon sensitivities of Fig.~\ref{fig:compar_red_bench2}, where the differing contributions of the HERA combined data to the MSHT20 and CT18$^\prime$ reduced fits (which use the same data sets) highlights the remaining differences between the reduced fits, including, for example, the functional form and heavy-quark scheme. 

%
\section{Sensitivities for CT18, CT18As, and CT18As\_Lat fits \label{sec:SFL2CT}}

As we summarized in Sec.~\ref{sec:CT18DataTheory}, the comprehensive CT18 NNLO study \cite{Hou:2019efy}, published by the CTEQ-TEA collaboration in early 2021, has extensively employed the $L_2$ sensitivities in the pre-fit and post-fit investigations of the constraints from the experiments. The CT18 website \cite{CT18L2Sensitivity} collects 340 plots of $L_2$ sensitivities for the individual experiments, PDF flavors, PDF combinations, and parton luminosities at the LHC in the CT18 and C18Z NNLO analyses. Many of these results are discussed in the CT18 publication.
Complementary insights were gained from visualizations of $L_1$ sensitivities of individual data points using the \texttt{PDFSense} program \cite{Wang:2018heo} and the CT14HERA2 NNLO ensemble (the immediate predecessor of CT18), with the resulting figures collected online at \cite{PDFSenseWebsite}. The \texttt{PDFSense} results were especially helpful for charting the global map of sensitivities of the fitted experiments to the PDF flavors \cite{Wang:2018heo} and to PDF combinations and Mellin moments that can be computed on the lattice \cite{Hobbs:2019gob}.

This article and its companion website \cite{L2website} reproduce many $S_{f,L2}$ plots from the CT18 NNLO analysis \cite{Hou:2019efy,CT18L2Sensitivity} in the format that facilitates the comparisons against the other PDF ensembles. Examples include Fig.~\ref{fig:compar_HERAcharm}
for HERA I charm production,
Fig.~\ref{fig:compar_bench_CMS8_TeV} for CMS 8 TeV inclusive jet production, Fig.~\ref{fig:compar_ttbar_CTMSHT} for ATLAS and CMS 8 TeV $t\bar t$ production (IDs=580 and 573), Fig.~\ref{fig:compar_HERA_Combined} for the combined HERA I+II DIS (ID=160), Fig.~\ref{fig:compar_ATLAS_WZ} for ATLAS 7 TeV $W, Z$ production (ID=248), and Fig.~\ref{fig:compar_E866} for E866 $pd/pp$ ratio; as well as plots for the LHCb $W, Z$ production, and D{Ø} charge asymmetry in the Supplemental Material. These figures are discussed later in their respective sections. 

\begin{figure*}[b]
\centering
\includegraphics[height=210pt]{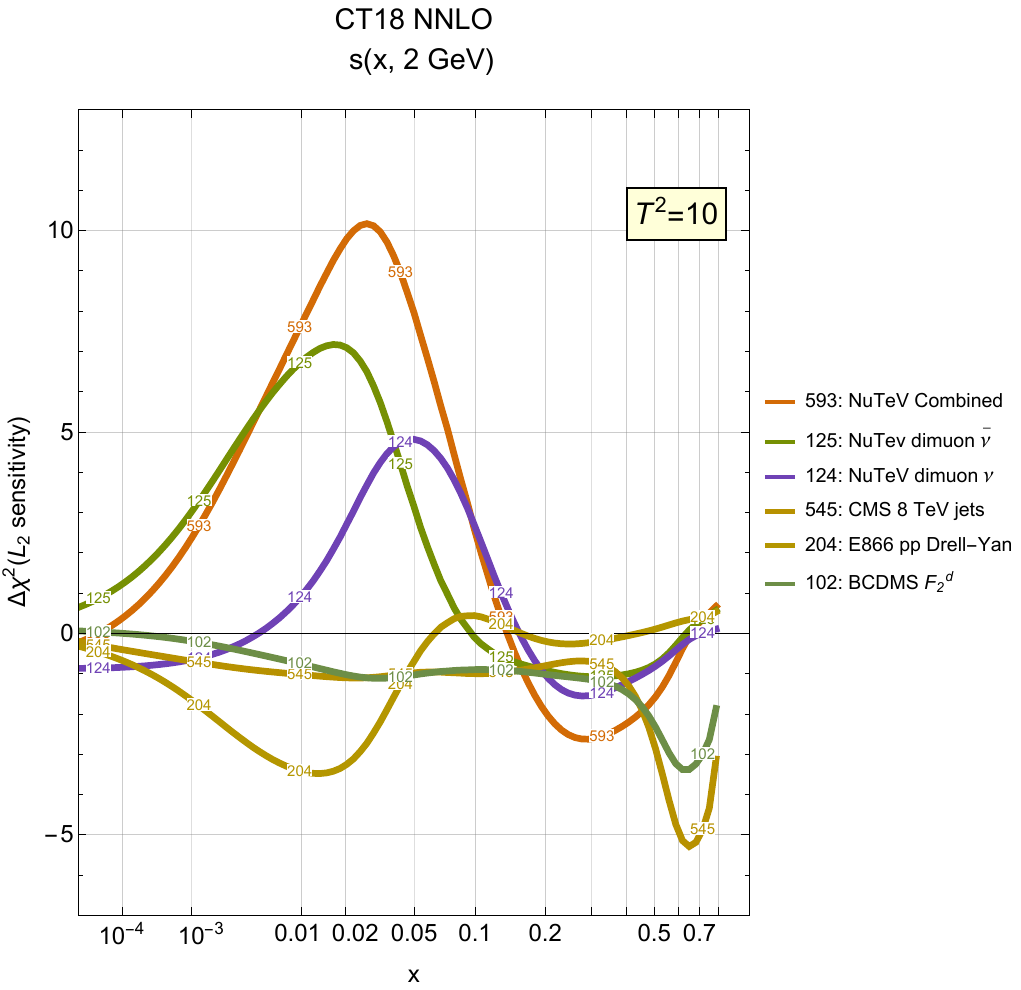}
\includegraphics[height=210pt]{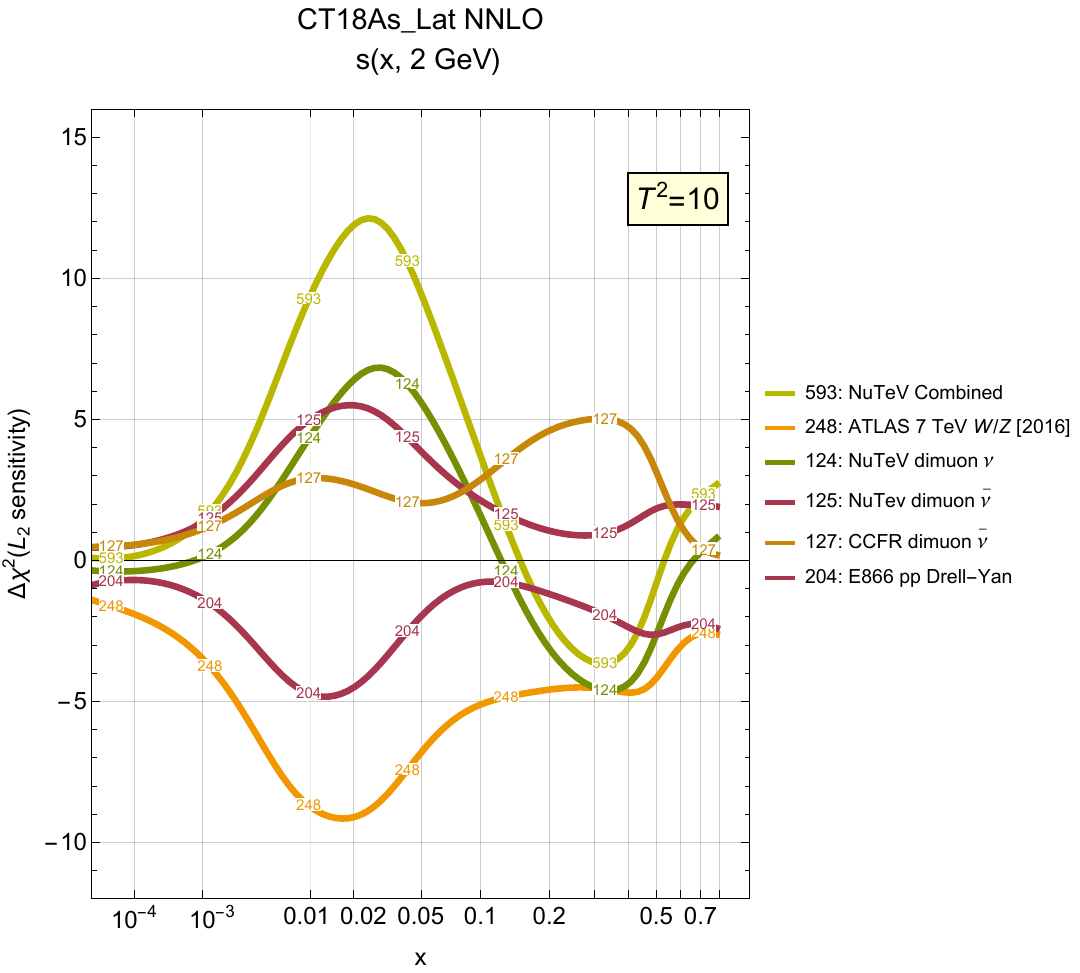}\\
\includegraphics[height=210pt]{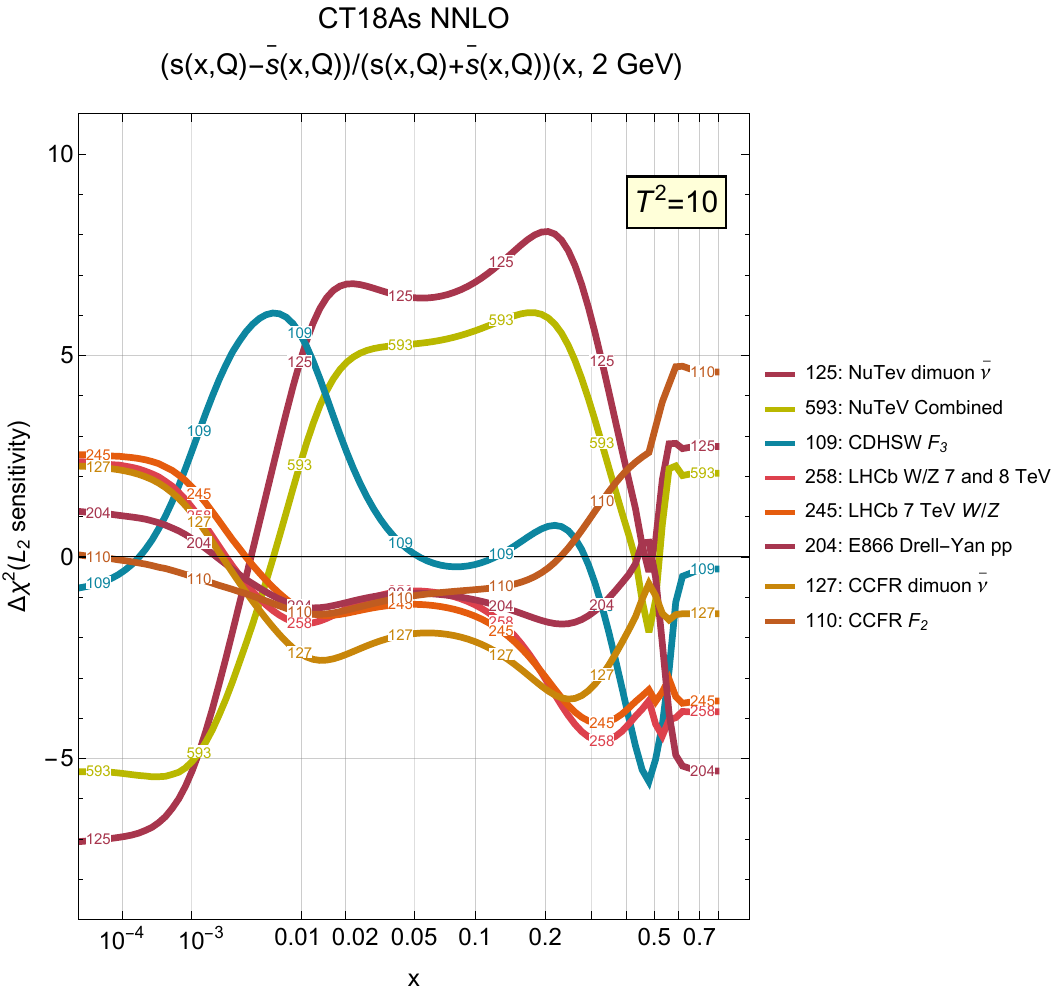}
\includegraphics[height=210pt]{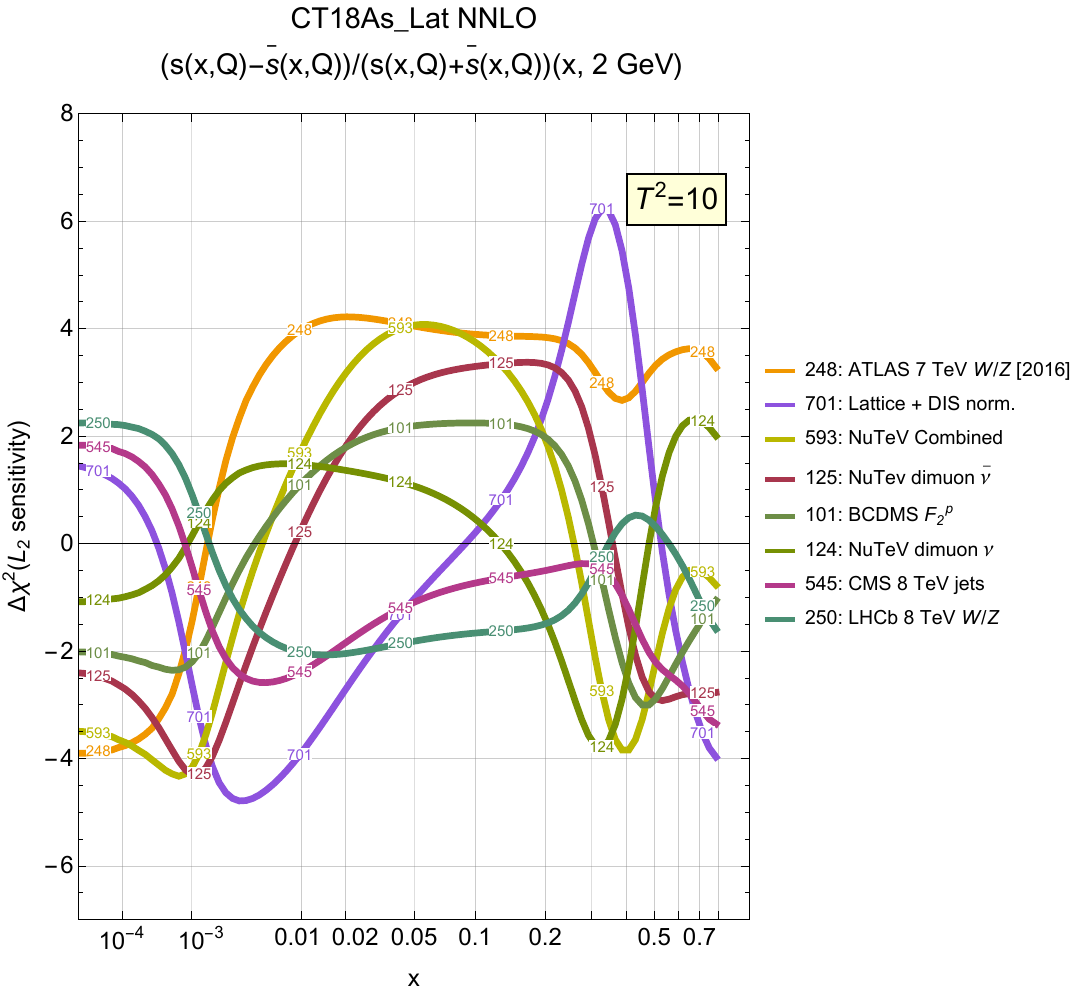}
\caption{Upper: Sensitivities to the strangeness PDF for CT18 and CT18As\_Lat NNLO fits  with $T^2=10$ at $Q=2$ GeV, shown for the six most sensitive experiments. Lower: Sensitivities to the strangeness asymmetry $(s-\bar{s})/(s+\bar{s})$  for  CT18As and CT18As\_Lat NNLO. [Notice that the color code and the ranges of the vertical scale vary between the plots.]}
\label{fig:compar_CT18_strange}
\end{figure*}

\begin{figure*}[bht]
\centering
\includegraphics[height=210pt]{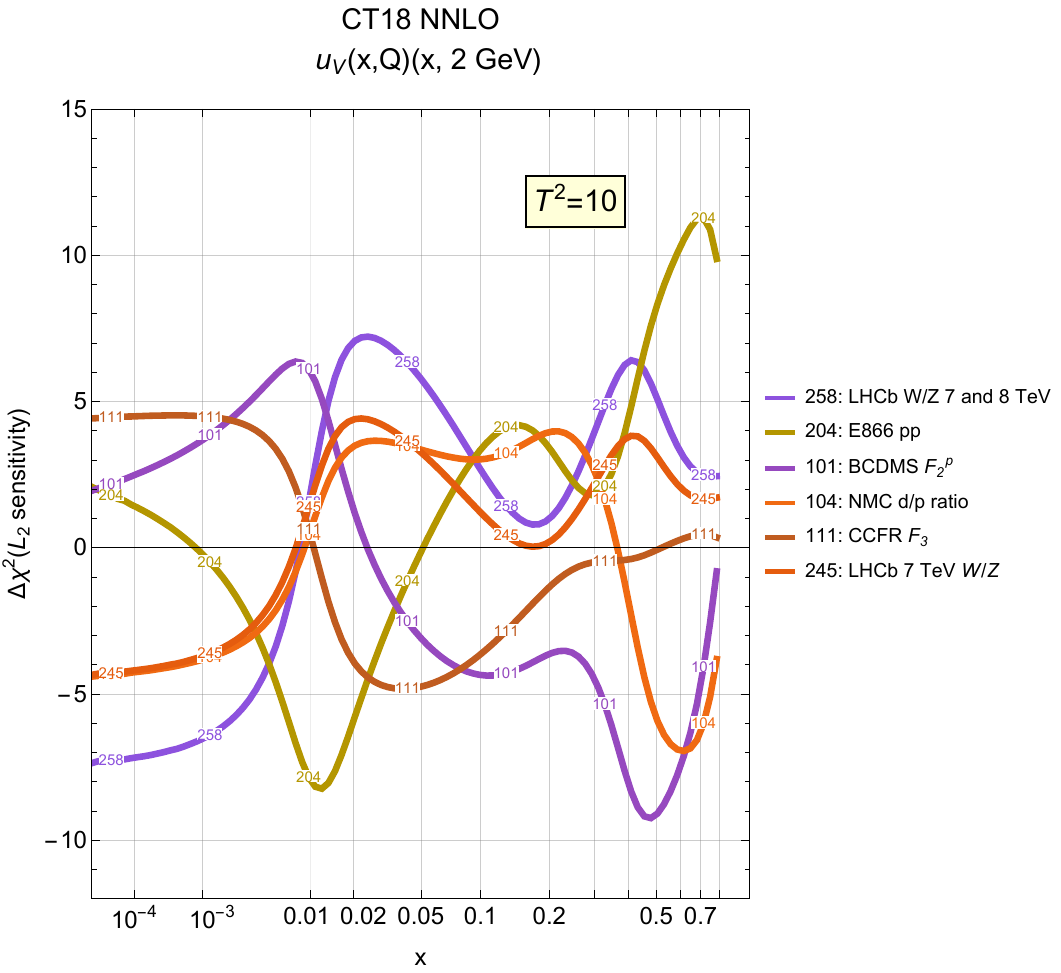}
\includegraphics[height=210pt]{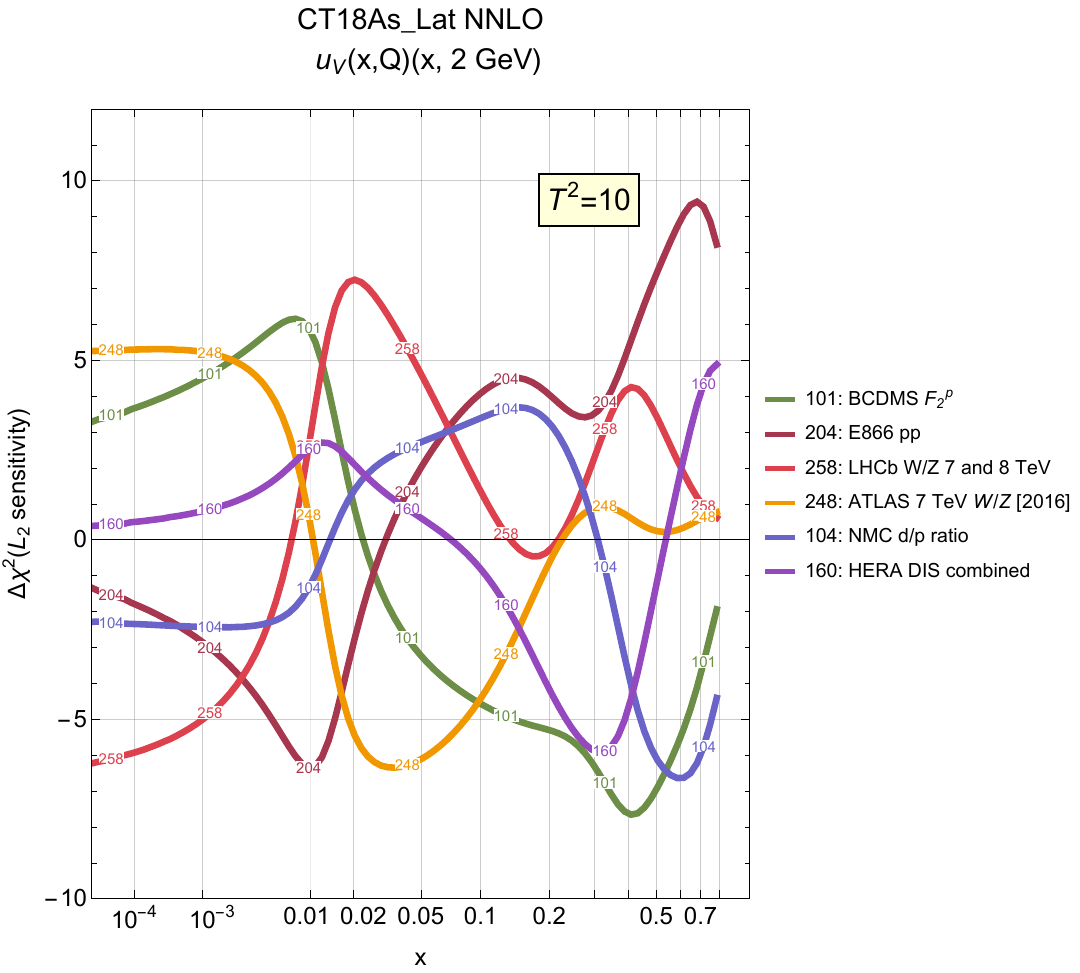}
\caption{Sensitivities to the valence PDF $u_V=u-\bar u$ for CT18 NNLO (left) and  CT18As\_Lat NNLO (right) at $Q=2$ GeV. }
\label{fig:compar_uv_CT}
\end{figure*}

\begin{figure*}[bht]
\centering
\includegraphics[height=210pt]{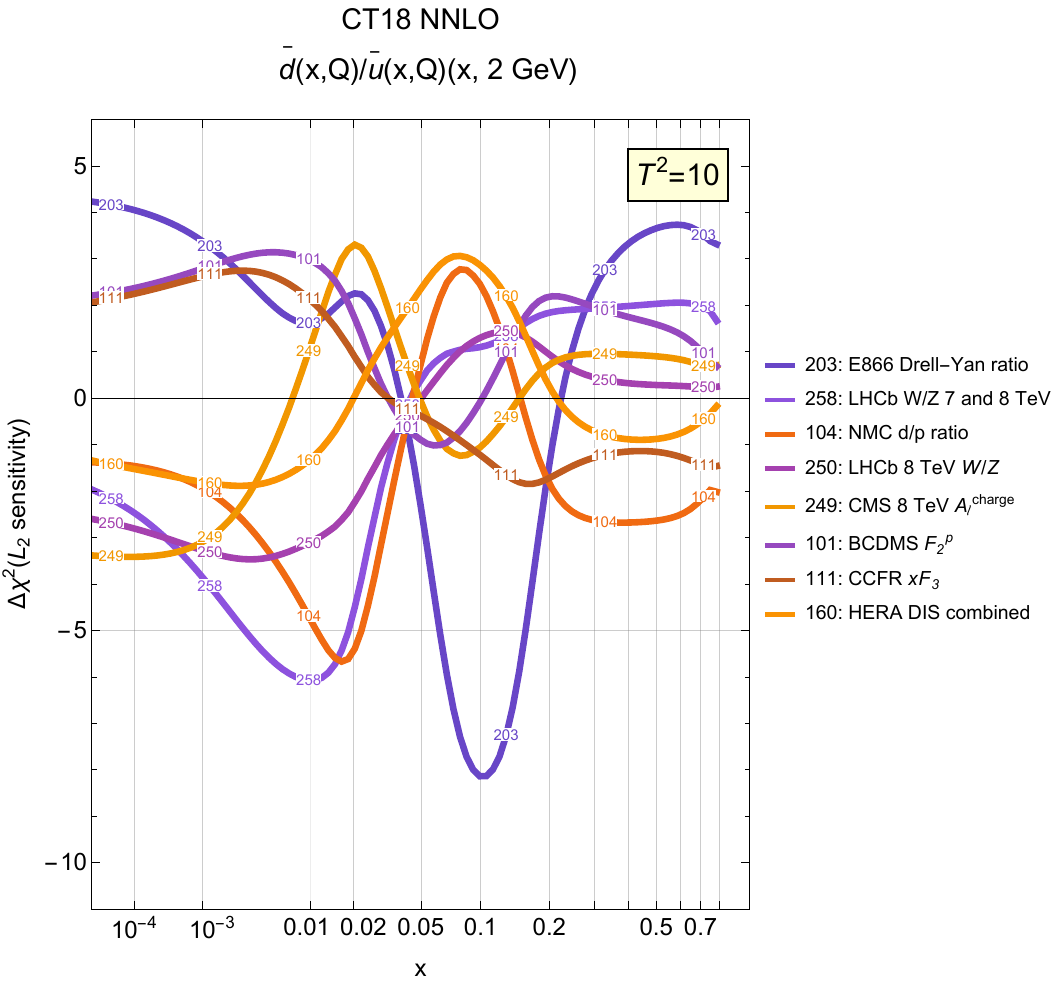}
\includegraphics[height=210pt]{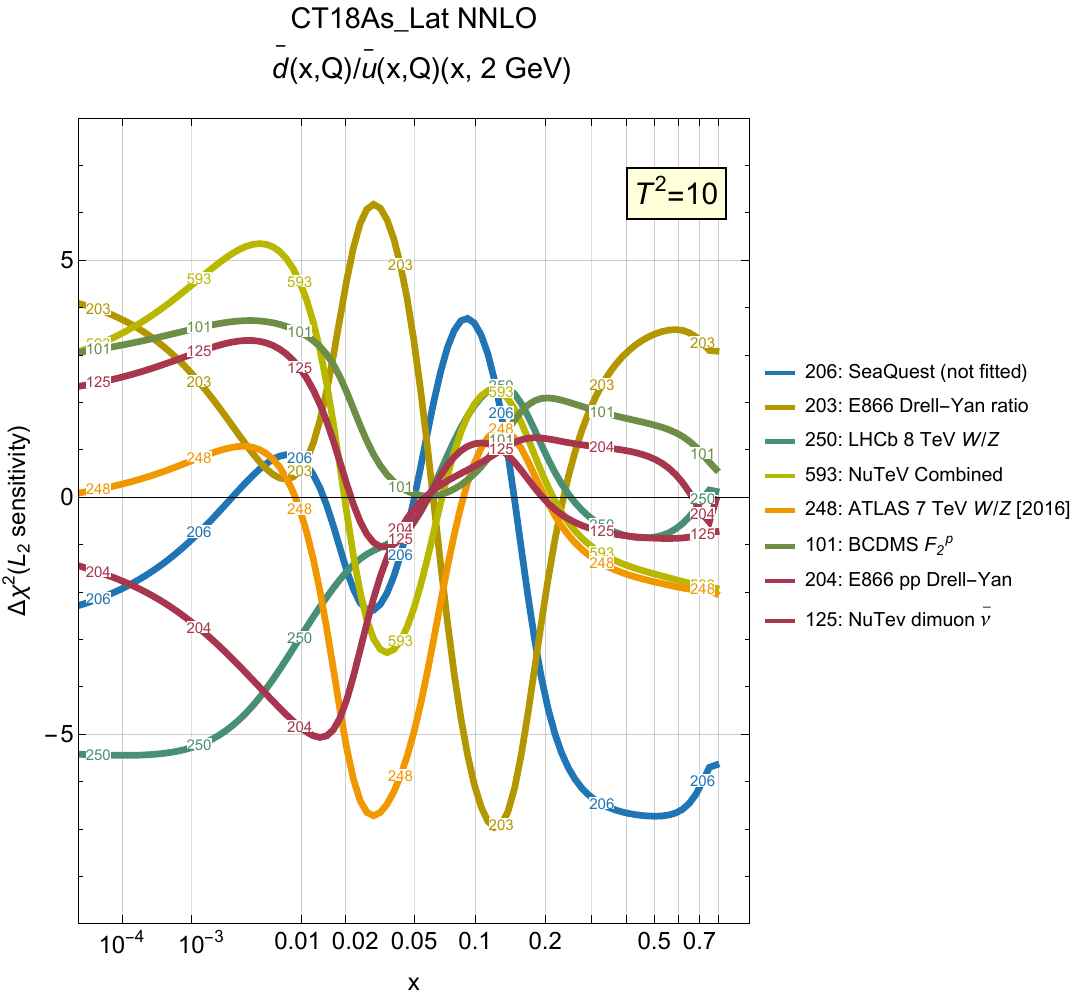}
\caption{Sensitivities to the ratio $\bar{d}/\bar{u}$  for CT18 (left) and  CT18As\_Lat (right) NNLO fits at $Q=2$ GeV. }
\label{fig:compar_db2ub_CT}
\end{figure*}

In addition to the experiment-by-experiment plots of $L_2$ sensitivities, we provide cumulative figures of sensitivities of the leading experiments to specific PDF flavors or PDF combinations, such as the valence PDFs $q_V(x,Q)\equiv q(x,Q)-\bar q(x,Q)$ and PDF ratios $f_a(x,Q)/f_b(x,Q)$. 
We already discussed such Fig.~\ref{fig:L2gluonCT18} for the CT18 NNLO sensitivities to the gluon.
As another illustration, consider Figs.~\ref{fig:compar_CT18_strange}-\ref{fig:compar_db2ub_CT} comparing $S_{f,L2}$ for quark and antiquark PDFs between the CT18, CT18As and CT18As\_Lat PDF ensembles. We expect, and the sensitivities reflect it, that the primary differences between CT18 and CT18As\_Lat arise in the (anti)quark flavor separation. The CT18A PDF ensemble is identical to CT18 but includes ATLAS 7 TeV W, Z precision data which show sizable tension with other data sets, such as the HERA I+II DIS combined data and NuTeV dimuon  data~\cite{Hou:2019efy}. On top of that, the CT18As fit also adds degrees of freedom in the strangeness parametrization to allow $s$ to be not equal to $\bar s$ at the initial scale $Q_0 =1.3$ GeV. This choice slightly reduces the above-mentioned tension~\cite{Hou:2022onq}: the $s(\bar{s})$ PDF error band becomes larger in comparison with the analogous CT18A PDFs. Through a feedback loop in the fit, the additional freedom in CT18As also enlarges the $d(\bar{d})$ error bands. The PDF ratio ${\bar d}/{\bar u}$ is also modified.

The CT18As\_Lat fit~\cite{Hou:2022onq} further includes lattice data~\cite{Zhang:2020dkn} to constrain the strangeness asymmetry in the large-$x$ region, $0.3<x<0.8$. The lattice $s_{-}$ data modifies both the central fit and the error bands of $R_s$ and $s_{-}(x)$. The $R_s$ of CT18As\_Lat gets higher (closer to that of MSHT20 and NNPDF4.0). The strangeness asymmetry $s_{-}$ of CT18As\_Lat is thus closer to zero at large $x$ than those of CT18As, MSHT20, and NNPDF4.0.

But how exactly are these changes distributed among the flavors, pulls of data sets, and as a function of partonic $x$? This question is answered by plotting the $L_2$ sensitivities. In the upper right panel in Fig.~\ref{fig:compar_CT18_strange}, we see that the ATLAS 7 TeV $W, Z$ data set (248) in the CT18As\_Lat fit prefers a higher $s$ PDF than the other data sets at $x\approx 0.02$ and actually along the whole $x$ range. This preference is discussed at length in Ref.~\cite{Hou:2019efy}. We further see that the pattern of leading sensitivities in CT18 NNLO shown in the upper left panel has been modified in CT18As\_Lat. For example, while in CT18 NNLO the NuTeV dimuon experiments (124 and 125) both prefer lower strangeness, but at somewhat different $x$, and the E866 $pp$ Drell-Yan production moderately prefers to increase it, in the CT18As\_Lat case the pulls of the NuTeV neutrino and antineutrino data sets become more alike at $x<0.1$; at $x>0.1$, the NuTeV $\nu$ dimuon data set (124) develops a preference for a higher $s$ that is largely canceled by the opposite pull by CCFR $\bar \nu$ dimuon set (127).

The CT18As and CT18As\_Lat sensitivities available from the website clarify how these modifications emerge through the triple effect of including the ATLAS $W, Z$ data set, releasing the $s\neq \bar s$ condition (which substantially modifies the flavor composition in CT18As), and including a lattice constraint on the magnitude of $s-\bar s$ (which partly offsets the previous effect). The lower two figures in Fig.~\ref{fig:compar_CT18_strange} compare $S_{f,L2}$ for the $(s-\bar s)/(s+\bar s)$ asymmetry of CT18As and CT18As\_Lat sets. In the CT18As case, without the lattice constraints on $s_-$,
the pulls on the strangeness asymmetry by CCFR inclusive (110) and $\bar\nu$ dimuon (127) data sets, together with the same-sign pulls of E866 $pp$ (204) and LHCb 7 and 8 TeV $W, Z$ (258), are counteracted by the strong pull by the NuTeV $\bar\nu$ dimuon data (125). For CT18As\_Lat, these sensitivities are substantially rebalanced by the lattice constraints on $s-\bar s$, which is included in the figures together with the contributions of normalizations for fixed-target experiments (ID=701). While the lattice+normalizations prefer a smaller asymmetry at $x>0.1$, they have to compensate by preferring a larger asymmetry at $x<0.1$ to satisfy the net absence of strangeness in the proton, $\int_0^1 s_-(x,Q)\ dx=0$. The pull of the lattice $s_-$ data is offset by the ATLAS $W, Z$, NuTeV, and BCDMS $F_2^p$ curves over a large span of $x$. [The sensitivity of the ATLAS 7 TeV $W, Z$ data to $s_-$ is small in the CT18As case without the lattice data.]

Figure~\ref{fig:compar_uv_CT} illustrates the six strongest $L_2$ pulls on the up-valence PDF at $Q=2$ GeV. In the CT18 NNLO fit (left panel), over most of the $x$ range we see the interplay of opposing pulls from the BCDMS $F_2^p$ (101) and E866 Drell-Yan $pp$ cross section (204), joined by a tug-of-war between CCFR $x F_3$ (111) and the LHCb $W, Z$ production (245 or 258) data sets over a wide $x$ range. When ATLAS 7 TeV $W, Z$ data are added in the CT18As\_Lat in the right panel, it partly offsets the pulls of LHCb (258) at $x<0.2$, with HERA DIS (160) and NMC $d/p$ ratio (104) offering additional opposing pulls. These figures can be compared with the $L_2$ sensitivities for MSHT20 $u_V$ PDFs in Fig.~\ref{fig:compar_uv_MSHT}, showing a similar behavior of the $L_2$ sensitivity for HERA DIS as in CT18As\_Lat, as well as more prominent roles of BCDMS $F_2^p$ and $F_2^d$ data sets (101 and 102), and of the D{\O} Run-2 $W$ boson-level asymmetry that is not included in CT18 fits.

In the same vein, the $\bar d/\bar u$ sensitivities for the CT18 and CT18As\_Lat fits shown in Fig.~\ref{fig:compar_db2ub_CT} can be compared against the MSHT20 ones in Fig.~\ref{fig:compar_db2ub_MSHT}.  Overall, the ratio of $pd$ and $pp$ Drell-Yan cross sections in the E866/NuSea experiment (ID=203 for CT and 13 for MSHT) prefers a lower $\bar d/\bar u$ ratio across the majority  of the $x$ range. The exception is the pronounced preference for a higher $\bar d/\bar u$ at $x=0.04-0.2$ in CT18 ($x=0.07-0.3$ in CT18As\_Lat). 
This E866 pull is balanced by a combination of HERA DIS, NMC $d/p$ ratio, BCDMS $F_2^p$, and LHC 8 TeV $W, Z$ pulls in the CT18 case.
For CT18As\_Lat, the E866 pull toward smaller $\bar d/\bar u$ at $0.01<x<0.07$ is amplified and nearly completely opposed by an upward preference for $\bar d/\bar u$ arising from the ATLAS 7 TeV $W, Z$ data set. In both the CT and MSHT cases, a variety of vector boson production experiments at the Tevatron and LHC have sensitivity to $\bar d/\bar u$. The (not fitted) E906/SeaQuest experiment included in the CT18As\_Lat analysis largely opposes the preference of the E866 ratio for diminished $\bar d/\bar u$ at $x>0.2$.

\newpage
\section{Sensitivities for the MSHT20 fits \label{sec:SFL2MSHT}}

\begin{figure*}[h]
\centering
\includegraphics[width=0.49\textwidth]{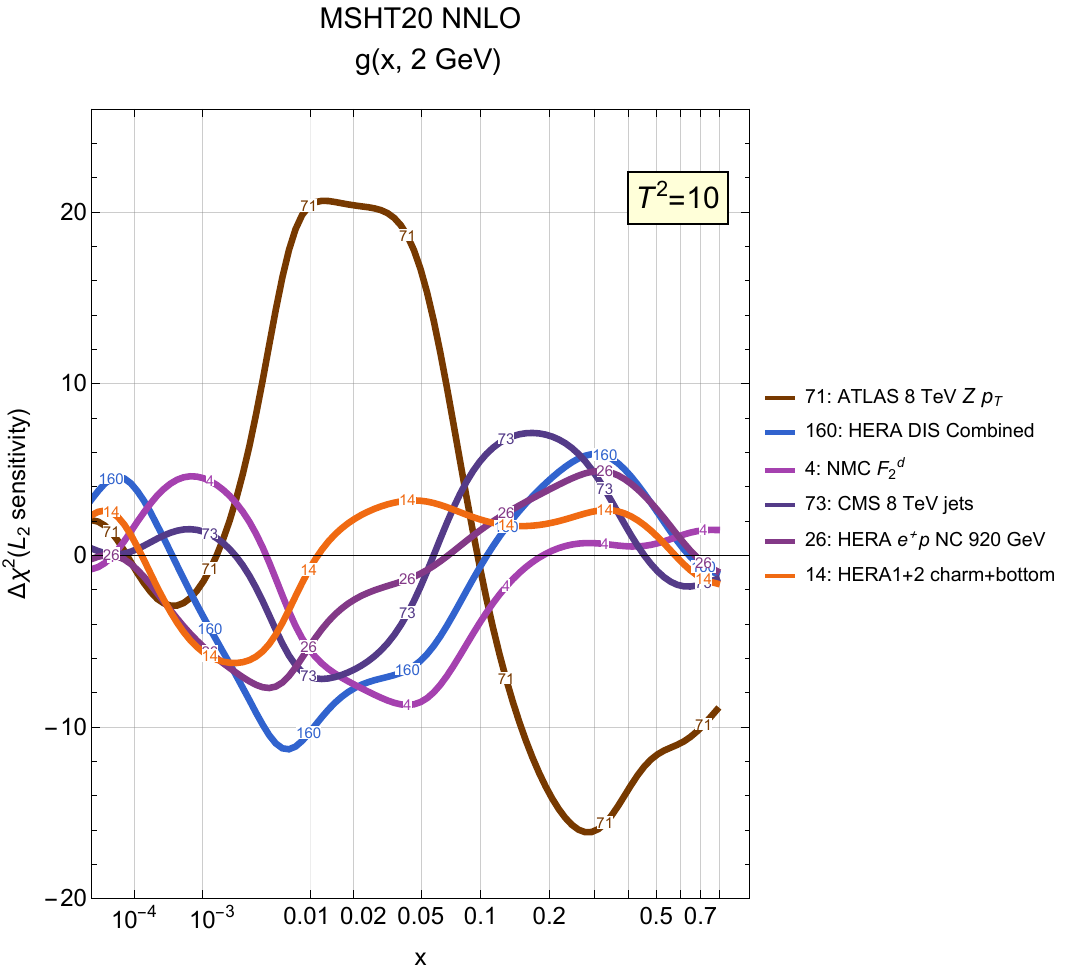}
\includegraphics[width=0.49\textwidth]{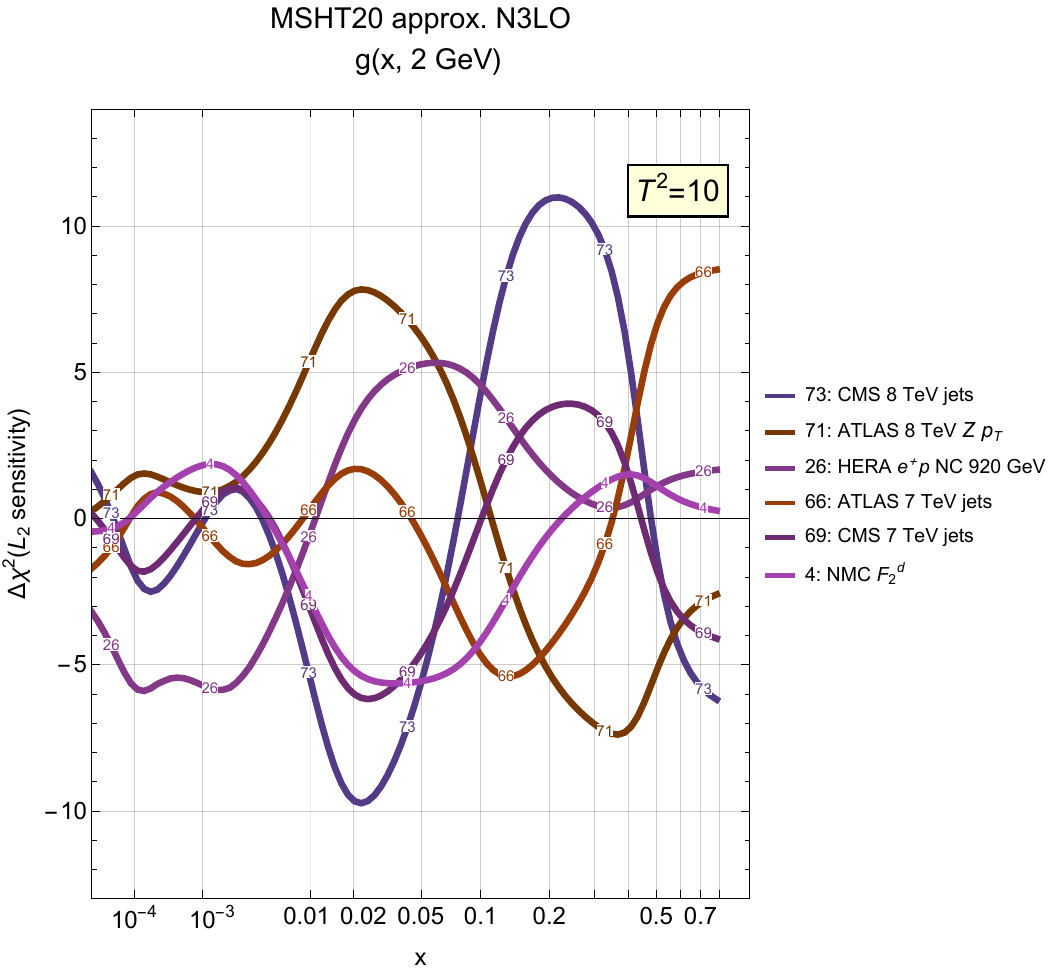}
\caption{$L_2$
sensitivities showing the six most sensitive data sets for the MSHT20 NNLO and MSHT20 aN3LO gluon PDFs.}
\label{fig:compar_MSHTNNLOMN3LOg}
\end{figure*}

In this section, we present some results comparing the $L_2$ sensitivities for the MSHT20 NNLO and aN3LO PDFs, showing the improvements that are obtained in some cases at aN3LO. First, in Fig.\ref{fig:compar_MSHTNNLOMN3LOg} we compare the sensitivities for the gluon. At NNLO we see this is dominated by the ATLAS $Z\ p_T$ data, which are also comparatively poorly fit according to the $\chi^2$ value presented in \cite{Bailey:2020ooq}. A similar pattern in the shape of the sensitivities is observed at aN3LO, although their amplitudes are substantially reduced, typically being less than a half of that at NNLO. It is also observed in \cite{McGowan:2022nag} that it has a substantially reduced $\chi^2$, corresponding to a better fit quality at aN3LO. At NNLO, there are large corrections for these precise ATLAS $Z\ p_T$ data so significant changes at aN3LO may well be expected. The sensitivity to the HERA combined structure function data at NNLO is also very much reduced at aN3LO, with this set no longer in the top six data sets shown in Fig.~\ref{fig:compar_MSHTNNLOMN3LOg}. 
Tensions between these two data sets at NNLO are largely eliminated at aN3LO. At the PDF level, the better agreement between the two experiments is accompanied by some rearrangement of the gluon and quark PDFs at aN3LO in the relevant $x$ region of about 0.01.
The sensitivities of the CMS 8 TeV jet data and NMC $F_2^d$ data do not change much and are even enhanced slightly when going to aN3LO (note the different axes scales), showing that inclusion of the  approximate N3LO corrections do not help with the tensions for these sets. 
However, overall, at NNLO the tension is overwhelmingly between ATLAS $Z$ $p_T$ data and other data sets, whereas at aN3LO the overall tension between different data sets is much reduced and more evenly spread, though ATLAS $Z\ p_T$ data and CMS 8 TeV jet data are still quite directly in conflict.  

\begin{figure*}[h]
\centering

\includegraphics[width=0.49\textwidth]{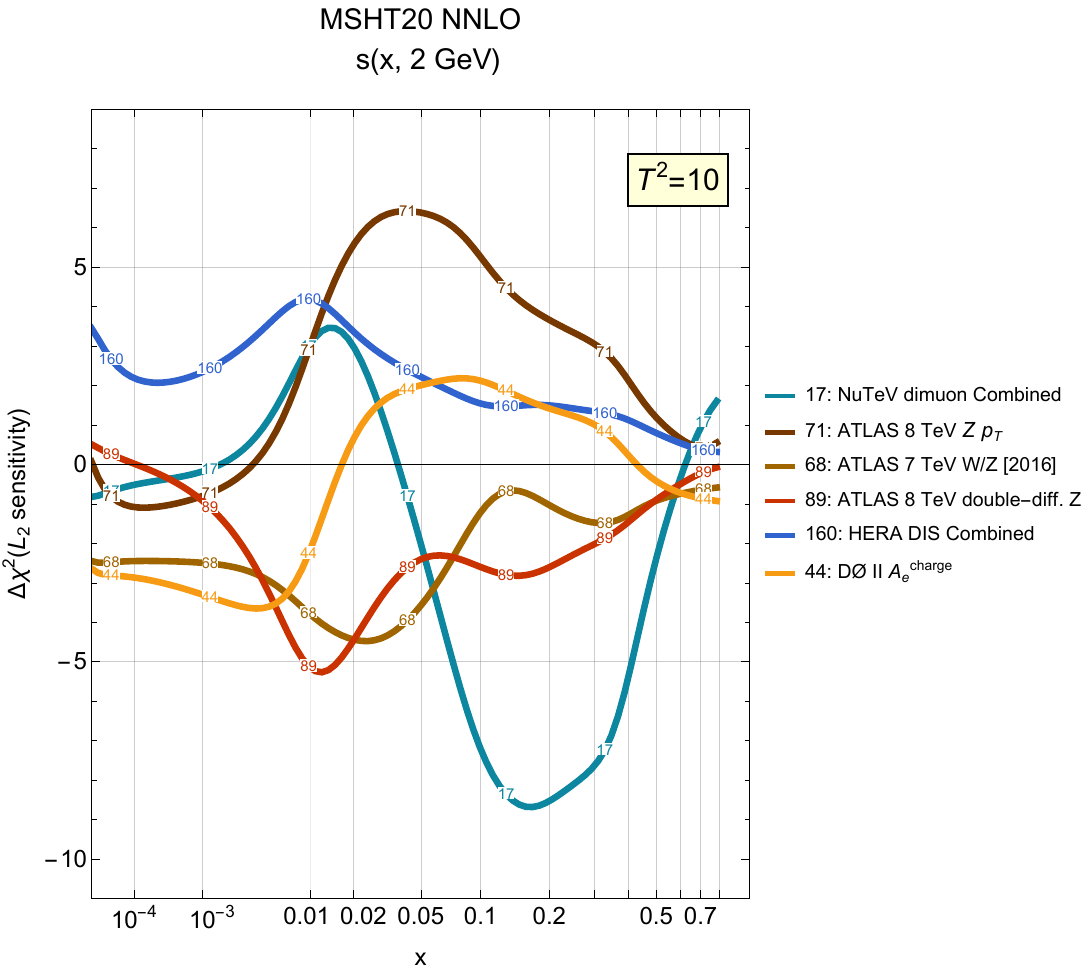}
\includegraphics[width=0.49\textwidth]{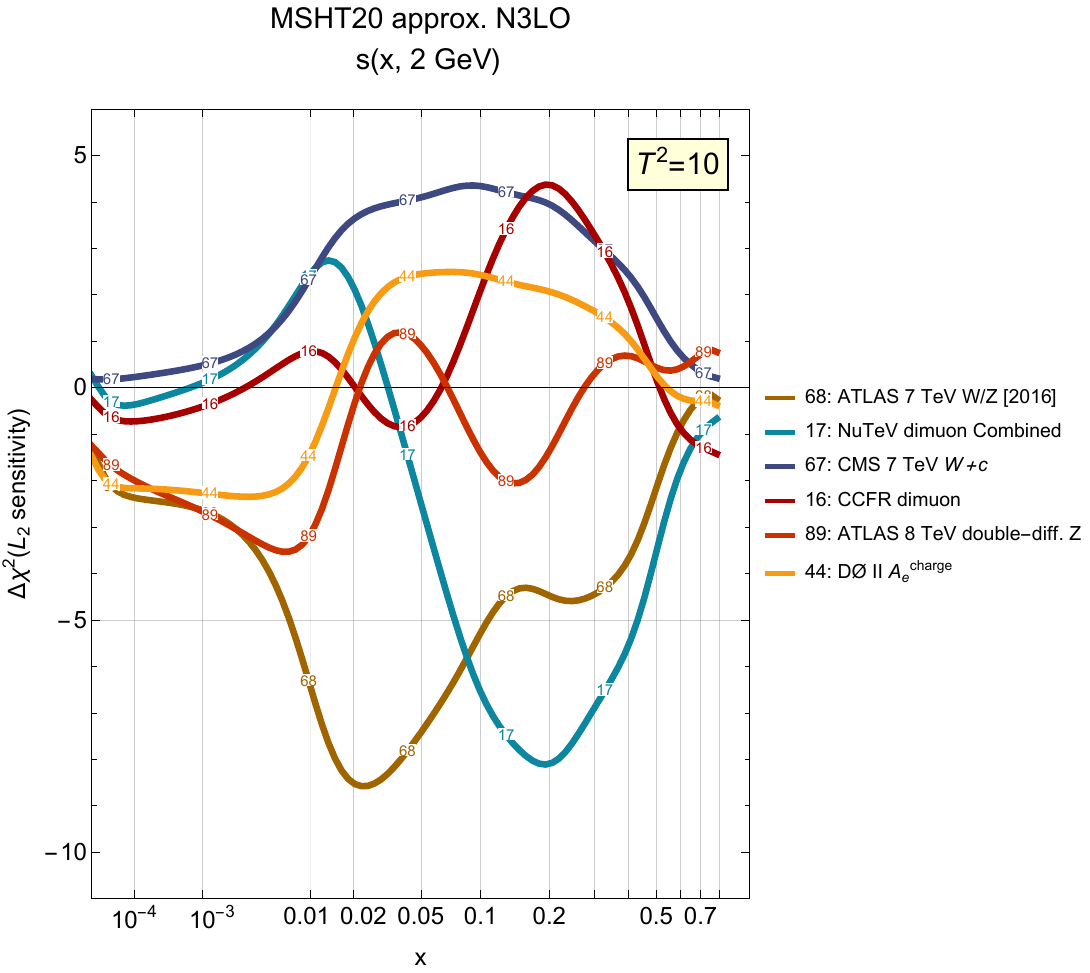}
\caption{$L_2$ sensitivities showing the six most sensitive data sets for the MSHT20 NNLO and aN3LO strange PDFs.}
\label{fig:fig:compar_str_MSHT}
\end{figure*}

Next, in Fig.~\ref{fig:fig:compar_str_MSHT} we compare the results for the strange quark. Here, despite no very obvious direct reason for sensitivity, we see the 
NNLO fit displays significant sensitivity again to the ATLAS $Z\ p_T$ data, showing the general issues in fitting these data at NNLO. This is absent at aN3LO. Similar sensitivity is seen in both PDFs for the NuTeV dimuon data and the ATLAS 7 TeV $W, Z$ data, which pull in the same direction at high $x$ but oppose at lower $x$. However, at NNLO these are in tension with ATLAS $Z\ p_T$ data, while at aN3LO it is largely with CMS $W +c$ data, which favors a slightly reduced strangeness over most of the $x$ range. This may be caused by minor issues in the aN3LO fit due to relatively little knowledge of the theory cross section beyond NLO, and also likely due to the larger strange quark at aN3LO. 

\begin{figure*}[h]
\centering
\includegraphics[width=0.49\textwidth]{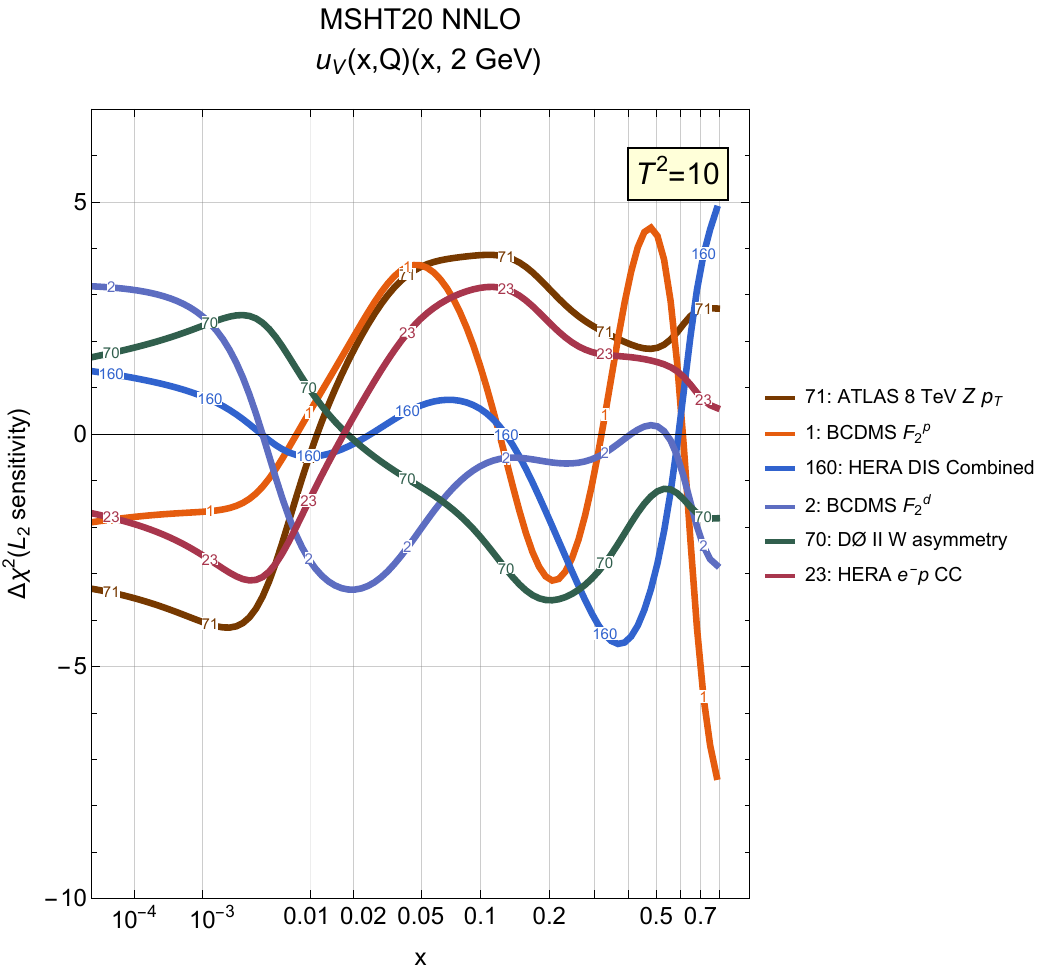}
\includegraphics[width=0.49\textwidth]{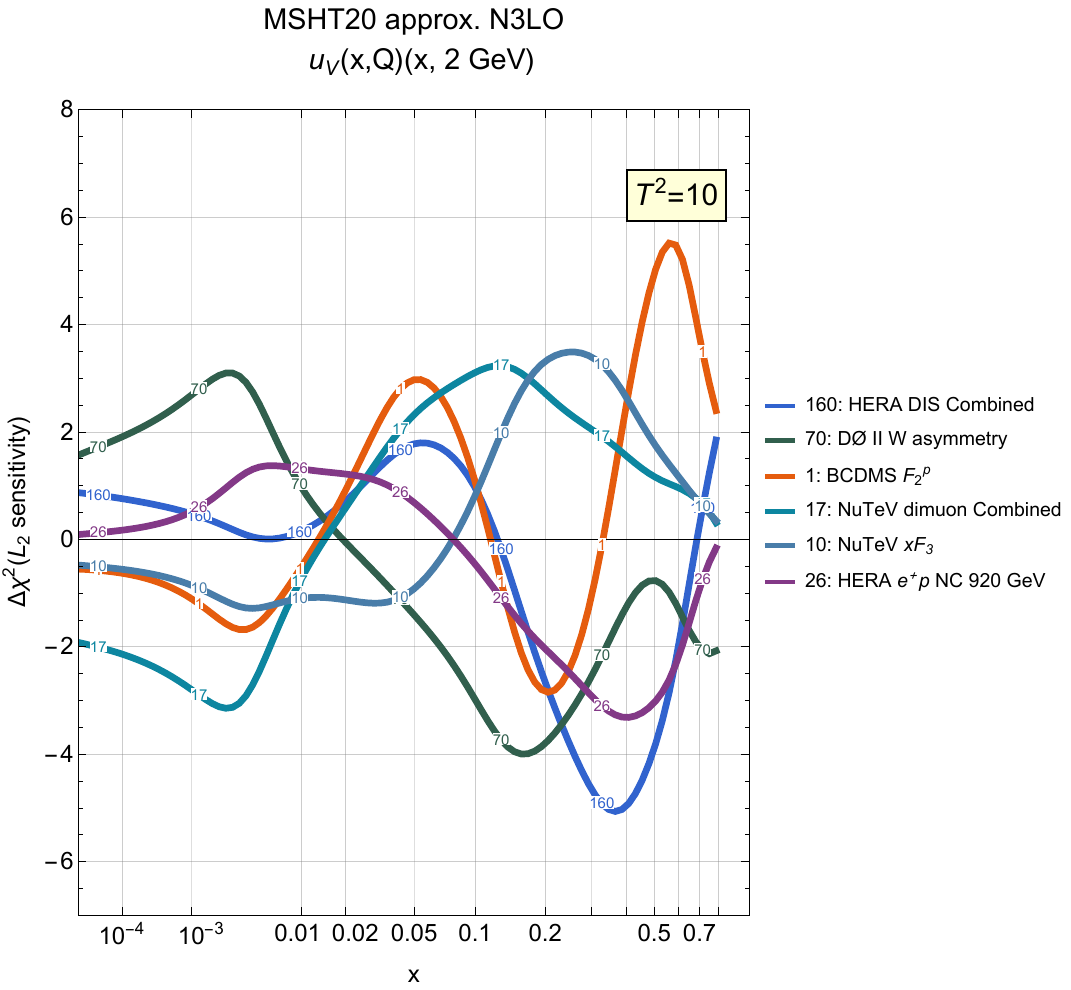}
\caption{ $L_2$
sensitivities showing the six most sensitive data sets for the MSHT20 NNLO and aN3LO $u_V$ PDFs.}
\label{fig:compar_uv_MSHT}
\end{figure*}

In Fig.~\ref{fig:compar_uv_MSHT}, we compare the results for the up valence quark. Again we see that the 
NNLO fit displays some significant sensitivity to the ATLAS $Z\ p_T$ data, which disappears at aN3LO. Beyond this, there are similar behaviors at NNLO and aN3LO for the HERA combined structure function data and the  BCDMS $F_2^p$ data, which are two of the most sensitive sets. This is because the aN3LO fit has smaller impact at high $x$ with respect to the NNLO fit and does not significantly alter the details of flavor decomposition. Overall the $u_V$ quark is well-constrained, with no very large tensions between data sets. Indeed, all sensitivities across the whole range of $x$ are $\lesssim 5$ units. The degree of tension is not obviously further reduced at aN3LO. 

\begin{figure*}[h]
\centering
\includegraphics[width=0.49\textwidth]{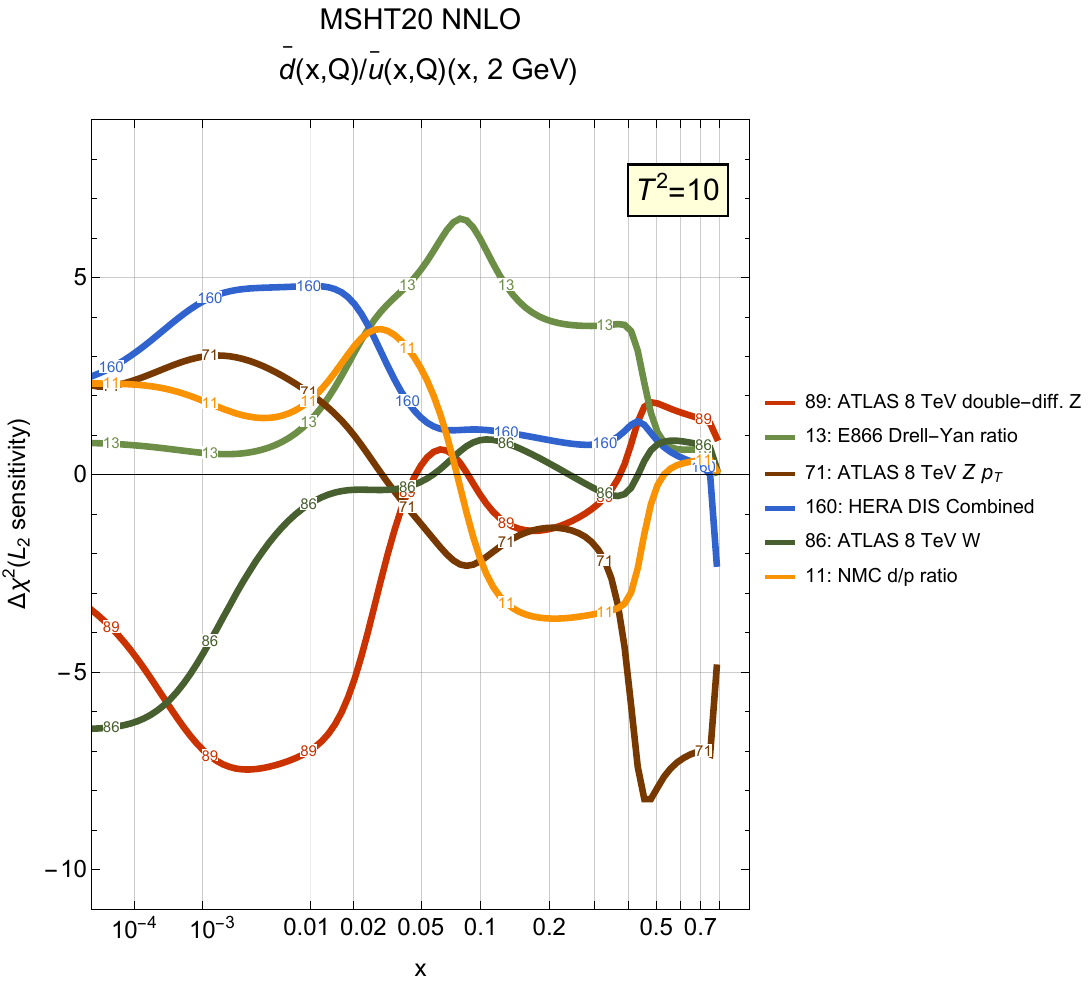}
\includegraphics[width=0.49\textwidth]{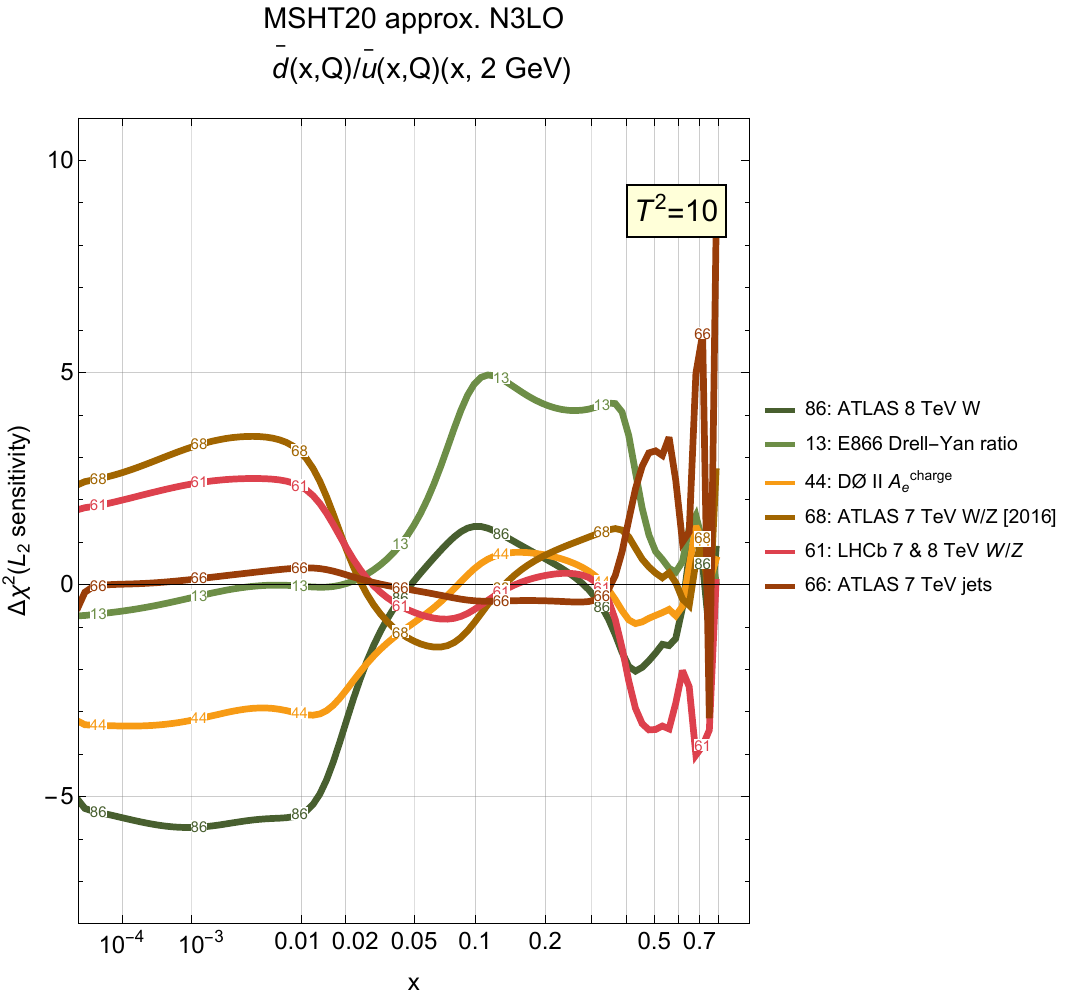}
\caption{ $L_2$ sensitivities showing the six most sensitive data sets for the MSHT20 NNLO and aN3LO $\bar d/\bar u$ PDF ratios.}
\label{fig:compar_db2ub_MSHT}
\end{figure*}

Finally, in Fig.~\ref{fig:compar_db2ub_MSHT} we compare the results for the $\bar d/\bar u$  antiquark ratio. Here we might expect little impact from aN3LO corrections. Indeed this is largely the case. In both cases the E866 Drell-Yan ratio data and the ATLAS 8 TeV $W$ data provide constraints at primarily large and small $x$, respectively. 
The former favors a reduced $\bar{d}/\bar{u}$ at large $x$ (i.e., a larger asymmetry with $\bar{d}<\bar{u}$ at large $x$), while the latter favors an enhanced $\bar{d}/\bar{u}$ at small $x$. This differs slightly from the observations made for CT18 in Fig.~\ref{fig:compar_db2ub_CT} where it was observed that the E866 Drell-Yan ratio data had sensitivity of -9 units (positive pull on the isospin asymmetry ratio) in the region $0.04 < x < 0.2$. These differences partly reflect differences in parametrization choices, with MSHT employing a more flexible parametrization at high $x$ and in turn obtaining a smaller $\bar{d}/\bar{u}$; see Fig.~\ref{fig:PDF_comp_DbovUb} and its discussion in the next section. However, a tension exists between these E866/NuSea Drell-Yan data and the (not fitted) E906 Seaquest data in pulls on $\bar{d}/\bar{u}$ at high $x$, which is visible in Fig.~\ref{fig:compar_db2ub_CT} (right), with the latter favoring instead an increased $\bar{d}/\bar{u}$. In MSHT20, the remaining data sets showing sensitivity to this PDF ratio include, once again, some ATLAS $Z\ p_T$ sensitivity at NNLO, which is absent at aN3LO. At NNLO, there is also some (small) tension between HERA structure function data and ATLAS 8 TeV $Z$ data at small $x$, with each showing pulls in opposite directions on the $\bar{d}/\bar{u}$ with sensitivities of around five units. This is absent at aN3LO, but replaced, to a lesser degree, by a smaller ATLAS 7 TeV $W, Z$ and D\O\ electron asymmetry data tension. In general, only small tensions ($\lesssim 5$ units) exist at each order, and at both orders the E866 Drell-Yan ratio data provides the dominant constraint at $x>0.1$.

\begin{figure*}[b]
\centering
\includegraphics[height=160pt,trim={ 0 0 1.55cm 0},clip]{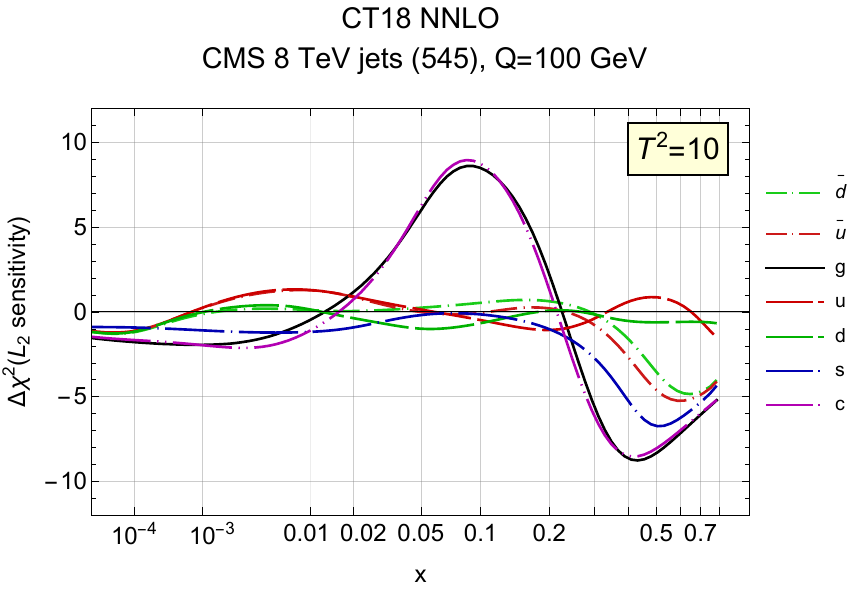}
\includegraphics[height=160pt,trim={ .8cm 0 0 0},clip]{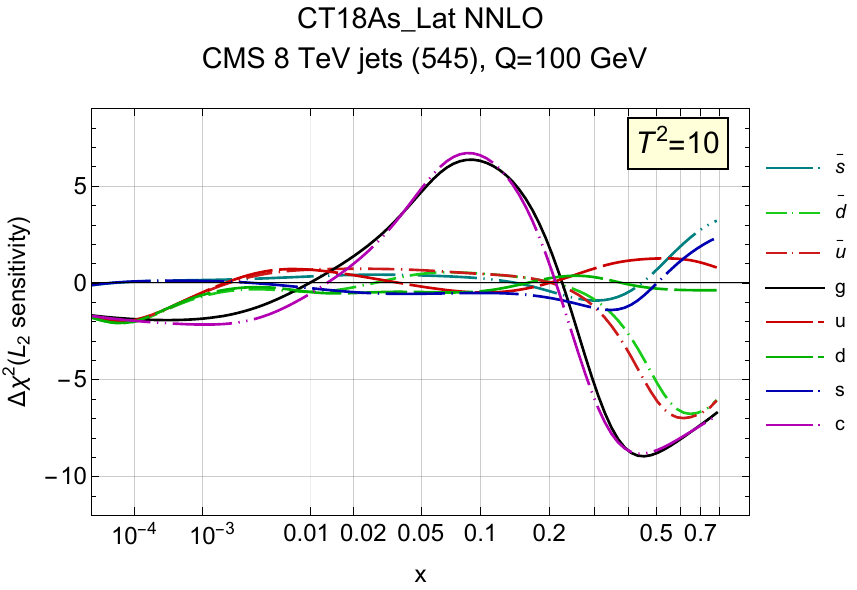}
\includegraphics[height=160pt,trim={ 0 0 1.55cm 0},clip]{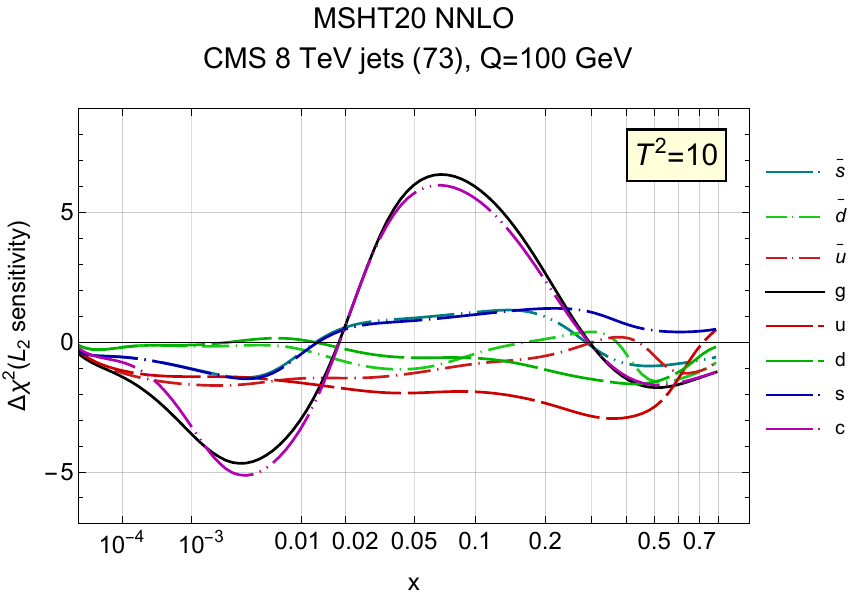}
\includegraphics[height=160pt,trim={ .8cm 0 0 0},clip]{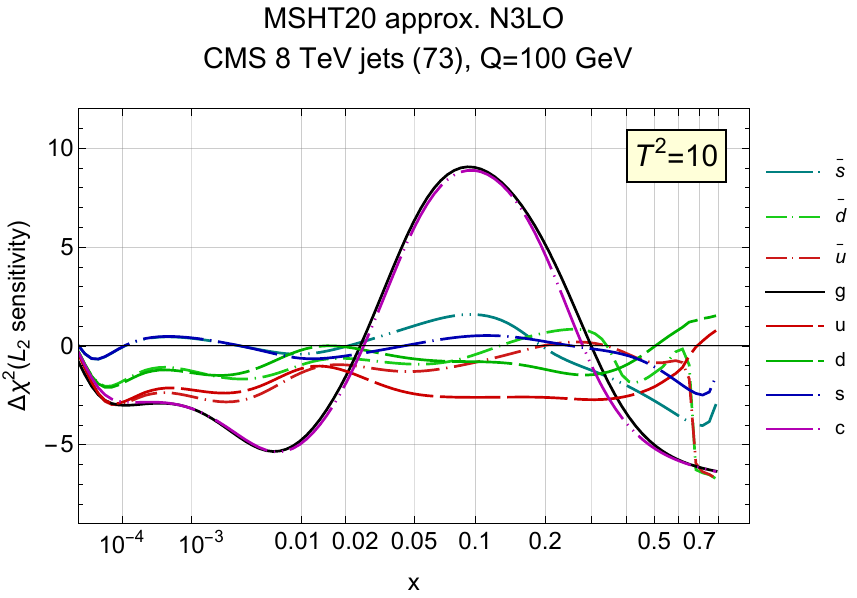}
\caption{Sensitivities of the CMS 8 TeV inclusive jet data set for the PDFs with $T^2=10$ at $Q=100$ GeV.  Top: CT18 and CT18As\_Lat at NNLO. Bottom: MSHT20 at NNLO and aN3LO. }
\label{fig:compar_bench_CMS8_TeV}
\end{figure*}

\section{Comparing sensitivities of experiments among different PDF fits \label{sec:SFL2global}}

Sections~\ref{sec:non-global}-\ref{sec:SFL2MSHT} provided examples of sensitivities computed for the PDF sets of each participating group. We saw that, among the PDFs of one group, the similarities were maintained for the most influential data sets. At the same time, various differences were also noted and could be readily explained in some cases. We also saw that the differences are even more pronounced when comparing sensitivities to the same PDF flavor among the different groups, such as those in Secs.~\ref{sec:SFL2CT} and \ref{sec:SFL2MSHT}. These differences partly reflect non-identical choices in the data selection and fitting methodologies reviewed in Sec.~\ref{sec:Particulars}. In this section, we will further elaborate on representative examples of differences and similarities among the PDFs of different groups, this time focusing on the plots of sensitivities to multiple PDF flavors for the shared experimental data sets. 

To start, we return to the example of the CMS 8 TeV jet data set, now showing $L_2$ sensitivities for CT18 and MSHT20 PDFs with $T^2=10$ at $Q=100$ GeV side-by-side in Fig.~\ref{fig:compar_bench_CMS8_TeV}.\footnote{The CT18 NNLO and MSHT20 NNLO sensitivities at 100 GeV, here compared against CT18As\_Lat NNLO and aN3LO, have been also compared against the reduced fits in App. D of Ref.~\cite{PDF4LHCWorkingGroup:2022cjn}.} 
In the figure, one readily observes that this data set plays an important role for the gluon distribution (the black solid curve) and the charm distribution (the magenta dot-dot-dashed curve) that closely follows the former. There is less sensitivity to (anti)quark PDFs, as reflected by much weaker magnitudes of the other curves.  We already remarked in Sec.~\ref{sec:MoreLML2Relation} that the CMS 8 TeV jet data set plays an important role in constraining the gluon distribution in the CT18 analysis. More remarkably, by  comparing the $S_{f,L2}$ patterns for the CT18As\_Lat, MSHT20 NNLO and aN3LO PDFs, we now confirm a very similar picture for these other PDF sets as well.

\begin{figure*}[b]
\centering
\includegraphics[height=160pt,trim={ 0 0 1.6cm 0},clip]{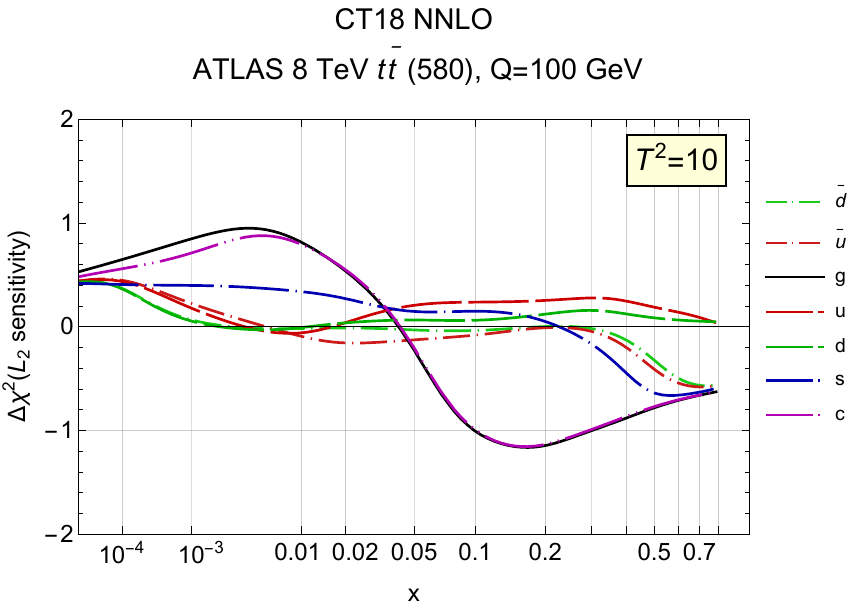}\quad\quad
\includegraphics[height=160pt,trim={ 1.cm 0 0 0},clip]{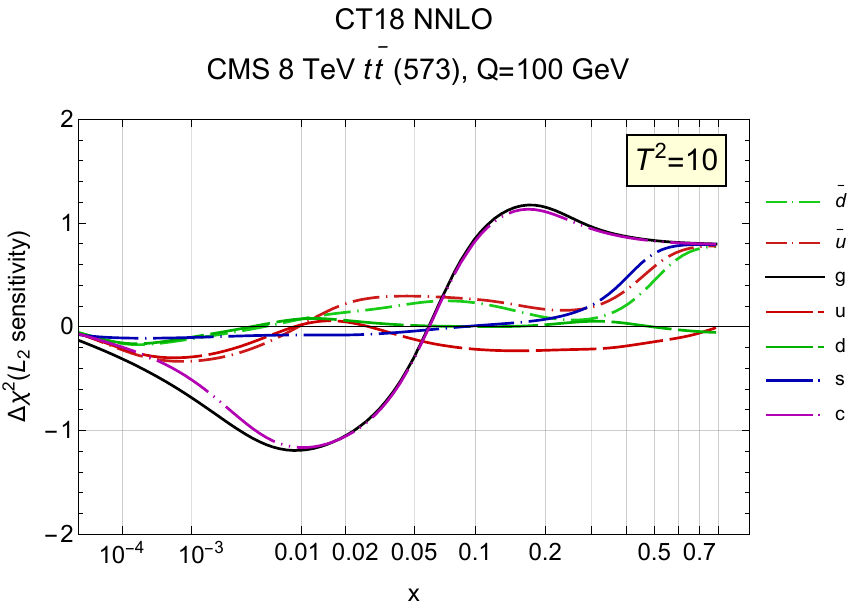}\\
\includegraphics[height=160pt,trim={ 0 0 1.6cm 0},clip]{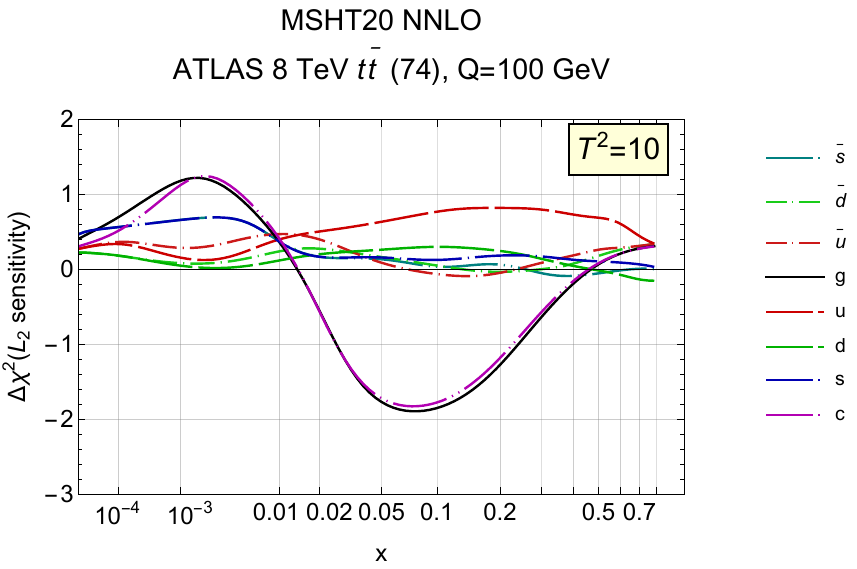}
\includegraphics[height=160pt,trim={ 1.cm 0 0 0},clip]{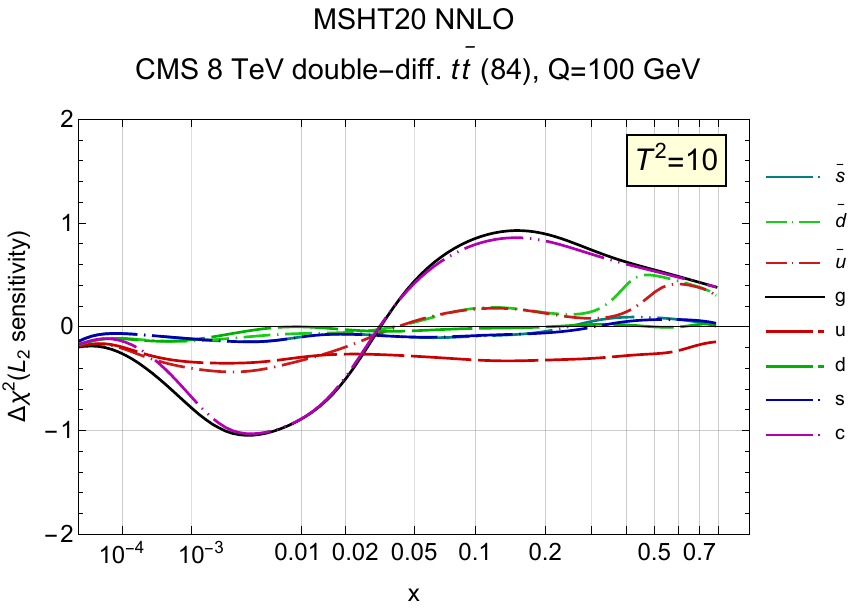}\\
\includegraphics[height=160pt,trim={ 0 0 1.6cm 0},clip]{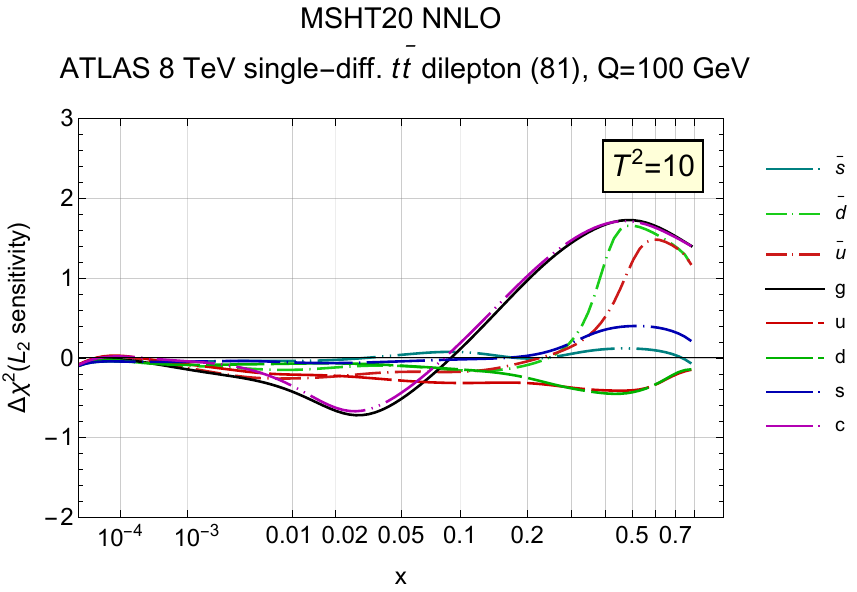}\quad
\includegraphics[height=160pt,trim={ 1.cm 0 0 0},clip]{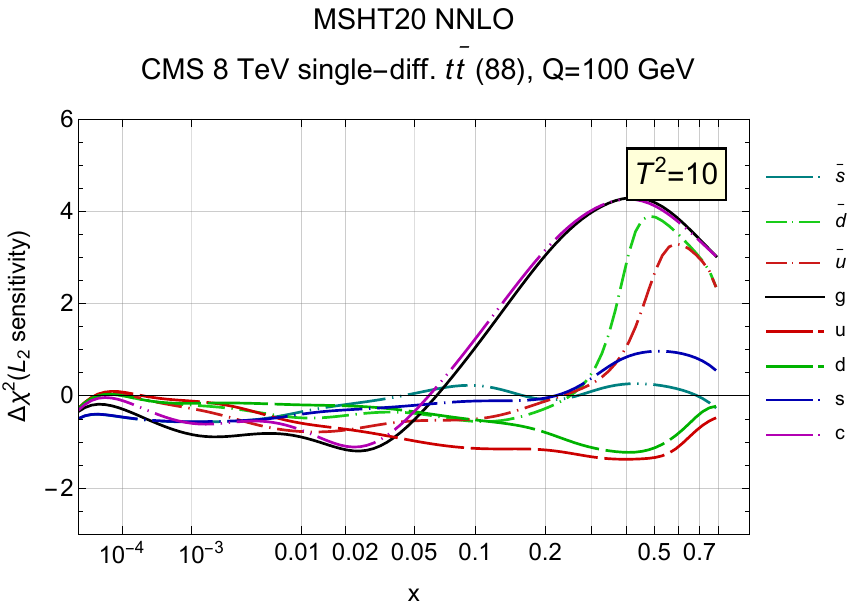}
\caption{
Sensitivities for ATLAS 8 TeV (left column) and CMS 8 TeV (right column) $t\bar t$ data sets in the CT18 (upper row) and MSHT20 (middle and lower rows) analyses at $Q=100$ GeV.}
\label{fig:compar_ttbar_CTMSHT}
\end{figure*}

The sensitivities tell us that, for all four PDF ensembles, at $x=0.1$ an increase of the value of the gluon PDF by one sigma would be disfavored by up to ten units of $\chi^2$. A smaller upward preference for the gluon at low $x$ is another common feature. 

This is one of the rare patterns where very good agreement is so apparent between the fitting groups. Note, however, that, for both CT and MSHT, the pull on the gluon from the CMS 8 TeV jet data set is actually very different to that seen in the reduced fits. For the latter, the pulls on the gluon by the CMS 8 TeV jets can be seen in Fig.~\ref{fig:compar_red_bench2}: the reduced fits both display the opposite pattern in the shape of the sensitivities across $x$, which in turn is tied to a different magnitude of the gluon PDF. The magnitude of the $L_2$ pulls is slightly smaller in the reduced sets, suggesting an improved agreement with the other data sets. This example illustrates that the tension can be enhanced in a global fit upon the addition of data sets to the reduced fit, and the patterns of the pulls can change, too.

Other interesting observations about the gluon come from the ATLAS and CMS $t\bar{t}$ data, displayed in Fig.~\ref{fig:compar_ttbar_CTMSHT}. The top and middle plots concern the lepton+jets $t\bar{t}$ data and reveal sensitivity to the gluon (and charm), at small and valence-like regions in $x$, though at a smaller magnitude {\it w.r.t.} the CMS 8 TeV jet data set. The patterns found by both CT18 and MSHT20 again agree between each other, but differ significantly between ATLAS and CMS: their pulls on the gluon distribution at $x>0.05$ are opposite (see the upper and middle rows). These $t\bar{t}$ data also affect the $\bar{u}$ and $\bar{d}$ distributions at large $x$, but with  weaker pulls than other data sets. On the other hand, consistency of the pulls by the dilepton $t\bar{t}$ data from ATLAS and CMS at 8 TeV on the the MSHT20 NNLO PDFs in the bottom row is quite good: the pulls are similar in shape, albeit with different magnitudes. Finally, the CT18
sensitivities to the gluon at $x\approx 0.01$ in the upper row of Fig.~\ref{fig:compar_ttbar_CTMSHT}
are in agreement, both in their magnitudes and signs, with the weak preferences for a lower (higher) gluon exhibited by the ATLAS (CMS) $t\bar t$ data sets in the LM scan in Fig.~\ref{fig:LMScan2}, where the respective curves have IDs 12 and 13.

\begin{figure*}[p]
\centering
\includegraphics[height=160pt,trim={ 0 0 1.6cm 0},clip]{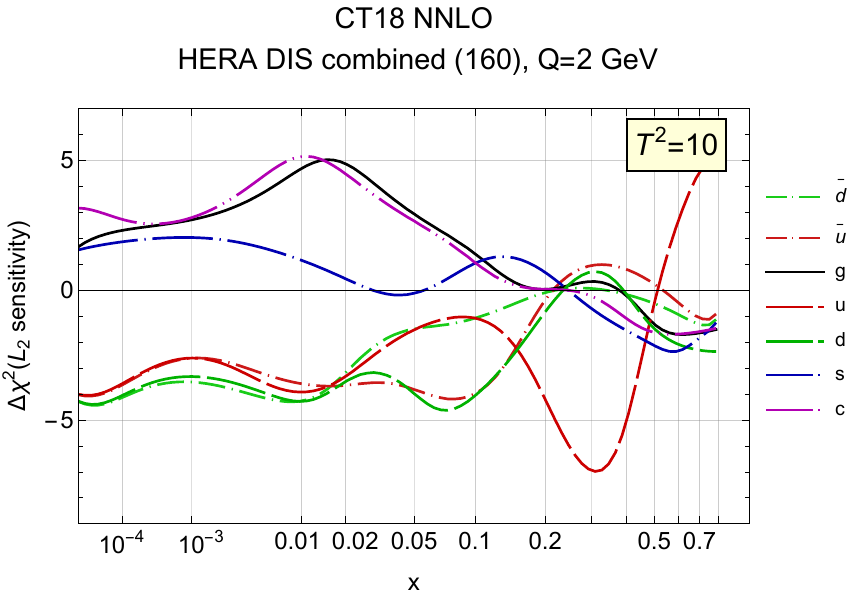}
\includegraphics[height=160pt,trim={ .8cm 0 0 0},clip]{plots/160_ATL21v2_SfL2_q2_1.pdf}
\includegraphics[height=160pt,trim={ 0 0 1.6cm 0},clip]{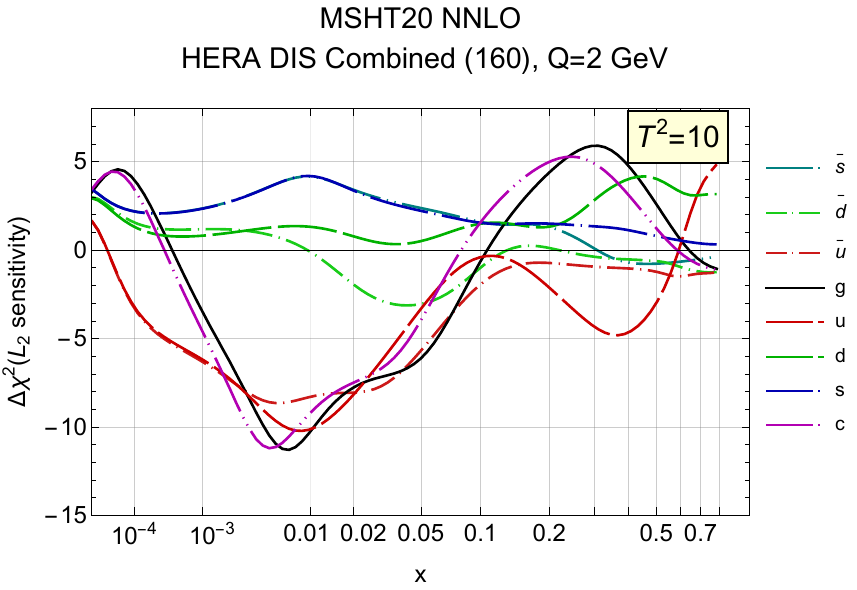}
\includegraphics[height=160pt,trim={ .8cm 0 0 0},clip]{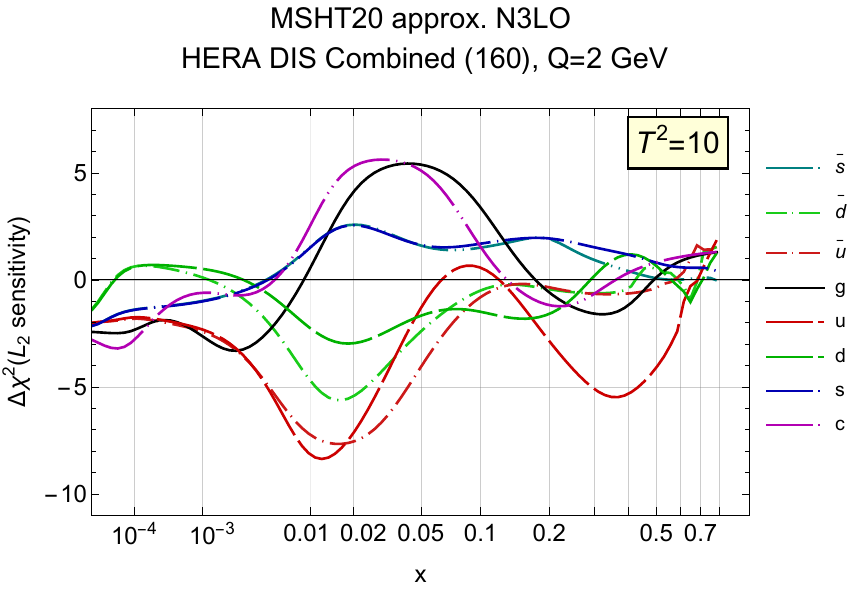}
\caption{Sensitivities  of the combined HERA DIS data. Top left:  CT18 NNLO, top right: ATLASpdf21 NNLO, bottom left: MSHT20 NNLO, bottom right: MSHT20 aN3LO.  }
\label{fig:compar_HERA_Combined}
\end{figure*}

The final data set influencing the gluon PDF that we will consider is the HERA combined DIS cross section data in Fig.~\ref{fig:compar_HERA_Combined}. 
As it happened with the 8 TeV CMS jet data, the pulls by HERA combined data on the gluon distribution are inverted when going from the reduced to the full CT18 NNLO set, whereas for MSHT20 they remain quite similar. Indeed, for the global fits the trends of the pull on the gluon from this data set for CT18 and MSHT20 are distinctly different, particularly at small $x$ values. As already noted, for MSHT aN3LO the trend is different to the NNLO fit and smaller in magnitude, due to the different and reduced tensions, while for the ATLASpdf21 NNLO set the pulls on the gluon from the HERA data are smaller, reflecting the dominant weight these data have in the ATLAS data set.  

\begin{figure*}[p]
  \begin{centering}
    \includegraphics[width=0.46\textwidth]{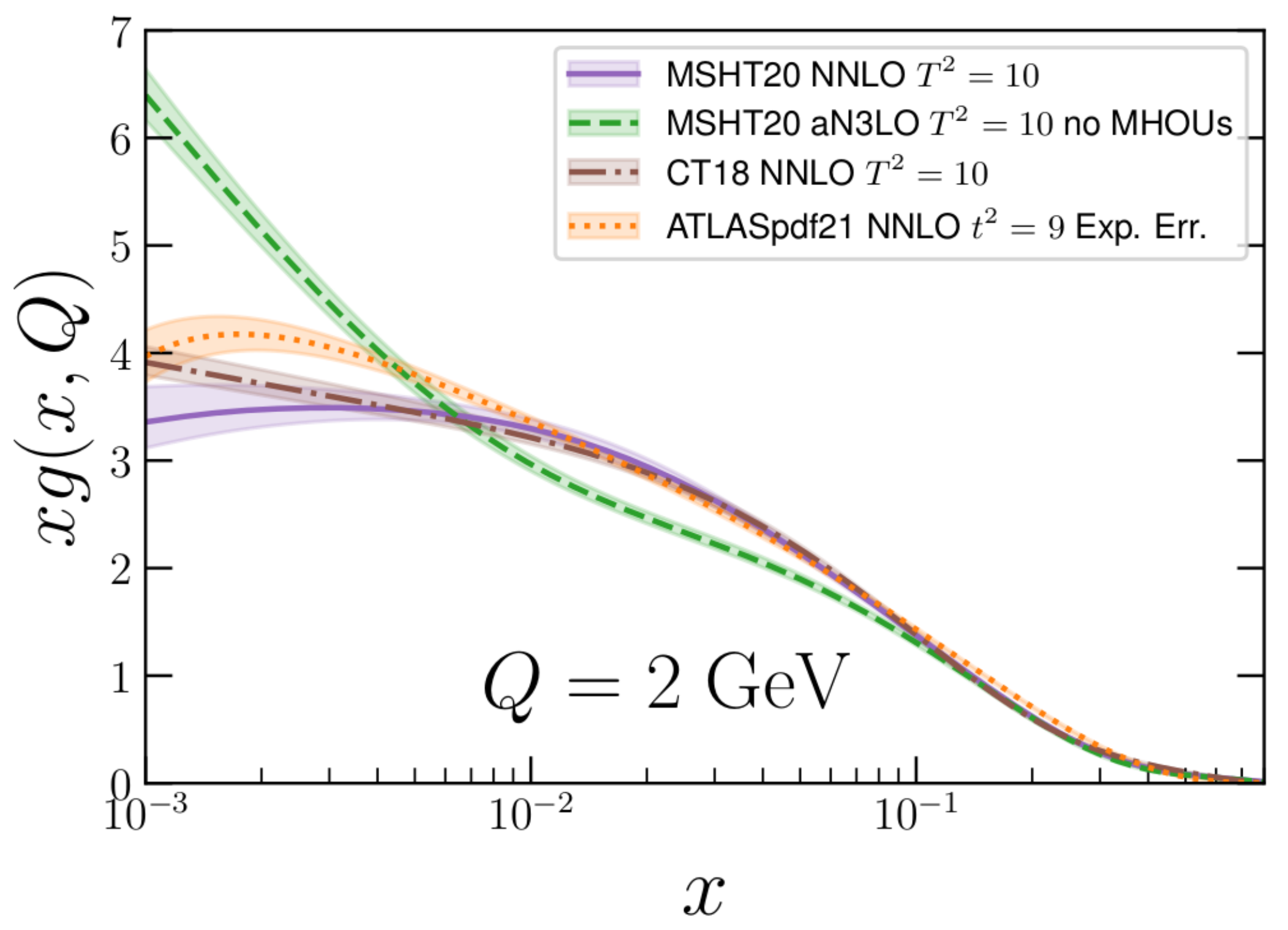}
    \includegraphics[width=0.50\textwidth]{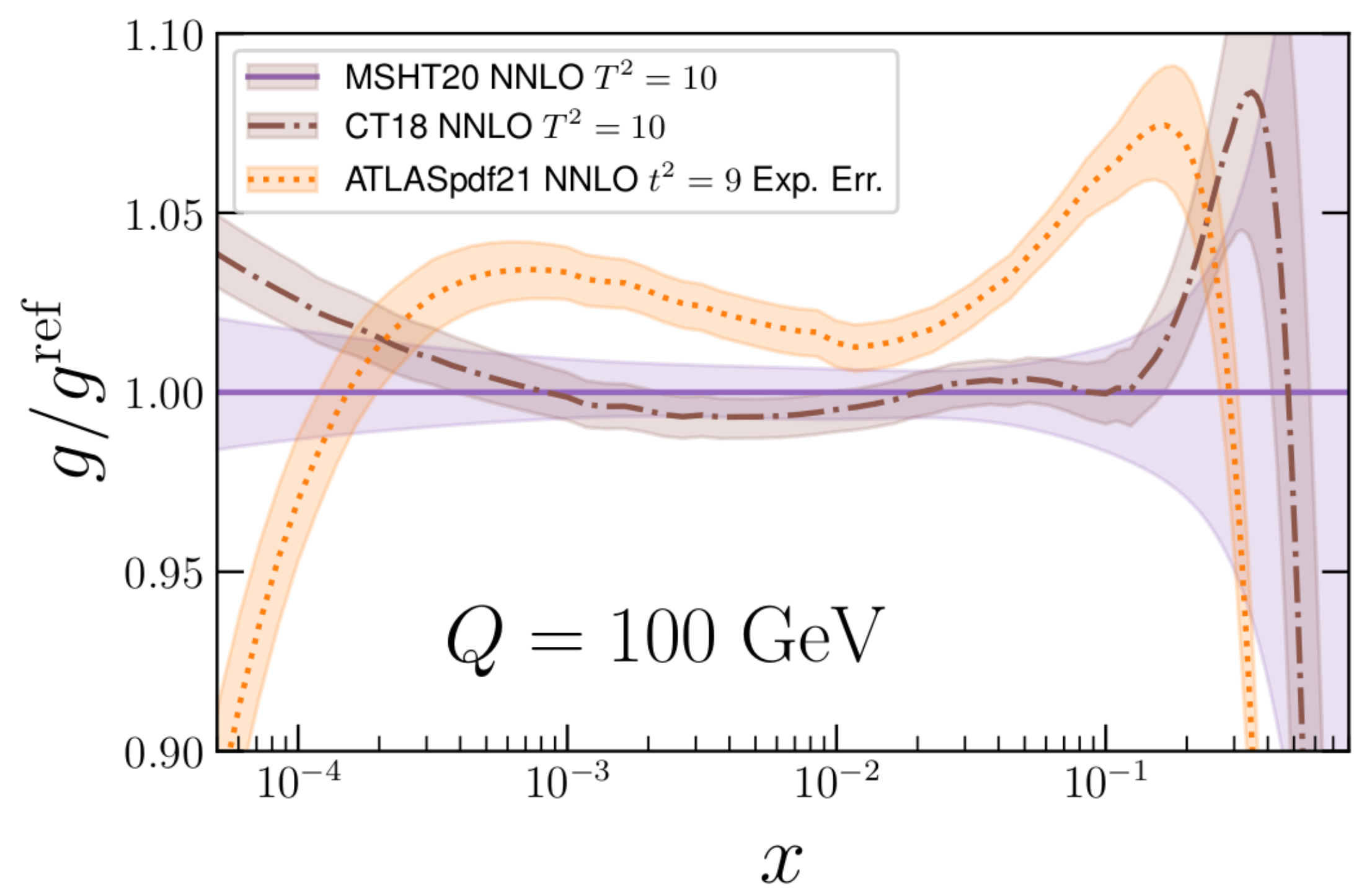}
    \caption{ Comparison of the gluon PDFs. Left: absolute PDFs at $Q=2 $ GeV shown for MSHT20 NNLO (purple, solid), MSHT20 aN3LO (green, dashed), CT18 NNLO (brown, dot-dashed) and ATLASpdf21 NNLO (orange, dotted). Right: 
    ratios to the central MSHT20 NNLO at $Q=100$ GeV. Notice that MSHT20 aN3LO and CT18As\_Lat are not displayed. All ensembles  are given for $T^2=10$ ($t^2=9$ for ATLAS).\label{fig:PDF_comp_gluon}}
  \end{centering}
\end{figure*}

Ultimately, the patterns of sensitivities to the gluon reflect several factors, e.g., the shapes and flexibility of the relevant PDFs, implementation of the theoretical cross sections, and implementation of systematic uncertainties for the considered experiment. Take, for instance, the gluon PDFs for the corresponding PDF sets, shown in Fig.~\ref{fig:PDF_comp_gluon} at $Q=2$ GeV in the left panel and as the ratios to the MSHT20 NNLO gluon PDF at $Q=100$ GeV in the right panel. Going back to the CMS 8 TeV jet data in Fig.~\ref{fig:compar_bench_CMS8_TeV}, we observed that NNLO PDFs predict very similar gluon sensitivities at $x>0.02$. In Fig.~\ref{fig:PDF_comp_gluon}, we observe that the NNLO gluon PDFs agree well at $x>0.02$, and hence it is not surprising that the sensitivities for the CMS 8 TeV jet data in the respective CT18 and MSHT20 analyses (which use close implementations of theoretical cross sections and systematic errors in this case) also agree. 

For the combined HERA DIS data in Fig.~\ref{fig:compar_HERA_Combined}, we pointed out large differences in the gluon sensitivities at $x<0.1$. For this experiment, the groups use different heavy-quark schemes and $\chi^2$ definitions, while the interplay of the pulls from various $x$ regions depends on the assumed parametrization forms. [Recall that the MSHT parametrization for the gluon is more flexible than the CT one.] As influential are the shapes of the central PDFs, which are distinctly apart at $x \lesssim 10^{-3}$ in Fig.~\ref{fig:PDF_comp_gluon}.
In the ATLASpdf21 case, the small-$x$ region is affected by a cut of $Q^2 > 10\mbox{ GeV}^2$ imposed on the fitted HERA data because of doubts about the adequacy of NNLO DGLAP to describe the HERA data at low $x$ and $Q$ below this cut. The ATLASpdf21 PDFs are designed for use at higher $x$, $x>10^{-4}$, and the absence of the low-$x$ data explains the reduction of the gluon at very small $x$ and its compensating enhancement at moderate $x$. The deviation of the ATLASpdf21 from the others at the lowest $x$ values demarcates the region where low-$x$ physics effects may need to be considered. 

Now turning to the sensitivities to (anti)quarks, it can be seen in Fig.~\ref{fig:compar_HERA_Combined} that the HERA DIS data also have significant pulls on PDFs other than the gluon. In particular, they affect the up and down distributions. For CT18, the data favors larger $u$ and $d$ PDFs at small- and mid/valence-$x$ values at $Q=2$ GeV, and smaller up distribution at very large $x$. While the trend for the $u$ PDF is also negative for MSHT20 NNLO (bottom left), the pulls on the $d$ distribution are smaller and positive, though there is now more pull on the strange quark. The magnitude of the pulls from HERA DIS data for the $u$ and $\bar{u}$ of the MSHT20 $L_2$ plots at small $x$ is similar to that of the gluon. The signs of pulls of MSHT20aN3LO align quite well to those of CT18 NNLO and MSHT20 NNLO, but are generally a little reduced. Overall, as with the gluon, ATLASpdf21 shows less tension with the HERA DIS data for the quarks, a consequence of its smaller sample of data sets.

\begin{figure*}[p]
\centering
\includegraphics[height=160pt,trim={ 0 0 1.6cm 0},clip]
{plots/68_ATL21v2_SfL2_q2_1.pdf}
\includegraphics[height=160pt,trim={ 0.8cm 0 0 0},clip]{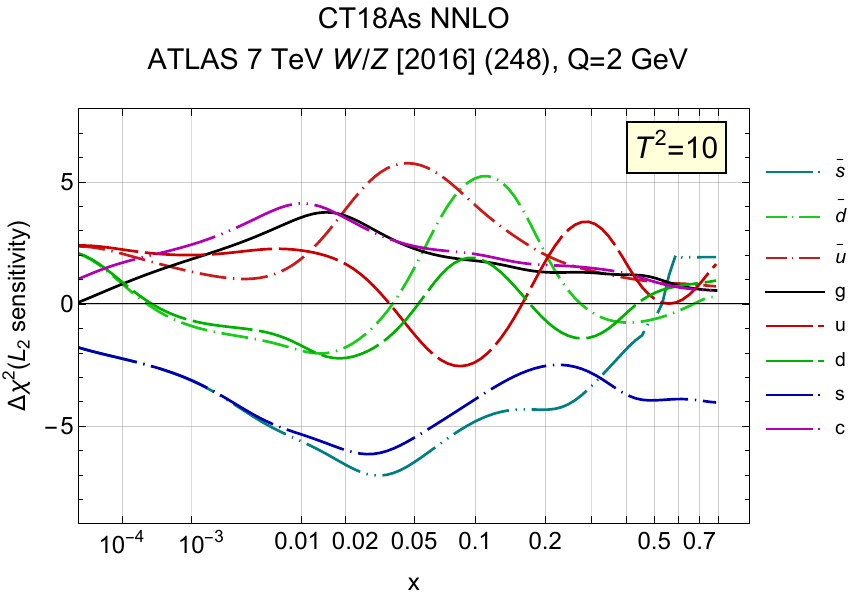}
\includegraphics[height=160pt,trim={ 0 0 1.6cm 0},clip]{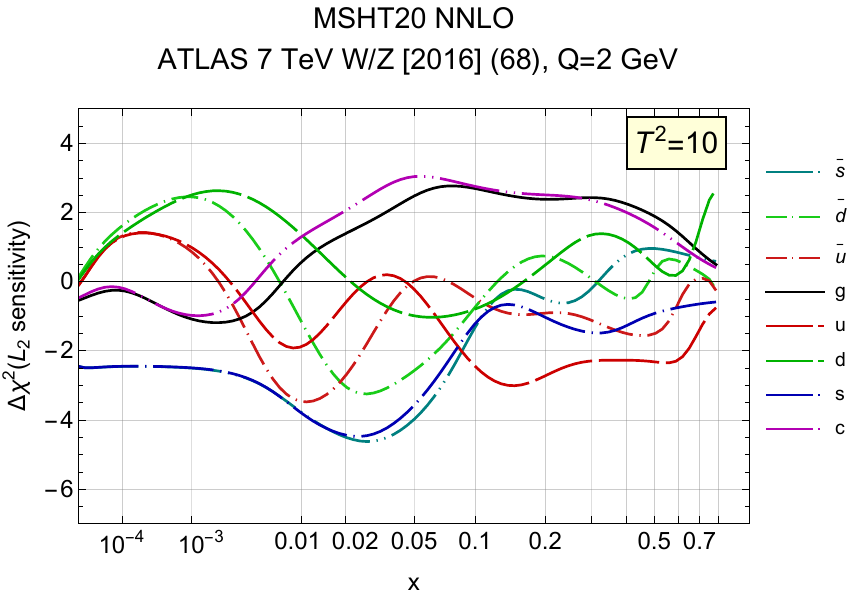}
\includegraphics[height=160pt,trim={ 0.8cm 0 0 0},clip]{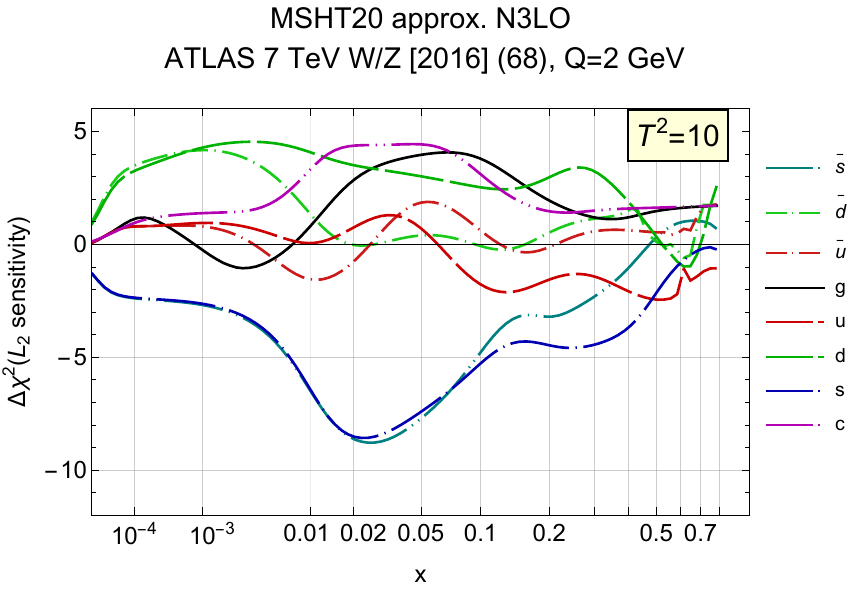}
\caption{Sensitivities of ATLAS 7 TeV $W, Z$ data at $Q=2$ GeV:  ATLASpdf21 NNLO (top left),  CT18As NNLO (top right), MSHT20 NNLO (bottom left), MSHT20 aN3LO (bottom right).
}
\label{fig:compar_ATLAS_WZ}
\end{figure*}
\begin{figure*}[p]
  \begin{centering}
    \includegraphics[height=150pt]{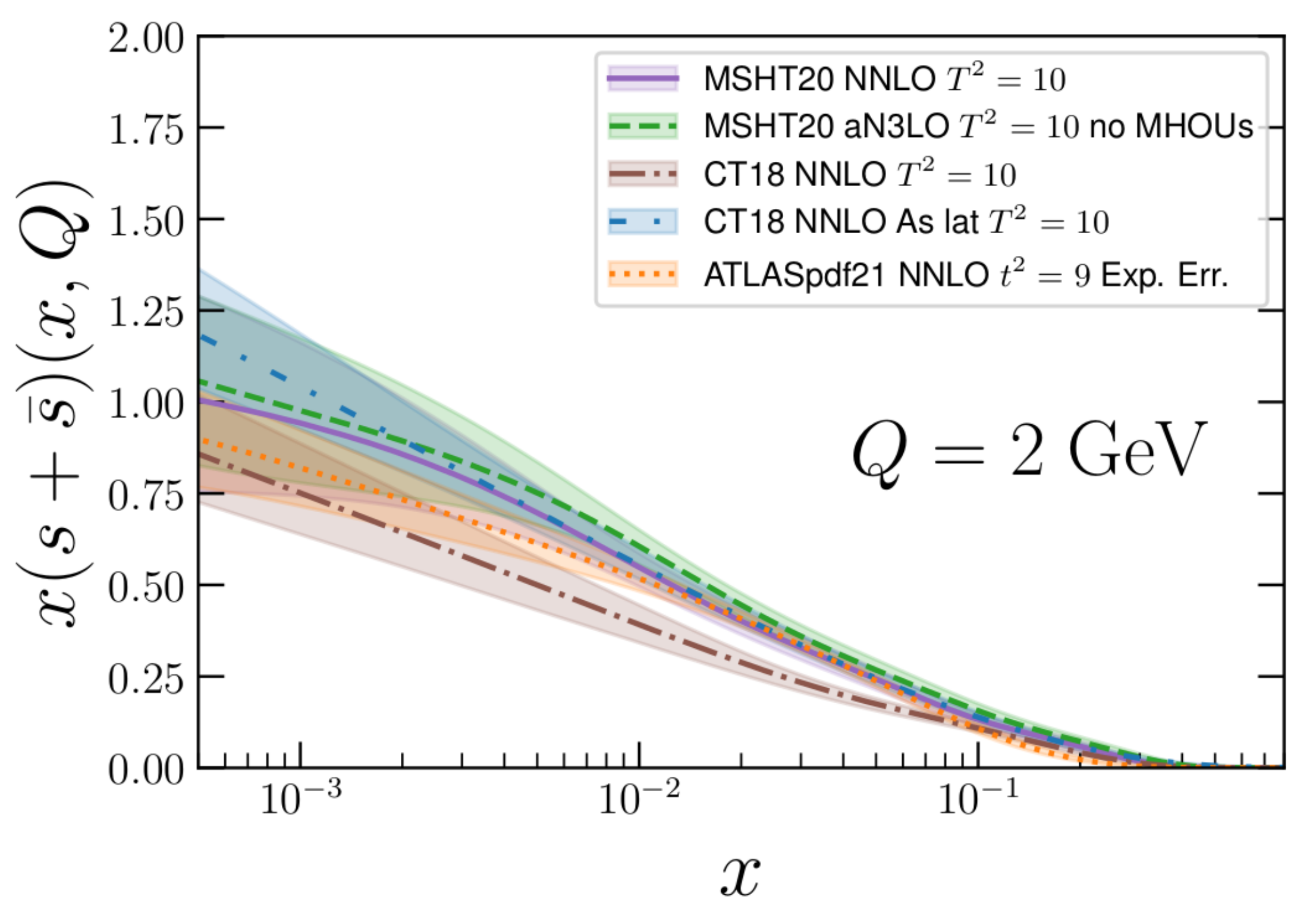}
    \includegraphics[height=150pt]{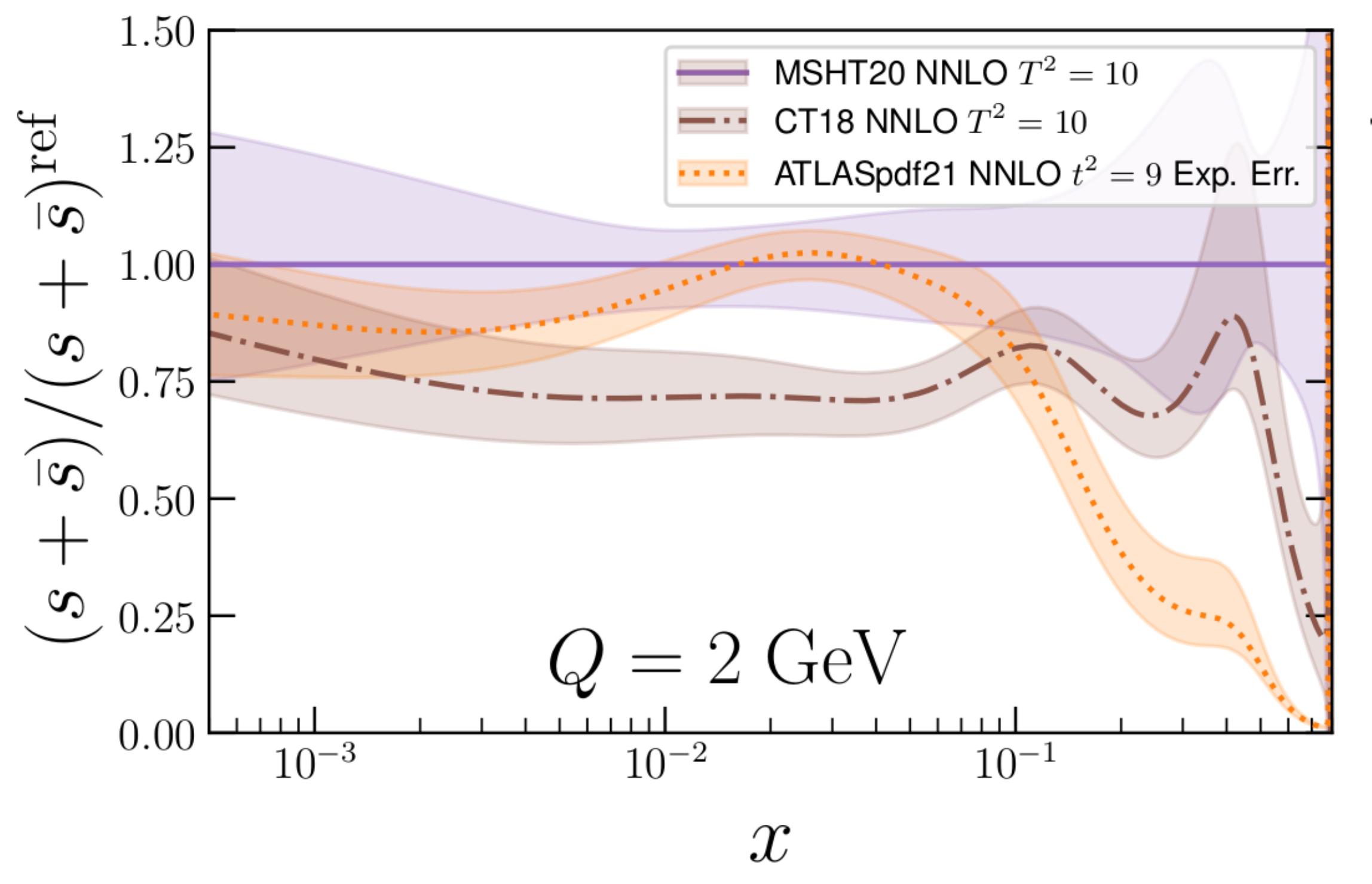}
    \caption{ Comparison of the strangeness PDFs. Left: $s+\bar{s}$ PDFs shown for MSHT20 NNLO (purple, solid), MSHT20 aN3LO (green, dashed), CT18 NNLO (brown, dot-dashed), CT18As\_Lat NNLO (blue, large dot-dashed)  and ATLASpdf21 NNLO (orange, dotted). Right: ratios to the central MSHT20 NNLO (MSHT20 aN3LO and CT18As\_Lat are not displayed, as their uncertainties are similar to their corresponding nominal sets). 
    All ensembles  are given for $T^2=10$ ($t^2=9$ for ATLAS). 
    \label{fig:PDF_comp_strange} } 
  \end{centering}
\end{figure*}

Another data set that is known to be strongly constraining the PDFs is the 
7 TeV ATLAS $W, Z$ precision data. Its $L_2$ sensitivity plots are shown in Fig.~\ref{fig:compar_ATLAS_WZ}. Since the rapidity dependence of these very precise data constrain the details of the $x$-dependence of 
all flavors of quarks, and hence, via evolution, also the gluon, there are significant pulls on a wide variety of PDFs. Indeed, the indirect gluon constraint does result in a noticeable set of pulls on this PDF, generally opposing that from HERA data at small $x$, particularly for ATLASpdf21. 
The very specific constraint this data set brings, though, is on the strange quark, which is not directly constrained by most of the other data sets in a global fit. 
In particular, it has long been known that NuTeV and CCFR combined dimuon data favor a smaller strangeness at $x\sim 0.02$, and historically this combination of data has been the main constraint on the strange quark PDF. However, this is now compensated by an upwards pull on the strangeness (corresponding to negative $L_2$ sensitivity)  from ATLAS 7 TeV $W, Z$ over a generally wide range of $x$-values. This is evident for both the CT18As PDF set (the default CT18 PDFs do not fit to this data set - as apparent  from its reduced strangeness seen in Fig.~\ref{fig:PDF_comp_strange}) and the MSHT20 sets (where there is also an opposing pull from CMS 7 TeV $W+c$ data for MSHT20aN3LO). The pull in the ATLASpdf21 sensitivity plot is much smaller at relatively high $x$ values, though is also negative at very small $x$ values, where these data are opposed by the 8 TeV Z3D data, which has opposite tendencies.

A comparison of the distributions for total strangeness is given in Fig.~\ref{fig:PDF_comp_strange}. All PDF sets are strongly influenced by the ATLAS $W, Z$ data, and hence agree quite well in the region covered by data. Noticeably  there is a smaller strangeness PDF for CT18 at $x\sim 0.01-0.1$, mostly due to the absence of the ATLAS $W, Z$ data set in the CT18 nominal fit, as opposed to the CT18A(Z) alternative sets.

\begin{figure*}[t]
\centering
\includegraphics[height=160pt,trim={ 0 0 1.6cm 0},clip]{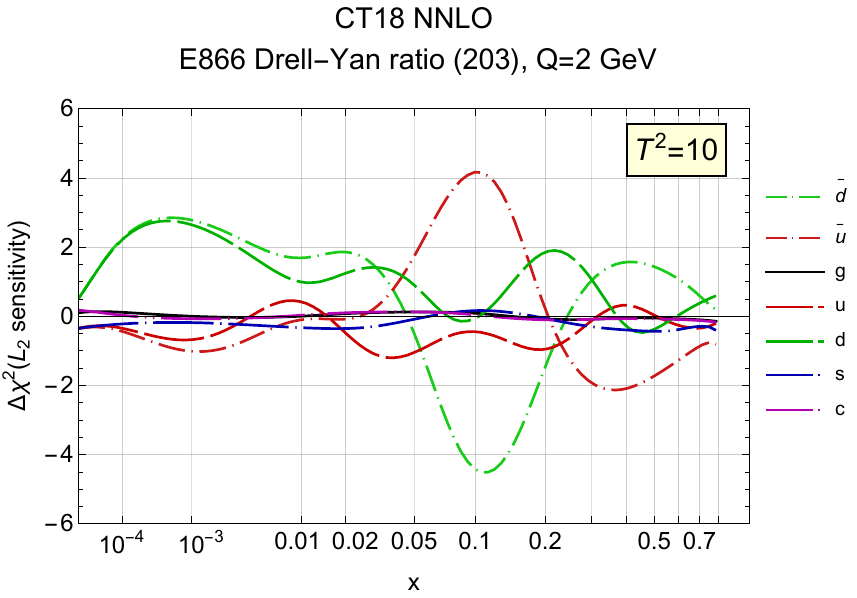}
\includegraphics[height=160pt,trim={ 0.8cm 0 0 0},clip]{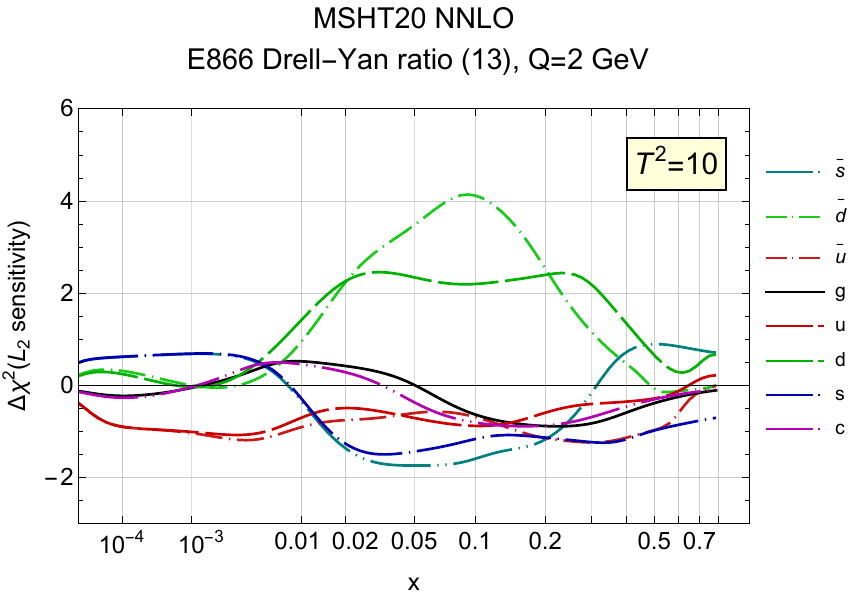}
\caption{ Sensitivities of the E866 $pd/pp$ Drell-Yan ratio at $Q=2$ GeV.  Left: CT18 NNLO, right: MSHT20 NNLO. }
\label{fig:compar_E866}
\end{figure*}

\begin{figure*}[t]
  \begin{centering}
    \includegraphics[height=150pt]{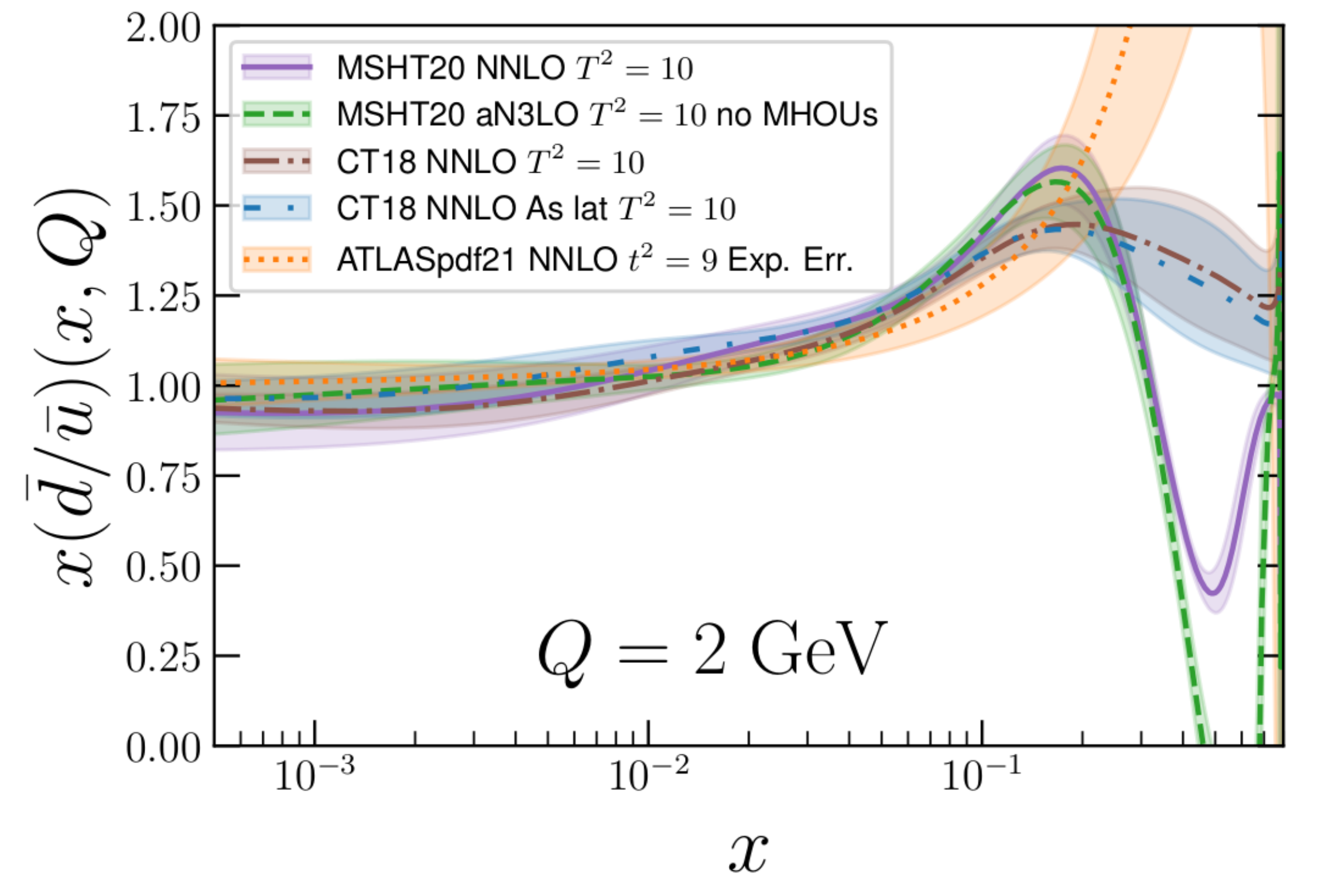}
    \includegraphics[height=150pt]{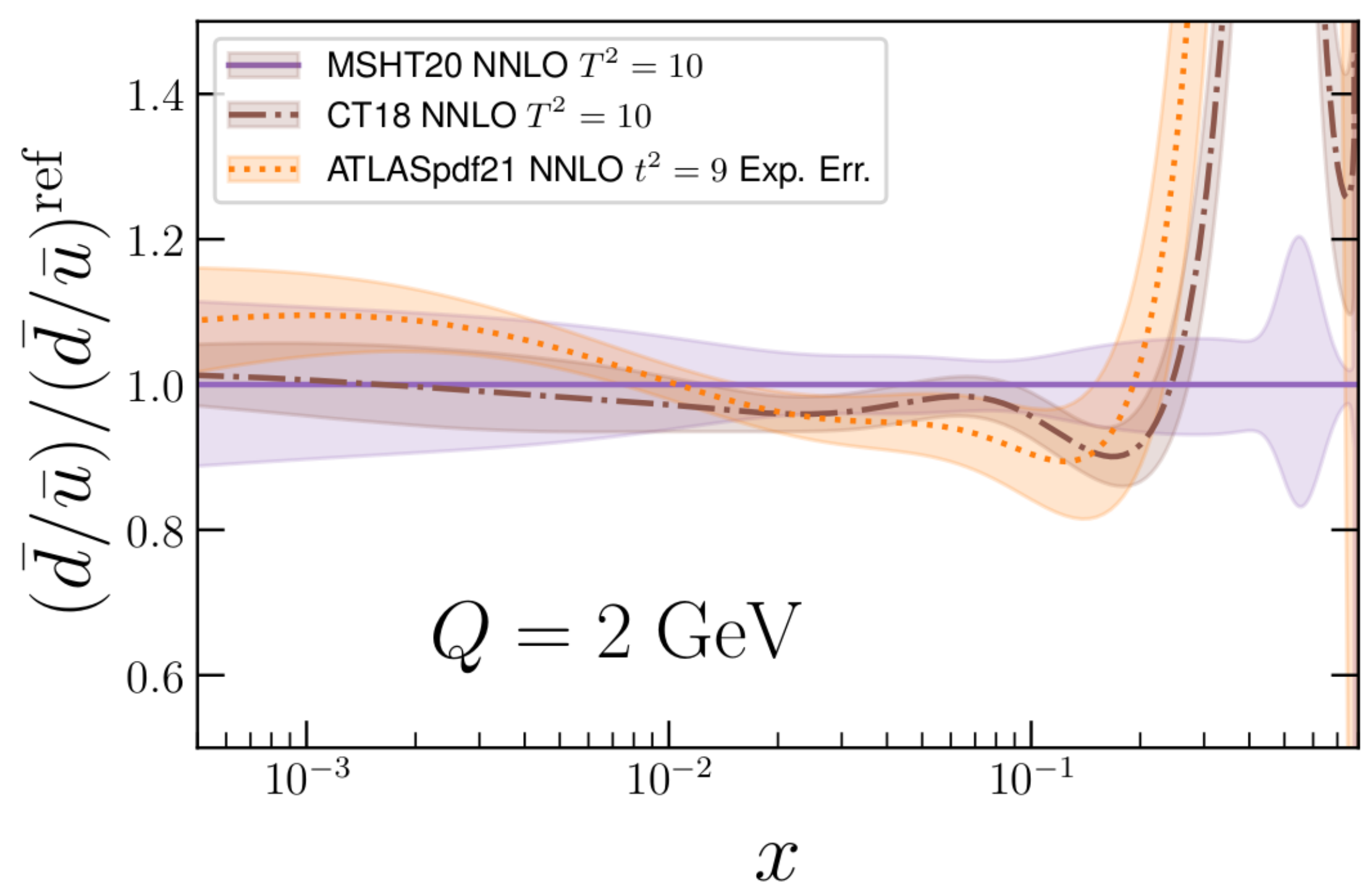}
    \caption{ Comparison of the $\bar{d}/\bar{u}$ PDF ratios. Left: absolute ratios of PDFs.  Right: ratios to central MSHT20 NNLO. Color code and legend as in Fig.~\ref{fig:PDF_comp_gluon}.
    \label{fig:PDF_comp_DbovUb} } 
  \end{centering}
\end{figure*}

The final data we consider are the E866 fixed-target Drell Yan cross section ratios, $\sigma^{pd}/\sigma^{pp}$, for which the corresponding sensitivity plots are shown in Fig.~\ref{fig:compar_E866}. The CTEQ-TEA publications singled out the E866 ratio as a small experiment that provides a particularly sensitive probe of flavor-symmetry breaking in the quark sea at high $x$, specifically by constraining $\bar d(x,Q)/ \bar u(x,Q)$  (or, alternatively, $\bar d (x,Q) - \bar u(x,Q)$) above $x=0.01$ and at $Q$ of a few GeV. Indeed, the \texttt{PDFSense} analyses in Refs.~\cite{Wang:2018heo,Hobbs:2019gob} noted that the E866 ratio is a top experiment in terms of the sensitivity per data point, and it is especially valuable for constraining a sea PDF combination that can be computed on the lattice~\cite{Hobbs:2019gob}. 

These observations remain partially true with respect to the total sensitivity of the E866 ratio to individual PDFs presented in Fig.~\ref{fig:compar_E866}.  
At $x>0.05$ --- the region of kinematic coverage by the experiment --- the pulls in CT18 are in opposing directions for the $\bar u$ and $\bar d$ flavors,
while MSHT20 very predominantly prefers lower values of both $d$ and $\bar d$, with weaker sensitivity to $u$ and $\bar u$. (The overall pulls are not large, but this is because this data set has few points and imposes a strong constraint on the central PDF, so little tension exists.) 
At least some of this difference is likely due to the PDF parametrizations. Different flavor combinations are parametrized at input by CT and MSHT, with the former parametrizing $\bar{u}$ and $\bar{d}$ separately and requiring the asymptotic $x\to 0$ or $x \to 1$ behavior of $\bar u$ and $\bar d$ to coincide~\cite[see App. C.~in Ref.][]{Hou:2019efy}. This places some constraint on $\bar{d}/\bar{u}$ at high $x$. Instead, MSHT effectively parametrizes the sum and ratio $\bar{d}/\bar{u}$, with the latter having a somewhat freer high-$x$ behavior as determined by the PDF fit. The plots of $\bar d /\bar u$ in Fig.~\ref{fig:PDF_comp_DbovUb} show that consequently there is close agreement between all PDFs below $x\!=\!0.2$, while substantial differences persist at larger $x$ with poor experimental constraints. The E906/SeaQuest ratio largely compensates the pulls of the E866 ratio in the CT18As\_Lat fit; see Fig.~\ref{fig:compar_db2ub_CT} and Fig.~SM.3 in the Supplemental Material.
Finally, we note that the E866 sensitivities also propagate to $x<0.01$ (beyond the actual kinematic reach), most likely through extrapolation according to the given functional form. Here,
both CT18 and MSHT PDF sets show a pull towards higher $\bar u$ compared to $\bar d$. However, this derives mainly from pull on $\bar d$ (and $d$) for CT18 and on $\bar u$ (and $u$) for MSHT20.  

The examples above illustrate how the sensitivities can reveal subtle dynamics among the experimental constraints in PDF fits. In the Supplemental material, we collect such comparisons for several other interesting experiments: LHCb production of $W, Z$ bosons at 8 TeV; D{\O} Run-2 charge asymmetry at the lepton and vector boson level; additional plots of PDF ratios for the E866 and E906 measurements of $\sigma_{pd}/\sigma_{pp}$ in Drell-Yan process; and $t\bar t$ production data sets in the ATLASpdf21 analysis.

\section{Conclusions \label{sec:Conclusions}}

In this article, we explored a new statistical technique to ``look inside'' the global fits that determine parton distribution functions for a variety of applications. 
Such examinations are important both for the PDF fitter and the PDF user, for instance,
to understand the role of each fitted data set and emergence of new constraints as more data sets are added. Tools such as the Lagrange Multiplier (LM) method have been extensively used by the PDF fitters for this purpose, but are not accessible to the PDF user. In addition, the LM scan is CPU-intensive and is limited to a single kinematic point at a time.  

To elucidate these issues, authors of the ATLASpdf, CT and MSHT NNLO PDF ensembles have presented a study using the $L_2$ sensitivity method to explore the relative importance of each data set used in their PDF fits. The $L_2$ sensitivity, or $S_{f, L2}$, can be easily computed using the Hessian error PDFs published along with each central PDF set. These error PDF sets are already extensively used in calculations and comparisons at the LHC. 

$S_{f, L2}$ estimates statistical pulls of one or several data sets on the best-fit PDF at a particular $x$ and $Q$ point, parton-parton luminosity, or another PDF-dependent quantity. It reflects both the power of that data set to constrain the PDF, i.e., its statistical power in the fit, as well as the correlation between the PDF and log-likelihood of the data set. Unlike the LM scan, the $L_2$ sensitivity covers the entire partonic $x$ range in one (quick) evaluation.  An examination of $S_{f, L2}$ indicates which experiments are important in a given $x$ range, and which data sets are competing or in ``tension'' with each other, i.e. one preferring to pull the PDF up (resulting in a negative $S_{f, L2}$), and another preferring to pull it down (a positive $S_{f, L2}$). Opposite pulls on the best-fit PDFs by the subsets of global data are present in any analysis. Excessive pulls indicate systematic disagreements, and $S_{f, L2}$ estimates the degree of such tensions. The sum of all pulls, i.e. the values of $S_{f,L2}$, should be close to zero according to the inequality in Eq.~(\ref{sumSFL2}). Checking this inequality in fact serves as a validation test of the Hessian error set. While in this study we focused on special PDFs obtained with the simplified global tolerance of about 10, the method works with the standard error definitions as well.

The $L_2$ sensitivity method is fast because it relies on the linear approximation for the $\chi^2$ functions of individual experiments in the close neighborhood of the global minimum. For the experiments that are not perfectly consistent, the linear pieces dominate their $\chi^2$ near the global minimum. That said, it is not uncommon that a highly constraining experiment agrees well with the best-fit PDF and renders a low $S_{f,L2}$ value in some $x$ region while still imposing strong constraints on the net PDF uncertainty via quadratic and higher-order terms. This has been illustrated by the influence of the CMS 8 TeV jets data set  on the gluon PDF at $x=0.01$ vs. $x=0.3$ at $Q=125$ GeV in Sec.~\ref{sec:MoreLML2Relation}. The $L_2$ sensitivity, in its role of the linearized indicator of the local agreement among the experiments, thus offers a tool that complements the methods testing quadratic dependence.

The nonlinear $\chi^2$ contributions become numerically important when some error sets substantially deviate from the global minimum. This happens when the $L_2$ sensitivity is computed using Monte Carlo replicas, as discussed in Sec.~\ref{sec:MCMethod}. The appendix shows how to organize sampling of Monte Carlo replicas in many parameter dimensions to prevent large deviations from the global minimum and obtain meaningful sensitivities.

In this work and on the companion website \cite{L2website}, we compare $L_2$ sensitivities from the non-global and global Hessian fits of three groups in the formats showing either the sensitivities of leading experiments to a given PDF flavor or PDF combination, or for all PDF flavors and a particular experiment. We demonstrate how the $L_2$ sensitivity confirms and expands conclusions made based on the traditional examination of $\chi^2$ for experiments and LM scans. 
The $L_2$ sensitivity can provide information on the impact of PDF parametrization choices and 
different perturbative orders. In this paper, this was examined in the context of comparing CT18, CT18As, and CT18As\_Lat NNLO PDFs (with the latter one including independent parametrizations of strange quark and antiquark PDFs together with the lattice QCD prediction for the large-$x$ strangeness asymmetry), as well as the MSHT20 NNLO and aN3LO PDFs. 
One interesting result is the examination of pulls on strangeness and other sea PDFs in the various fits using the same $L_2$ metric. Another is the evidence of reduced tensions in the MSHT aN3LO PDFs, compared to the corresponding NNLO ones; this is most clearly seen in the strong decrease of the value of $S_{f,L2}$ for the gluon distribution for the ATLAS Z $p_T$ distribution from NNLO to aN3LO, perhaps indicating a release of tension by transitioning to the (approximate) higher order in the PDF fit. 

To conclude, the $L_2$ sensitivities offer any user of Hessian PDFs valuable insights using the tools outside of the PDF fit. Additional comparisons are provided on the website \cite{L2website} and in the Supplemental material. The website also provides a C++ program that computes or plots sensitivities using tabulated vectors of $\chi^2$ and error PDFs in either LHAPDF or tabulated format.

\section*{Acknowledgments}
We thank Yao Fu for help with some shown PDF sets, Jun Gao for information about the Lagrange Multiplier scans, and A. Accardi, A. Deiana-McCarn, R. Gal, E. Nocera, F. Olness, and members of the PDF4LHC21 group for stimulating exchanges. 
A.C. is supported by the UNAM Grant No. DGAPA-PAPIIT IN111222 and CONACyT Ciencia de Frontera
2019 No. 51244 (FORDECYT-PRONACES). 
A. C-S. wishes to thank the Leverhulme Trust for support.
T.C. acknowledges that his work and this project has received funding from the European Research Council (ERC) under the European Union's Horizon 2020 research and innovation programme (Grant agreement No. 101002090 COLORFREE).
The work of T.J.H. at Argonne National Laboratory was supported by the U.S.~Department of Energy, Office of Science, under Contract No.~DE-AC02-06CH11357.
L. H.-L. and R.S.T. thank STFC for support via grant
award ST/T000856/1. 
X.J. and P.M.N. are partially supported by the U.S. Department of Energy under Grant No.~DE-SC0010129. P.M.N. is also supported by the Fermilab URA award, using the resources of the Fermi National Accelerator Laboratory (Fermilab), a U.S. Department of Energy, Office of Science, HEP User Facility. Fermilab is managed by Fermi Research Alliance, LLC (FRA), acting under Contract No. DE-AC02-07CH11359.
The work of K.X. is supported by the U.S. Department
of Energy under grant No. DE-SC0007914, the U.S. National Science Foundation under Grants No. PHY-1820760 and also in part by PITT PACC.
C.-P.Y. was supported by the U.S. National Science Foundation under Grant No.~PHY-2013791 as well as the Wu-Ki Tung endowed chair in particle physics.
%

\clearpage \newpage
\begin{appendix}

\section{From a Hessian to a Monte-Carlo PDF ensemble \label{sec:Hessian2MC}}

In this article, we focused on the exploration of the $L_2$ sensitivity using error PDF sets in the Hessian approximation. We pointed out in Sec.~\ref{sec:MCMethod} that the $L_2$ sensitivity can also be defined in the MC method, while its calculation in this method requires reducing the impact of non-linearities.  
     
If $X(\vec R)$ is a non-linear function of the PDF parameters $\vec R$, the cumulative probability ${\cal P}(X(\vec R))$ is generally not the same as ${\cal P} (\vec R)$, implying that the PDF parameters at a certain confidence level for $\vec R$ correspond to a different confidence level for $X(\vec R)$ \cite{Hou:2016sho}. As a flip side, a linear function $X(\vec R)$ of the normally distributed $\vec R$ is also normally distributed. Consequently, the error PDFs work the best when the variations of $X(\vec R)$ from the best-fit value $X_0$ are approximately linear.  An MC replica ensemble that is generated from a Hessian eigenvector ensemble by following one of the available methods  \cite{Watt:2012tq,Hou:2016sho} then closely reproduces the Hessian PDF uncertainty and correlations. 
     
In some situations the normally sampled MC replicas lead to unphysical results, for example, by predicting negative cross sections with a non-zero probability at large Bjorken $x$. In this case, MC sampling according to an alternative, e.g., log-normal, probability distribution is needed to get the physical behavior \cite{Hou:2016sho}. 
     
The log-likelihood function $\chi^2_E$, with its quadratic and higher-order terms, is particularly prone to non-linear effects. Indeed, nearly all MC replicas $f^{(k)}$ generated with a standard method deviate from the best fit by several standard deviations in some directions because of the large number of PDF parameters $D \sim 30$ in a typical fit \cite{Hou:2016sho}. For these replicas, the $\chi^2_E(f^{(k)})$ functions are large and non-linear. To demonstrate the equivalence of the Hessian and MC sensitivities, one must resort to an alternative method that stays close enough to the global minimum to suppress the quadratic component.

When we use the normal sampling, we consider $N_{{\rm rep}}$ sets of the PDF vectors $\vec{R}^{(k)}\equiv \{R_1^{(k)},...,R_D^{(k)}\}$ that are  randomly sampled from the $D$-dimensional standard normal distribution ${\cal N}_D(0,1)$ associated with the probability density $\mathcal{P}(\vec{R})= (2\pi)^{-D/2}\,\exp\left(-\frac{1}{2}\sum_{i=1}^{D}R_i^2\right)$.
For a PDF $f(\vec R)$, the MC replica can be constructed as\footnote{In some replica generation prescriptions \cite{Hou:2016sho}, the random variation for the MC replica also contains a constant shift from the central value that vanishes at $N_{\rm rep}\to \infty$. This shift does not modify our conclusions.} 
\begin{equation}
    f^{(k)} \equiv  f_0 + \sum_{i=1}^{D} \frac{f_{+i} -f_{-i}}{2}R_i^{(k)},
\end{equation}
with only the linear term kept on the right-hand side. 
For $N_{\rm rep}\to \infty$, the MC estimators for the mean PDF $\langle f\rangle$, the PDF uncertainty $\delta_{\rm MC} f$, and PDF-PDF correlation $C_{\rm MC}(f_1,f_2)$,  given respectively in Eqs.~(\ref{X0MC}-\ref{CorrMC}), coincide with the Hessian $f_0$, $\delta_{\rm H}f$ and $C_{\rm H}(f_1,f_2)$
in Eqs.~(\ref{dHessian}) and (\ref{CorrHessian}) \cite{JingPhdthesis2021}. 
To demonstrate this, we use 
\begin{align}
\lim_{N_{\rm rep}\to \infty} \langle f\rangle =
\lim_{N_{\rm rep}\to \infty} \frac{1}{N_{\rm rep}} \sum_{k=1}^{N_{\rm rep}}f^{(k)} = \int d^D R\ f(\vec R)\  {\cal P} (\vec R) 
\end{align}
and 
\begin{equation}
\langle R_i \rangle = 0, \quad \langle R_i^2 \rangle = 1, \quad \langle \vec R^2\rangle = \sum_{i=1}^D \langle  R_i^2 \rangle = D \mbox{ for } \vec R\sim {\cal N}_D(0,1). 
\end{equation}

The $\chi^2_E$ function for such replicas contains a non-linear component ${\cal O}(R^2)$ that is at least comparable to the linear term if $D$ is large:
\begin{equation}
  \chi^{2\ (k)}_E\equiv \chi^2_E(f^{(k)})\approx \chi^2_E(f_0) + \sum_{i=1}^{D} \frac{\chi^2_{E,+i} -\chi^2_{E,-i}}{2}R_i^{(k)}+{\cal O}(R^2).
\end{equation}
We wish to suppress the spurious non-linear component for the reasons discussed above. To stay in the close vicinity of the global minimum, where the  ${\cal O}(R^2)$ component is small, we can generate the MC replicas according to the uniform distribution ${\cal U}_D(R^2=1)$ on the tolerance hypersphere instead of ${\cal N}_D(0,1)$. The probability density associated with ${\cal U}_D(R^2=1)$ is ${\cal P}=1/\Omega_D$, where $\Omega_D=2\pi^{D/2}/\Gamma(D/2)$ is the full solid angle in $D$ dimensions. We also have
\begin{equation}
\langle R_i \rangle = 0, \quad \langle R_i^2 \rangle = 1/D, \quad \langle \vec R^2\rangle  = 1 \mbox{ for } \vec R\sim {\cal U}_D(R^2=1),  
\end{equation}
from which it follows that the uncertainty values on the two probability distributions are related as 
\begin{equation}
\left. \delta_{\rm MC} f \right|_{{\cal N}_D(0,1)} = \sqrt{D} \left. \delta_{\rm MC} f \right|_{{\cal U}_D(R^2=1)} + {\cal O}(R^2).
\end{equation}
We now discard the ${\cal O}(R^2)$ terms on the right-hand side, retaining only the sought linear part of the expectation value. 
This prescription gives us a trustworthy  MC formula for the $L_2$ sensitivity,
\begin{equation}\label{SFL2MC2}
S_{f,L2}^{\rm MC}(E) = \sqrt{D} \,\,\left( \delta_{\rm MC} \chi^2_E \right)\,\, C_{\rm MC}(f,\chi^2_E),
\end{equation}
where the MC replicas are generated from the uniform distribution ${\cal U}_D(R^2=1)$ on the tolerance hypersphere. 

The MC sensitivity in Eq.~(\ref{SFL2MC2}) is close to the Hessian one because (a) the means of the linear terms in Eq.~(\ref{SFL2MC2}) are related by a constant factor, since both ${\cal N}_D(0,1)$ and ${\cal U}_D(R^2=1)$ distributions are spherically symmetric; (b) the non-linear terms can be suppressed on the tolerance hypersphere by choosing a small $T^2$. 

\begin{figure*}[th]
\centering
\includegraphics[scale=0.5]{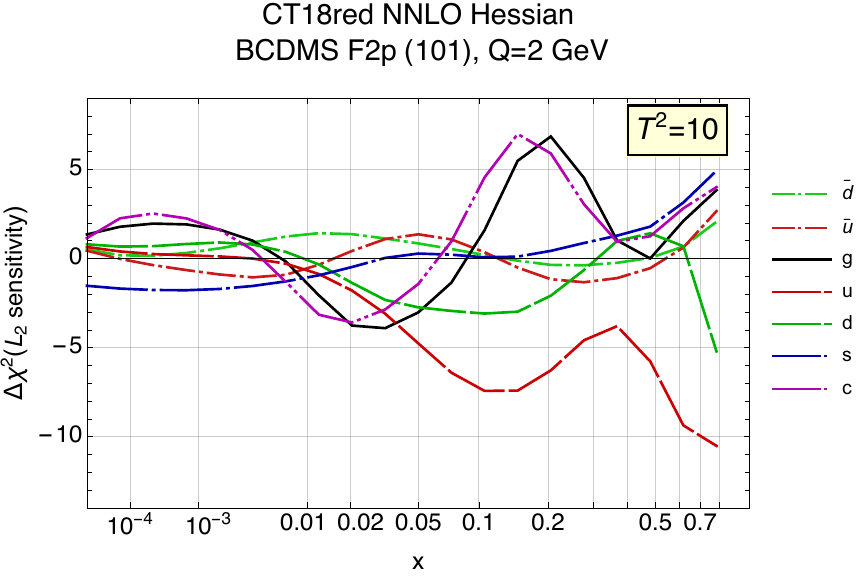}
\includegraphics[scale=0.5]{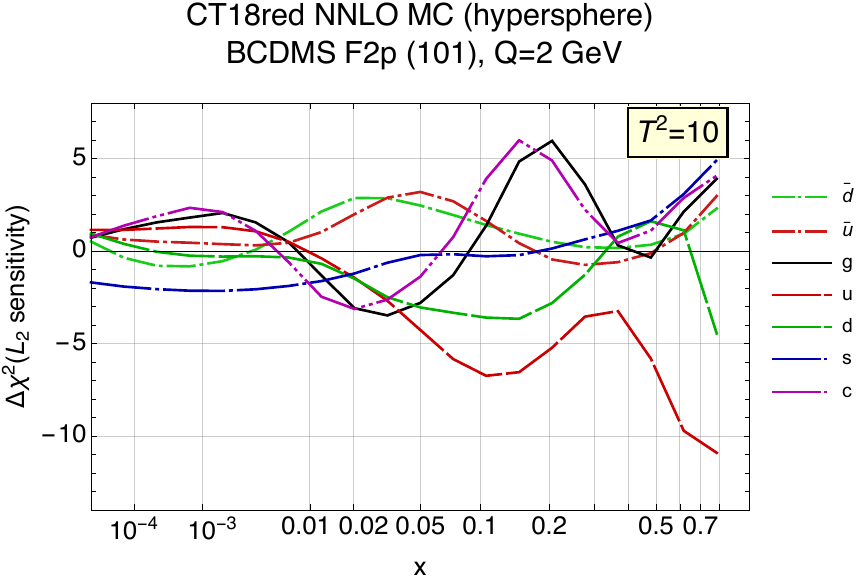}\\
 {\small (a)} \hspace{2.6in} {\small (b)}\\
\includegraphics[scale=0.5]{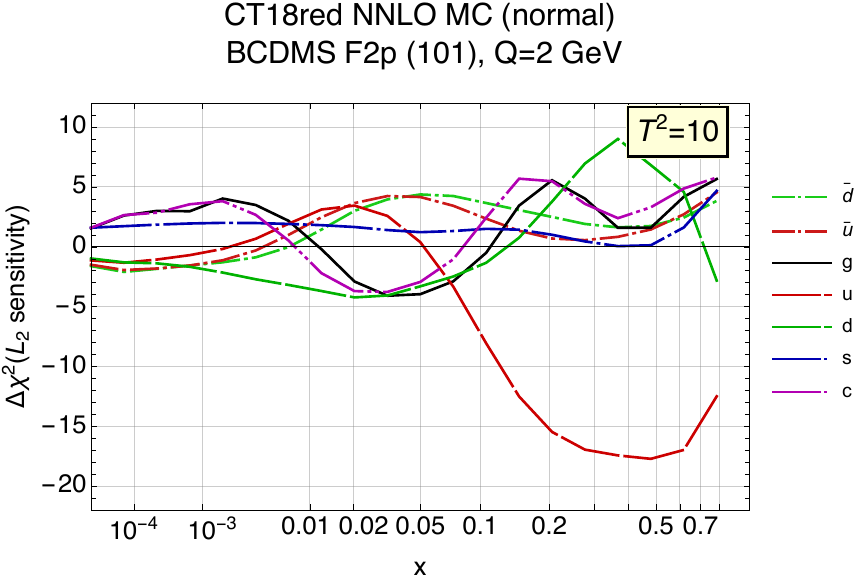}\\
 {\small (c)}
\caption{$L_2$ sensitivities for the BCDMS $F_2^p(x,Q^2)$ data set and CT18 NNLO PDFs with $T^2=10$ in (a) the Hessian representation; (b,c) MC representations generated with the uniform distribution ${\cal U}_D(R^2=1)$ on the tolerance hypersphere and the standard normal distribution ${\cal N}_D(0,1)$.
}
\label{fig:Hessian_to_MC_sensitivity}
\end{figure*}

To illustrate that this prescription is successful, we compare the Hessian $L_2$ sensitivities for CT18 NNLO and those computed with the MC replicas that were generated using either ${\cal U}_D(R^2=1)$ or ${\cal N}_D(0,1)$. Figure~\ref{fig:Hessian_to_MC_sensitivity} presents such comparison for the BCDMS $F_2^p(x,Q^2)$ data set [ID=101]. The MC representation obtained with ${\cal U}_D(R^2=1)$ in Fig.~\ref{fig:Hessian_to_MC_sensitivity}(b) closely reproduces the Hessian sensitivity in Fig.~\ref{fig:Hessian_to_MC_sensitivity}(a). Similar level of agreement has been observed for the $L_2$ sensitivities of the other experiments. 

In contrast,  the sensitivities for the MC representation based on ${\cal N}_D(0,1)$ in Fig.~\ref{fig:Hessian_to_MC_sensitivity}(c) show larger deviations from the Hessian result, with especially manifest differences present in the $d$- and $u$-quark sensitivities at large $x$. These discrepancies reflect the large total number $D$ of the PDF parameters, which should be contrasted with the good agreement between the Hessian and MC sensitivities in the 1-dimensional case demonstrated in Sec.~\ref{sec:SFL2OneDim}. 

Estimation of the $L_2$ sensitivity with the MC replicas therefore requires a sampling algorithm that populates well the close vicinity of the global minimum and follows an isotropic probability distribution that can be related to ${\cal N}_D(0,1)$ in the linear approximation. These conditions are not automatic when using the commonly available MC replicas. We have shown how to realize them when converting a Hessian error set into an MC one. The form of the $\chi^2$ or cost function during training of MC replicas may differ from the one used in the post-fit computations of the $L_2$ sensitivity. The dependence on the $\chi^2$ definition is not insignificant \cite{Courtoy:2022ocu}.

\end{appendix}
\clearpage \newpage


\begin{thebibliography}{164}
\expandafter\ifx\csname natexlab\endcsname\relax\def\natexlab#1{#1}\fi
\expandafter\ifx\csname bibnamefont\endcsname\relax
  \def\bibnamefont#1{#1}\fi
\expandafter\ifx\csname bibfnamefont\endcsname\relax
  \def\bibfnamefont#1{#1}\fi
\expandafter\ifx\csname citenamefont\endcsname\relax
  \def\citenamefont#1{#1}\fi
\expandafter\ifx\csname url\endcsname\relax
  \def\url#1{\texttt{#1}}\fi
\expandafter\ifx\csname urlprefix\endcsname\relax\def\urlprefix{URL }\fi
\providecommand{\bibinfo}[2]{#2}
\providecommand{\eprint}[2][]{\url{#2}}

\bibitem[{\citenamefont{Pumplin et~al.}(2001)\citenamefont{Pumplin, Stump,
  Brock, Casey, Huston, Kalk, Lai, and Tung}}]{Pumplin:2001ct}
\bibinfo{author}{\bibfnamefont{J.}~\bibnamefont{Pumplin}},
  \bibinfo{author}{\bibfnamefont{D.}~\bibnamefont{Stump}},
  \bibinfo{author}{\bibfnamefont{R.}~\bibnamefont{Brock}},
  \bibinfo{author}{\bibfnamefont{D.}~\bibnamefont{Casey}},
  \bibinfo{author}{\bibfnamefont{J.}~\bibnamefont{Huston}},
  \bibinfo{author}{\bibfnamefont{J.}~\bibnamefont{Kalk}},
  \bibinfo{author}{\bibfnamefont{H.-L.} \bibnamefont{Lai}}, \bibnamefont{and}
  \bibinfo{author}{\bibfnamefont{W.-K.} \bibnamefont{Tung}},
  \bibinfo{journal}{Phys. Rev.} \textbf{\bibinfo{volume}{D65}},
  \bibinfo{pages}{014013} (\bibinfo{year}{2001}), \eprint{hep-ph/0101032}.

\bibitem[{\citenamefont{Pumplin et~al.}(2002)\citenamefont{Pumplin, Stump,
  Huston, Lai, Nadolsky, and Tung}}]{Pumplin:2002vw}
\bibinfo{author}{\bibfnamefont{J.}~\bibnamefont{Pumplin}},
  \bibinfo{author}{\bibfnamefont{D.~R.} \bibnamefont{Stump}},
  \bibinfo{author}{\bibfnamefont{J.}~\bibnamefont{Huston}},
  \bibinfo{author}{\bibfnamefont{H.~L.} \bibnamefont{Lai}},
  \bibinfo{author}{\bibfnamefont{P.~M.} \bibnamefont{Nadolsky}},
  \bibnamefont{and} \bibinfo{author}{\bibfnamefont{W.-K.} \bibnamefont{Tung}},
  \bibinfo{journal}{JHEP} \textbf{\bibinfo{volume}{07}}, \bibinfo{pages}{012}
  (\bibinfo{year}{2002}), \eprint{hep-ph/0201195}.

\bibitem[{\citenamefont{Giele and Keller}(1998)}]{Giele:1998gw}
\bibinfo{author}{\bibfnamefont{W.~T.} \bibnamefont{Giele}} \bibnamefont{and}
  \bibinfo{author}{\bibfnamefont{S.}~\bibnamefont{Keller}},
  \bibinfo{journal}{Phys. Rev. D} \textbf{\bibinfo{volume}{58}},
  \bibinfo{pages}{094023} (\bibinfo{year}{1998}), \eprint{hep-ph/9803393}.

\bibitem[{\citenamefont{Nadolsky et~al.}(2008)\citenamefont{Nadolsky, Lai, Cao,
  Huston, Pumplin, Stump, Tung, and Yuan}}]{Nadolsky:2008zw}
\bibinfo{author}{\bibfnamefont{P.~M.} \bibnamefont{Nadolsky}},
  \bibinfo{author}{\bibfnamefont{H.-L.} \bibnamefont{Lai}},
  \bibinfo{author}{\bibfnamefont{Q.-H.} \bibnamefont{Cao}},
  \bibinfo{author}{\bibfnamefont{J.}~\bibnamefont{Huston}},
  \bibinfo{author}{\bibfnamefont{J.}~\bibnamefont{Pumplin}},
  \bibinfo{author}{\bibfnamefont{D.}~\bibnamefont{Stump}},
  \bibinfo{author}{\bibfnamefont{W.-K.} \bibnamefont{Tung}}, \bibnamefont{and}
  \bibinfo{author}{\bibfnamefont{C.~P.} \bibnamefont{Yuan}},
  \bibinfo{journal}{Phys. Rev.} \textbf{\bibinfo{volume}{D78}},
  \bibinfo{pages}{013004} (\bibinfo{year}{2008}), \eprint{0802.0007}.

\bibitem[{\citenamefont{Wang et~al.}(2018)\citenamefont{Wang, Hobbs, Doyle,
  Gao, Hou, Nadolsky, and Olness}}]{Wang:2018heo}
\bibinfo{author}{\bibfnamefont{B.-T.} \bibnamefont{Wang}},
  \bibinfo{author}{\bibfnamefont{T.~J.} \bibnamefont{Hobbs}},
  \bibinfo{author}{\bibfnamefont{S.}~\bibnamefont{Doyle}},
  \bibinfo{author}{\bibfnamefont{J.}~\bibnamefont{Gao}},
  \bibinfo{author}{\bibfnamefont{T.-J.} \bibnamefont{Hou}},
  \bibinfo{author}{\bibfnamefont{P.~M.} \bibnamefont{Nadolsky}},
  \bibnamefont{and} \bibinfo{author}{\bibfnamefont{F.~I.}
  \bibnamefont{Olness}}, \bibinfo{journal}{Phys. Rev.}
  \textbf{\bibinfo{volume}{D98}}, \bibinfo{pages}{094030}
  (\bibinfo{year}{2018}), \eprint{1803.02777}.

\bibitem[{\citenamefont{Hobbs et~al.}(2019)\citenamefont{Hobbs, Wang, Nadolsky,
  and Olness}}]{Hobbs:2019gob}
\bibinfo{author}{\bibfnamefont{T.~J.} \bibnamefont{Hobbs}},
  \bibinfo{author}{\bibfnamefont{B.-T.} \bibnamefont{Wang}},
  \bibinfo{author}{\bibfnamefont{P.~M.} \bibnamefont{Nadolsky}},
  \bibnamefont{and} \bibinfo{author}{\bibfnamefont{F.~I.}
  \bibnamefont{Olness}}, \bibinfo{journal}{Phys. Rev. D}
  \textbf{\bibinfo{volume}{100}}, \bibinfo{pages}{094040}
  (\bibinfo{year}{2019}), \eprint{1904.00022}.

\bibitem[{\citenamefont{Collins and Pumplin}(2001)}]{Collins:2001es}
\bibinfo{author}{\bibfnamefont{J.~C.} \bibnamefont{Collins}} \bibnamefont{and}
  \bibinfo{author}{\bibfnamefont{J.}~\bibnamefont{Pumplin}}
  (\bibinfo{year}{2001}), \eprint{hep-ph/0105207}.

\bibitem[{\citenamefont{Martin et~al.}(2009)\citenamefont{Martin, Stirling,
  Thorne, and Watt}}]{Martin:2009iq}
\bibinfo{author}{\bibfnamefont{A.~D.} \bibnamefont{Martin}},
  \bibinfo{author}{\bibfnamefont{W.~J.} \bibnamefont{Stirling}},
  \bibinfo{author}{\bibfnamefont{R.~S.} \bibnamefont{Thorne}},
  \bibnamefont{and} \bibinfo{author}{\bibfnamefont{G.}~\bibnamefont{Watt}},
  \bibinfo{journal}{Eur. Phys. J.} \textbf{\bibinfo{volume}{C63}},
  \bibinfo{pages}{189} (\bibinfo{year}{2009}), \eprint{0901.0002}.

\bibitem[{\citenamefont{Kova\v{r}\'\i{}k
  et~al.}(2020)\citenamefont{Kova\v{r}\'\i{}k, Nadolsky, and
  Soper}}]{Kovarik:2019xvh}
\bibinfo{author}{\bibfnamefont{K.}~\bibnamefont{Kova\v{r}\'\i{}k}},
  \bibinfo{author}{\bibfnamefont{P.~M.} \bibnamefont{Nadolsky}},
  \bibnamefont{and} \bibinfo{author}{\bibfnamefont{D.~E.} \bibnamefont{Soper}},
  \bibinfo{journal}{Rev. Mod. Phys.} \textbf{\bibinfo{volume}{92}},
  \bibinfo{pages}{045003} (\bibinfo{year}{2020}), \eprint{1905.06957}.

\bibitem[{\citenamefont{Stump et~al.}(2001)\citenamefont{Stump, Pumplin, Brock,
  Casey, Huston, Kalk, Lai, and Tung}}]{Stump:2001gu}
\bibinfo{author}{\bibfnamefont{D.}~\bibnamefont{Stump}},
  \bibinfo{author}{\bibfnamefont{J.}~\bibnamefont{Pumplin}},
  \bibinfo{author}{\bibfnamefont{R.}~\bibnamefont{Brock}},
  \bibinfo{author}{\bibfnamefont{D.}~\bibnamefont{Casey}},
  \bibinfo{author}{\bibfnamefont{J.}~\bibnamefont{Huston}},
  \bibinfo{author}{\bibfnamefont{J.}~\bibnamefont{Kalk}},
  \bibinfo{author}{\bibfnamefont{H.-L.} \bibnamefont{Lai}}, \bibnamefont{and}
  \bibinfo{author}{\bibfnamefont{W.-K.} \bibnamefont{Tung}},
  \bibinfo{journal}{Phys. Rev.} \textbf{\bibinfo{volume}{D65}},
  \bibinfo{pages}{014012} (\bibinfo{year}{2001}), \eprint{hep-ph/0101051}.

\bibitem[{\citenamefont{Paukkunen and Zurita}(2014)}]{Paukkunen:2014zia}
\bibinfo{author}{\bibfnamefont{H.}~\bibnamefont{Paukkunen}} \bibnamefont{and}
  \bibinfo{author}{\bibfnamefont{P.}~\bibnamefont{Zurita}},
  \bibinfo{journal}{JHEP} \textbf{\bibinfo{volume}{12}}, \bibinfo{pages}{100}
  (\bibinfo{year}{2014}), \eprint{1402.6623}.

\bibitem[{\citenamefont{Schmidt et~al.}(2018)\citenamefont{Schmidt, Pumplin,
  and Yuan}}]{Schmidt:2018hvu}
\bibinfo{author}{\bibfnamefont{C.}~\bibnamefont{Schmidt}},
  \bibinfo{author}{\bibfnamefont{J.}~\bibnamefont{Pumplin}}, \bibnamefont{and}
  \bibinfo{author}{\bibfnamefont{C.-P.} \bibnamefont{Yuan}},
  \bibinfo{journal}{Phys. Rev.} \textbf{\bibinfo{volume}{D98}},
  \bibinfo{pages}{094005} (\bibinfo{year}{2018}), \eprint{1806.07950}.

\bibitem[{\citenamefont{Hou et~al.}(2019)\citenamefont{Hou, Yu, Dulat, Schmidt,
  and Yuan}}]{Hou:2019gfw}
\bibinfo{author}{\bibfnamefont{T.-J.} \bibnamefont{Hou}},
  \bibinfo{author}{\bibfnamefont{Z.}~\bibnamefont{Yu}},
  \bibinfo{author}{\bibfnamefont{S.}~\bibnamefont{Dulat}},
  \bibinfo{author}{\bibfnamefont{C.}~\bibnamefont{Schmidt}}, \bibnamefont{and}
  \bibinfo{author}{\bibfnamefont{C.~P.} \bibnamefont{Yuan}},
  \bibinfo{journal}{Phys. Rev. D} \textbf{\bibinfo{volume}{100}},
  \bibinfo{pages}{114024} (\bibinfo{year}{2019}), \eprint{1907.12177}.

\bibitem[{\citenamefont{Ball et~al.}(2011)\citenamefont{Ball, Bertone, Cerutti,
  Del~Debbio, Forte, Guffanti, Latorre, Rojo, and Ubiali}}]{Ball:2010gb}
\bibinfo{author}{\bibfnamefont{R.~D.} \bibnamefont{Ball}},
  \bibinfo{author}{\bibfnamefont{V.}~\bibnamefont{Bertone}},
  \bibinfo{author}{\bibfnamefont{F.}~\bibnamefont{Cerutti}},
  \bibinfo{author}{\bibfnamefont{L.}~\bibnamefont{Del~Debbio}},
  \bibinfo{author}{\bibfnamefont{S.}~\bibnamefont{Forte}},
  \bibinfo{author}{\bibfnamefont{A.}~\bibnamefont{Guffanti}},
  \bibinfo{author}{\bibfnamefont{J.~I.} \bibnamefont{Latorre}},
  \bibinfo{author}{\bibfnamefont{J.}~\bibnamefont{Rojo}}, \bibnamefont{and}
  \bibinfo{author}{\bibfnamefont{M.}~\bibnamefont{Ubiali}}
  (\bibinfo{collaboration}{NNPDF}), \bibinfo{journal}{Nucl. Phys. B}
  \textbf{\bibinfo{volume}{849}}, \bibinfo{pages}{112} (\bibinfo{year}{2011}),
  \bibinfo{note}{[Erratum: Nucl.Phys.B 854, 926--927 (2012), Erratum:
  Nucl.Phys.B 855, 927--928 (2012)]}, \eprint{1012.0836}.

\bibitem[{\citenamefont{Ball et~al.}(2012)\citenamefont{Ball, Bertone, Cerutti,
  Del~Debbio, Forte, Guffanti, Hartland, Latorre, Rojo, and
  Ubiali}}]{Ball:2011gg}
\bibinfo{author}{\bibfnamefont{R.~D.} \bibnamefont{Ball}},
  \bibinfo{author}{\bibfnamefont{V.}~\bibnamefont{Bertone}},
  \bibinfo{author}{\bibfnamefont{F.}~\bibnamefont{Cerutti}},
  \bibinfo{author}{\bibfnamefont{L.}~\bibnamefont{Del~Debbio}},
  \bibinfo{author}{\bibfnamefont{S.}~\bibnamefont{Forte}},
  \bibinfo{author}{\bibfnamefont{A.}~\bibnamefont{Guffanti}},
  \bibinfo{author}{\bibfnamefont{N.~P.} \bibnamefont{Hartland}},
  \bibinfo{author}{\bibfnamefont{J.~I.} \bibnamefont{Latorre}},
  \bibinfo{author}{\bibfnamefont{J.}~\bibnamefont{Rojo}}, \bibnamefont{and}
  \bibinfo{author}{\bibfnamefont{M.}~\bibnamefont{Ubiali}},
  \bibinfo{journal}{Nucl. Phys. B} \textbf{\bibinfo{volume}{855}},
  \bibinfo{pages}{608} (\bibinfo{year}{2012}), \eprint{1108.1758}.

\bibitem[{\citenamefont{Sato et~al.}(2014)\citenamefont{Sato, Owens, and
  Prosper}}]{Sato:2013ika}
\bibinfo{author}{\bibfnamefont{N.}~\bibnamefont{Sato}},
  \bibinfo{author}{\bibfnamefont{J.}~\bibnamefont{Owens}}, \bibnamefont{and}
  \bibinfo{author}{\bibfnamefont{H.}~\bibnamefont{Prosper}},
  \bibinfo{journal}{Phys. Rev. D} \textbf{\bibinfo{volume}{89}},
  \bibinfo{pages}{114020} (\bibinfo{year}{2014}), \eprint{1310.1089}.

\bibitem[{\citenamefont{Hou et~al.}(2021)}]{Hou:2019efy}
\bibinfo{author}{\bibfnamefont{T.-J.} \bibnamefont{Hou}} \bibnamefont{et~al.},
  \bibinfo{journal}{Phys. Rev. D} \textbf{\bibinfo{volume}{103}},
  \bibinfo{pages}{014013} (\bibinfo{year}{2021}), \eprint{1912.10053}.

\bibitem[{\citenamefont{Ball et~al.}(2022)}]{PDF4LHCWorkingGroup:2022cjn}
\bibinfo{author}{\bibfnamefont{R.~D.} \bibnamefont{Ball}} \bibnamefont{et~al.}
  (\bibinfo{collaboration}{PDF4LHC Working Group}), \bibinfo{journal}{J. Phys.
  G} \textbf{\bibinfo{volume}{49}}, \bibinfo{pages}{080501}
  (\bibinfo{year}{2022}), \eprint{2203.05506}.

\bibitem[{\citenamefont{Accardi et~al.}(2021)\citenamefont{Accardi, Hobbs,
  Jing, and Nadolsky}}]{Accardi:2021ysh}
\bibinfo{author}{\bibfnamefont{A.}~\bibnamefont{Accardi}},
  \bibinfo{author}{\bibfnamefont{T.~J.} \bibnamefont{Hobbs}},
  \bibinfo{author}{\bibfnamefont{X.}~\bibnamefont{Jing}}, \bibnamefont{and}
  \bibinfo{author}{\bibfnamefont{P.~M.} \bibnamefont{Nadolsky}},
  \bibinfo{journal}{Eur. Phys. J. C} \textbf{\bibinfo{volume}{81}},
  \bibinfo{pages}{603} (\bibinfo{year}{2021}), \eprint{2102.01107}.

\bibitem[{\citenamefont{Aad et~al.}(2022)}]{ATLAS:2021vod}
\bibinfo{author}{\bibfnamefont{G.}~\bibnamefont{Aad}} \bibnamefont{et~al.}
  (\bibinfo{collaboration}{ATLAS}), \bibinfo{journal}{Eur. Phys. J. C}
  \textbf{\bibinfo{volume}{82}}, \bibinfo{pages}{438} (\bibinfo{year}{2022}),
  \eprint{2112.11266}.

\bibitem[{\citenamefont{Bailey et~al.}(2021)\citenamefont{Bailey, Cridge,
  Harland-Lang, Martin, and Thorne}}]{Bailey:2020ooq}
\bibinfo{author}{\bibfnamefont{S.}~\bibnamefont{Bailey}},
  \bibinfo{author}{\bibfnamefont{T.}~\bibnamefont{Cridge}},
  \bibinfo{author}{\bibfnamefont{L.~A.} \bibnamefont{Harland-Lang}},
  \bibinfo{author}{\bibfnamefont{A.~D.} \bibnamefont{Martin}},
  \bibnamefont{and} \bibinfo{author}{\bibfnamefont{R.~S.}
  \bibnamefont{Thorne}}, \bibinfo{journal}{Eur. Phys. J. C}
  \textbf{\bibinfo{volume}{81}}, \bibinfo{pages}{341} (\bibinfo{year}{2021}),
  \eprint{2012.04684}.

\bibitem[{\citenamefont{Hou et~al.}(2023)\citenamefont{Hou, Lin, Yan, and
  Yuan}}]{Hou:2022onq}
\bibinfo{author}{\bibfnamefont{T.-J.} \bibnamefont{Hou}},
  \bibinfo{author}{\bibfnamefont{H.-W.} \bibnamefont{Lin}},
  \bibinfo{author}{\bibfnamefont{M.}~\bibnamefont{Yan}}, \bibnamefont{and}
  \bibinfo{author}{\bibfnamefont{C.~P.} \bibnamefont{Yuan}},
  \bibinfo{journal}{Phys. Rev. D} \textbf{\bibinfo{volume}{107}},
  \bibinfo{pages}{076018} (\bibinfo{year}{2023}), \eprint{2211.11064}.

\bibitem[{\citenamefont{McGowan et~al.}(2022)\citenamefont{McGowan, Cridge,
  Harland-Lang, and Thorne}}]{McGowan:2022nag}
\bibinfo{author}{\bibfnamefont{J.}~\bibnamefont{McGowan}},
  \bibinfo{author}{\bibfnamefont{T.}~\bibnamefont{Cridge}},
  \bibinfo{author}{\bibfnamefont{L.~A.} \bibnamefont{Harland-Lang}},
  \bibnamefont{and} \bibinfo{author}{\bibfnamefont{R.~S.} \bibnamefont{Thorne}}
  (\bibinfo{year}{2022}), \eprint{2207.04739}.

\bibitem[{\citenamefont{Hou et~al.}(2017{\natexlab{a}})}]{Hou:2016sho}
\bibinfo{author}{\bibfnamefont{T.-J.} \bibnamefont{Hou}} \bibnamefont{et~al.},
  \bibinfo{journal}{JHEP} \textbf{\bibinfo{volume}{03}}, \bibinfo{pages}{099}
  (\bibinfo{year}{2017}{\natexlab{a}}), \eprint{1607.06066}.

\bibitem[{\citenamefont{Nadolsky and Sullivan}(2001)}]{Nadolsky:2001yg}
\bibinfo{author}{\bibfnamefont{P.~M.} \bibnamefont{Nadolsky}} \bibnamefont{and}
  \bibinfo{author}{\bibfnamefont{Z.}~\bibnamefont{Sullivan}},
  \bibinfo{journal}{eConf} \textbf{\bibinfo{volume}{C010630}},
  \bibinfo{pages}{P510} (\bibinfo{year}{2001}), \eprint{hep-ph/0110378}.

\bibitem[{\citenamefont{Watt and Thorne}(2012)}]{Watt:2012tq}
\bibinfo{author}{\bibfnamefont{G.}~\bibnamefont{Watt}} \bibnamefont{and}
  \bibinfo{author}{\bibfnamefont{R.~S.} \bibnamefont{Thorne}},
  \bibinfo{journal}{JHEP} \textbf{\bibinfo{volume}{08}}, \bibinfo{pages}{052}
  (\bibinfo{year}{2012}), \eprint{1205.4024}.

\bibitem[{\citenamefont{Gao and Nadolsky}(2014)}]{Gao:2013bia}
\bibinfo{author}{\bibfnamefont{J.}~\bibnamefont{Gao}} \bibnamefont{and}
  \bibinfo{author}{\bibfnamefont{P.}~\bibnamefont{Nadolsky}},
  \bibinfo{journal}{JHEP} \textbf{\bibinfo{volume}{07}}, \bibinfo{pages}{035}
  (\bibinfo{year}{2014}), \eprint{1401.0013}.

\bibitem[{\citenamefont{Carrazza et~al.}(2015)\citenamefont{Carrazza, Forte,
  Kassabov, Latorre, and Rojo}}]{Carrazza:2015aoa}
\bibinfo{author}{\bibfnamefont{S.}~\bibnamefont{Carrazza}},
  \bibinfo{author}{\bibfnamefont{S.}~\bibnamefont{Forte}},
  \bibinfo{author}{\bibfnamefont{Z.}~\bibnamefont{Kassabov}},
  \bibinfo{author}{\bibfnamefont{J.~I.} \bibnamefont{Latorre}},
  \bibnamefont{and} \bibinfo{author}{\bibfnamefont{J.}~\bibnamefont{Rojo}},
  \bibinfo{journal}{Eur. Phys. J. C} \textbf{\bibinfo{volume}{75}},
  \bibinfo{pages}{369} (\bibinfo{year}{2015}), \eprint{1505.06736}.

\bibitem[{\citenamefont{Carrazza et~al.}(2016)\citenamefont{Carrazza, Forte,
  Kassabov, and Rojo}}]{Carrazza:2016htc}
\bibinfo{author}{\bibfnamefont{S.}~\bibnamefont{Carrazza}},
  \bibinfo{author}{\bibfnamefont{S.}~\bibnamefont{Forte}},
  \bibinfo{author}{\bibfnamefont{Z.}~\bibnamefont{Kassabov}}, \bibnamefont{and}
  \bibinfo{author}{\bibfnamefont{J.}~\bibnamefont{Rojo}},
  \bibinfo{journal}{Eur. Phys. J. C} \textbf{\bibinfo{volume}{76}},
  \bibinfo{pages}{205} (\bibinfo{year}{2016}), \eprint{1602.00005}.

\bibitem[{L2w()}]{L2website}
\bibinfo{howpublished}{\protect{The online plotter of ATLAS21, CT18, and MSHT20
  sensitivities}, \url{https://metapdf.hepforge.org/L2}}.

\bibitem[{\citenamefont{Abramowicz et~al.}(2013)}]{H1:2012xnw}
\bibinfo{author}{\bibfnamefont{H.}~\bibnamefont{Abramowicz}}
  \bibnamefont{et~al.} (\bibinfo{collaboration}{H1, ZEUS}),
  \bibinfo{journal}{Eur. Phys. J. C} \textbf{\bibinfo{volume}{73}},
  \bibinfo{pages}{2311} (\bibinfo{year}{2013}), \eprint{1211.1182}.

\bibitem[{\citenamefont{Abramowicz et~al.}(2018)}]{H1:2018flt}
\bibinfo{author}{\bibfnamefont{H.}~\bibnamefont{Abramowicz}}
  \bibnamefont{et~al.} (\bibinfo{collaboration}{H1, ZEUS}),
  \bibinfo{journal}{Eur. Phys. J. C} \textbf{\bibinfo{volume}{78}},
  \bibinfo{pages}{473} (\bibinfo{year}{2018}), \eprint{1804.01019}.

\bibitem[{\citenamefont{Guzzi et~al.}(2022{\natexlab{a}})\citenamefont{Guzzi,
  Xie, Hou, Nadolsky, Schmidt, Yan, and Yuan}}]{Guzzi:2021rvo}
\bibinfo{author}{\bibfnamefont{M.}~\bibnamefont{Guzzi}},
  \bibinfo{author}{\bibfnamefont{K.}~\bibnamefont{Xie}},
  \bibinfo{author}{\bibfnamefont{T.-J.} \bibnamefont{Hou}},
  \bibinfo{author}{\bibfnamefont{P.}~\bibnamefont{Nadolsky}},
  \bibinfo{author}{\bibfnamefont{C.}~\bibnamefont{Schmidt}},
  \bibinfo{author}{\bibfnamefont{M.}~\bibnamefont{Yan}}, \bibnamefont{and}
  \bibinfo{author}{\bibfnamefont{C.~P.} \bibnamefont{Yuan}},
  \bibinfo{journal}{PoS} \textbf{\bibinfo{volume}{EPS-HEP2021}},
  \bibinfo{pages}{370} (\bibinfo{year}{2022}{\natexlab{a}}),
  \eprint{2110.11495}.

\bibitem[{\citenamefont{Soper and Collins}(1994)}]{Soper:1994km}
\bibinfo{author}{\bibfnamefont{D.~E.} \bibnamefont{Soper}} \bibnamefont{and}
  \bibinfo{author}{\bibfnamefont{J.~C.} \bibnamefont{Collins}}
  (\bibinfo{year}{1994}), \eprint{hep-ph/9411214}.

\bibitem[{\citenamefont{Abramowicz et~al.}(2015)}]{Abramowicz:2015mha}
\bibinfo{author}{\bibfnamefont{H.}~\bibnamefont{Abramowicz}}
  \bibnamefont{et~al.} (\bibinfo{collaboration}{H1, ZEUS}),
  \bibinfo{journal}{Eur. Phys. J.} \textbf{\bibinfo{volume}{C75}},
  \bibinfo{pages}{580} (\bibinfo{year}{2015}), \eprint{1506.06042}.

\bibitem[{\citenamefont{Aaboud et~al.}(2017{\natexlab{a}})}]{ATLAS:2016nqi}
\bibinfo{author}{\bibfnamefont{M.}~\bibnamefont{Aaboud}} \bibnamefont{et~al.}
  (\bibinfo{collaboration}{ATLAS}), \bibinfo{journal}{Eur. Phys. J. C}
  \textbf{\bibinfo{volume}{77}}, \bibinfo{pages}{367}
  (\bibinfo{year}{2017}{\natexlab{a}}), \eprint{1612.03016}.

\bibitem[{\citenamefont{Aaboud et~al.}(2017{\natexlab{b}})}]{ATLAS:2017rue}
\bibinfo{author}{\bibfnamefont{M.}~\bibnamefont{Aaboud}} \bibnamefont{et~al.}
  (\bibinfo{collaboration}{ATLAS}), \bibinfo{journal}{JHEP}
  \textbf{\bibinfo{volume}{12}}, \bibinfo{pages}{059}
  (\bibinfo{year}{2017}{\natexlab{b}}), \eprint{1710.05167}.

\bibitem[{\citenamefont{Aad et~al.}(2019{\natexlab{a}})}]{ATLAS:2019fgb}
\bibinfo{author}{\bibfnamefont{G.}~\bibnamefont{Aad}} \bibnamefont{et~al.}
  (\bibinfo{collaboration}{ATLAS}), \bibinfo{journal}{Eur. Phys. J. C}
  \textbf{\bibinfo{volume}{79}}, \bibinfo{pages}{760}
  (\bibinfo{year}{2019}{\natexlab{a}}), \eprint{1904.05631}.

\bibitem[{\citenamefont{Aaboud et~al.}(2018)}]{ATLAS:2017irc}
\bibinfo{author}{\bibfnamefont{M.}~\bibnamefont{Aaboud}} \bibnamefont{et~al.}
  (\bibinfo{collaboration}{ATLAS}), \bibinfo{journal}{JHEP}
  \textbf{\bibinfo{volume}{05}}, \bibinfo{pages}{077} (\bibinfo{year}{2018}),
  \bibinfo{note}{[Erratum: JHEP 10, 048 (2020)]}, \eprint{1711.03296}.

\bibitem[{\citenamefont{Aad et~al.}(2019{\natexlab{b}})}]{ATLAS:2019bsa}
\bibinfo{author}{\bibfnamefont{G.}~\bibnamefont{Aad}} \bibnamefont{et~al.}
  (\bibinfo{collaboration}{ATLAS}), \bibinfo{journal}{Eur. Phys. J. C}
  \textbf{\bibinfo{volume}{79}}, \bibinfo{pages}{847}
  (\bibinfo{year}{2019}{\natexlab{b}}), \eprint{1907.06728}.

\bibitem[{\citenamefont{Aad et~al.}(2016{\natexlab{a}})}]{ATLAS:2015lsn}
\bibinfo{author}{\bibfnamefont{G.}~\bibnamefont{Aad}} \bibnamefont{et~al.}
  (\bibinfo{collaboration}{ATLAS}), \bibinfo{journal}{Eur. Phys. J. C}
  \textbf{\bibinfo{volume}{76}}, \bibinfo{pages}{538}
  (\bibinfo{year}{2016}{\natexlab{a}}), \eprint{1511.04716}.

\bibitem[{\citenamefont{Aaboud et~al.}(2016)}]{ATLAS:2016pal}
\bibinfo{author}{\bibfnamefont{M.}~\bibnamefont{Aaboud}} \bibnamefont{et~al.}
  (\bibinfo{collaboration}{ATLAS}), \bibinfo{journal}{Phys. Rev. D}
  \textbf{\bibinfo{volume}{94}}, \bibinfo{pages}{092003}
  (\bibinfo{year}{2016}), \bibinfo{note}{[Addendum: Phys.Rev.D 101, 119901
  (2020)]}, \eprint{1607.07281}.

\bibitem[{\citenamefont{Aad et~al.}(2019{\natexlab{c}})}]{ATLAS:2019hxz}
\bibinfo{author}{\bibfnamefont{G.}~\bibnamefont{Aad}} \bibnamefont{et~al.}
  (\bibinfo{collaboration}{ATLAS}), \bibinfo{journal}{Eur. Phys. J. C}
  \textbf{\bibinfo{volume}{79}}, \bibinfo{pages}{1028}
  (\bibinfo{year}{2019}{\natexlab{c}}), \bibinfo{note}{[Erratum: Eur.Phys.J.C
  80, 1092 (2020)]}, \eprint{1908.07305}.

\bibitem[{\citenamefont{Aaboud et~al.}(2019)}]{ATLAS:2019drj}
\bibinfo{author}{\bibfnamefont{M.}~\bibnamefont{Aaboud}} \bibnamefont{et~al.}
  (\bibinfo{collaboration}{ATLAS}), \bibinfo{journal}{JHEP}
  \textbf{\bibinfo{volume}{04}}, \bibinfo{pages}{093} (\bibinfo{year}{2019}),
  \eprint{1901.10075}.

\bibitem[{\citenamefont{Aaboud et~al.}(2017{\natexlab{c}})}]{ATLAS:2017kux}
\bibinfo{author}{\bibfnamefont{M.}~\bibnamefont{Aaboud}} \bibnamefont{et~al.}
  (\bibinfo{collaboration}{ATLAS}), \bibinfo{journal}{JHEP}
  \textbf{\bibinfo{volume}{09}}, \bibinfo{pages}{020}
  (\bibinfo{year}{2017}{\natexlab{c}}), \eprint{1706.03192}.

\bibitem[{\citenamefont{Aaboud et~al.}()}]{ATLAS:2018ttbar}
\bibinfo{author}{\bibfnamefont{M.}~\bibnamefont{Aaboud}} \bibnamefont{et~al.}
  (\bibinfo{collaboration}{ATLAS}) (????),
  \urlprefix\url{https://cds.cern.ch/record/2633819}.

\bibitem[{\citenamefont{Aad et~al.}(2021)}]{ATLAS:2021qnl}
\bibinfo{author}{\bibfnamefont{G.}~\bibnamefont{Aad}} \bibnamefont{et~al.}
  (\bibinfo{collaboration}{ATLAS}), \bibinfo{journal}{JHEP}
  \textbf{\bibinfo{volume}{07}}, \bibinfo{pages}{223} (\bibinfo{year}{2021}),
  \eprint{2101.05095}.

\bibitem[{\citenamefont{Catani and Grazzini}(2007)}]{Catani:2007vq}
\bibinfo{author}{\bibfnamefont{S.}~\bibnamefont{Catani}} \bibnamefont{and}
  \bibinfo{author}{\bibfnamefont{M.}~\bibnamefont{Grazzini}},
  \bibinfo{journal}{Phys. Rev. Lett.} \textbf{\bibinfo{volume}{98}},
  \bibinfo{pages}{222002} (\bibinfo{year}{2007}), \eprint{hep-ph/0703012}.

\bibitem[{\citenamefont{Catani et~al.}(2009)\citenamefont{Catani, Cieri,
  Ferrera, de~Florian, and Grazzini}}]{Catani:2009sm}
\bibinfo{author}{\bibfnamefont{S.}~\bibnamefont{Catani}},
  \bibinfo{author}{\bibfnamefont{L.}~\bibnamefont{Cieri}},
  \bibinfo{author}{\bibfnamefont{G.}~\bibnamefont{Ferrera}},
  \bibinfo{author}{\bibfnamefont{D.}~\bibnamefont{de~Florian}},
  \bibnamefont{and} \bibinfo{author}{\bibfnamefont{M.}~\bibnamefont{Grazzini}},
  \bibinfo{journal}{Phys. Rev. Lett.} \textbf{\bibinfo{volume}{103}},
  \bibinfo{pages}{082001} (\bibinfo{year}{2009}), \eprint{0903.2120}.

\bibitem[{\citenamefont{Gavin et~al.}(2011)\citenamefont{Gavin, Li, Petriello,
  and Quackenbush}}]{Gavin:2010az}
\bibinfo{author}{\bibfnamefont{R.}~\bibnamefont{Gavin}},
  \bibinfo{author}{\bibfnamefont{Y.}~\bibnamefont{Li}},
  \bibinfo{author}{\bibfnamefont{F.}~\bibnamefont{Petriello}},
  \bibnamefont{and}
  \bibinfo{author}{\bibfnamefont{S.}~\bibnamefont{Quackenbush}},
  \bibinfo{journal}{Comput. Phys. Commun.} \textbf{\bibinfo{volume}{182}},
  \bibinfo{pages}{2388} (\bibinfo{year}{2011}), \eprint{1011.3540}.

\bibitem[{\citenamefont{Gavin et~al.}(2013)\citenamefont{Gavin, Li, Petriello,
  and Quackenbush}}]{Gavin:2012sy}
\bibinfo{author}{\bibfnamefont{R.}~\bibnamefont{Gavin}},
  \bibinfo{author}{\bibfnamefont{Y.}~\bibnamefont{Li}},
  \bibinfo{author}{\bibfnamefont{F.}~\bibnamefont{Petriello}},
  \bibnamefont{and}
  \bibinfo{author}{\bibfnamefont{S.}~\bibnamefont{Quackenbush}},
  \bibinfo{journal}{Comput. Phys. Commun.} \textbf{\bibinfo{volume}{184}},
  \bibinfo{pages}{208} (\bibinfo{year}{2013}), \eprint{1201.5896}.

\bibitem[{\citenamefont{Li and Petriello}(2012)}]{Li:2012wna}
\bibinfo{author}{\bibfnamefont{Y.}~\bibnamefont{Li}} \bibnamefont{and}
  \bibinfo{author}{\bibfnamefont{F.}~\bibnamefont{Petriello}},
  \bibinfo{journal}{Phys. Rev.} \textbf{\bibinfo{volume}{D86}},
  \bibinfo{pages}{094034} (\bibinfo{year}{2012}), \eprint{1208.5967}.

\bibitem[{\citenamefont{Currie et~al.}(2017)\citenamefont{Currie, Glover, and
  Pires}}]{Currie:2016bfm}
\bibinfo{author}{\bibfnamefont{J.}~\bibnamefont{Currie}},
  \bibinfo{author}{\bibfnamefont{E.~W.~N.} \bibnamefont{Glover}},
  \bibnamefont{and} \bibinfo{author}{\bibfnamefont{J.}~\bibnamefont{Pires}},
  \bibinfo{journal}{Phys. Rev. Lett.} \textbf{\bibinfo{volume}{118}},
  \bibinfo{pages}{072002} (\bibinfo{year}{2017}), \eprint{1611.01460}.

\bibitem[{\citenamefont{Gehrmann-De~Ridder
  et~al.}(2018)\citenamefont{Gehrmann-De~Ridder, Gehrmann, Glover, Huss, and
  Walker}}]{Gehrmann-DeRidder:2017mvr}
\bibinfo{author}{\bibfnamefont{A.}~\bibnamefont{Gehrmann-De~Ridder}},
  \bibinfo{author}{\bibfnamefont{T.}~\bibnamefont{Gehrmann}},
  \bibinfo{author}{\bibfnamefont{E.~W.~N.} \bibnamefont{Glover}},
  \bibinfo{author}{\bibfnamefont{A.}~\bibnamefont{Huss}}, \bibnamefont{and}
  \bibinfo{author}{\bibfnamefont{D.~M.} \bibnamefont{Walker}},
  \bibinfo{journal}{Phys. Rev. Lett.} \textbf{\bibinfo{volume}{120}},
  \bibinfo{pages}{122001} (\bibinfo{year}{2018}), \eprint{1712.07543}.

\bibitem[{\citenamefont{Gehrmann-De~Ridder
  et~al.}(2016)\citenamefont{Gehrmann-De~Ridder, Gehrmann, Glover, Huss, and
  Morgan}}]{Gehrmann-DeRidder:2015wbt}
\bibinfo{author}{\bibfnamefont{A.}~\bibnamefont{Gehrmann-De~Ridder}},
  \bibinfo{author}{\bibfnamefont{T.}~\bibnamefont{Gehrmann}},
  \bibinfo{author}{\bibfnamefont{E.~W.~N.} \bibnamefont{Glover}},
  \bibinfo{author}{\bibfnamefont{A.}~\bibnamefont{Huss}}, \bibnamefont{and}
  \bibinfo{author}{\bibfnamefont{T.~A.} \bibnamefont{Morgan}},
  \bibinfo{journal}{Phys. Rev. Lett.} \textbf{\bibinfo{volume}{117}},
  \bibinfo{pages}{022001} (\bibinfo{year}{2016}), \eprint{1507.02850}.

\bibitem[{\citenamefont{Czakon et~al.}(2017{\natexlab{a}})\citenamefont{Czakon,
  Heymes, and Mitov}}]{Czakon:2017dip}
\bibinfo{author}{\bibfnamefont{M.}~\bibnamefont{Czakon}},
  \bibinfo{author}{\bibfnamefont{D.}~\bibnamefont{Heymes}}, \bibnamefont{and}
  \bibinfo{author}{\bibfnamefont{A.}~\bibnamefont{Mitov}}
  (\bibinfo{year}{2017}{\natexlab{a}}), \eprint{1704.08551}.

\bibitem[{\citenamefont{Czakon et~al.}(2017{\natexlab{b}})\citenamefont{Czakon,
  Hartland, Mitov, Nocera, and Rojo}}]{Czakon:2016olj}
\bibinfo{author}{\bibfnamefont{M.}~\bibnamefont{Czakon}},
  \bibinfo{author}{\bibfnamefont{N.~P.} \bibnamefont{Hartland}},
  \bibinfo{author}{\bibfnamefont{A.}~\bibnamefont{Mitov}},
  \bibinfo{author}{\bibfnamefont{E.~R.} \bibnamefont{Nocera}},
  \bibnamefont{and} \bibinfo{author}{\bibfnamefont{J.}~\bibnamefont{Rojo}},
  \bibinfo{journal}{JHEP} \textbf{\bibinfo{volume}{04}}, \bibinfo{pages}{044}
  (\bibinfo{year}{2017}{\natexlab{b}}), \eprint{1611.08609}.

\bibitem[{\citenamefont{Campbell et~al.}(2017)\citenamefont{Campbell, Ellis,
  and Williams}}]{Campbell:2016lzl}
\bibinfo{author}{\bibfnamefont{J.~M.} \bibnamefont{Campbell}},
  \bibinfo{author}{\bibfnamefont{R.~K.} \bibnamefont{Ellis}}, \bibnamefont{and}
  \bibinfo{author}{\bibfnamefont{C.}~\bibnamefont{Williams}},
  \bibinfo{journal}{Phys. Rev. Lett.} \textbf{\bibinfo{volume}{118}},
  \bibinfo{pages}{222001} (\bibinfo{year}{2017}), \bibinfo{note}{[Erratum:
  Phys.Rev.Lett. 124, 259901 (2020)]}, \eprint{1612.04333}.

\bibitem[{\citenamefont{Czakon et~al.}(2017{\natexlab{c}})\citenamefont{Czakon,
  Heymes, Mitov, Pagani, Tsinikos, and Zaro}}]{Czakon:2017wor}
\bibinfo{author}{\bibfnamefont{M.}~\bibnamefont{Czakon}},
  \bibinfo{author}{\bibfnamefont{D.}~\bibnamefont{Heymes}},
  \bibinfo{author}{\bibfnamefont{A.}~\bibnamefont{Mitov}},
  \bibinfo{author}{\bibfnamefont{D.}~\bibnamefont{Pagani}},
  \bibinfo{author}{\bibfnamefont{I.}~\bibnamefont{Tsinikos}}, \bibnamefont{and}
  \bibinfo{author}{\bibfnamefont{M.}~\bibnamefont{Zaro}},
  \bibinfo{journal}{JHEP} \textbf{\bibinfo{volume}{10}}, \bibinfo{pages}{186}
  (\bibinfo{year}{2017}{\natexlab{c}}), \eprint{1705.04105}.

\bibitem[{\citenamefont{Becher and Garcia~i Tormo}(2013)}]{Becher:2013zua}
\bibinfo{author}{\bibfnamefont{T.}~\bibnamefont{Becher}} \bibnamefont{and}
  \bibinfo{author}{\bibfnamefont{X.}~\bibnamefont{Garcia~i Tormo}},
  \bibinfo{journal}{Phys. Rev. D} \textbf{\bibinfo{volume}{88}},
  \bibinfo{pages}{013009} (\bibinfo{year}{2013}), \eprint{1305.4202}.

\bibitem[{\citenamefont{Dittmaier et~al.}(2012)\citenamefont{Dittmaier, Huss,
  and Speckner}}]{Dittmaier:2012kx}
\bibinfo{author}{\bibfnamefont{S.}~\bibnamefont{Dittmaier}},
  \bibinfo{author}{\bibfnamefont{A.}~\bibnamefont{Huss}}, \bibnamefont{and}
  \bibinfo{author}{\bibfnamefont{C.}~\bibnamefont{Speckner}},
  \bibinfo{journal}{JHEP} \textbf{\bibinfo{volume}{11}}, \bibinfo{pages}{095}
  (\bibinfo{year}{2012}), \eprint{1210.0438}.

\bibitem[{\citenamefont{Hou et~al.}(2017{\natexlab{b}})\citenamefont{Hou,
  Dulat, Gao, Guzzi, Huston, Nadolsky, Pumplin, Schmidt, Stump, and
  Yuan}}]{Hou:2016nqm}
\bibinfo{author}{\bibfnamefont{T.-J.} \bibnamefont{Hou}},
  \bibinfo{author}{\bibfnamefont{S.}~\bibnamefont{Dulat}},
  \bibinfo{author}{\bibfnamefont{J.}~\bibnamefont{Gao}},
  \bibinfo{author}{\bibfnamefont{M.}~\bibnamefont{Guzzi}},
  \bibinfo{author}{\bibfnamefont{J.}~\bibnamefont{Huston}},
  \bibinfo{author}{\bibfnamefont{P.}~\bibnamefont{Nadolsky}},
  \bibinfo{author}{\bibfnamefont{J.}~\bibnamefont{Pumplin}},
  \bibinfo{author}{\bibfnamefont{C.}~\bibnamefont{Schmidt}},
  \bibinfo{author}{\bibfnamefont{D.}~\bibnamefont{Stump}}, \bibnamefont{and}
  \bibinfo{author}{\bibfnamefont{C.~P.} \bibnamefont{Yuan}},
  \bibinfo{journal}{Phys. Rev.} \textbf{\bibinfo{volume}{D95}},
  \bibinfo{pages}{034003} (\bibinfo{year}{2017}{\natexlab{b}}),
  \eprint{1609.07968}.

\bibitem[{\citenamefont{CTEQ-TEA}()}]{CT18L2Sensitivity}
\bibinfo{author}{\bibnamefont{CTEQ-TEA}},
  \bibinfo{howpublished}{\url{https://ct.hepforge.org/PDFs/ct18/figures/L2Sensitivity/}}.

\bibitem[{\citenamefont{Xie et~al.}(2022)\citenamefont{Xie, Hobbs, Hou,
  Schmidt, Yan, and Yuan}}]{Xie:2021equ}
\bibinfo{author}{\bibfnamefont{K.}~\bibnamefont{Xie}},
  \bibinfo{author}{\bibfnamefont{T.~J.} \bibnamefont{Hobbs}},
  \bibinfo{author}{\bibfnamefont{T.-J.} \bibnamefont{Hou}},
  \bibinfo{author}{\bibfnamefont{C.}~\bibnamefont{Schmidt}},
  \bibinfo{author}{\bibfnamefont{M.}~\bibnamefont{Yan}}, \bibnamefont{and}
  \bibinfo{author}{\bibfnamefont{C.~P.} \bibnamefont{Yuan}}
  (\bibinfo{collaboration}{CTEQ-TEA}), \bibinfo{journal}{Phys. Rev. D}
  \textbf{\bibinfo{volume}{105}}, \bibinfo{pages}{054006}
  (\bibinfo{year}{2022}), \eprint{2106.10299}.

\bibitem[{\citenamefont{Xie et~al.}(2023{\natexlab{a}})\citenamefont{Xie, Zhou,
  and Hobbs}}]{Xie:2023qbn}
\bibinfo{author}{\bibfnamefont{K.}~\bibnamefont{Xie}},
  \bibinfo{author}{\bibfnamefont{B.}~\bibnamefont{Zhou}}, \bibnamefont{and}
  \bibinfo{author}{\bibfnamefont{T.~J.} \bibnamefont{Hobbs}}
  (\bibinfo{year}{2023}{\natexlab{a}}), \eprint{2305.10497}.

\bibitem[{\citenamefont{Gao et~al.}(2022)\citenamefont{Gao, Hobbs, Nadolsky,
  Sun, and Yuan}}]{Gao:2021fle}
\bibinfo{author}{\bibfnamefont{J.}~\bibnamefont{Gao}},
  \bibinfo{author}{\bibfnamefont{T.~J.} \bibnamefont{Hobbs}},
  \bibinfo{author}{\bibfnamefont{P.~M.} \bibnamefont{Nadolsky}},
  \bibinfo{author}{\bibfnamefont{C.}~\bibnamefont{Sun}}, \bibnamefont{and}
  \bibinfo{author}{\bibfnamefont{C.~P.} \bibnamefont{Yuan}},
  \bibinfo{journal}{Phys. Rev. D} \textbf{\bibinfo{volume}{105}},
  \bibinfo{pages}{L011503} (\bibinfo{year}{2022}), \eprint{2107.00460}.

\bibitem[{\citenamefont{Xie et~al.}(2023{\natexlab{b}})\citenamefont{Xie, Gao,
  Hobbs, Stump, and Yuan}}]{Xie:2023suk}
\bibinfo{author}{\bibfnamefont{K.}~\bibnamefont{Xie}},
  \bibinfo{author}{\bibfnamefont{J.}~\bibnamefont{Gao}},
  \bibinfo{author}{\bibfnamefont{T.~J.} \bibnamefont{Hobbs}},
  \bibinfo{author}{\bibfnamefont{D.~R.} \bibnamefont{Stump}}, \bibnamefont{and}
  \bibinfo{author}{\bibfnamefont{C.~P.} \bibnamefont{Yuan}}
  (\bibinfo{year}{2023}{\natexlab{b}}), \eprint{2303.13607}.

\bibitem[{\citenamefont{Courtoy and Nadolsky}(2021)}]{Courtoy:2020fex}
\bibinfo{author}{\bibfnamefont{A.}~\bibnamefont{Courtoy}} \bibnamefont{and}
  \bibinfo{author}{\bibfnamefont{P.~M.} \bibnamefont{Nadolsky}},
  \bibinfo{journal}{Phys. Rev. D} \textbf{\bibinfo{volume}{103}},
  \bibinfo{pages}{054029} (\bibinfo{year}{2021}), \eprint{2011.10078}.

\bibitem[{\citenamefont{Guzzi et~al.}(2022{\natexlab{b}})\citenamefont{Guzzi,
  Hobbs, Xie, Huston, Nadolsky, and Yuan}}]{Guzzi:2022rca}
\bibinfo{author}{\bibfnamefont{M.}~\bibnamefont{Guzzi}},
  \bibinfo{author}{\bibfnamefont{T.~J.} \bibnamefont{Hobbs}},
  \bibinfo{author}{\bibfnamefont{K.}~\bibnamefont{Xie}},
  \bibinfo{author}{\bibfnamefont{J.}~\bibnamefont{Huston}},
  \bibinfo{author}{\bibfnamefont{P.}~\bibnamefont{Nadolsky}}, \bibnamefont{and}
  \bibinfo{author}{\bibfnamefont{C.~P.} \bibnamefont{Yuan}}
  (\bibinfo{year}{2022}{\natexlab{b}}), \eprint{2211.01387}.

\bibitem[{\citenamefont{Sitiwaldi et~al.}(2023)\citenamefont{Sitiwaldi, Xie,
  Ablat, Dulat, Hou, and Yuan}}]{Sitiwaldi:2023jjp}
\bibinfo{author}{\bibfnamefont{I.}~\bibnamefont{Sitiwaldi}},
  \bibinfo{author}{\bibfnamefont{K.}~\bibnamefont{Xie}},
  \bibinfo{author}{\bibfnamefont{A.}~\bibnamefont{Ablat}},
  \bibinfo{author}{\bibfnamefont{S.}~\bibnamefont{Dulat}},
  \bibinfo{author}{\bibfnamefont{T.-J.} \bibnamefont{Hou}}, \bibnamefont{and}
  \bibinfo{author}{\bibfnamefont{C.-P.} \bibnamefont{Yuan}}
  (\bibinfo{year}{2023}), \eprint{2305.10733}.

\bibitem[{\citenamefont{Kluge et~al.}(2006)\citenamefont{Kluge, Rabbertz, and
  Wobisch}}]{Kluge:2006xs}
\bibinfo{author}{\bibfnamefont{T.}~\bibnamefont{Kluge}},
  \bibinfo{author}{\bibfnamefont{K.}~\bibnamefont{Rabbertz}}, \bibnamefont{and}
  \bibinfo{author}{\bibfnamefont{M.}~\bibnamefont{Wobisch}}
  (\bibinfo{year}{2006}), pp. \bibinfo{pages}{483--486},
  \eprint{hep-ph/0609285},
  \urlprefix\url{http://lss.fnal.gov/cgi-bin/find_paper.pl?conf-06-352}.

\bibitem[{\citenamefont{Wobisch et~al.}(2011)\citenamefont{Wobisch, Britzger,
  Kluge, Rabbertz, and Stober}}]{Wobisch:2011ij}
\bibinfo{author}{\bibfnamefont{M.}~\bibnamefont{Wobisch}},
  \bibinfo{author}{\bibfnamefont{D.}~\bibnamefont{Britzger}},
  \bibinfo{author}{\bibfnamefont{T.}~\bibnamefont{Kluge}},
  \bibinfo{author}{\bibfnamefont{K.}~\bibnamefont{Rabbertz}}, \bibnamefont{and}
  \bibinfo{author}{\bibfnamefont{F.}~\bibnamefont{Stober}}
  (\bibinfo{collaboration}{fastNLO}) (\bibinfo{year}{2011}),
  \eprint{1109.1310}.

\bibitem[{\citenamefont{Britzger et~al.}(2012)\citenamefont{Britzger, Rabbertz,
  Stober, and Wobisch}}]{Britzger:2012bs}
\bibinfo{author}{\bibfnamefont{D.}~\bibnamefont{Britzger}},
  \bibinfo{author}{\bibfnamefont{K.}~\bibnamefont{Rabbertz}},
  \bibinfo{author}{\bibfnamefont{F.}~\bibnamefont{Stober}}, \bibnamefont{and}
  \bibinfo{author}{\bibfnamefont{M.}~\bibnamefont{Wobisch}}
  (\bibinfo{collaboration}{fastNLO}) (\bibinfo{year}{2012}), pp.
  \bibinfo{pages}{217--221}, \eprint{1208.3641}.

\bibitem[{\citenamefont{Carli et~al.}(2010)\citenamefont{Carli, Clements,
  Cooper-Sarkar, Gwenlan, Salam, Siegert, Starovoitov, and
  Sutton}}]{Carli:2010rw}
\bibinfo{author}{\bibfnamefont{T.}~\bibnamefont{Carli}},
  \bibinfo{author}{\bibfnamefont{D.}~\bibnamefont{Clements}},
  \bibinfo{author}{\bibfnamefont{A.}~\bibnamefont{Cooper-Sarkar}},
  \bibinfo{author}{\bibfnamefont{C.}~\bibnamefont{Gwenlan}},
  \bibinfo{author}{\bibfnamefont{G.~P.} \bibnamefont{Salam}},
  \bibinfo{author}{\bibfnamefont{F.}~\bibnamefont{Siegert}},
  \bibinfo{author}{\bibfnamefont{P.}~\bibnamefont{Starovoitov}},
  \bibnamefont{and} \bibinfo{author}{\bibfnamefont{M.}~\bibnamefont{Sutton}},
  \bibinfo{journal}{Eur. Phys. J.} \textbf{\bibinfo{volume}{C66}},
  \bibinfo{pages}{503} (\bibinfo{year}{2010}), \eprint{0911.2985}.

\bibitem[{\citenamefont{Campbell and Ellis}(2010)}]{Campbell:2010ff}
\bibinfo{author}{\bibfnamefont{J.~M.} \bibnamefont{Campbell}} \bibnamefont{and}
  \bibinfo{author}{\bibfnamefont{R.~K.} \bibnamefont{Ellis}},
  \bibinfo{journal}{Nucl. Phys. Proc. Suppl.}
  \textbf{\bibinfo{volume}{205-206}}, \bibinfo{pages}{10}
  (\bibinfo{year}{2010}), \eprint{1007.3492}.

\bibitem[{\citenamefont{Boughezal et~al.}(2017)\citenamefont{Boughezal,
  Campbell, Ellis, Focke, Giele, Liu, Petriello, and
  Williams}}]{Boughezal:2016wmq}
\bibinfo{author}{\bibfnamefont{R.}~\bibnamefont{Boughezal}},
  \bibinfo{author}{\bibfnamefont{J.~M.} \bibnamefont{Campbell}},
  \bibinfo{author}{\bibfnamefont{R.~K.} \bibnamefont{Ellis}},
  \bibinfo{author}{\bibfnamefont{C.}~\bibnamefont{Focke}},
  \bibinfo{author}{\bibfnamefont{W.}~\bibnamefont{Giele}},
  \bibinfo{author}{\bibfnamefont{X.}~\bibnamefont{Liu}},
  \bibinfo{author}{\bibfnamefont{F.}~\bibnamefont{Petriello}},
  \bibnamefont{and} \bibinfo{author}{\bibfnamefont{C.}~\bibnamefont{Williams}},
  \bibinfo{journal}{Eur. Phys. J.} \textbf{\bibinfo{volume}{C77}},
  \bibinfo{pages}{7} (\bibinfo{year}{2017}), \eprint{1605.08011}.

\bibitem[{\citenamefont{Campbell et~al.}(2011)\citenamefont{Campbell, Ellis,
  and Williams}}]{MCFM6}
\bibinfo{author}{\bibfnamefont{J.~M.} \bibnamefont{Campbell}},
  \bibinfo{author}{\bibfnamefont{R.~K.} \bibnamefont{Ellis}}, \bibnamefont{and}
  \bibinfo{author}{\bibfnamefont{C.}~\bibnamefont{Williams}},
  \bibinfo{journal}{JHEP} \textbf{\bibinfo{volume}{07}}, \bibinfo{pages}{018}
  (\bibinfo{year}{2011}), \eprint{1105.0020}.

\bibitem[{\citenamefont{Campbell et~al.}(2015)\citenamefont{Campbell, Ellis,
  and Giele}}]{MCFM8}
\bibinfo{author}{\bibfnamefont{J.~M.} \bibnamefont{Campbell}},
  \bibinfo{author}{\bibfnamefont{R.~K.} \bibnamefont{Ellis}}, \bibnamefont{and}
  \bibinfo{author}{\bibfnamefont{W.~T.} \bibnamefont{Giele}},
  \bibinfo{journal}{Eur. Phys. J.} \textbf{\bibinfo{volume}{C75}},
  \bibinfo{pages}{246} (\bibinfo{year}{2015}), \eprint{1503.06182}.

\bibitem[{\citenamefont{Nagy}(2003)}]{Nagy:2003tz}
\bibinfo{author}{\bibfnamefont{Z.}~\bibnamefont{Nagy}}, \bibinfo{journal}{Phys.
  Rev.} \textbf{\bibinfo{volume}{D68}}, \bibinfo{pages}{094002}
  (\bibinfo{year}{2003}), \eprint{hep-ph/0307268}.

\bibitem[{\citenamefont{Bertone et~al.}(2014)\citenamefont{Bertone, Frederix,
  Frixione, Rojo, and Sutton}}]{Bertone:2014zva}
\bibinfo{author}{\bibfnamefont{V.}~\bibnamefont{Bertone}},
  \bibinfo{author}{\bibfnamefont{R.}~\bibnamefont{Frederix}},
  \bibinfo{author}{\bibfnamefont{S.}~\bibnamefont{Frixione}},
  \bibinfo{author}{\bibfnamefont{J.}~\bibnamefont{Rojo}}, \bibnamefont{and}
  \bibinfo{author}{\bibfnamefont{M.}~\bibnamefont{Sutton}},
  \bibinfo{journal}{JHEP} \textbf{\bibinfo{volume}{08}}, \bibinfo{pages}{166}
  (\bibinfo{year}{2014}), \eprint{1406.7693}.

\bibitem[{fas()}]{fastnnlo:grids}
\bibinfo{howpublished}{\url{http://www.precision.hep.phy.cam.ac.uk/results/ttbar-fastnlo/}}.

\bibitem[{\citenamefont{Zhang et~al.}(2021)\citenamefont{Zhang, Lin, and
  Yoon}}]{Zhang:2020dkn}
\bibinfo{author}{\bibfnamefont{R.}~\bibnamefont{Zhang}},
  \bibinfo{author}{\bibfnamefont{H.-W.} \bibnamefont{Lin}}, \bibnamefont{and}
  \bibinfo{author}{\bibfnamefont{B.}~\bibnamefont{Yoon}},
  \bibinfo{journal}{Phys. Rev. D} \textbf{\bibinfo{volume}{104}},
  \bibinfo{pages}{094511} (\bibinfo{year}{2021}), \eprint{2005.01124}.

\bibitem[{\citenamefont{Benvenuti et~al.}(1989)}]{Benvenuti:1989rh}
\bibinfo{author}{\bibfnamefont{A.~C.} \bibnamefont{Benvenuti}}
  \bibnamefont{et~al.} (\bibinfo{collaboration}{BCDMS}),
  \bibinfo{journal}{Phys. Lett.} \textbf{\bibinfo{volume}{B223}},
  \bibinfo{pages}{485} (\bibinfo{year}{1989}).

\bibitem[{\citenamefont{Arneodo et~al.}(1997{\natexlab{a}})}]{Arneodo:1996qe}
\bibinfo{author}{\bibfnamefont{M.}~\bibnamefont{Arneodo}} \bibnamefont{et~al.}
  (\bibinfo{collaboration}{New Muon}), \bibinfo{journal}{Nucl. Phys.}
  \textbf{\bibinfo{volume}{B483}}, \bibinfo{pages}{3}
  (\bibinfo{year}{1997}{\natexlab{a}}), \eprint{hep-ph/9610231}.

\bibitem[{\citenamefont{Arneodo et~al.}(1997{\natexlab{b}})}]{NewMuon:1996uwk}
\bibinfo{author}{\bibfnamefont{M.}~\bibnamefont{Arneodo}} \bibnamefont{et~al.}
  (\bibinfo{collaboration}{New Muon}), \bibinfo{journal}{Nucl. Phys. B}
  \textbf{\bibinfo{volume}{487}}, \bibinfo{pages}{3}
  (\bibinfo{year}{1997}{\natexlab{b}}), \eprint{hep-ex/9611022}.

\bibitem[{\citenamefont{Whitlow et~al.}(1992)\citenamefont{Whitlow, Riordan,
  Dasu, Rock, and Bodek}}]{Whitlow:1991uw}
\bibinfo{author}{\bibfnamefont{L.~W.} \bibnamefont{Whitlow}},
  \bibinfo{author}{\bibfnamefont{E.~M.} \bibnamefont{Riordan}},
  \bibinfo{author}{\bibfnamefont{S.}~\bibnamefont{Dasu}},
  \bibinfo{author}{\bibfnamefont{S.}~\bibnamefont{Rock}}, \bibnamefont{and}
  \bibinfo{author}{\bibfnamefont{A.}~\bibnamefont{Bodek}},
  \bibinfo{journal}{Phys. Lett. B} \textbf{\bibinfo{volume}{282}},
  \bibinfo{pages}{475} (\bibinfo{year}{1992}).

\bibitem[{\citenamefont{Whitlow et~al.}(1990)\citenamefont{Whitlow, Rock,
  Bodek, Riordan, and Dasu}}]{Whitlow:1990gk}
\bibinfo{author}{\bibfnamefont{L.~W.} \bibnamefont{Whitlow}},
  \bibinfo{author}{\bibfnamefont{S.}~\bibnamefont{Rock}},
  \bibinfo{author}{\bibfnamefont{A.}~\bibnamefont{Bodek}},
  \bibinfo{author}{\bibfnamefont{E.~M.} \bibnamefont{Riordan}},
  \bibnamefont{and} \bibinfo{author}{\bibfnamefont{S.}~\bibnamefont{Dasu}},
  \bibinfo{journal}{Phys. Lett. B} \textbf{\bibinfo{volume}{250}},
  \bibinfo{pages}{193} (\bibinfo{year}{1990}).

\bibitem[{\citenamefont{Adams et~al.}(1996)}]{E665:1996mob}
\bibinfo{author}{\bibfnamefont{M.~R.} \bibnamefont{Adams}} \bibnamefont{et~al.}
  (\bibinfo{collaboration}{E665}), \bibinfo{journal}{Phys. Rev. D}
  \textbf{\bibinfo{volume}{54}}, \bibinfo{pages}{3006} (\bibinfo{year}{1996}).

\bibitem[{\citenamefont{Aaron et~al.}(2010)}]{H1:2009pze}
\bibinfo{author}{\bibfnamefont{F.~D.} \bibnamefont{Aaron}} \bibnamefont{et~al.}
  (\bibinfo{collaboration}{H1, ZEUS}), \bibinfo{journal}{JHEP}
  \textbf{\bibinfo{volume}{01}}, \bibinfo{pages}{109} (\bibinfo{year}{2010}),
  \eprint{0911.0884}.

\bibitem[{\citenamefont{Aktas et~al.}(2005)}]{Aktas:2004az}
\bibinfo{author}{\bibfnamefont{A.}~\bibnamefont{Aktas}} \bibnamefont{et~al.}
  (\bibinfo{collaboration}{H1}), \bibinfo{journal}{Eur. Phys. J.}
  \textbf{\bibinfo{volume}{C40}}, \bibinfo{pages}{349} (\bibinfo{year}{2005}),
  \eprint{hep-ex/0411046}.

\bibitem[{\citenamefont{Aaron et~al.}(2011)}]{H1:2010fzx}
\bibinfo{author}{\bibfnamefont{F.~D.} \bibnamefont{Aaron}} \bibnamefont{et~al.}
  (\bibinfo{collaboration}{H1}), \bibinfo{journal}{Eur. Phys. J. C}
  \textbf{\bibinfo{volume}{71}}, \bibinfo{pages}{1579} (\bibinfo{year}{2011}),
  \eprint{1012.4355}.

\bibitem[{\citenamefont{Aaron et~al.}(2008)}]{H1:2008rkk}
\bibinfo{author}{\bibfnamefont{F.~D.} \bibnamefont{Aaron}} \bibnamefont{et~al.}
  (\bibinfo{collaboration}{H1}), \bibinfo{journal}{Phys. Lett. B}
  \textbf{\bibinfo{volume}{665}}, \bibinfo{pages}{139} (\bibinfo{year}{2008}),
  \eprint{0805.2809}.

\bibitem[{\citenamefont{Chekanov et~al.}(2009)}]{ZEUS:2009nwk}
\bibinfo{author}{\bibfnamefont{S.}~\bibnamefont{Chekanov}} \bibnamefont{et~al.}
  (\bibinfo{collaboration}{ZEUS}), \bibinfo{journal}{Phys. Lett. B}
  \textbf{\bibinfo{volume}{682}}, \bibinfo{pages}{8} (\bibinfo{year}{2009}),
  \eprint{0904.1092}.

\bibitem[{\citenamefont{Berge et~al.}(1991)}]{Berge:1989hr}
\bibinfo{author}{\bibfnamefont{J.~P.} \bibnamefont{Berge}}
  \bibnamefont{et~al.}, \bibinfo{journal}{Z. Phys.}
  \textbf{\bibinfo{volume}{C49}}, \bibinfo{pages}{187} (\bibinfo{year}{1991}).

\bibitem[{\citenamefont{Yang et~al.}(2001)}]{Yang:2000ju}
\bibinfo{author}{\bibfnamefont{U.-K.} \bibnamefont{Yang}} \bibnamefont{et~al.}
  (\bibinfo{collaboration}{CCFR/NuTeV}), \bibinfo{journal}{Phys. Rev. Lett.}
  \textbf{\bibinfo{volume}{86}}, \bibinfo{pages}{2742} (\bibinfo{year}{2001}),
  \eprint{hep-ex/0009041}.

\bibitem[{\citenamefont{Seligman et~al.}(1997)}]{Seligman:1997mc}
\bibinfo{author}{\bibfnamefont{W.~G.} \bibnamefont{Seligman}}
  \bibnamefont{et~al.}, \bibinfo{journal}{Phys. Rev. Lett.}
  \textbf{\bibinfo{volume}{79}}, \bibinfo{pages}{1213} (\bibinfo{year}{1997}),
  \eprint{hep-ex/9701017}.

\bibitem[{\citenamefont{Onengut et~al.}(2006)}]{CHORUS:2005cpn}
\bibinfo{author}{\bibfnamefont{G.}~\bibnamefont{Onengut}} \bibnamefont{et~al.}
  (\bibinfo{collaboration}{CHORUS}), \bibinfo{journal}{Phys. Lett. B}
  \textbf{\bibinfo{volume}{632}}, \bibinfo{pages}{65} (\bibinfo{year}{2006}).

\bibitem[{\citenamefont{Tzanov et~al.}(2006)}]{NuTeV:2005wsg}
\bibinfo{author}{\bibfnamefont{M.}~\bibnamefont{Tzanov}} \bibnamefont{et~al.}
  (\bibinfo{collaboration}{NuTeV}), \bibinfo{journal}{Phys. Rev. D}
  \textbf{\bibinfo{volume}{74}}, \bibinfo{pages}{012008}
  (\bibinfo{year}{2006}), \eprint{hep-ex/0509010}.

\bibitem[{\citenamefont{Goncharov et~al.}(2001)}]{Goncharov:2001qe}
\bibinfo{author}{\bibfnamefont{M.}~\bibnamefont{Goncharov}}
  \bibnamefont{et~al.} (\bibinfo{collaboration}{NuTeV}),
  \bibinfo{journal}{Phys. Rev.} \textbf{\bibinfo{volume}{D64}},
  \bibinfo{pages}{112006} (\bibinfo{year}{2001}), \eprint{hep-ex/0102049}.

\bibitem[{\citenamefont{Mason}(2006)}]{Mason:2006qa}
\bibinfo{author}{\bibfnamefont{D.~A.} \bibnamefont{Mason}}, Ph.D. thesis,
  \bibinfo{school}{Oregon U.} (\bibinfo{year}{2006}).

\bibitem[{\citenamefont{Aad et~al.}(2012)}]{Aad:2011dm}
\bibinfo{author}{\bibfnamefont{G.}~\bibnamefont{Aad}} \bibnamefont{et~al.}
  (\bibinfo{collaboration}{ATLAS}), \bibinfo{journal}{Phys. Rev.}
  \textbf{\bibinfo{volume}{D85}}, \bibinfo{pages}{072004}
  (\bibinfo{year}{2012}), \eprint{1109.5141}.

\bibitem[{\citenamefont{Aad et~al.}(2013)}]{Aad:2013iua}
\bibinfo{author}{\bibfnamefont{G.}~\bibnamefont{Aad}} \bibnamefont{et~al.}
  (\bibinfo{collaboration}{ATLAS}), \bibinfo{journal}{Phys. Lett.}
  \textbf{\bibinfo{volume}{B725}}, \bibinfo{pages}{223} (\bibinfo{year}{2013}),
  \eprint{1305.4192}.

\bibitem[{\citenamefont{Aad et~al.}(2016{\natexlab{b}})}]{ATLAS:2016gic}
\bibinfo{author}{\bibfnamefont{G.}~\bibnamefont{Aad}} \bibnamefont{et~al.}
  (\bibinfo{collaboration}{ATLAS}), \bibinfo{journal}{JHEP}
  \textbf{\bibinfo{volume}{08}}, \bibinfo{pages}{009}
  (\bibinfo{year}{2016}{\natexlab{b}}), \eprint{1606.01736}.

\bibitem[{\citenamefont{Aad et~al.}(2016{\natexlab{c}})}]{ATLAS:2015iiu}
\bibinfo{author}{\bibfnamefont{G.}~\bibnamefont{Aad}} \bibnamefont{et~al.}
  (\bibinfo{collaboration}{ATLAS}), \bibinfo{journal}{Eur. Phys. J. C}
  \textbf{\bibinfo{volume}{76}}, \bibinfo{pages}{291}
  (\bibinfo{year}{2016}{\natexlab{c}}), \eprint{1512.02192}.

\bibitem[{\citenamefont{Abe et~al.}(1998)}]{CDF:1998uzn}
\bibinfo{author}{\bibfnamefont{F.}~\bibnamefont{Abe}} \bibnamefont{et~al.}
  (\bibinfo{collaboration}{CDF}), \bibinfo{journal}{Phys. Rev. Lett.}
  \textbf{\bibinfo{volume}{81}}, \bibinfo{pages}{5754} (\bibinfo{year}{1998}),
  \eprint{hep-ex/9809001}.

\bibitem[{\citenamefont{Acosta et~al.}(2005)}]{CDF:2005cgc}
\bibinfo{author}{\bibfnamefont{D.}~\bibnamefont{Acosta}} \bibnamefont{et~al.}
  (\bibinfo{collaboration}{CDF}), \bibinfo{journal}{Phys. Rev. D}
  \textbf{\bibinfo{volume}{71}}, \bibinfo{pages}{051104}
  (\bibinfo{year}{2005}), \eprint{hep-ex/0501023}.

\bibitem[{\citenamefont{Aaltonen et~al.}(2009)}]{CDF:2009cjw}
\bibinfo{author}{\bibfnamefont{T.}~\bibnamefont{Aaltonen}} \bibnamefont{et~al.}
  (\bibinfo{collaboration}{CDF}), \bibinfo{journal}{Phys. Rev. Lett.}
  \textbf{\bibinfo{volume}{102}}, \bibinfo{pages}{181801}
  (\bibinfo{year}{2009}), \eprint{0901.2169}.

\bibitem[{\citenamefont{Aaltonen et~al.}(2010)}]{Aaltonen:2010zza}
\bibinfo{author}{\bibfnamefont{T.~A.} \bibnamefont{Aaltonen}}
  \bibnamefont{et~al.} (\bibinfo{collaboration}{CDF}), \bibinfo{journal}{Phys.
  Lett.} \textbf{\bibinfo{volume}{B692}}, \bibinfo{pages}{232}
  (\bibinfo{year}{2010}), \eprint{0908.3914}.

\bibitem[{\citenamefont{Chatrchyan et~al.}(2011)}]{CMS:2011bet}
\bibinfo{author}{\bibfnamefont{S.}~\bibnamefont{Chatrchyan}}
  \bibnamefont{et~al.} (\bibinfo{collaboration}{CMS}), \bibinfo{journal}{JHEP}
  \textbf{\bibinfo{volume}{04}}, \bibinfo{pages}{050} (\bibinfo{year}{2011}),
  \eprint{1103.3470}.

\bibitem[{\citenamefont{Chatrchyan
  et~al.}(2012{\natexlab{a}})}]{Chatrchyan:2012xt}
\bibinfo{author}{\bibfnamefont{S.}~\bibnamefont{Chatrchyan}}
  \bibnamefont{et~al.} (\bibinfo{collaboration}{CMS}), \bibinfo{journal}{Phys.
  Rev. Lett.} \textbf{\bibinfo{volume}{109}}, \bibinfo{pages}{111806}
  (\bibinfo{year}{2012}{\natexlab{a}}), \eprint{1206.2598}.

\bibitem[{\citenamefont{Chatrchyan
  et~al.}(2014{\natexlab{a}})}]{Chatrchyan:2013mza}
\bibinfo{author}{\bibfnamefont{S.}~\bibnamefont{Chatrchyan}}
  \bibnamefont{et~al.} (\bibinfo{collaboration}{CMS}), \bibinfo{journal}{Phys.
  Rev.} \textbf{\bibinfo{volume}{D90}}, \bibinfo{pages}{032004}
  (\bibinfo{year}{2014}{\natexlab{a}}), \eprint{1312.6283}.

\bibitem[{\citenamefont{Chatrchyan et~al.}(2012{\natexlab{b}})}]{CMS:2011wyd}
\bibinfo{author}{\bibfnamefont{S.}~\bibnamefont{Chatrchyan}}
  \bibnamefont{et~al.} (\bibinfo{collaboration}{CMS}), \bibinfo{journal}{Phys.
  Rev. D} \textbf{\bibinfo{volume}{85}}, \bibinfo{pages}{032002}
  (\bibinfo{year}{2012}{\natexlab{b}}), \eprint{1110.4973}.

\bibitem[{\citenamefont{Chatrchyan et~al.}(2013)}]{CMS:2013zfg}
\bibinfo{author}{\bibfnamefont{S.}~\bibnamefont{Chatrchyan}}
  \bibnamefont{et~al.} (\bibinfo{collaboration}{CMS}), \bibinfo{journal}{JHEP}
  \textbf{\bibinfo{volume}{12}}, \bibinfo{pages}{030} (\bibinfo{year}{2013}),
  \eprint{1310.7291}.

\bibitem[{\citenamefont{Khachatryan et~al.}(2016{\natexlab{a}})}]{CMS:2016qqr}
\bibinfo{author}{\bibfnamefont{V.}~\bibnamefont{Khachatryan}}
  \bibnamefont{et~al.} (\bibinfo{collaboration}{CMS}), \bibinfo{journal}{Eur.
  Phys. J. C} \textbf{\bibinfo{volume}{76}}, \bibinfo{pages}{469}
  (\bibinfo{year}{2016}{\natexlab{a}}), \eprint{1603.01803}.

\bibitem[{\citenamefont{Moreno et~al.}(1991)}]{Moreno:1990sf}
\bibinfo{author}{\bibfnamefont{G.}~\bibnamefont{Moreno}} \bibnamefont{et~al.},
  \bibinfo{journal}{Phys. Rev.} \textbf{\bibinfo{volume}{D43}},
  \bibinfo{pages}{2815} (\bibinfo{year}{1991}).

\bibitem[{\citenamefont{Towell et~al.}(2001)}]{Towell:2001nh}
\bibinfo{author}{\bibfnamefont{R.~S.} \bibnamefont{Towell}}
  \bibnamefont{et~al.} (\bibinfo{collaboration}{NuSea}),
  \bibinfo{journal}{Phys. Rev.} \textbf{\bibinfo{volume}{D64}},
  \bibinfo{pages}{052002} (\bibinfo{year}{2001}), \eprint{hep-ex/0103030}.

\bibitem[{\citenamefont{Webb et~al.}(2003)}]{Webb:2003ps}
\bibinfo{author}{\bibfnamefont{J.~C.} \bibnamefont{Webb}} \bibnamefont{et~al.}
  (\bibinfo{collaboration}{NuSea}) (\bibinfo{year}{2003}),
  \eprint{hep-ex/0302019}.

\bibitem[{\citenamefont{Webb}(2003)}]{Webb:2003bj}
\bibinfo{author}{\bibfnamefont{J.~C.} \bibnamefont{Webb}}, Ph.D. thesis,
  \bibinfo{school}{New Mexico State U.} (\bibinfo{year}{2003}),
  \eprint{hep-ex/0301031}.

\bibitem[{\citenamefont{Dove et~al.}(2021)}]{SeaQuest:2021zxb}
\bibinfo{author}{\bibfnamefont{J.}~\bibnamefont{Dove}} \bibnamefont{et~al.}
  (\bibinfo{collaboration}{SeaQuest}), \bibinfo{journal}{Nature}
  \textbf{\bibinfo{volume}{590}}, \bibinfo{pages}{561} (\bibinfo{year}{2021}),
  \bibinfo{note}{[Erratum: Nature 604, E26 (2022)]}, \eprint{2103.04024}.

\bibitem[{\citenamefont{Abazov et~al.}(2008{\natexlab{a}})}]{D0:2007pcy}
\bibinfo{author}{\bibfnamefont{V.~M.} \bibnamefont{Abazov}}
  \bibnamefont{et~al.} (\bibinfo{collaboration}{D0}), \bibinfo{journal}{Phys.
  Rev. D} \textbf{\bibinfo{volume}{77}}, \bibinfo{pages}{011106}
  (\bibinfo{year}{2008}{\natexlab{a}}), \eprint{0709.4254}.

\bibitem[{\citenamefont{Abazov et~al.}(2013)}]{D0:2013xqc}
\bibinfo{author}{\bibfnamefont{V.~M.} \bibnamefont{Abazov}}
  \bibnamefont{et~al.} (\bibinfo{collaboration}{D0}), \bibinfo{journal}{Phys.
  Rev. D} \textbf{\bibinfo{volume}{88}}, \bibinfo{pages}{091102}
  (\bibinfo{year}{2013}), \eprint{1309.2591}.

\bibitem[{\citenamefont{Abazov et~al.}(2015)}]{D0:2014kma}
\bibinfo{author}{\bibfnamefont{V.~M.} \bibnamefont{Abazov}}
  \bibnamefont{et~al.} (\bibinfo{collaboration}{D0}), \bibinfo{journal}{Phys.
  Rev.} \textbf{\bibinfo{volume}{D91}}, \bibinfo{pages}{032007}
  (\bibinfo{year}{2015}), \bibinfo{note}{[Erratum: Phys.
  Rev.D91,no.7,079901(2015)]}, \eprint{1412.2862}.

\bibitem[{\citenamefont{Abazov et~al.}(2008{\natexlab{b}})}]{D0:2008cgv}
\bibinfo{author}{\bibfnamefont{V.~M.} \bibnamefont{Abazov}}
  \bibnamefont{et~al.} (\bibinfo{collaboration}{D0}), \bibinfo{journal}{Phys.
  Rev. Lett.} \textbf{\bibinfo{volume}{101}}, \bibinfo{pages}{211801}
  (\bibinfo{year}{2008}{\natexlab{b}}), \eprint{0807.3367}.

\bibitem[{\citenamefont{Abazov et~al.}(2007)}]{D0:2007djv}
\bibinfo{author}{\bibfnamefont{V.~M.} \bibnamefont{Abazov}}
  \bibnamefont{et~al.} (\bibinfo{collaboration}{D0}), \bibinfo{journal}{Phys.
  Rev. D} \textbf{\bibinfo{volume}{76}}, \bibinfo{pages}{012003}
  (\bibinfo{year}{2007}), \eprint{hep-ex/0702025}.

\bibitem[{\citenamefont{Abazov et~al.}(2014)}]{D0:2013lql}
\bibinfo{author}{\bibfnamefont{V.~M.} \bibnamefont{Abazov}}
  \bibnamefont{et~al.} (\bibinfo{collaboration}{D0}), \bibinfo{journal}{Phys.
  Rev. Lett.} \textbf{\bibinfo{volume}{112}}, \bibinfo{pages}{151803}
  (\bibinfo{year}{2014}), \bibinfo{note}{[Erratum: Phys.Rev.Lett. 114, 049901
  (2015)]}, \eprint{1312.2895}.

\bibitem[{\citenamefont{Aaij et~al.}(2012)}]{Aaij:2012vn}
\bibinfo{author}{\bibfnamefont{R.}~\bibnamefont{Aaij}} \bibnamefont{et~al.}
  (\bibinfo{collaboration}{LHCb}), \bibinfo{journal}{JHEP}
  \textbf{\bibinfo{volume}{06}}, \bibinfo{pages}{058} (\bibinfo{year}{2012}),
  \eprint{1204.1620}.

\bibitem[{\citenamefont{Aaij et~al.}(2013)}]{LHCb:2012gii}
\bibinfo{author}{\bibfnamefont{R.}~\bibnamefont{Aaij}} \bibnamefont{et~al.}
  (\bibinfo{collaboration}{LHCb}), \bibinfo{journal}{JHEP}
  \textbf{\bibinfo{volume}{02}}, \bibinfo{pages}{106} (\bibinfo{year}{2013}),
  \eprint{1212.4620}.

\bibitem[{\citenamefont{Aaij et~al.}(2015{\natexlab{a}})}]{LHCb:2015okr}
\bibinfo{author}{\bibfnamefont{R.}~\bibnamefont{Aaij}} \bibnamefont{et~al.}
  (\bibinfo{collaboration}{LHCb}), \bibinfo{journal}{JHEP}
  \textbf{\bibinfo{volume}{08}}, \bibinfo{pages}{039}
  (\bibinfo{year}{2015}{\natexlab{a}}), \eprint{1505.07024}.

\bibitem[{\citenamefont{Aaij et~al.}(2015{\natexlab{b}})}]{LHCb:2015kwa}
\bibinfo{author}{\bibfnamefont{R.}~\bibnamefont{Aaij}} \bibnamefont{et~al.}
  (\bibinfo{collaboration}{LHCb}), \bibinfo{journal}{JHEP}
  \textbf{\bibinfo{volume}{05}}, \bibinfo{pages}{109}
  (\bibinfo{year}{2015}{\natexlab{b}}), \eprint{1503.00963}.

\bibitem[{\citenamefont{Aaij et~al.}(2016)}]{LHCb:2015mad}
\bibinfo{author}{\bibfnamefont{R.}~\bibnamefont{Aaij}} \bibnamefont{et~al.}
  (\bibinfo{collaboration}{LHCb}), \bibinfo{journal}{JHEP}
  \textbf{\bibinfo{volume}{01}}, \bibinfo{pages}{155} (\bibinfo{year}{2016}),
  \eprint{1511.08039}.

\bibitem[{\citenamefont{Aad et~al.}(2015)}]{ATLAS:2014riz}
\bibinfo{author}{\bibfnamefont{G.}~\bibnamefont{Aad}} \bibnamefont{et~al.}
  (\bibinfo{collaboration}{ATLAS}), \bibinfo{journal}{JHEP}
  \textbf{\bibinfo{volume}{02}}, \bibinfo{pages}{153} (\bibinfo{year}{2015}),
  \bibinfo{note}{[Erratum: JHEP 09, 141 (2015)]}, \eprint{1410.8857}.

\bibitem[{\citenamefont{Aaltonen et~al.}(2008)}]{CDF:2008hmn}
\bibinfo{author}{\bibfnamefont{T.}~\bibnamefont{Aaltonen}} \bibnamefont{et~al.}
  (\bibinfo{collaboration}{CDF}), \bibinfo{journal}{Phys. Rev. D}
  \textbf{\bibinfo{volume}{78}}, \bibinfo{pages}{052006}
  (\bibinfo{year}{2008}), \bibinfo{note}{[Erratum: Phys.Rev.D 79, 119902
  (2009)]}, \eprint{0807.2204}.

\bibitem[{\citenamefont{Abulencia et~al.}(2007)}]{CDF:2007bvv}
\bibinfo{author}{\bibfnamefont{A.}~\bibnamefont{Abulencia}}
  \bibnamefont{et~al.} (\bibinfo{collaboration}{CDF}), \bibinfo{journal}{Phys.
  Rev. D} \textbf{\bibinfo{volume}{75}}, \bibinfo{pages}{092006}
  (\bibinfo{year}{2007}), \bibinfo{note}{[Erratum: Phys.Rev.D 75, 119901
  (2007)]}, \eprint{hep-ex/0701051}.

\bibitem[{\citenamefont{Khachatryan et~al.}(2016{\natexlab{b}})}]{CMS:2015jdl}
\bibinfo{author}{\bibfnamefont{V.}~\bibnamefont{Khachatryan}}
  \bibnamefont{et~al.} (\bibinfo{collaboration}{CMS}), \bibinfo{journal}{Eur.
  Phys. J. C} \textbf{\bibinfo{volume}{76}}, \bibinfo{pages}{265}
  (\bibinfo{year}{2016}{\natexlab{b}}), \eprint{1512.06212}.

\bibitem[{\citenamefont{Chatrchyan et~al.}(2014{\natexlab{b}})}]{CMS:2014nvq}
\bibinfo{author}{\bibfnamefont{S.}~\bibnamefont{Chatrchyan}}
  \bibnamefont{et~al.} (\bibinfo{collaboration}{CMS}), \bibinfo{journal}{Phys.
  Rev. D} \textbf{\bibinfo{volume}{90}}, \bibinfo{pages}{072006}
  (\bibinfo{year}{2014}{\natexlab{b}}), \eprint{1406.0324}.

\bibitem[{\citenamefont{Khachatryan et~al.}(2017)}]{CMS:2016lna}
\bibinfo{author}{\bibfnamefont{V.}~\bibnamefont{Khachatryan}}
  \bibnamefont{et~al.} (\bibinfo{collaboration}{CMS}), \bibinfo{journal}{JHEP}
  \textbf{\bibinfo{volume}{03}}, \bibinfo{pages}{156} (\bibinfo{year}{2017}),
  \eprint{1609.05331}.

\bibitem[{\citenamefont{Abazov et~al.}(2008{\natexlab{c}})}]{D0:2008nou}
\bibinfo{author}{\bibfnamefont{V.~M.} \bibnamefont{Abazov}}
  \bibnamefont{et~al.} (\bibinfo{collaboration}{D0}), \bibinfo{journal}{Phys.
  Rev. Lett.} \textbf{\bibinfo{volume}{101}}, \bibinfo{pages}{062001}
  (\bibinfo{year}{2008}{\natexlab{c}}), \eprint{0802.2400}.

\bibitem[{\citenamefont{Abazov et~al.}(2012)}]{D0:2011jpq}
\bibinfo{author}{\bibfnamefont{V.~M.} \bibnamefont{Abazov}}
  \bibnamefont{et~al.} (\bibinfo{collaboration}{D0}), \bibinfo{journal}{Phys.
  Rev. D} \textbf{\bibinfo{volume}{85}}, \bibinfo{pages}{052006}
  (\bibinfo{year}{2012}), \eprint{1110.3771}.

\bibitem[{\citenamefont{Chatrchyan
  et~al.}(2014{\natexlab{c}})}]{Chatrchyan:2013uja}
\bibinfo{author}{\bibfnamefont{S.}~\bibnamefont{Chatrchyan}}
  \bibnamefont{et~al.} (\bibinfo{collaboration}{CMS}), \bibinfo{journal}{JHEP}
  \textbf{\bibinfo{volume}{02}}, \bibinfo{pages}{013}
  (\bibinfo{year}{2014}{\natexlab{c}}), \eprint{1310.1138}.

\bibitem[{\citenamefont{Aaltonen et~al.}(2014)}]{CDF:2013hmv}
\bibinfo{author}{\bibfnamefont{T.~A.} \bibnamefont{Aaltonen}}
  \bibnamefont{et~al.} (\bibinfo{collaboration}{CDF, D0}),
  \bibinfo{journal}{Phys. Rev. D} \textbf{\bibinfo{volume}{89}},
  \bibinfo{pages}{072001} (\bibinfo{year}{2014}), \eprint{1309.7570}.

\bibitem[{\citenamefont{Chatrchyan et~al.}(2014{\natexlab{d}})}]{CMS:2013hon}
\bibinfo{author}{\bibfnamefont{S.}~\bibnamefont{Chatrchyan}}
  \bibnamefont{et~al.} (\bibinfo{collaboration}{CMS}), \bibinfo{journal}{JHEP}
  \textbf{\bibinfo{volume}{02}}, \bibinfo{pages}{024}
  (\bibinfo{year}{2014}{\natexlab{d}}), \bibinfo{note}{[Erratum: JHEP 02, 102
  (2014)]}, \eprint{1312.7582}.

\bibitem[{\citenamefont{Sirunyan et~al.}(2017)}]{CMS:2017iqf}
\bibinfo{author}{\bibfnamefont{A.~M.} \bibnamefont{Sirunyan}}
  \bibnamefont{et~al.} (\bibinfo{collaboration}{CMS}), \bibinfo{journal}{Eur.
  Phys. J. C} \textbf{\bibinfo{volume}{77}}, \bibinfo{pages}{459}
  (\bibinfo{year}{2017}), \eprint{1703.01630}.

\bibitem[{\citenamefont{Khachatryan et~al.}(2015)}]{CMS:2015rld}
\bibinfo{author}{\bibfnamefont{V.}~\bibnamefont{Khachatryan}}
  \bibnamefont{et~al.} (\bibinfo{collaboration}{CMS}), \bibinfo{journal}{Eur.
  Phys. J. C} \textbf{\bibinfo{volume}{75}}, \bibinfo{pages}{542}
  (\bibinfo{year}{2015}), \eprint{1505.04480}.

\bibitem[{\citenamefont{Harland-Lang et~al.}(2015)\citenamefont{Harland-Lang,
  Martin, Motylinski, and Thorne}}]{Harland-Lang:2014zoa}
\bibinfo{author}{\bibfnamefont{L.~A.} \bibnamefont{Harland-Lang}},
  \bibinfo{author}{\bibfnamefont{A.~D.} \bibnamefont{Martin}},
  \bibinfo{author}{\bibfnamefont{P.}~\bibnamefont{Motylinski}},
  \bibnamefont{and} \bibinfo{author}{\bibfnamefont{R.~S.}
  \bibnamefont{Thorne}}, \bibinfo{journal}{Eur. Phys. J.}
  \textbf{\bibinfo{volume}{C75}}, \bibinfo{pages}{204} (\bibinfo{year}{2015}),
  \eprint{1412.3989}.

\bibitem[{\citenamefont{Martin et~al.}(2013)\citenamefont{Martin, Mathijssen,
  Stirling, Thorne, Watt, and Watt}}]{Martin:2012da}
\bibinfo{author}{\bibfnamefont{A.~D.} \bibnamefont{Martin}},
  \bibinfo{author}{\bibfnamefont{A.~J. T.~M.} \bibnamefont{Mathijssen}},
  \bibinfo{author}{\bibfnamefont{W.~J.} \bibnamefont{Stirling}},
  \bibinfo{author}{\bibfnamefont{R.~S.} \bibnamefont{Thorne}},
  \bibinfo{author}{\bibfnamefont{B.~J.~A.} \bibnamefont{Watt}},
  \bibnamefont{and} \bibinfo{author}{\bibfnamefont{G.}~\bibnamefont{Watt}},
  \bibinfo{journal}{Eur. Phys. J. C} \textbf{\bibinfo{volume}{73}},
  \bibinfo{pages}{2318} (\bibinfo{year}{2013}), \eprint{1211.1215}.

\bibitem[{\citenamefont{Harland-Lang et~al.}(2018)\citenamefont{Harland-Lang,
  Martin, and Thorne}}]{Harland-Lang:2017ytb}
\bibinfo{author}{\bibfnamefont{L.~A.} \bibnamefont{Harland-Lang}},
  \bibinfo{author}{\bibfnamefont{A.~D.} \bibnamefont{Martin}},
  \bibnamefont{and} \bibinfo{author}{\bibfnamefont{R.~S.}
  \bibnamefont{Thorne}}, \bibinfo{journal}{Eur. Phys. J.}
  \textbf{\bibinfo{volume}{C78}}, \bibinfo{pages}{248} (\bibinfo{year}{2018}),
  \eprint{1711.05757}.

\bibitem[{\citenamefont{Bailey and Harland-Lang}(2020)}]{Bailey:2019yze}
\bibinfo{author}{\bibfnamefont{S.}~\bibnamefont{Bailey}} \bibnamefont{and}
  \bibinfo{author}{\bibfnamefont{L.}~\bibnamefont{Harland-Lang}},
  \bibinfo{journal}{Eur. Phys. J. C} \textbf{\bibinfo{volume}{80}},
  \bibinfo{pages}{60} (\bibinfo{year}{2020}), \eprint{1909.10541}.

\bibitem[{\citenamefont{Thorne et~al.}(2019)\citenamefont{Thorne, Bailey,
  Cridge, Harland-Lang, Martin, and Nathvani}}]{Thorne:2019mpt}
\bibinfo{author}{\bibfnamefont{R.~S.} \bibnamefont{Thorne}},
  \bibinfo{author}{\bibfnamefont{S.}~\bibnamefont{Bailey}},
  \bibinfo{author}{\bibfnamefont{T.}~\bibnamefont{Cridge}},
  \bibinfo{author}{\bibfnamefont{L.~A.} \bibnamefont{Harland-Lang}},
  \bibinfo{author}{\bibfnamefont{A.~D.} \bibnamefont{Martin}},
  \bibnamefont{and} \bibinfo{author}{\bibfnamefont{R.}~\bibnamefont{Nathvani}}
  (\bibinfo{year}{2019}), vol. \bibinfo{volume}{DIS2019}, p.
  \bibinfo{pages}{036}, \eprint{1907.08147}.

\bibitem[{\citenamefont{Thorne et~al.}(2022)\citenamefont{Thorne, Bailey,
  Cridge, Harland-Lang, and Martin}}]{Thorne:2022abv}
\bibinfo{author}{\bibfnamefont{R.}~\bibnamefont{Thorne}},
  \bibinfo{author}{\bibfnamefont{S.}~\bibnamefont{Bailey}},
  \bibinfo{author}{\bibfnamefont{T.}~\bibnamefont{Cridge}},
  \bibinfo{author}{\bibfnamefont{L.}~\bibnamefont{Harland-Lang}},
  \bibnamefont{and} \bibinfo{author}{\bibfnamefont{A.~D.}
  \bibnamefont{Martin}}, \bibinfo{journal}{SciPost Phys. Proc.}
  \textbf{\bibinfo{volume}{8}}, \bibinfo{pages}{018} (\bibinfo{year}{2022}).

\bibitem[{\citenamefont{Boughezal et~al.}(2015)\citenamefont{Boughezal, Focke,
  Liu, and Petriello}}]{Boughezal:2015dva}
\bibinfo{author}{\bibfnamefont{R.}~\bibnamefont{Boughezal}},
  \bibinfo{author}{\bibfnamefont{C.}~\bibnamefont{Focke}},
  \bibinfo{author}{\bibfnamefont{X.}~\bibnamefont{Liu}}, \bibnamefont{and}
  \bibinfo{author}{\bibfnamefont{F.}~\bibnamefont{Petriello}},
  \bibinfo{journal}{Phys. Rev. Lett.} \textbf{\bibinfo{volume}{115}},
  \bibinfo{pages}{062002} (\bibinfo{year}{2015}), \eprint{1504.02131}.

\bibitem[{\citenamefont{Cridge et~al.}(2022)\citenamefont{Cridge, Harland-Lang,
  Martin, and Thorne}}]{Cridge:2021pxm}
\bibinfo{author}{\bibfnamefont{T.}~\bibnamefont{Cridge}},
  \bibinfo{author}{\bibfnamefont{L.~A.} \bibnamefont{Harland-Lang}},
  \bibinfo{author}{\bibfnamefont{A.~D.} \bibnamefont{Martin}},
  \bibnamefont{and} \bibinfo{author}{\bibfnamefont{R.~S.}
  \bibnamefont{Thorne}}, \bibinfo{journal}{Eur. Phys. J. C}
  \textbf{\bibinfo{volume}{82}}, \bibinfo{pages}{90} (\bibinfo{year}{2022}),
  \eprint{2111.05357}.

\bibitem[{\citenamefont{Cridge et~al.}(2021)\citenamefont{Cridge, Harland-Lang,
  Martin, and Thorne}}]{Cridge:2021qfd}
\bibinfo{author}{\bibfnamefont{T.}~\bibnamefont{Cridge}},
  \bibinfo{author}{\bibfnamefont{L.~A.} \bibnamefont{Harland-Lang}},
  \bibinfo{author}{\bibfnamefont{A.~D.} \bibnamefont{Martin}},
  \bibnamefont{and} \bibinfo{author}{\bibfnamefont{R.~S.}
  \bibnamefont{Thorne}}, \bibinfo{journal}{Eur. Phys. J. C}
  \textbf{\bibinfo{volume}{81}}, \bibinfo{pages}{744} (\bibinfo{year}{2021}),
  \eprint{2106.10289}.

\bibitem[{\citenamefont{Nathvani et~al.}(2018)\citenamefont{Nathvani, Thorne,
  Harland-Lang, and Martin}}]{Nathvani:2018pys}
\bibinfo{author}{\bibfnamefont{R.}~\bibnamefont{Nathvani}},
  \bibinfo{author}{\bibfnamefont{R.}~\bibnamefont{Thorne}},
  \bibinfo{author}{\bibfnamefont{L.}~\bibnamefont{Harland-Lang}},
  \bibnamefont{and} \bibinfo{author}{\bibfnamefont{A.}~\bibnamefont{Martin}},
  \bibinfo{journal}{PoS} \textbf{\bibinfo{volume}{DIS2018}},
  \bibinfo{pages}{029} (\bibinfo{year}{2018}), \eprint{1807.07846}.

\bibitem[{\citenamefont{Cridge}(2022)}]{Cridge:2021qjj}
\bibinfo{author}{\bibfnamefont{T.}~\bibnamefont{Cridge}}
  (\bibinfo{collaboration}{PDF4LHC21 combination group}),
  \bibinfo{journal}{SciPost Phys. Proc.} \textbf{\bibinfo{volume}{8}},
  \bibinfo{pages}{101} (\bibinfo{year}{2022}), \eprint{2108.09099}.

\bibitem[{\citenamefont{Lepage et~al.}(2002)\citenamefont{Lepage, Clark,
  Davies, Hornbostel, Mackenzie, Morningstar, and Trottier}}]{Lepage:2001ym}
\bibinfo{author}{\bibfnamefont{G.~P.} \bibnamefont{Lepage}},
  \bibinfo{author}{\bibfnamefont{B.}~\bibnamefont{Clark}},
  \bibinfo{author}{\bibfnamefont{C.~T.~H.} \bibnamefont{Davies}},
  \bibinfo{author}{\bibfnamefont{K.}~\bibnamefont{Hornbostel}},
  \bibinfo{author}{\bibfnamefont{P.~B.} \bibnamefont{Mackenzie}},
  \bibinfo{author}{\bibfnamefont{C.}~\bibnamefont{Morningstar}},
  \bibnamefont{and} \bibinfo{author}{\bibfnamefont{H.}~\bibnamefont{Trottier}}
  (\bibinfo{collaboration}{HPQCD}), \bibinfo{journal}{Nucl. Phys. B Proc.
  Suppl.} \textbf{\bibinfo{volume}{106}}, \bibinfo{pages}{12}
  (\bibinfo{year}{2002}), \eprint{hep-lat/0110175}.

\bibitem[{\citenamefont{Amoroso et~al.}(2022)}]{Amoroso:2022eow}
\bibinfo{author}{\bibfnamefont{S.}~\bibnamefont{Amoroso}} \bibnamefont{et~al.},
  \bibinfo{journal}{Acta Phys. Polon. B} \textbf{\bibinfo{volume}{53}},
  \bibinfo{pages}{A1} (\bibinfo{year}{2022}), \eprint{2203.13923}.

\bibitem[{\citenamefont{D'Agostini}(1994)}]{DAgostini:1993arp}
\bibinfo{author}{\bibfnamefont{G.}~\bibnamefont{D'Agostini}},
  \bibinfo{journal}{Nucl. Instrum. Meth. A} \textbf{\bibinfo{volume}{346}},
  \bibinfo{pages}{306} (\bibinfo{year}{1994}).

\bibitem[{\citenamefont{D'Agostini}(1999)}]{DAgostini:1999gfj}
\bibinfo{author}{\bibfnamefont{G.}~\bibnamefont{D'Agostini}},
  \emph{\bibinfo{title}{{Bayesian reasoning in high-energy physics: Principles
  and applications}}}, \bibinfo{howpublished}{\protect{CERN-99-03,
  CERN-YELLOW-99-03}} (\bibinfo{year}{1999}).

\bibitem[{\citenamefont{Ball et~al.}(2010)\citenamefont{Ball, Del~Debbio,
  Forte, Guffanti, Latorre, Rojo, and Ubiali}}]{Ball:2009qv}
\bibinfo{author}{\bibfnamefont{R.~D.} \bibnamefont{Ball}},
  \bibinfo{author}{\bibfnamefont{L.}~\bibnamefont{Del~Debbio}},
  \bibinfo{author}{\bibfnamefont{S.}~\bibnamefont{Forte}},
  \bibinfo{author}{\bibfnamefont{A.}~\bibnamefont{Guffanti}},
  \bibinfo{author}{\bibfnamefont{J.~I.} \bibnamefont{Latorre}},
  \bibinfo{author}{\bibfnamefont{J.}~\bibnamefont{Rojo}}, \bibnamefont{and}
  \bibinfo{author}{\bibfnamefont{M.}~\bibnamefont{Ubiali}}
  (\bibinfo{collaboration}{NNPDF}), \bibinfo{journal}{JHEP}
  \textbf{\bibinfo{volume}{05}}, \bibinfo{pages}{075} (\bibinfo{year}{2010}),
  \eprint{0912.2276}.

\bibitem[{\citenamefont{Ball et~al.}(2013)}]{Ball:2012wy}
\bibinfo{author}{\bibfnamefont{R.~D.} \bibnamefont{Ball}} \bibnamefont{et~al.},
  \bibinfo{journal}{JHEP} \textbf{\bibinfo{volume}{04}}, \bibinfo{pages}{125}
  (\bibinfo{year}{2013}), \eprint{1211.5142}.

\bibitem[{\citenamefont{Gao et~al.}(2014)\citenamefont{Gao, Guzzi, Huston, Lai,
  Li, Nadolsky, Pumplin, Stump, and Yuan}}]{Gao:2013xoa}
\bibinfo{author}{\bibfnamefont{J.}~\bibnamefont{Gao}},
  \bibinfo{author}{\bibfnamefont{M.}~\bibnamefont{Guzzi}},
  \bibinfo{author}{\bibfnamefont{J.}~\bibnamefont{Huston}},
  \bibinfo{author}{\bibfnamefont{H.-L.} \bibnamefont{Lai}},
  \bibinfo{author}{\bibfnamefont{Z.}~\bibnamefont{Li}},
  \bibinfo{author}{\bibfnamefont{P.}~\bibnamefont{Nadolsky}},
  \bibinfo{author}{\bibfnamefont{J.}~\bibnamefont{Pumplin}},
  \bibinfo{author}{\bibfnamefont{D.}~\bibnamefont{Stump}}, \bibnamefont{and}
  \bibinfo{author}{\bibfnamefont{C.~P.} \bibnamefont{Yuan}},
  \bibinfo{journal}{Phys. Rev.} \textbf{\bibinfo{volume}{D89}},
  \bibinfo{pages}{033009} (\bibinfo{year}{2014}), \eprint{1302.6246}.

\bibitem[{\citenamefont{Courtoy et~al.}(2023)\citenamefont{Courtoy, Huston,
  Nadolsky, Xie, Yan, and Yuan}}]{Courtoy:2022ocu}
\bibinfo{author}{\bibfnamefont{A.}~\bibnamefont{Courtoy}},
  \bibinfo{author}{\bibfnamefont{J.}~\bibnamefont{Huston}},
  \bibinfo{author}{\bibfnamefont{P.}~\bibnamefont{Nadolsky}},
  \bibinfo{author}{\bibfnamefont{K.}~\bibnamefont{Xie}},
  \bibinfo{author}{\bibfnamefont{M.}~\bibnamefont{Yan}}, \bibnamefont{and}
  \bibinfo{author}{\bibfnamefont{C.~P.} \bibnamefont{Yuan}},
  \bibinfo{journal}{Phys. Rev. D} \textbf{\bibinfo{volume}{107}},
  \bibinfo{pages}{034008} (\bibinfo{year}{2023}), \eprint{2205.10444}.

\bibitem[{PDF()}]{PDFSenseWebsite}
\bibinfo{howpublished}{\protect{PDFSense: visualization of experimental
  constraints on the nucleon structure},
  \url{https://metapdf.hepforge.org/PDFSense/}}.

\bibitem[{\citenamefont{Jing}(2021)}]{JingPhdthesis2021}
\bibinfo{author}{\bibfnamefont{X.}~\bibnamefont{Jing}}, Ph.D. thesis,
  \bibinfo{school}{Southern Methodist University}, \bibinfo{address}{Dallas,
  TX} (\bibinfo{year}{2021}).

\end{thebibliography}

\clearpage \newpage

\section*{Supplemental material \label{sec:supplement}}
\setcounter{figure}{0}
\renewcommand{\thefigure}{SM.\arabic{figure}}
\setcounter{page}{1}
\renewcommand{\thepage}{SM.\arabic{page}}

\begin{figure*}[h]
\begin{flushleft}
\includegraphics[height=160pt,trim={ 0 0 1.6cm 0},clip]{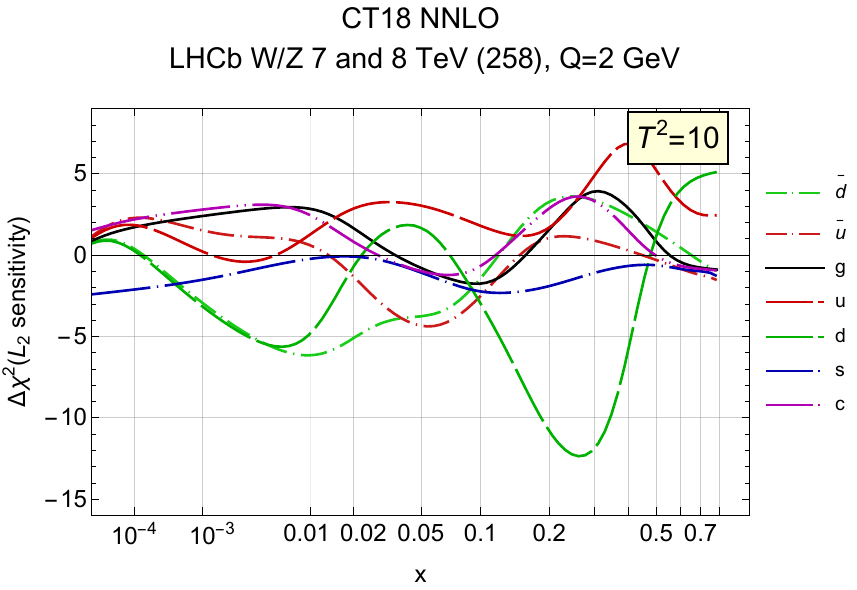}
\includegraphics[height=160pt,trim={ 0.8cm 0 0 0},clip]{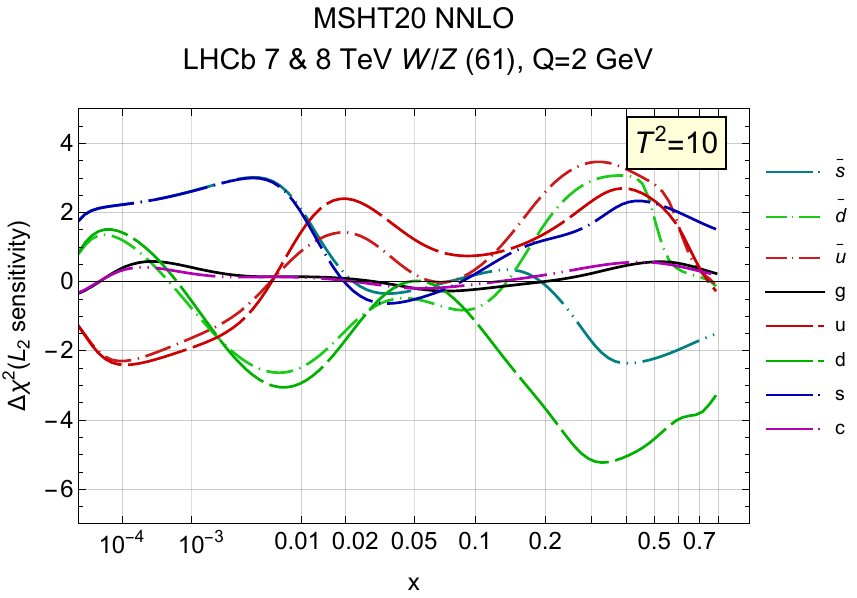}\\
\includegraphics[height=150pt,trim={ 0 0 6cm 0},clip]{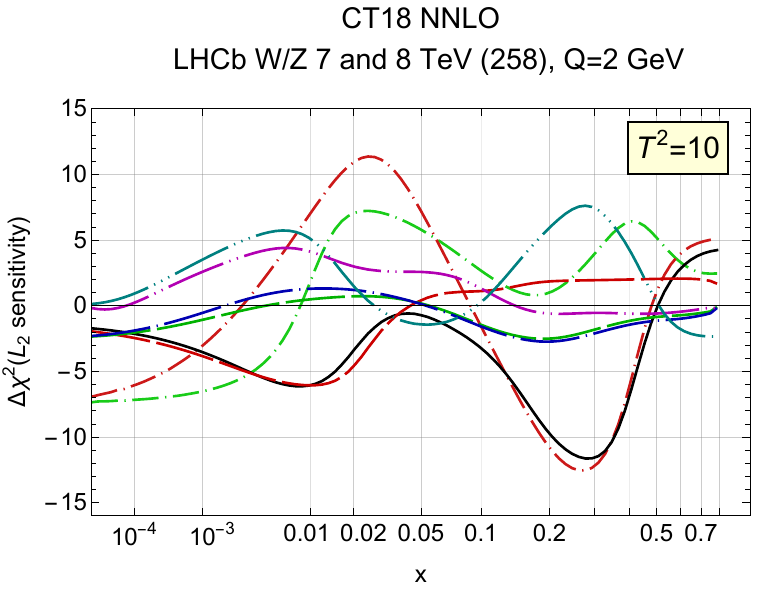}
\includegraphics[height=150pt,trim={ 0.8cm 0 0 0},clip]{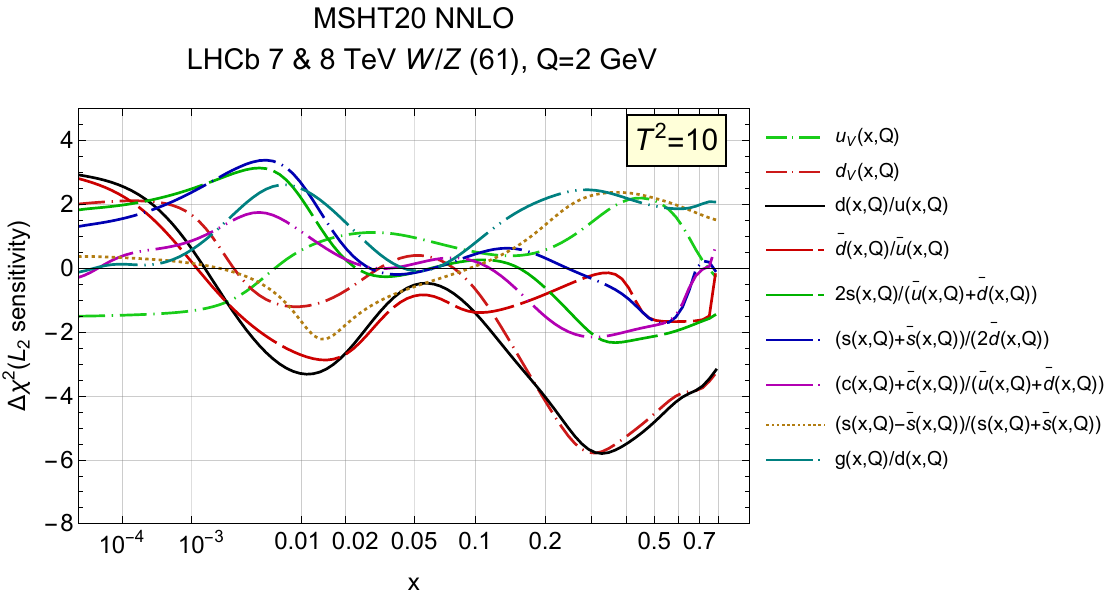}\\
\includegraphics[height=150pt,trim={ 0 0 6cm 0},clip]{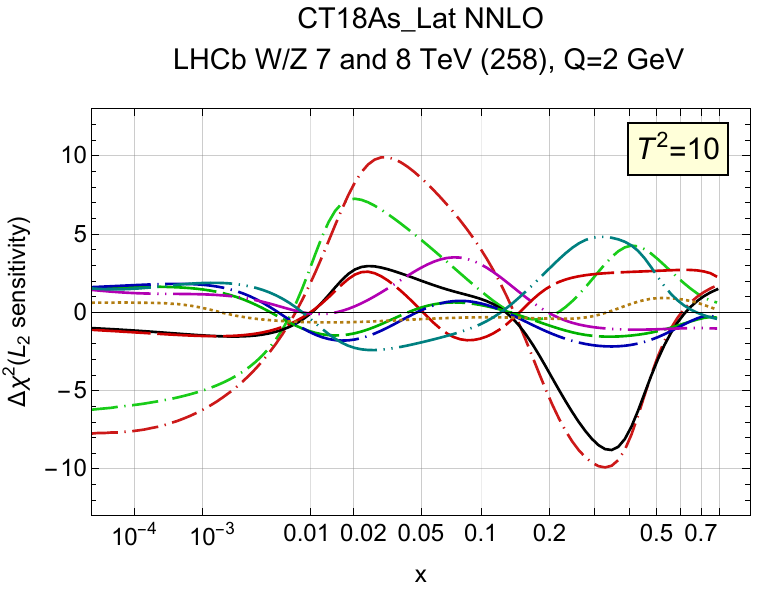}
\includegraphics[height=150pt,trim={ 0.8cm 0 0 0},clip]{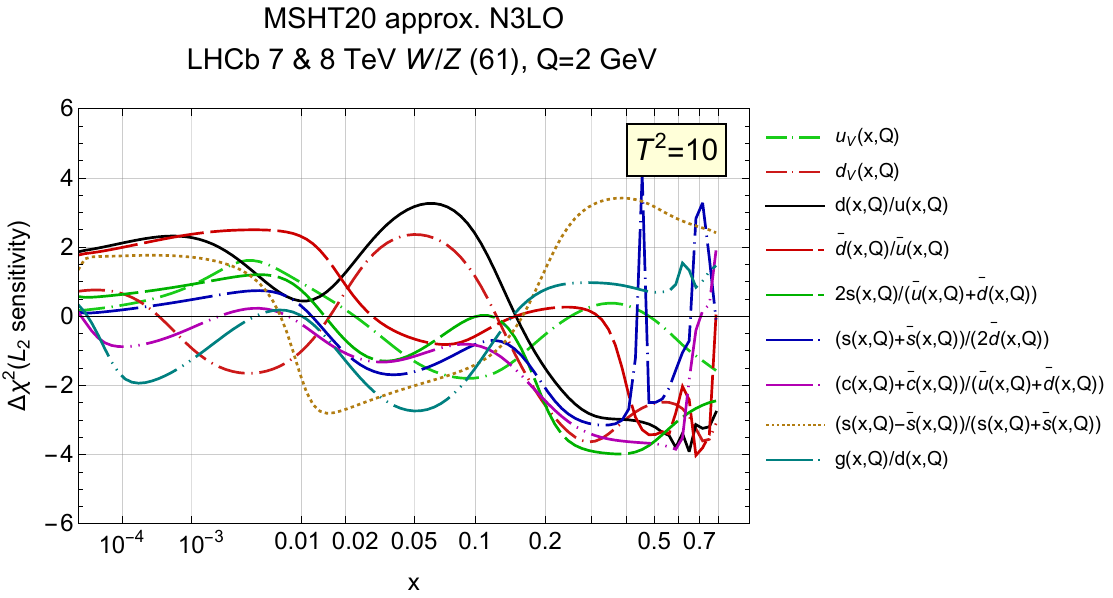}
\end{flushleft}
\caption{ Sensitivities for the full fit with $T^2=10$ for LHCb 7+8 TeV data set at $Q=2$ GeV.  Left: CT18 NNLO and CT18As\_Lat NNLO. Right: MSHT20 NNLO and aN3LO.}
\label{fig:compar_LHCb_WZ}
\end{figure*}

\begin{figure*}[p]
\centering
\includegraphics[height=160pt,trim={ 0 0 6cm 0},clip]{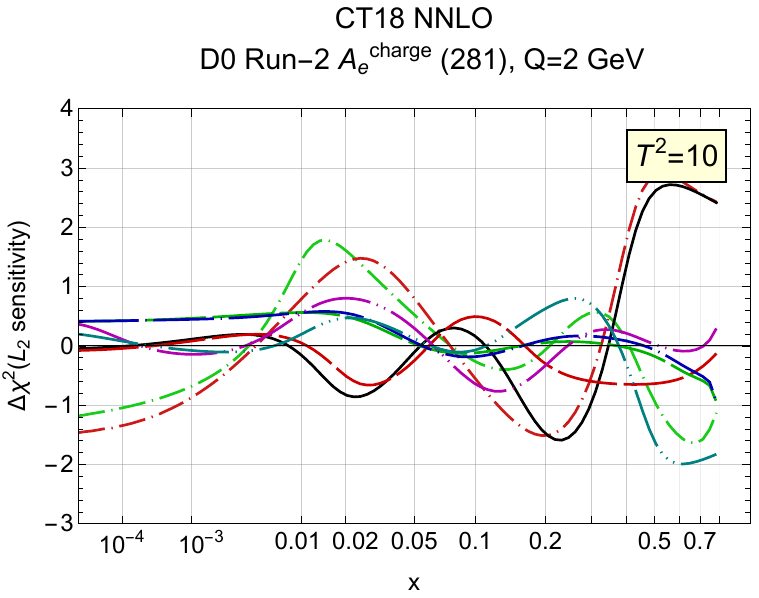}
\includegraphics[height=160pt,trim={ 0.8cm 0 6cm 0},clip]{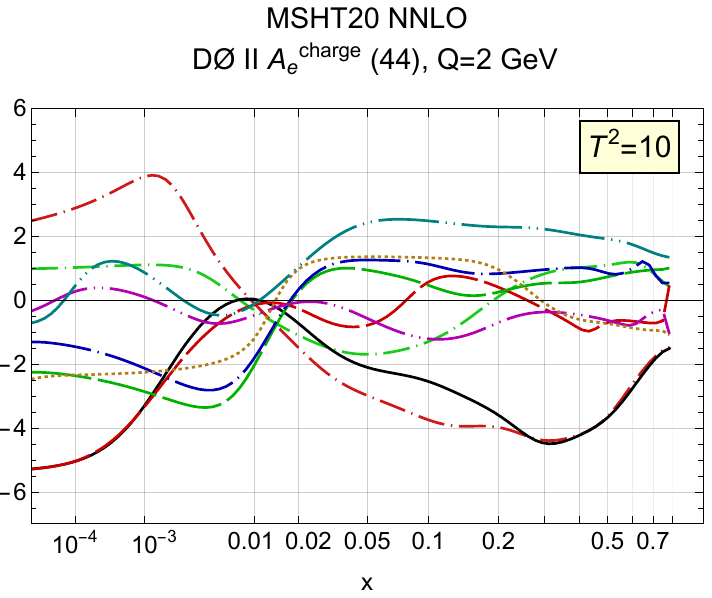}\hspace{0.2\textwidth}\quad \\
\includegraphics[height=160pt,trim={ 0 0 0 0},clip]{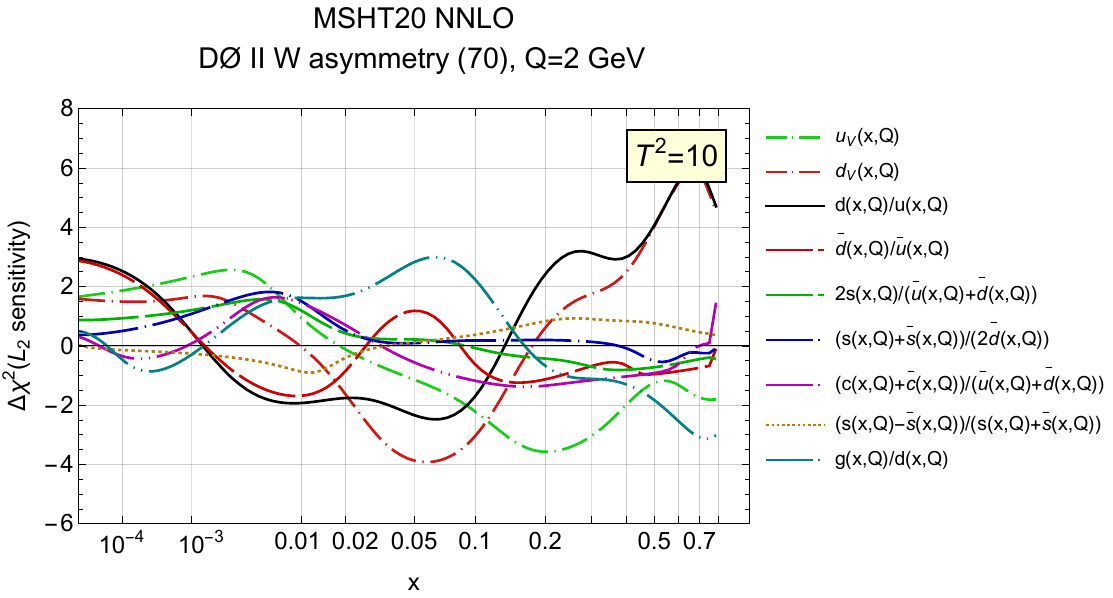}\hspace{95pt}\quad
\caption{
Sensitivities for the D{\O} Run-2 electron charge asymmetries in the CT18 (upper left) and MSHT20 (upper right) fits; as well as the reconstructed $W$ charge asymmetry in the MSHT20 fit (lower left) at $Q=2$ GeV.}
\label{fig:compar_D0asym}
\end{figure*}

\begin{figure*}[p]
\centering
\includegraphics[height=150pt,trim={ 0 0 6cm 0},clip]{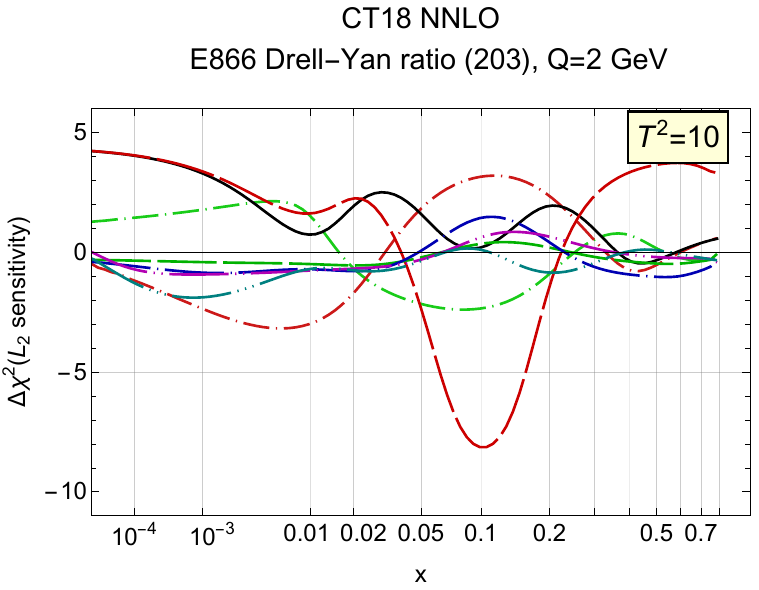}
\includegraphics[height=150pt,trim={ 0.8cm 0 0 0},clip]{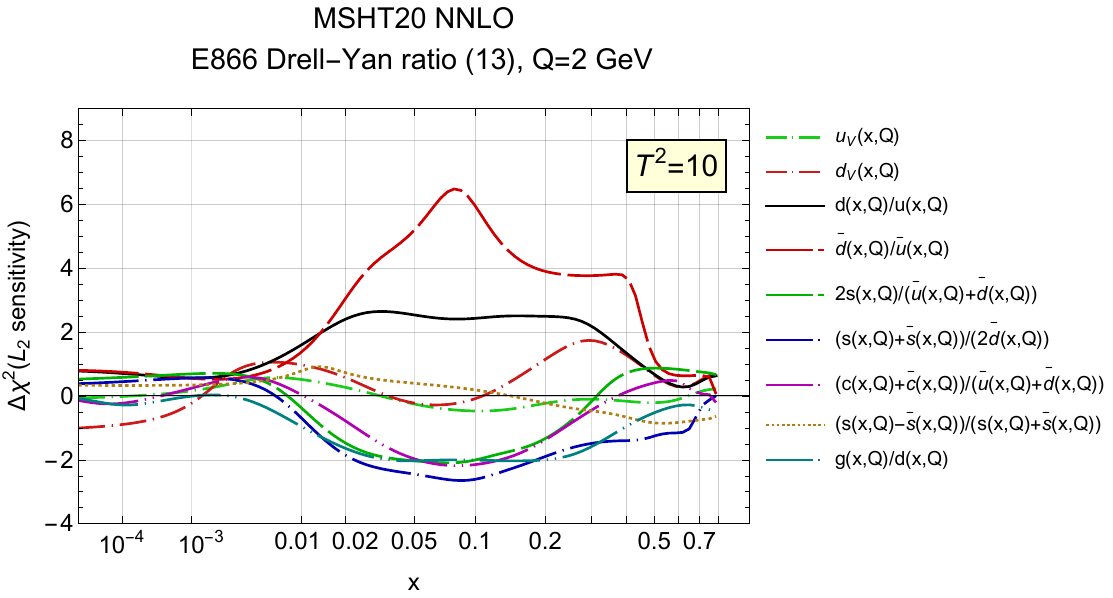}\\
\includegraphics[height=150pt,trim={ 0 0 6cm 0},clip]{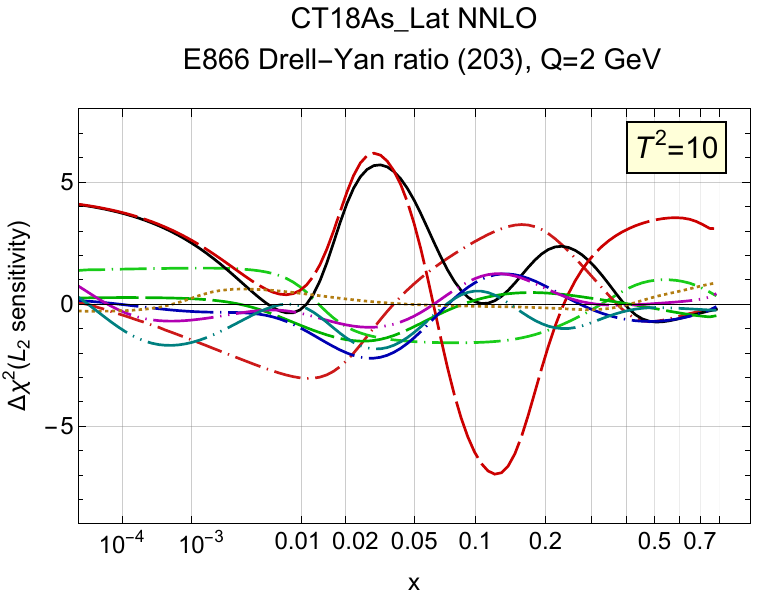}
\includegraphics[height=150pt,trim={ 0.8cm 0 0 0},clip]{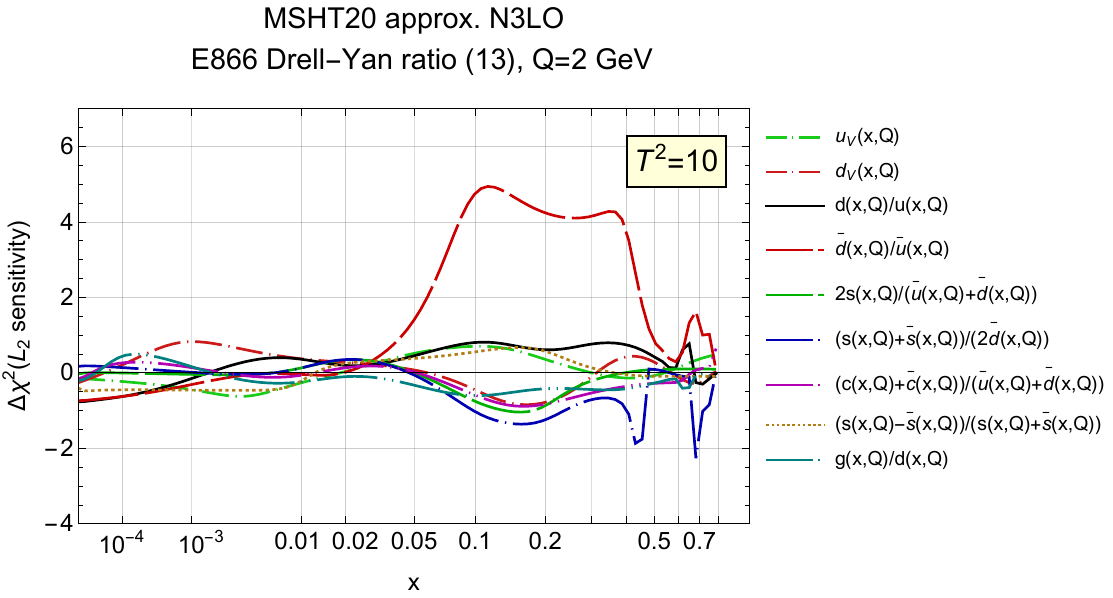}\\
\includegraphics[height=150pt,trim={ 0 0 0 0},clip]{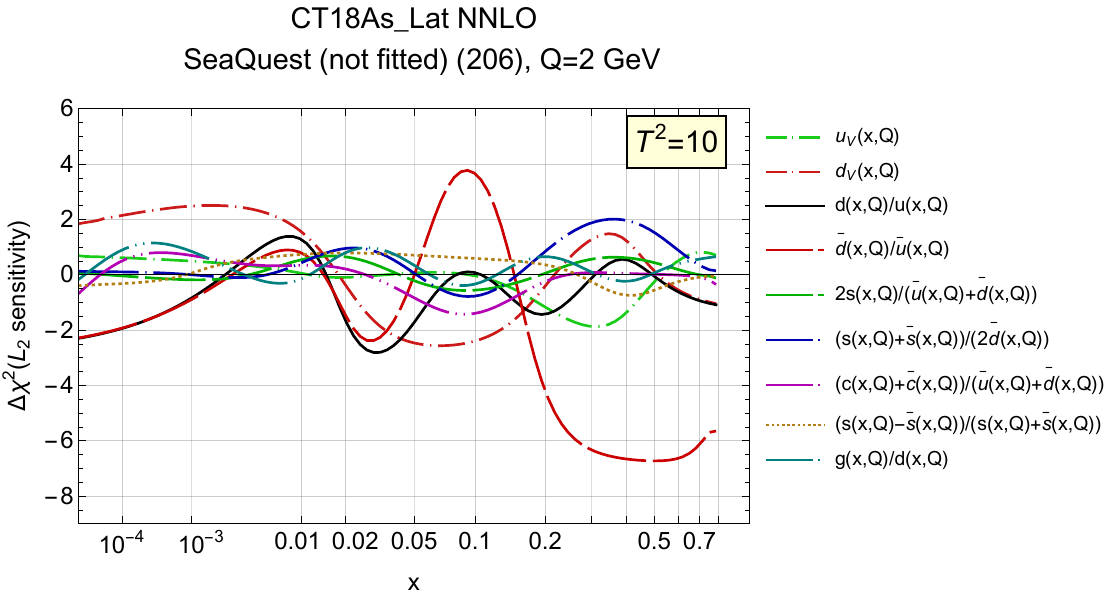} \hfill\quad
\caption{ Sensitivities for the full fit with $T^2=10$ for E866 and E906 experiments plotted as PDF combinations and ratios at $Q=2$ GeV.  Left: CT18 NNLO and CT18As\_Lat. Right: MSHT20 NNLO and aN3LO.}
\label{fig:compar_E866_rat}
\end{figure*}

\begin{figure*}[b]
\centering
\includegraphics[height=160pt,trim={ 0 0 1.6cm 0},clip]{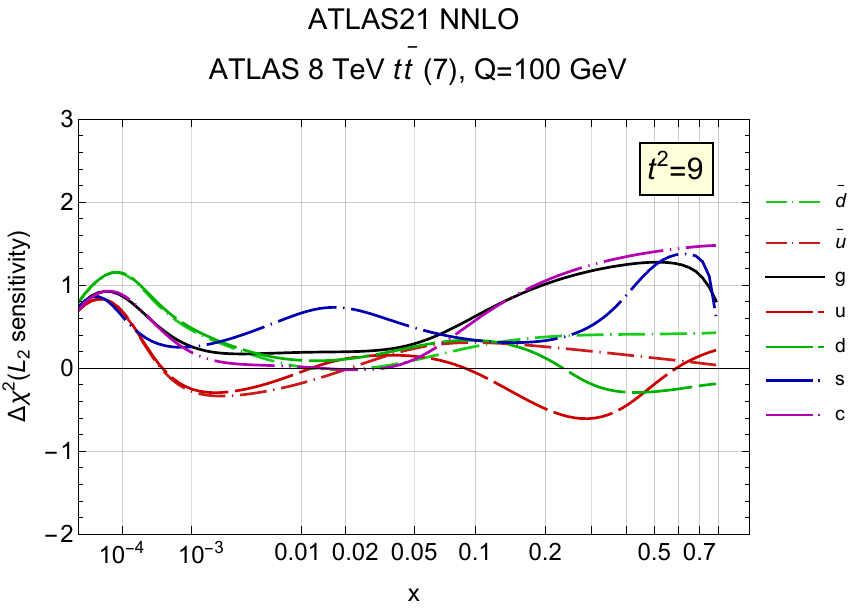}
\includegraphics[height=160pt,trim={ 0.8cm 0 0 0},clip]{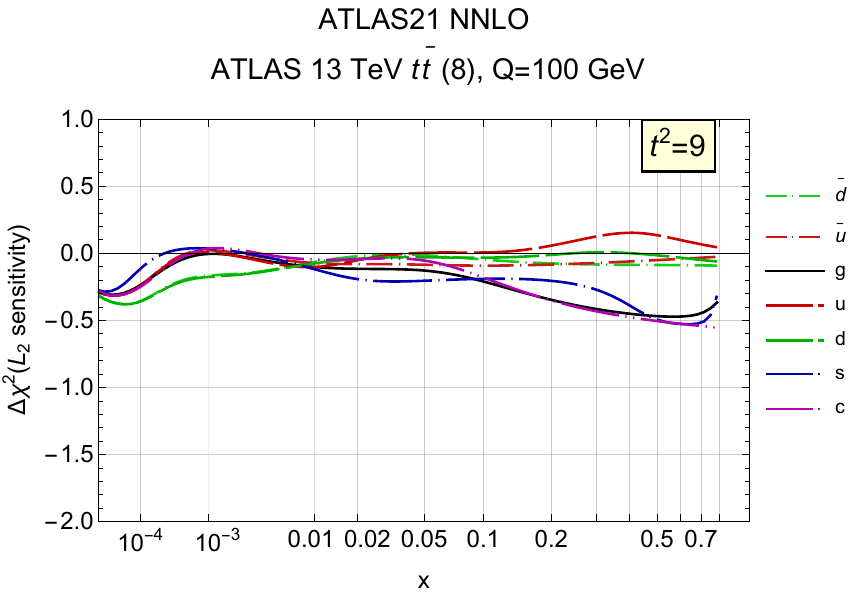}
\caption{
Sensitivities for ATLAS 8 TeV and 13 TeV $t\bar t$ data sets in the ATLASpdf21 analysis at $Q=100$ GeV.}
\label{fig:compar_ttbar_ATLAS}
\end{figure*}

\end{document}